\documentclass[11pt]{article}
\usepackage{jheppub}
\usepackage{amsthm}
\usepackage{tikz-cd}
\usepackage{mathrsfs}
\usepackage{mathtools}
\usepackage{dsfont}
\usepackage{blkarray}
\usepackage{todonotes}
\usepackage[T1]{fontenc}

\definecolor{darkyellow}{rgb}{0.5, 0.5, 0.0}
\definecolor{darkpurple}{rgb}{0.5, 0.2, 0.8}
\definecolor{darkblue}{rgb}{0.0, 0.0, 0.8}
\definecolor{darkgreen}{rgb}{0.0, 0.5, 0.0}
\definecolor{darkred}{rgb}{0.5, 0.0, 0.0}
\definecolor{newdarkblue}{HTML}{0044B3}
\definecolor{olddarkgreen}{rgb}{0.0, 0.4, 0.0}
\definecolor{etacol3}{RGB}{137,138,60}
\definecolor{etacol2}{RGB}{137,60,138}
\definecolor{etacol}{RGB}{60,137,138}
\definecolor{etacol4}{RGB}{127,23,14}
\definecolor{polecol}{RGB}{127,140,200}
\hypersetup{
    linktocpage,
    colorlinks,
    citecolor=olddarkgreen,
    linkcolor= olddarkgreen,
    urlcolor=olddarkgreen
}

\graphicspath{{Figures}}

\usepackage{tikz-3dplot}
\usetikzlibrary{patterns}
\usetikzlibrary{knots}
\usetikzlibrary{decorations}
\usetikzlibrary{decorations.pathreplacing}
\usetikzlibrary{calc,math,angles,quotes}
\usetikzlibrary{decorations.markings,arrows.meta}

\newcommand{\circleleft}{
\begin{tikzpicture}
 \draw[
        decoration={markings, mark=at position 0.3 with   {\arrow[scale=1.4]{>}}},
        postaction={decorate}
        ] (-0.01,0.1) -- (-0.011,0.1);
\draw[] (0.05,0) circle (0.1);
\end{tikzpicture}
}

\tikzset{->-/.style={decoration={
  markings,
  mark=at position 1 with {\arrow[scale=1.3]{>}}},postaction={decorate}}}

\tikzset{-<-/.style={decoration={
  markings,
  mark=at position .4 with {\arrow[scale=1.3]{<}}},postaction={decorate}}}

\newcommand{\eps}{\ensuremath{\varepsilon}}
\newcommand{\xrightarrowdbl}[2][]{%
  \xrightarrow[#1]{#2}\mathrel{\mkern-14mu}\rightarrow
}

\newcommand{\red}[1]{ {\color{darkred}{#1}}}

\renewcommand{\circlearrowleft}{\circleleft}
\newcommand{\kappab}{{\overline{\kappa}}}
\newcommand{\kappas}{{\overline{\kappa}}}
\newcommand{\kappaps}{{\overline{\kappap}}}
\newcommand{\kappapps}{{\overline{\kappapp}}}
\newcommand{\kappap}{{\kappa^\prime}}
\newcommand{\kappapp}{{\kappa^{\prime\prime}}}

\numberwithin{equation}{section}
\newtheorem{definition}{Definition}
\newtheorem*{definition*}{Definition}

\newtheorem{remark}{Remark}
\newtheorem{theorem}{Theorem}
\newtheorem*{theorem*}{Theorem}
\newtheorem{lemma}{Lemma}
\newtheorem*{lemma*}{Lemma}

\newcommand{\eq}[1]{\begin{equation}\begin{aligned}#1\end{aligned}\end{equation}}
\newcommand{\disc}{\text{Disc}}
\newcommand{\bbone} { {\mathds 1}}
\newcommand{\bbR}{\ensuremath{\mathbb{R}}}
\newcommand{\bbC}{\ensuremath{\mathbb{C}}}

\newcommand{\cA}{{\mathcal A}}
\newcommand{\cM}{{\mathcal M}}

\newcommand{\EE}{{\mathcal E}}
\newcommand{\I}{I}
\renewcommand{\S}{{\mathcal{S}}}
\newcommand{\LL}{\mathcal{L}}
\renewcommand{\P}{\mathcal{P}}
\newcommand{\MM}{\eta}
\newcommand{\monM}{\mathscr{M}}
\renewcommand{\gg}{\gamma}
\newcommand{\Eint}{E_{\mathrm{int}}}
\renewcommand{\leq}{\leqslant}
\renewcommand{\geq}{\geqslant}
\newcommand{\B}{\P}

\newcommand{\rd} {\mathrm{d}}

\newcommand{\nex}{n_{\text{ext}}}
\newcommand{\nint}{n_{\text{int}}}
\newcommand{\Chat}{\widehat{C}}

\renewcommand{\Re}{\ensuremath{\mathrm{Re}}}
\renewcommand{\Im}{\ensuremath{\mathrm{Im}}}

\newcommand{\Up}{U_p}
\newcommand{\vt}{{\tilde{v}}}
\newcommand{\EX}{\mathcal{E}}
\newcommand{\Xs}{X}
\newcommand{\Xp}{{X_p}}

\usetikzlibrary{arrows}

\definecolor{col5}{HTML}{008b8b}
\definecolor{col6}{rgb}{0.9,0.0,0.0}
\definecolor{col3}{rgb}{0.0, 0.4, 0.0}
\definecolor{col4}{rgb}{0.5, 0.0, 0.0}
\definecolor{englishviolet}{HTML}{62466B}
\definecolor{newDarkGreen}{rgb}{0.0,0.5,0.0}
\definecolor{darkcyan}{HTML}{00A1A1}
\definecolor{royalblue}{HTML}{5576d1}
\definecolor{olivegreen}{HTML}{423B0B}
\definecolor{queenpink}{HTML}{ED6B86}
\definecolor{mulberry}{HTML}{AF467C}
\definecolor{newdarkblue}{HTML}{027098}
\definecolor{newdarkblue2}{HTML}{00008B}
\definecolor{darkorange}{HTML}{ED6B86}

\newcommand{\trit}{{
\begin{tikzpicture}[line width=0.8]
\draw [black] (-0.16,-0.08) -- (-0.16,0.08) -- (0,0) -- cycle;
\end{tikzpicture}
}}

\newcommand{\tritcol}{{
\begin{tikzpicture}[line width=0.8]
\draw [black] (-0.16,-0.08) -- (-0.16,0.08) -- (0,0) -- cycle;
\draw [newdarkblue2] (-0.16,-0.08) -- (-0.16,0.08);
\draw [darkred] (-0.16,0.08) -- (0,0);
\draw [olddarkgreen] (0,0) -- (-0.16,-0.08);
\end{tikzpicture}
}}

\newcommand{\squaret}{{
\begin{tikzpicture}[line width=0.8]
\draw [black] (-0.1,-0.1) -- (-0.1,0.1) -- (0.1,0.1) -- (0.1,-0.1) -- cycle;
\end{tikzpicture}
}}

\newcommand{\squaretcol}{{
\begin{tikzpicture}[line width=0.8]
\draw [black] (-0.1,-0.1) -- (-0.1,0.1) -- (0.1,0.1) -- (0.1,-0.1) -- cycle;
\draw [newdarkblue2] (-0.1,-0.1) -- (-0.1,0.1);
\draw[olddarkgreen] (-0.1,0.1) -- (0.1,0.1);
\draw[darkred] (0.1,0.1)-- (0.1,-0.1);
\draw[darkorange] (0.1,-0.1) -- (-0.1,-0.1);
\end{tikzpicture}
}}

\newcommand{\squaretcolii}{{
\begin{tikzpicture}[line width=0.8]
\draw [black] (-0.1,-0.1) -- (-0.1,0.1) -- (0.1,0.1) -- (0.1,-0.1) -- cycle;
\draw [newdarkblue2] (-0.1,-0.1) -- (-0.1,0.1);
\draw[darkred] (-0.1,0.1) -- (0.1,0.1);
\draw[darkorange] (0.1,0.1)-- (0.1,-0.1);
\draw[olddarkgreen] (0.1,-0.1) -- (-0.1,-0.1);
\end{tikzpicture}
}}

\newcommand{\squaresquare}{{
\begin{tikzpicture}[line width=0.8]
\draw [darkred] (-0.1,-0.1) -- (-0.1,0.1) -- (0.1,0.1) -- (0.1,-0.1) -- cycle;
\draw [etacol] (0.1,-0.1) -- (0.1,0.1) -- (0.3,0.1) -- (0.3,-0.1) -- cycle;
\draw [black] (0.1,-0.1) -- (0.1,0.1);
\end{tikzpicture}
}}

\newcommand{\triright}{{
\begin{tikzpicture}[line width=0.8,scale=1.2]
\draw [darkred] (-0.08,-0.06) -- (-0.08,0.06) -- (0,0) -- cycle;
\end{tikzpicture}
}}

\newcommand{\trileft}{{
\begin{tikzpicture}[line width=0.8,scale=1.2]
\draw [etacol] (0.08,0.06) -- (0.08,-0.06) -- (0,0) -- cycle;
\end{tikzpicture}
}}

\newcommand{\bubt}{{
\begin{tikzpicture}[line width=0.8]
\path [newdarkblue2,out=40,in=140] (-0.12,0) edge (0.12,0);
\path [newdarkblue2,out=-40,in=220] (-0.12,0) edge (0.12,0);
\end{tikzpicture}
}}

\newcommand{\bubrgt}{{
\begin{tikzpicture}[line width=0.8]
\path [darkred,out=40,in=140] (-0.15,0) edge (0.15,0);
\path [olddarkgreen,out=-40,in=220] (-0.15,0) edge (0.15,0);
\end{tikzpicture}
}}
\newcommand{\bubrgT}{{
\begin{tikzpicture}[line width=0.8]
\draw[darkred] (0.1,0) arc (0:180:0.1);
\draw[olddarkgreen] (0.1,0) arc (0:-180:0.1);
\end{tikzpicture}
}}

\newcommand{\sunt}{{
\begin{tikzpicture}[line width=0.6]
\path [newdarkblue2,out=-60,in=240] (-0.12,0) edge (0.12,0);
\path [darkred,out=60,in=120] (-0.12,0) edge (0.12,0);
\draw [darkred] (-0.12,0) -- (0.12,0);
\end{tikzpicture}
}}
\newcommand{\icet}{{
\begin{tikzpicture}[scale=1.5,line width=0.73]
\path [darkred,out=130,in=230] (0,-0.06,0) edge (0,0.06);
\path [darkred,out=50,in=-40] (0,-0.06,0) edge (0,0.06);
\draw [newdarkblue2] (0,0.06) -- (0.15,0);
\draw [newdarkblue2] (0,-0.06) -- (0.15,0);
\end{tikzpicture}
}}
\newcommand{\dubbubt}{
{\begin{tikzpicture}
  \draw[darkred,line width=1.1] (-0.1,0) circle (0.1);
  \draw[newdarkblue2,line width=1.1] (0.1,0) circle (0.1);
\end{tikzpicture}}
}
\newcommand{\bubbluet}{{
\begin{tikzpicture}
  \draw[newdarkblue2,line width=1.1] (0,0) circle (0.09);
\end{tikzpicture}
}}
\newcommand{\bubredt}{{
\begin{tikzpicture}
  \draw[darkred,line width=1.1] (0,0) circle (0.09);
\end{tikzpicture}
}}

\newcommand{\Gbub}{G_{\bubt}}
\newcommand{\Gsun}{G_{\sunt}}
\newcommand{\Gice}{G_{\icet}}
\newcommand{\kappabubp}{\kappas_{\bubt}}
\newcommand{\kappasunp}{\kappas_{\sunt}}
\newcommand{\kappabub}{\kappa_{\bubt}}
\newcommand{\kappasun}{\kappa_{\sunt}}

\tikzset{
  on each segment/.style={
    decorate,
    decoration={
      show path construction,
      moveto code={},
      lineto code={
        \path [#1]
        (\tikzinputsegmentfirst) -- (\tikzinputsegmentlast);
      },
      curveto code={
        \path [#1] (\tikzinputsegmentfirst)
        .. controls
        (\tikzinputsegmentsupporta) and (\tikzinputsegmentsupportb)
        ..
        (\tikzinputsegmentlast);
      },
      closepath code={
        \path [#1]
        (\tikzinputsegmentfirst) -- (\tikzinputsegmentlast);
      },
    },
  },
  mid arrow/.style={postaction={decorate,decoration={
        markings,
        mark=at position .5 with {\arrow[#1]{stealth}}
      }}},
}

\title{%
Constraints on Sequential Discontinuities from the Geometry of On-shell Spaces
}

\author[a, b]{Holmfridur~S.~Hannesdottir,}
\author[c, d]{Andrew~J.~McLeod,}
\author[b]{Matthew~D.~Schwartz,}
\author[e]{Cristian~Vergu}

\affiliation[a]{Institute for Advanced Study, Einstein Drive, Princeton, NJ 08540, USA}
\affiliation[b]{Department of Physics, Harvard University, Cambridge, MA 02138, USA}
\affiliation[c]{CERN, Theoretical Physics Department, 1211 Geneva 23, Switzerland}
\affiliation[d]{Mani L. Bhaumik Institute for Theoretical Physics, Department of Physics and Astronomy,\\UCLA, Los Angeles, CA 90095, USA}
\affiliation[e]{Niels Bohr International Academy and Discovery Center, Niels Bohr Institute,\\University of Copenhagen, Blegdamsvej 17, DK-2100, Copenhagen \O, Denmark}

\emailAdd{hofie@ias.edu}
\emailAdd{a.mcleod@cern.ch}
\emailAdd{schwartz@g.harvard.edu}
\emailAdd{c.vergu@gmail.com}

\preprint{CERN-TH-2022-189}

\abstract{
We present several classes of constraints on the discontinuities of Feynman integrals that go beyond the Steinmann relations. These constraints follow from a geometric formulation of the Landau equations that was advocated by Pham, in which the singularities of Feynman integrals correspond to critical points of maps between on-shell spaces. To establish our results, we review elements of Picard--Lefschetz theory, which connect the homotopy properties of the space of complexified external momenta to the homology of the combined space of on-shell internal and external momenta. An important concept that emerges from this analysis is the question of whether or not a pair of Landau singularities is compatible---namely, whether or not the Landau equations for the two singularities can be satisfied simultaneously. Under conditions we describe, sequential discontinuities with respect to non-compatible Landau singularities must vanish. Although we only rigorously prove results for Feynman integrals with generic masses in this paper, we expect the geometric and algebraic insights that we gain will also assist in the analysis of more general Feynman integrals.
}

\begin{document}
\maketitle

  \section{Introduction}

One of the main goals of the $S$-matrix program of the nineteen-sixties was to  completely understand the analytic structure of scattering amplitudes~\cite{Karplus:1958zz,Karplus:1959zz,mandelstam1959analytic,nakanishi1959,Bjorken:1959fd,landau1959,Mandelstam:1960zz,Cutkosky:1960sp,PhysRevLett.5.213,PhysRev.119.1763,PhysRev.120.1514,PhysRev.121.1567,Nakanishi1961,boyling1964hermitian,Bros:1964iho,BarucchiRegge:1964,Bros:1965kbd,Polkinghorne1966,Bloxham1966,boyling:1966,Landshoffolive:1966,ELOP,Hwa:102287,pham,boyling1968homological,Risk1968,Goddard:1969ci,Bloxham:1969cm,Bloxham:1969cp,Bloxham:1969cq,Ponzano:1969tk,Sommer:1970mr,Ponzano:1970ch,Goddard:1970dz,todorov:1971,Bros:1972jh,Regge:1972ns,Streater:1975vw}.  Although this dream still seems far from being realized, significant advances have been made in our understanding of the analytic structure of certain classes of amplitudes. Much of this progress has come from studying specific examples in the planar limit of maximally supersymmetric Yang-Mills theory, where amplitudes involving fewer than ten particles are believed to be expressible in terms of multiple polylogarithms. As the analytic structure of multiple polylogarithms can be fully characterized (up to algebraic identities) using the motivic coaction~\cite{Goncharov:2001iea,Brown:2009qja,Goncharov:2010jf,Brown1102.1312}, the analytic structure of these amplitudes has become increasingly well understood. This has, in turn, facilitated calculations at incredibly high loop order and large particle multiplicity~\cite{Caron-Huot:2011zgw,Caron-Huot:2018dsv,McLeod:2020dxg,Caron-Huot:2020bkp,He:2020vob,He:2020lcu,Dixon:2020bbt,Golden:2021ggj,Li:2021bwg,Dixon:2021nzr}, and led to the discovery of surprising new connections between amplitudes, cluster algebras, and tropical geometry~\cite{Arkani-Hamed:2012zlh,Golden:2013xva,Golden:2014xqa,Golden:2014pua,Abreu:2017enx,Drummond:2017ssj,Drummond:2018dfd,Golden:2018gtk,Golden:2019kks,Drummond:2019qjk,Arkani-Hamed:2019rds,Henke:2019hve,Abreu:2019wzk,Drummond:2019cxm,Henke:2020dfp,Drummond:2020kqg,Mago:2020kmp,Mago:2020nuv,Ren:2021ztg,Mago:2021luw,Abreu:2021vhb,Herderschee:2021dez,Arkani-Hamed:2022cqe}. 

The singularities and discontinuities of amplitudes can be studied using the Landau equations~\cite{nakanishi1959,landau1959,Bjorken:1959fd} and absorption integrals (often also called cut integrals)~\cite{Cutkosky:1960sp}. In recent years, these topics have received renewed attention~\cite{Brown:2009ta,Bloch:2015efx,Kreimer:2016tqq,Abreu:2014cla,Abreu:2017ptx,Arkani-Hamed:2017ahv,Collins:2020euz,Bourjaily:2020wvq,Hannesdottir:2021kpd,Klausen:2021yrt,Muhlbauer:2020kut,Mizera:2021fap,Correia:2021etg,Bourjaily:2022vti,Flieger:2022xyq}, including in connection with the motivic coaction~\cite{Goncharov:2001iea,Brown:2009qja,Goncharov:2010jf,Brown1102.1312,Abreu:2014cla} and in special theories such as planar maximally supersymmetric Yang-Mills theory~\cite{Dennen:2015bet,Dennen:2016mdk,Prlina:2017azl,Prlina:2017tvx,Prlina:2018ukf,Gurdogan:2020tip} and integrable theories~\cite{Dorey:2022fvs}. 
In this paper, we contribute to this literature by studying the locations at which the discontinuities of Feynman integrals can develop branch cuts, and the ways in which sequences of branch points can be accessed via analytic continuation~\cite{pham}.

The Landau equations impose two types of restrictions on the internal and external momenta of Feynman integrals. First, they require that either the internal momenta associated with each line is on-shell or that the corresponding line be contracted to a point and effectively removed from the graph.
These on-shell conditions identify where singularities can occur in the {\it integrand} in a Feynman integral, which constitutes a necessary but not a sufficient condition for a singularity to develop in the full {\it integral}. Second, the Landau loop equations identify points at which the singularities in the integrand can pinch the integration contour. If the integration contour passes through this pinch (and there are no cancellations due to numerators), the on-shell and loop equations comprise sufficient conditions for the integral to develop a singularity or branch point.

A physical way of understanding the Landau equations, which was first appreciated by Coleman and Norton~\cite{Coleman:1965xm}, is that these equations pick out configurations of scattering particles in which each loop forms a closed path in spacetime. The loop equations therefore identify scattering configurations that can occur classically. However, while physically appealing, the Coleman-Norton interpretation does not immediately yield insight into singularities that occur outside of the physical region. 

A less widely appreciated interpretation of the Landau loop equations is that these equations identify the critical points of differential maps between spaces of on-shell momenta~\cite{pham}. This geometric interpretation holds both inside and outside of the physical region, and was expounded at length by Pham in his thesis~\cite{pham} (see also Ref.~\cite{boyling:1966} for related earlier work). In particular, Pham showed that at special points in the space of external momenta, namely at the solutions of the Landau equations, the topology of the on-shell surface in the space of internal momenta changes.
Mathematically, the Landau equations are satisfied by the critical points of projection maps from the combined space of internal and external momenta to the space of external momenta. Thus, by studying the geometry of these spaces one can gain insights into the properties and locations of singularities. Generalizing this construction to maps between different on-shell spaces, one can also study the locations of sequential discontinuities. We will explain how this is done in Sec.~\ref{sec:deform_contours}.

Studying the locations at which the discontinuities of Feynman integrals can themselves become singular requires us to generalize the graphical way in which we identify these singularities. Following Pham, we generalize the usual association of singularities with individual graphs (usually referred to as Landau diagrams) to the association of singularities with contractions between graphs. More specifically, we can consider two graphs $G_1$ and $G_2$, where $G_1$ is given by {\it contracting} some of the edges of $G_2$ to a point. For example:
\begin{equation}
\resizebox{6.5cm}{!}{
\begin{tikzpicture}[line width=1.1,baseline=(current bounding box.center)]
\node at (0,0) {
\begin{tikzpicture}[baseline=(current bounding box.center),line width=1.15]
\path [darkred,out=130,in=230] (0,-0.6,0) edge (0,0.6);
\path [darkred,out=50,in=-40] (0,-0.6,0) edge (0,0.6);
\node[darkred] at (0.5,0) {$q_3$};
\node[darkred] at (-0.5,0) {$q_4$};
\draw [newdarkblue2] (0,0.6) -- (1.5,0);
\draw [newdarkblue2] (0,-0.6) -- (1.5,0);
\node [newdarkblue2] at (0.75,0.55) {$q_1$};
\node [newdarkblue2] at (0.75,-0.55) {$q_2$};
\draw[black] (0,-0.6) -- ++(-60:0.5) ;
\draw[black] (0,-0.6) -- ++(-120:0.5) ;
\draw[black] (0,0.6) -- ++(120:0.5) ;
\draw[black] (0,0.6) -- ++(60:0.5) ;
\draw[black] (1.5,0) -- ++(30:0.5) ;
\draw[black] (1.5,0) -- ++(-30:0.5) ;
\end{tikzpicture}
};
\node at (5,0) {
\begin{tikzpicture}[baseline=(current bounding box.center),line width=1.2]
\path [newdarkblue2,out=60,in=120] (-1,0) edge (0,0);
\path [newdarkblue2,out=-60,in=240] (-1,0) edge (0,0);
\node[newdarkblue2] at (-0.5,0.55) {$q_1$};
\node[newdarkblue2] at (-0.5,-0.55) {$q_2$};
\draw[black] (-1,0) -- ++(60:0.5);
\draw[black] (-1,0) -- ++(120:0.5);
\draw[black] (-1,0) -- ++(-60:0.5);
\draw[black] (-1,0) -- ++(-120:0.5);
\draw[black] (0,0) -- ++(45:0.5);
\draw[black] (0,0) -- ++(-45:0.5);
\end{tikzpicture}
};
\node[scale=1.2] at (5,1.5) {$G_{1}$};
\node[scale=1.2] at (0,1.5) {$G_{2}$};
\draw[->>,black!50] (2,0) -- (3.5,0) ;
\end{tikzpicture}
}
\end{equation}
Algebraically, the equations associated with each such graph contraction are the Landau equations associated with $G_2$, except that a Landau loop equation is only included for each loop that can be constructed entirely out of edges that have been contracted. The solutions to this set of equations describe where singularities can appear after one computes a discontinuity with respect to the singularity associated with the Landau diagram $G_1$. The standard way of thinking about the singularities of Feynman integrals as corresponding to a single graph is recaptured by taking $G_1$ to be the graph of an elementary process, in which all external lines meeting at a single vertex. This corresponds to contracting all of the internal lines of $G_2$.

In this graphical language, the study of sequential discontinuities becomes the study of sequences of graph contractions. One can then ask questions such as: when do two sequential discontinuities commute? That is, when will the same function result from computing the same pair of discontinuities in the opposite order? As we will see in Sec.~\ref{sec:codim2}, the discontinuities of Feynman integrals in generic kinematics commute if and only if the graph contractions are {\it compatible}, meaning that the Landau equations associated to each of the two singularities can be solved simultaneously.
Intuitively, this corresponds to the observation that the first discontinuity of a Feynman integral can be computed as a cut integral, in which some of the particles in a diagram are put on shell. This cut integral only has support for loop momenta that are consistent with the corresponding solution to the Landau equations. Thus, if the Landau equations corresponding to the second discontinuity force the loop momenta to take different values from those that were required by the first discontinuity, the sequential discontinuity must vanish.  In Sec.~\ref{sec:codim2} we also show that one can formulate this compatibility graphically for Feynman integrals involving generic masses, following Pham.

To establish these results, and their generalizations outside the physical region and/or for non-generic masses, we use techniques provided by Picard--Lefschetz theory and Leray's multivariate residue calculus. Although these mathematical tools have been described elsewhere in the physics literature (see for example~\cite{MR214101,Hwa:102287,boyling:1966,boyling1968homological,Abreu:2017ptx}), they have not yet been widely adopted. We thus provide an introduction to these methods by means of examples in Sec.~\ref{sec:PicardLefschetz}, before applying them to study sequential discontinuities in the remainder of the paper. For recent applications of Picard--Lefschetz theory in the physics literature, see e.g.~Refs.~\cite{Bogner:2017vim,Bonisch:2021yfw,Broedel:2021zij}.

To understand the discontinuities described by longer sequences of graph contractions, it is sufficient in the generic-mass case to separately study each sequential pair of contractions. As we will describe in Sec.~\ref{sec:sequential_disc} and Sec.~\ref{sec:codim2},  key properties of pairs of such contractions can be studied by embedding them in larger diagrams that we call Pham diagrams. These Pham diagrams take a different form depending on whether the combined pair of contractions identifies a codimension-one or codimension-two locus. 
In the codimension-one case, where each contraction encodes a single kinematic constraint, Pham diagrams depict an identity between the double discontinuity represented by the pair of contractions and the single discontinuity represented by the combined contraction. We write this as
\begin{equation}
\Big(\bbone - 
  \monM_{\P_\kappap} \Big) \Big(\bbone - \monM_{\P_\kappa} \Big) \I_G(p) 
  = \Big(\bbone -   \monM_{\P_\kappap} \Big) \I_G(p)
\end{equation}
where $\monM_{\P}$ is the monodromy operator that analytically continues a function around the singular surface $\P$ in the space of external kinematics, and $I_G(p)$ is a Feynman integral.
When the combined contraction has codimension two, Pham diagrams instead allow us to diagnose whether or not the sequence of discontinuities represented by this pair of contractions vanishes:
\begin{equation}
    \Big(\bbone - 
   \monM_{\P_\kappap} \Big) 
   \Big(\bbone - \monM_{\P_\kappa}\Big) \I_G(p)
=
   0 \, .
\end{equation}
These two types of relations can provide highly non-trivial constraints on the structure of scattering amplitudes. 

Our motivation for studying the analytic structure of scattering amplitudes comes, in part, from the strong constraints that the Steinmann relations~\cite{Steinmann,Steinmann2,araki:1961,Cahill:1973qp} and their recent generalizations~\cite{Drummond:2017ssj,Caron-Huot:2019bsq,Bourjaily:2020wvq,Benincasa:2020aoj,Dixon:2021tdw} have been seen to put on the discontinuity structure of amplitudes and Feynman integrals. As we will show, the codimension-two Pham diagram analysis implies the usual Steinmann relations for amplitudes with generic masses. However, Pham diagrams also go beyond the Steinmann relations---which only apply to normal thresholds---and put new restrictions on the sequential discontinuities of Feynman integrals. This type of information can be fed directly into bootstrap approaches (as exemplified, for instance, in~\cite{Dixon:2016nkn,Almelid:2017qju,Drummond:2018caf,Henn:2018cdp,Caron-Huot:2019vjl,Dixon:2022rse}).

In order to motivate and illustrate the constraints that follow from Pham diagrams, we consider a number of examples throughout the paper. Many of these examples require computing sequential discontinuities of Feynman integrals, which requires the careful construction of explicit analytic continuation contours.  We deal with this difficulty by requiring the contours to stay real as much as possible, and introduce complex detours only to avoid singularities in the integrand. As we will show, this then implies a choice for these detours in the space of external kinematics that stays consistent with causality, which allows us to construct the appropriate analytic continuation paths.

In the majority of this paper, we restrict our attention to integrals that involve generic internal and external masses. In these cases, we expect to be able to isolate contributions from different solutions of the Landau equations in the space of external variables. We also focus on first-type Landau singularities, in which all momenta take finite values; for this class of singularities, there is a well-defined notion of a threshold in the space of external variables (for discussions of homological methods applied to second-type singularities~\cite{Fairlie:1962secondtype}, see~Refs.~\cite{MR214101,boyling1968homological,Abreu:2017ptx}). When some masses are equal, additional technical complications arise (such as the possibility of elastic scattering), although sometimes these complications can be overcome.  Massless particles also lead to technical complications due to the possibility of collinear and soft singularities, as well as IR singularities.\footnote{For massless particles, the bubble and triangle Landau loop equations inevitably lead to collinear kinematics, and therefore to singular configurations. However, some solutions to the box Landau loop equations are well-behaved, even though the internal particles are massless. To study such singular configurations arising in Feynman integrals with massless particles, a more careful analysis is needed.}  Although we believe that the general theory described herein can be adapted to the study of massless particles, the generic mass case is intricate enough that we focus almost entirely on it in this paper.

We structure the paper as follows. In Sec.~\ref{sec:review} we discuss Feynman integrals, introduce a number of crucial definitions, and discuss the Landau equations.  In Sec.~\ref{sec:PicardLefschetz} we discuss how the Picard--Lefschetz theorem can be used to understand monodromies of Feynman integrals in both Feynman-parameter space and momentum space.  We also discuss how to determine the correct complex detours of the integration contour of Feynman integrals, and illustrate how these detours can be represented as vector fields.  We then use these techniques to provide a sketch of proof of Cutkosky's formula.  In Sec.~\ref{sec:iterated} we discuss absorption integrals, and where they can develop singularities and discontinuities.  In particular, we illustrate the application of the Picard--Lefschetz theorem to absorption integrals in some examples.  In Sec.~\ref{sec:deform_contours} we review Pham's interpretation of Landau singularities as critical values of projection maps between on-shell spaces.  We work out a number of examples of these on-shell spaces and the projections between them, in order to see explicitly that the critical points of these projections correspond to solutions to the Landau equations.  We also present a discussion of principal branches of the Landau variety, which are those for which there is a well-defined notion of being below the threshold region, and for which Cutkosky's formula exists.  In Sec.~\ref{sec:sequential_disc} we present constraints on hierarchical sequential discontinuities, while in Sec.~\ref{sec:codim2} we present constraints on non-hierarchical sequential discontinuities.  Finally, in Sec.~\ref{sec:exceptional_kinematics} we briefly illustrate new phenomena that can arise in exceptional kinematics.  In Sec.~\ref{sec:summary} we summarize the main results in the paper, and provide some concluding remarks in Sec.~\ref{sec:conclusion}.

We also include a number of appendices, in which we review salient concepts related to graph theory (Sec.~\ref{sec:graph_theory}), Kronecker indices (Sec.~\ref{sec:kronecker_index}), Poincar\'e duality (Sec.~\ref{sec:signs}), the nature of critical points from a Lagrange multiplier perspective (Sec.~\ref{sec:hessians}), and the homotopy of fibrations (Sec.~\ref{sec:homotopy}).  We present some details of the proof of Theorem~\ref{thm:pham} in Sec.~\ref{sec:sing_absorption}.  %

  \section{Branch Points in Feynman Integrals}
\label{sec:review}

We being by reviewing some of the tools that have been developed for studying the singularities of Feynman integrals. In particular, our goal will be to understand the singularities of $\I_G(p)$, the scalar Feynman integral whose propagator structure is given by the graph $G$, in terms of other algebraic and geometric objects that can be constructed using $G$, such as on-shell spaces, Landau varieties, 
and so on.\footnote{When nontrivial numerators are included in Feynman integrals, the analysis changes in two ways.  First, these numerators may cancel some of the singularities that appear in the scalar integrals, and second they may lead to extra singularities at infinity.}  Useful references for this material include~\cite{MR0494126,pham}. Our notation will make use of a number of notions from graph theory, such as the definitions of fundamental cycles and incidence matrices. We have collected these definitions in Appendix~\ref{sec:graph_theory} for easy reference.

\subsection{Feynman Integrals and the Landau Equations}
\label{sec:landau_review}

Feynman integrals describe the contributions to scattering processes that come from different sets of virtual interactions. These contributions can each be represented graphically as a Feynman diagram $G$. In particular, to each graph $G$, we associate a scalar \textbf{Feynman integral}
\begin{equation}
 \label{eq:Feynman_integral_def}
    \I_G (p) = \int \prod_{c \in \Chat(G)} \rd^d k_c \prod_{e \in \Eint(G)} \frac{1}{\left[q_e(k, p)\right]^2-m_e^2 + i\varepsilon} \,,
\end{equation}
where $k$ denotes the set of independent momenta flowing through each loop, which are in one-to-one correspondence with a set of fundamental cycles $\Chat(G)$ of $G$, $p$ denotes the set of external momenta, and $d$ is the (integer) space-time dimension.  We use $E(G)$ to denote the set of edges of the graph $G$, and $V(G)$ to denote its vertices.
In general, we will employ $p$ to denote external momenta, $q$ to denote momenta associated with internal lines, and $k$ to denote a basis of loop momenta being integrated over; however, we will sometimes also use $q_e(k, p)$ to denote both internal and external edges of the graph, where $e \in E(G)$.
We moreover use $\Eint(G)$ to denote the set of internal edges in $G$ and $\nint = |\Eint(G)|$ to denote their number. The number of external edges $|E(G)| - \nint$ will similarly be denoted by $\nex$. In general, we assume that the integral $I_G(p)$ is free of UV and IR divergences, and that the external kinematics $p$ can be deformed away from the singularities of $I_G(p)$.

In this graph theory language, momentum conservation in $I_G(p)$ corresponds to the constraint
\begin{equation}
\label{eq:landau_2}
\sum_{e \in E(G)} a_{v e}(G) q_e^\mu = 0
\end{equation}
for each vertex $v \in V(G)$, where $a_{v e}(G)$ is the incidence matrix of the graph. Namely, $a_{v e}$ is the $|V(G)| \times |E(G)|$ matrix $a$ whose entries are given by 
\begin{equation}
a_{v e} = \begin{cases} \ 1 \quad \text{if the edge $e$ starts on the vertex $v$,} \\ 
 -1 \quad \text{if the edge $e$ ends on the vertex $v$,} \\
\  0 \quad \text{otherwise.} \end{cases}
\end{equation}
Since we will always work in kinematics where momentum is conserved, we henceforth assume this set of equations is satisfied, and leave them implicit.  We denote the space of internal and external momenta of a graph $G$ where momentum is conserved by $\EE(G)$:\footnote{The dimension of $\EE(G)$ is $d (\lvert E(G)\rvert - \lvert V(G)\rvert)$.  On this space we have an action of the orthogonal group $O(d-1, 1)$ so we can eliminate $\frac{d (d-1)}{2}$ variables by picking a special frame.  This assumes that the action of the Lorentz group is free, that is there are no non-trivial Lorentz transformations which leave the configuration of momenta invariant.}
\begin{equation}
    \label{eq:eg_set}
    \EE(G) = \Bigl\{(q_e)_{e \in E(G)} \in \mathbb{C}^{d \times \lvert E(G)\rvert} \mid \sum_{e' \in E(G)} a_{v e'} q_{e'} = 0, \quad \forall v \in V(G)\Bigr\}.
\end{equation}

To study the singularities of $\I_G(p)$ it can be helpful to introduce Feynman parameters, which allow us to rewrite $\I_G(p)$ as
\begin{equation}
\label{eq:landauform}
    \I_G(p) = (\nint-1)!  
    \int_0^\infty \prod_{e \in \Eint(G)} \rd\alpha_e \int \prod_{c \in \Chat(G)} \rd^d k_c \frac{1}{(\ell  +i\varepsilon)^{\nint}}
    \delta\Big(1 -\!\! \sum_{e \in \Eint(G)} \alpha_e\Big) \, ,
\end{equation}
 where 
 \begin{equation}
     \ell = \sum_{e \in \Eint(G)} \alpha_e (q_e^2-m_e^2) \, ,
     \label{elldef}
\end{equation}    
and where the integral over each Feynman parameter $\alpha_e$ is evaluated along the positive real axis, from zero to infinity. Although the delta function constrains each of these Feynman parameters to be less than or equal to one, 
we have the freedom to rescale all of the variables $\alpha_e$ uniformly; this will change the integral by a constant, but not affect its singularities.  It is possible to use this $\textrm{GL}(1)$ covariance to fix $\sum_{e' \in E'} \alpha_{e'}$ to some convenient value, where $E'$ is any subset of the edges in $E(G)$.

In the Feynman-parametrized form in Eq.~\eqref{eq:landauform}, the integrand is singular whenever $\ell=0$. This can happen if the \textbf{on-shell equation}
\begin{equation}
q_e^2 = m_e^2 \,  \label{eq:landau_1}
\end{equation}
is satisfied for any subset of the internal edges and the Feynman parameters $\alpha_e$ associated with the remaining edges are set to zero. It follows that there will be  different regions of the $\ell=0$ surface in which different subsets of the internal edges are on-shell; we will characterize these different regions in terms of graph contractions in Sec.~\ref{sec:branches}.

We define the \textbf{on-shell variety} of the graph $G$ to be the set of points in the space of internal and external kinematics for which \emph{all} edges are on-shell:

\begin{equation}
\label{onshellvariety}
    \S(G) = \Bigl\{(q_e)_{e \in E(G)} \in \EE(G) \mid q_e^2 - m_e^2 = 0, \quad \forall e \in E(G)\Bigr\}.
\end{equation}
As mentioned above, we leave momentum conservation implicit, so $\S(G) \subset \EE(G)$.\footnote{We will use both the real and complex versions of the spaces $\S(G)$ and $\EE(G)$.  The real versions are required to study so-called principal singularities, which we will discuss in Sec.~\ref{sec:principal}.  The complex versions are used when deforming contours around singularities and in the application of the Picard--Lefschetz theorem.  We do not distinguish between these usages in our notation, but hope that which one is meant is clear from the context.
} We will find it useful to use $G_0$ to denote the elementary graph in which all of the external lines in $G$ meet at a single vertex; this allows us to denote the space of on-shell external momenta associated with the graph $G$ by $\S(G_0)$.

Not all singularities in the integrand will lead to singularities of the integral $\I_G(p)$. For a singularity to develop at the level of the full integral, it must also be the case that the integration contour cannot be deformed to avoid the relevant integrand-level singularities. 
In the physics literature, this condition is generally described as the contour being pinched between singularities of the integrand (as $\varepsilon \to 0$), or as resulting from the integrand becoming singular on the boundary of the integration region (since the contour cannot be deformed to avoid this point). While singularities at integration endpoints are easy to identify, it is useful to have a set of equations that describe when the contour can become pinched. Just as two circles are tangent when their normal vectors at the intersection point are parallel, there will be a contour pinch when the gradients of the on-shell constraints with respect to the independent loop momenta are linearly dependent. This happens where there exists some set of values $\alpha_e$, not all vanishing, such that for each fundamental cycle $c$ associated with the independent loop momentum $k_c$,
\begin{equation}
 \sum_{e \in \Eint(G^\kappa)} \alpha_e \frac{\partial}{\partial k_c} (q_e^2 -m_e^2) = 0 \, .
 \label{partialform}
\end{equation}
Using that $\frac{\partial q_e}{\partial k_c} = b_{ce}$, we find the 
\textbf{Landau loop equation} for each independent $k_c$: 
\begin{equation}
\sum_{e \in \Eint(G^\kappa)} b_{c e} \alpha_e q_e^\mu 
=  0 \, ,
\label{eq:landau_3}
\end{equation}
where $b_{c e}$ is the circuit matrix of the graph, defined as
\begin{equation}
    b_{c e} = \begin{cases} \ 1 \quad \text{if the edge $e$ is in the circuit $c$ and is oriented in the same way as $c$,} \\ 
     -1 \quad \text{if the edge $e$ is in the circuit $c$ and is oriented in the opposite way to $c$,} \\
    \  0 \quad \text{otherwise.} \end{cases}
\end{equation}
A more rigorous derivation of the pinch condition can be found in Refs.~\cite{pham2011singularities, pham, Hwa:102287}.\footnote{The Landau equations were first derived by Landau, Bjorken, and Nakanishi~\cite{landau1959, Bjorken:1959fd, nakanishi1959}. Earlier work on this topic includes Refs.~\cite{Nambu:1957shl, Karplus:1958zz, Karplus:1959zz}, and Ref.~\cite{BSMF_1959__87__81_0} provides a more mathematical point of view.}

The solutions to the Landau loop equations and the corresponding set of on-shell conditions describe the set of locations where $I_G(p)$ can become singular at specific kinematic points $p \in \S(G_0)$.\footnote{When studying massless scattering amplitudes as in $\mathcal{N} = 4$ super-Yang--Mills theory, some singularities seemingly can not be explained by a Landau singularity analysis.  However, this case involves integrals which need regularization and whose external kinematics is non-generic.  Such cases require an extension of the methods we present in this paper.}
We call this variety the \textbf{Landau variety} and denote it by $\LL_G$. That is, the Landau variety is defined to be
 \begin{equation}
 \label{LV}
\LL_G \colon \qquad
 p \in \S(G_0) \, \Big|
\bigcap_{c \in \Chat(G)} 
\!\!\Big\{ \sum_{e \in \Eint(G)} b_{c e} \, \alpha_e q_e^\mu =0 \Big\}
    \bigcap_{e\in \Eint(G)} \!\!
\Big\{ \alpha_e (q_e^2 - m_e^2) = 0 
\Big\} 
\end{equation}
As discussed above, the momenta $q_e$ in these equations are implicit functions of the external momenta, and we assume that at least one variable $\alpha_e$ is nonzero.
We emphasize that if $\alpha_e \ne 0$ for an internal line, then the line must be on-shell; however, if $\alpha_e = 0$ the line is allowed to be off-shell or on-shell.\footnote{The Landau equations in Eq.~\eqref{LV} can also describe singularities at infinite loop momenta (sometimes referred to as second-type singularities), after compactifying the kinematic space. Such singularities do not arise in the physical region, and we will not consider these types of singularities in this paper.}

The Landau loop equations can also be neatly encoded as the vanishing of a differential form, as follows. Consider the on-shell variety associated with the graph $G$.
The normals to the hypersurfaces defined by the on-shell conditions $q_e^2 - m_e^2 = 0$ are given by the differentials $\rd (q_e^2 - m_e^2)\rvert_{p}$, where $\rvert_{p}$ means that we keep the external momenta fixed (in addition to the internal masses, which are always held fixed).
The pinch condition 
is equivalent to requiring that these differential forms be linearly dependent.
That is, there must exist some $\alpha_e$, not all zero, such that
\begin{equation}
\label{formvanish}
    \sum_{e \in \Eint(G)} \alpha_e \rd (q_e^2 - m_e^2)\rvert_{p} = 0\, .
\end{equation}
This is leads to the same condition as in Eq.~\eqref{partialform}.
By decomposing this differential form over a basis of fundamental cycles, we recover the Landau equations in Eq.~\eqref{eq:landau_3}; however, this formulation is more invariant since we do not have to pick a basis of fundamental cycles.

 \begin{figure}[t]
    \centering
\resizebox{10cm}{!}{%
\begin{tikzpicture}
[decoration={markings,
 mark=between positions 0.05 and 1 step 0.11 with {\arrow[line width=1pt]{>}}}]
\draw [ultra thick,etacol] plot [smooth] coordinates { (0,0) (4,1) (8,0) (12,1)};
\draw[-, ultra thick,etacol] (0,0) -- (-2,4);
\draw [ultra thick,etacol] plot [smooth] coordinates { (-2,4) (2,5) (6,4) (10,5)};
\draw[-, ultra thick,etacol] (12,1) -- (10,5);
\draw[etacol,fill=etacol!3] (12,1) -- (10,5) -- plot [smooth] coordinates {(10,5)(6,4) (2,5) (-2,4)  } -- (-2,4)  -- (0,0) -- plot [smooth] coordinates { (0,0) (4,1) (8,0) (12,1)} ;
\draw [ultra thick,etacol2,dashed] (5,2.5) ellipse[x radius = 4, y radius=1, rotate=0] ellipse[x radius = 2, y radius=0.5, rotate=0];
\draw[ultra thick,olddarkgreen] (1,7) circle[radius=1];
\draw[ultra thick,darkred] (1,8.5) circle[radius=0.5];
\draw[-latex,line width=3] (1,5.5) -- (1,2.5);
\draw[ultra thick,olddarkgreen] (2.5,9) circle[radius=1];
\draw[ultra thick,darkred] (2.5,10) circle[radius=0.5];
\draw[-latex,line width=3] (2.5,7.5) -- (2.5,2.7);
\draw[ultra thick,olddarkgreen] (5,7.8) circle[radius=1];
\draw[ultra thick,darkred] (5,8) circle[radius=0.5];
\draw[-latex,line width=3] (5,6.2) -- (5,2.3);
\draw[ultra thick,olddarkgreen] (7,6) circle[radius=1];
\draw[ultra thick,darkred] (7,6.5) circle[radius=0.5];
\draw[-latex,line width=3] (7,4.5) -- (7,2.5);
\draw[ultra thick,olddarkgreen] (8.9,8.7) circle[radius=1];
\draw[ultra thick,darkred] (8.8,7) circle[radius=0.5];
\draw[-latex,line width=3] (8.8,6) -- (8.8,4);
\node[ultra thick,darkred] at (-1.3,8.6) {\Huge $q_2^2 = m_2^2$};
\node[ultra thick,olddarkgreen,left] at (0,6.3) {\Huge  $q_1^2 = m_1^2$};
\draw[black,fill=black] (1,8) circle(0.1);
\draw[black,fill=black] (2.05,9.9) circle(0.1);
\draw[black,fill=black] (2.95,9.9) circle(0.1);
\draw[black,fill=black] (7,7) circle(0.1);
\node[black] at (6,10) {\Huge $\S(G)$};
\draw[-latex,ultra thick, black] (5.7,9.5) to[out=220,in=-30] (3,10);
\draw[-latex,ultra thick, black] (5,10) to[out=150,in=120] (2,10);
\draw[-latex,ultra thick, black] (6,9.5) to[out=-30,in=90] (7,7.3);
\node[etacol,ultra thick,scale=1] at (-4,3.5) {\Huge $\S(G_0)$};
\draw[-latex,etacol,line width=2] (-2.8,3.6) to[out=30,in=120] (-1,3.5);
\node[ultra thick,etacol2] at (1,1) {\Huge $\LL_G(p)$};
\draw[-latex,ultra thick,etacol2] (1,1.7) -- (1.2,2.0);
\draw[-latex,ultra thick,etacol2] (1.9,1.45) -- (3.17,2.18);
\draw[ultra thick, darkred] (17,5) circle[radius=2];
\draw[ultra thick, olddarkgreen] (17,7) circle[radius=1];
\draw[black,fill=black] (16.05,6.8) circle(0.1);
\draw[black,fill=black] (17.95,6.8) circle(0.1);
\draw[ultra thick, black,dotted, postaction=decorate] (17,6.5) .. controls (14,9.5) and (14,5) .. (17,6.5) .. controls (20,9) and (20,5) .. (17,6.5);
\node[black] at (15,8) {\Huge $h$};
\end{tikzpicture}%
}
    \caption{We illustrate  the relation that can hold between pairs of on-shell equations $q_e^2=m_e^2$, each of which we depict as a circle. The intersection of all the on-shell equations is the on-shell locus $\S(G)$ for the graph $G$, which lives in the embedding space ${\mathcal E}(G)$ where only momentum-conservation equations are imposed. When the circles become tangent, the homology group of the space ${\mathcal E}(G)\setminus \S(G)$ shrinks. The vanishing cycle is labeled $h$ in the inset on the right: as the points coincide $h$ becomes trivial.
    The Landau variety $\LL_G(p)$ is the projection of these special tangent points in the on-shell space $S(G_0)$ of the external momenta.
    }
\label{fig:fibration}
\end{figure}
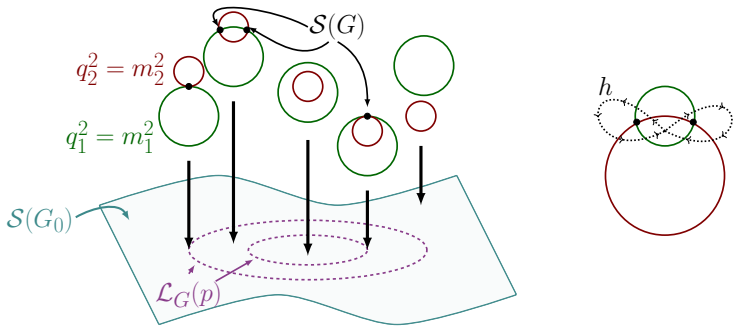

At a point of intersection where the normals to two surfaces are proportional, the two surfaces are tangent. To appreciate this, consider again the on-shell conditions $q_e^2 = m_e^2$. Each such condition constitutes a single constraint on a $d$-dimensional vector. For instance, in the Euclidean $d=2$ case, each such surface is a circle. Not all of the edge momenta are independent, however, due to momentum conservation. 
Thus, when multiple on-shell conditions are imposed, the circles can intersect either at two points, at one point (in which case they are tangent) or at no points. As the momenta are varied, these circles move around relative to each other. At some exceptional points, the circles are tangent and the Landau equations are satisfied. 
This is shown schematically in Fig.~\ref{fig:fibration}. We can also characterize these points of tangency by noting that they correspond to points where the homology group (of the complexification of the complement of the on-shell space) shrinks: there is a vanishing homology cycle.
This geometric understanding of pinch conditions will be essential for deriving the constraints we put on sequential discontinuities in the later sections of this paper. We will explain the vanishing cycle picture in more detail in Sec.~\ref{sec:PicardLefschetz}.

\subsection{Branches of the Landau Variety} \label{sec:branches}

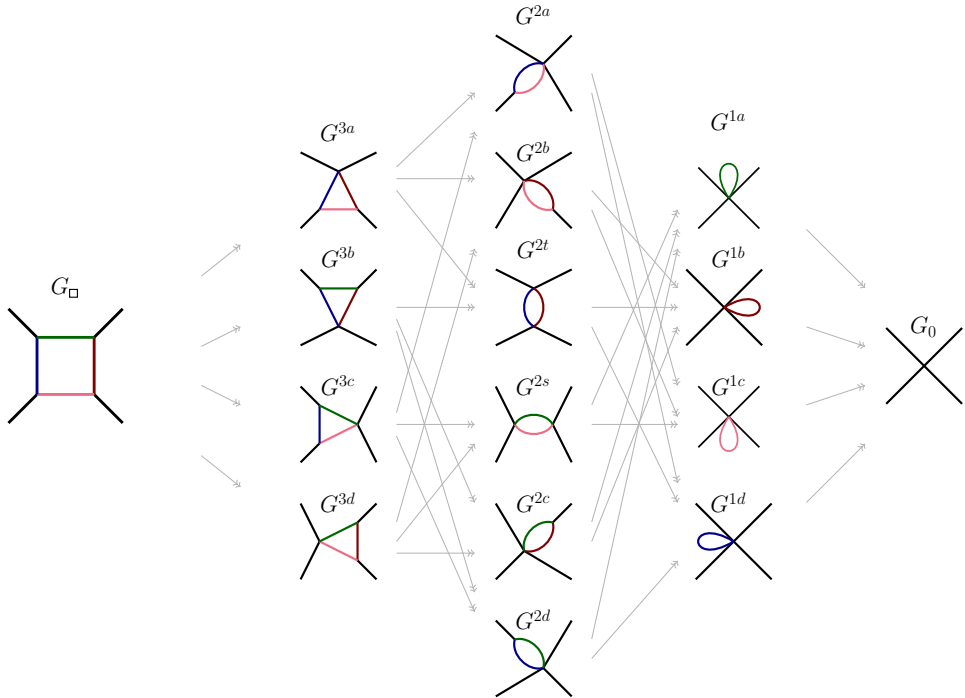
\begin{figure}[t!]
\vspace{30pt}
\centering
\resizebox{13cm}{!}{
\begin{tikzpicture}
 \node[black!100] at (-2,2) {\LARGE $G_\squaret$};
 \node[black!100] at (5,6) {\LARGE $G^{3a}$};
 \node[black!100] at (5,2.8) {\LARGE $G^{3b}$};
 \node[black!100] at (5,-0.5) {\LARGE $G^{3c}$};
 \node[black!100] at (5,-3.5) {\LARGE $G^{3d}$};
 \node[black!100] at (10,9) {\LARGE $G^{2a}$};
 \node[black!100] at (10,5.5) {\LARGE $G^{2b}$};
 \node[black!100] at (10,3) {\LARGE $G^{2t}$};
 \node[black!100] at (10,-0.5) {\LARGE $G^{2s}$};
 \node[black!100] at (10,-3.5) {\LARGE $G^{2c}$};
 \node[black!100] at (10,-6.5) {\LARGE $G^{2d}$};
 \node[black!100] at (15,6.3) {\LARGE $G^{1a}$};
 \node[black!100] at (15,2.8) {\LARGE $G^{1b}$};
 \node[black!100] at (15,-0.5) {\LARGE $G^{1c}$};
 \node[black!100] at (15,-3.5) {\LARGE $G^{1d}$};
\node[black!100] at (20,1) {\LARGE $G_0$};
 \draw[->>,black!30] (1.5,2.3) -- (2.5,3.1);
 \draw[->>,black!30] (1.5,0.5) -- (2.5,1);
 \draw[->>,black!30] (1.5,-0.5) -- (2.5,-1);
 \draw[->>,black!30] (1.5,-2.3) -- (2.5,-3.1);
 \draw[->>,black!30] (6.5,4.5) -- (8.5,2);
 \draw[->>,black!30] (6.5,4.8) -- (8.5,4.8);
 \draw[->>,black!30] (6.5,5.1) -- (8.5,7);
 \draw[->>,black!30] (6.5,1.5) -- (8.5,1.5);
 \draw[->>,black!30] (6.5,1.2) -- (8.5,-3.5);
 \draw[->>,black!30] (6.5,0.9) -- (8.5,-5.8);
 \draw[->>,black!30] (6.5,-1.2) -- (8.5,6);
 \draw[->>,black!30] (6.5,-1.5) -- (8.5,-1.5);
 \draw[->>,black!30] (6.5,-1.8) -- (8.5,-6.3);
 \draw[->>,black!30] (6.5,-4.8) -- (8.5,-4.8);
 \draw[->>,black!30] (6.5,-4.5) -- (8.5,-2);
 \draw[->>,black!30] (6.5,-4) -- (8.5,3);
  \draw[->>,black!30] (11.5,7.5) -- (13.7,-0.5);
  \draw[->>,black!30] (11.5,7) -- (13.7,-3);
  \draw[->>,black!30] (11.5,4.5) -- (13.7,2);
  \draw[->>,black!30] (11.5,4) -- (13.7,-1.3);
  \draw[->>,black!30] (11.5,1.5) -- (13.7,1.5);
  \draw[->>,black!30] (11.5,1) -- (13.7,-3.5);
  \draw[->>,black!30] (11.5,-1) -- (13.7,4);
  \draw[->>,black!30] (11.5,-1.5) -- (13.7,-1.5);
  \draw[->>,black!30] (11.5,-4) -- (13.7,3.5);
  \draw[->>,black!30] (11.5,-4.5) -- (13.7,1);
  \draw[->>,black!30] (11.5,-7) -- (13.7,3);
  \draw[->>,black!30] (11.5,-7.5) -- (13.7,-5);
 \draw[->>,black!30] (17,3.5) -- (18.5,2);
 \draw[->>,black!30] (17,1) -- (18.5,0.5);
 \draw[->>,black!30] (17,-1) -- (18.5,-0.5);
 \draw[->>,black!30] (17,-3.5) -- (18.5,-2);
\node at (-2,0) {
\resizebox{!}{3cm}{
\begin{tikzpicture}[line width=5.7]
    \draw[black] (-4,-4)   -- (-2,-2);
    \draw[black] (-4,4) -- (-2,2);
    \draw[black] (4,4) -- (2,2);
    \draw[black] (4,-4) -- (2,-2);
    \draw[darkred] (2,2) -- (2,-2);
    \draw[darkorange] (2,-2) -- (-2,-2);
    \draw[newdarkblue2] (-2,-2) -- (-2,2);
    \draw[olddarkgreen] (-2,2) -- (2,2);
\end{tikzpicture}}};
\node at (5,4.5) {
\resizebox{!}{2cm}{
\begin{tikzpicture}[line width=6.5]
    \draw[black] (-4,-4)   -- (-2,-2);
    \draw[black] (-4,4) -- (0,2);
    \draw[black] (4,4) -- (0,2);
    \draw[black] (4,-4) -- (2,-2);
    \draw[darkred] (0,2) -- (2,-2);
    \draw[darkorange] (2,-2) -- (-2,-2);
    \draw[newdarkblue2] (-2,-2) -- (0,2);
\end{tikzpicture}}};
\node at (5,1.5) {
\resizebox{!}{2cm}{
\begin{tikzpicture}[line width=6.5]
    \draw[black] (-4,-4)   -- (0,-2);
    \draw[black] (-4,4) -- (-2,2);
    \draw[black] (4,4) -- (2,2);
    \draw[black] (4,-4) -- (0,-2);
    \draw[darkred] (2,2) -- (0,-2);
    \draw[newdarkblue2] (0,-2) -- (-2,2);
    \draw[olddarkgreen] (-2,2) -- (2,2);
\end{tikzpicture}}};
\node at (5,-1.5) {
\resizebox{!}{2cm}{
\begin{tikzpicture}[line width=6.5]
    \draw[black] (-4,-4)   -- (-2,-2);
    \draw[black] (-4,4) -- (-2,2);
    \draw[black] (4,4) -- (2,0);
    \draw[black] (4,-4) -- (2,0);
    \draw[darkorange] (2,0) -- (-2,-2);
    \draw[newdarkblue2] (-2,-2) -- (-2,2);
    \draw[olddarkgreen] (-2,2) -- (2,0);
\end{tikzpicture}}};
\node at (5,-4.5) {
\resizebox{!}{2cm}{
\begin{tikzpicture}[line width=6.5]
    \draw[black] (-4,-4)   -- (-2,0);
    \draw[black] (-4,4) -- (-2,0);
    \draw[black] (4,4) -- (2,2);
    \draw[black] (4,-4) -- (2,-2);
    \draw[darkred] (2,2) -- (2,-2);
    \draw[darkorange] (2,-2) -- (-2,0);
    \draw[olddarkgreen] (-2,0) -- (2,2);
\end{tikzpicture}}};
\node at (10,7.5) {
\resizebox{!}{2cm}{
\begin{tikzpicture}[line width=6.5]
    \draw[black] (-4,-4)   -- (-2,-2);
    \draw[black] (-4,4) -- (1,1);
    \draw[black] (4,4) -- (1,1);
    \draw[black] (4,-4) -- (1,1);
    \draw[darkorange] (-2,-2) to[out=45-60,in=45-120] (1,1);
    \draw[newdarkblue2] (-2,-2) to[out=45+60,in=45+120] (1,1);
\end{tikzpicture}}};
\node at (10,4.5) {
\resizebox{!}{2cm}{
\begin{tikzpicture}[line width=6.5]
    \draw[black] (-4,-4) -- (-1,1);
    \draw[black] (-4,4) -- (-1,1);
    \draw[black] (4,4) -- (-1,1);
    \draw[black] (4,-4) -- (2,-2);
    \draw[darkred] (2,-2) to[out=45-60+90,in=45-120+90] (-1,1);
    \draw[darkorange] (2,-2)  to[out=45+60+90,in=45+120+90] (-1,1);
\end{tikzpicture}}};
\node at (10,1.5) {
\resizebox{!}{2cm}{
\begin{tikzpicture}[line width=6.5]
    \draw[black] (-4,-4)   -- (0,-2);
    \draw[black] (-4,4) -- (0,2);
    \draw[black] (4,4) -- (0,2);
    \draw[black] (4,-4) -- (0,-2);
     \draw[darkred] (0,2) to[out=60-90,in=120-90] (0,-2);
     \draw[newdarkblue2] (0,2) to[out=-60-90,in=-120-90] (0,-2);
\end{tikzpicture}}};
\node at (10,-1.5) {
\resizebox{!}{2cm}{
\begin{tikzpicture}[line width=6.5]
    \draw[black] (-4,-4)   -- (-2,0);
    \draw[black] (-4,4) -- (-2,0);
    \draw[black] (4,4) -- (2,0);
    \draw[black] (4,-4) -- (2,0);
    \draw[darkorange] (-2,0) to[out=-60,in=-120] (2,0);
    \draw[olddarkgreen] (-2,0) to[out=60,in=120] (2,0);
\end{tikzpicture}}};
\node at (10,-4.5) {
\resizebox{!}{2cm}{
\begin{tikzpicture}[line width=6.5]
    \draw[black] (-4,-4)   -- (-1,-1);
    \draw[black] (-4,4) -- (-1,-1);
    \draw[black] (4,4) -- (2,2);
    \draw[black] (4,-4) -- (-1,-1);
    \draw[darkred] (2,2)  to[out=45+60+180,in=45+120+180](-1,-1);
    \draw[olddarkgreen] (2,2) to[out=45-60+180,in=45-120+180] (-1,-1);
\end{tikzpicture}}};
\node at (10,-7.5) {
\resizebox{!}{2cm}{
\begin{tikzpicture}[line width=6.5]
    \draw[black] (-4,-4)   -- (1,-1);
    \draw[black] (-4,4) -- (-2,2);
    \draw[black] (4,4) -- (1,-1);
    \draw[black] (4,-4) -- (1,-1);
    \draw[newdarkblue2] (-2,2) to[out=45-60-90,in=45-120-90] (1,-1);
    \draw[olddarkgreen] (-2,2) to[out=45+60-90,in=45+120-90](1,-1);
\end{tikzpicture}}};
\node at (15,4.5) {
\resizebox{!}{2cm}{
\begin{tikzpicture}[line width=6.5]
    \draw[black] (-4,-4)   -- (0,0);
    \draw[black] (-4,4) -- (0,0);
    \draw[black] (4,4) -- (0,0);
    \draw[black] (4,-4) -- (0,0);
    \draw[olddarkgreen] (0,0).. controls (-4,6) and (4,6) .. (0,0);
\end{tikzpicture}}};
\node at (15,1.5) {
\resizebox{!}{2cm}{
\begin{tikzpicture}[line width=6.5]
    \draw[black] (-4,-4)   -- (0,0);
    \draw[black] (-4,4) -- (0,0);
    \draw[black] (4,4) -- (0,0);
    \draw[black] (4,-4) -- (0,0);
    \draw[darkred] (0,0).. controls (5,-3) and (5,3) .. (0,0);
\end{tikzpicture}}};
\node at (15,-1.5) {
\resizebox{!}{2cm}{
\begin{tikzpicture}[line width=6.5]
    \draw[black] (-4,-4)   -- (0,0);
    \draw[black] (-4,4) -- (0,0);
    \draw[black] (4,4) -- (0,0);
    \draw[black] (4,-4) -- (0,0);
    \draw[darkorange] (0,0).. controls (-4,-6) and (4,-6) .. (0,0);
\end{tikzpicture}}};
\node at (15,-4.5) {
\resizebox{!}{2cm}{
\begin{tikzpicture}[line width=6.5]
    \draw[black] (-4,-4)   -- (0,0);
    \draw[black] (-4,4) -- (0,0);
    \draw[black] (4,4) -- (0,0);
    \draw[black] (4,-4) -- (0,0);
    \draw[newdarkblue2] (0,0).. controls (-5,-3) and (-5,3) .. (0,0);
\end{tikzpicture}}};
\node at (20,0) {
\resizebox{!}{2cm}{
\begin{tikzpicture}[line width=6.5]
    \draw[black] (-4,-4)   -- (0,0);
    \draw[black] (-4,4) -- (0,0);
    \draw[black] (4,4) -- (0,0);
    \draw[black] (4,-4) -- (0,0);
\end{tikzpicture}}};
\end{tikzpicture}
}
\vspace{-10pt}
\caption{All the possible contractions of the box diagram. Each contraction gives a Landau diagram which represents a possible singularity of the box integral.
\label{fig:boxcontractions}}
\end{figure}

In the later sections of this paper, we will be interested in distinguishing between singularities that appear on different branches of the Landau variety $\ell=0$, where these branches are differentiated by having different propagators on-shell. To characterize these branches, we introduce the notion of a \textbf{graph contraction} $\kappa$, which is a map between graphs
\begin{equation} 
\label{eq:G2_to_G1_contraction}
\begin{tikzcd}
\kappa \colon G_1 \arrow[r, twoheadrightarrow, ""] & G_2  \, ,
  \end{tikzcd}
\end{equation}
where the graph $G_2$ is formed by contracting a subset of the edges in $G_1$. (Contracting an edge here means replacing the vertices that were previously connected by this edge by a single vertex and deleting the contracted edge from the graph.) 
As an example, the possible graph contractions of the box diagram are shown in Fig.~\ref{fig:boxcontractions}. Any composition of contractions also constitutes a valid contraction.

Every contraction of the graph $G$ can be embedded in a sequence of contractions to the elementary graph $G_0$ in which all internal lines have been contracted, for instance as
\begin{equation}
 G\xrightarrowdbl{~\kappas~} G^\kappa \xrightarrowdbl{~\kappa~} G_0 \, ,
\end{equation}
where $\kappa$ contracts all of the internal lines that are not contracted by $\kappas$. In our notation, we will often put a bar over a contraction when its image is not the elementary graph, as done here. 
We can then decompose the Landau variety $\LL_G$ into a number of different branches $\P_\kappa$, associated with different possible contractions. Namely, we can write it as
\begin{equation}
 \label{LV2}
    \LL_G = \bigcup_{ G^\kappa \subset G} \P_\kappa \, ,
\end{equation}
where $\P_\kappa$ is defined as 
\begin{equation}
\label{LGKdef}
     \P_\kappa  : \qquad  p \in \S(G_0) \, \Big|
    \bigcap_{c \in \Chat(G^\kappa)} 
\!\!\Big\{ \sum_{e \in \Eint(G^\kappa)} b_{c e} \, \alpha_e q_e^\mu =0 \Big\}
    \bigcap_{e\in \Eint(G^\kappa)} \!\!
\Big\{ q_e^2 = m_e^2\Big\} ,
\end{equation}
where we assume that at least one variable $\alpha_e$ is nonzero.
In Ref.~\cite{pham}, Pham calls $\P_\kappa$ the ``lieu de Landau'', which translates to Landau locus. However, to disambiguate it both from the Landau variety, and from one of its subspaces in which a specified set of Feynman parameters vanish, we call it the \textbf{Pham locus}. Let us emphasize that the Pham locus $\P_\kappa$ classifies different branches of the Landau variety by which lines of the original Feynman diagram are put \textit{on shell}, regardless of whether the corresponding variables $\alpha_e$ are zero or not.
An important result of Pham's is that  $\P_\kappa$ corresponds to the set of critical points of the projection map among on-shell surfaces $\pi: \S(G^\kappa) \to \S(G_0)$~\cite{pham}. We will return to this perspective in Sec.~\ref{sec:critical}. 

It is sometimes helpful to refer to the branches of the Landau variety using \textbf{Landau diagrams}.
A Landau diagram can be associated with any graph, but unlike Feynman diagrams, all propagators in Landau diagrams are understood to be on shell. Thus, each of the branches $\P_\kappa$ of $\LL_G$ can be represented by the Landau diagram associated with the graph $G^\kappa$.
Note that in some of the $S$-matrix literature, a Landau diagram $G^\kappa$ is instead associated with the part of the Landau variety in which all of the Feynman parameters in $G^\kappa$ are nonzero, while the remaining Feynman parameters in $G$ vanish.  Indeed, solutions to the $\alpha_e (q_e^2-m_e^2) =0$ condition in Eq.~\eqref{LV} are often discussed as having an $\alpha_e= 0$ branch and a $q_e^2 = m_e^2$ branch, as if these choices were mutually exclusive. However, to understand the singularity structure of Feynman integrals, it is also important to consider the case where both $\alpha_e = 0$ and $q_e^2 = m_e^2$ are satisfied simultaneously.
Our definition of $\P_\kappa$ in Eq.~\eqref{LGKdef} (which matches the definition adopted by Pham~\cite{pham}) allows the Feynman parameters in $G^\kappa$ to vanish.
To connect to previous literature, we will use the term \textbf{leading singularity} to describe the part of the Pham locus where $\alpha_e \ne 0$ for all edges in a given Landau diagram. For example, when Cutkosky refers to a Landau diagram, he means what we call the leading singularity~\cite{Cutkosky:1960sp}. We remark, however, that in some of the more recent literature, the leading singularity is instead defined as the one at which all lines in the Landau diagram are on shell, regardless of whether the $\alpha_e$ are nonzero~\cite{Prlina:2018ukf}. In this work, we refer to that as $\P_\kappa$, where $\kappa$ is the contraction from the original diagram $G$ to $G_0$.

An important property of Pham loci is that they are not ordered. That is, if we have a sequence of graphs $G \twoheadrightarrow  G^\kappap \twoheadrightarrow G^\kappa \twoheadrightarrow  G_0$ such that $G^\kappa$ is a contraction of $G^\kappap$, then it does not automatically follow that $\P_\kappa \subset \P_\kappap$. This is due to the fact that, while the on-shell conditions in $\P_\kappap$ include all of those in $\P_\kappa$, the loop equations in $\P_\kappap$ involve more Feynman parameters than the loop equations in $\P_\kappa$. For example, when $G^\kappap$ is the triangle diagram in three dimensions and $G^\kappa$ is a bubble diagram arising from contracting a line in $G^\kappap$, then $\P_\kappap$ and $\P_\kappa$ are both two-dimensional surfaces in the three-dimensional space $\S(G_0)$ of external kinematics associated with the triangle diagram. (We will discuss this example at more length in Sec.~\ref{sec:iterated}, and the intersection of this pair of Pham loci is illustrated in Figure~\ref{fig:triangletangent}.)

We highlight that our definitions of the Landau and Pham loci do not require all Feynman parameters to be real and non-negative, as 
is true in the integration contour in Eq.~\eqref{eq:landauform}.
Nonetheless, whether or not a point in these varieties involves only positive Feynman parameters is important. In particular, solutions to the Landau equations that require some Feynman parameters to be either negative or complex do not correspond to singularities of $\I_G(p)$ that can be accessed with real on-shell external momenta. We thus define the
\textbf{$\boldsymbol{\alpha}$-positive} Landau variety (Pham locus) to be the subspace of the Landau variety (Pham locus) in which all $\alpha_e \ge 0$.\footnote{Note that $\alpha$-positive really means all Feynman parameters are real and non-negative, so the terminology is slightly imprecise.}

The singularities of $\I_G(p)$ that correspond to points in the $\alpha$-positive Landau variety arise on the physical sheet. To make this precise, we first define the \textbf{physical region} to be the subspace of the space of external momenta that are real and on shell, with positive energy for some choice of which momenta are incoming and outgoing. 
Note that this definition of the physical region does not refer to the analytic structure of any specific function; thus, when $\I_G(p)$ is multivalued, one can be in the physical region on different Riemann sheets. 
We then define the \textbf{physical sheet} for a given Feynman integral $\I_G(p)$ to consist of all the complex points in $\mathcal{S}(G_0)$ that are accessible by analytic continuation from the physical region, other than those on the branch cuts of $\I_G(p)$.\footnote{Since all of the kinematic dependence of $\I_G(p)$ is encoded in the integrand of Eq.~\eqref{eq:Feynman_integral_def}, the branch cuts of these integrals are uniquely specified by where the integrand is singular on the integration contour (for $\eps=0$). One can write down different integral representations in which these branch cuts are deformed, as long as the physical values of $\I_G(p)$ remain intact~\cite{Binoth:2000ps,Binoth:2003ak,Hannesdottir:2022bmo}. However, such rotations can change the sheet structure of these integrals, and thereby also the analysis of which singularities appear on the physical sheet.}
It follows that $\I_G(p)$ is single-valued when restricted to the physical sheet.

Singularities that occur in connection with Feynman parameters that take negative or complex values are not on the physical sheet. However, $\I_G(p)$ can develop a singularity at any Pham locus (whether $\alpha$-positive or not) when it is analytically continued to sheets other than the physical one. 
While the parts of the Landau variety that are not $\alpha$-positive are important for understanding the full analytic structure of these integrals, we in large part focus in this paper on $\alpha$-positive singularities. In particular, we will see that many relatively simple theorems can be proved for singularities in the \textbf{principal Pham locus}, which is defined to be the union of all $\alpha$-positive branches in $\LL_G$ of codimension one that have at least one non-vanishing $\alpha$ in each loop. We will introduce the properties of principal Pham loci in more detail in Sec.~\ref{sec:principal}.

\subsection{Absorption Integrals} \label{sec:absorption_integrals}

At the points where $\I_G(p)$ becomes singular, it can develop branch cuts and therefore discontinuities. As first shown by Cutkosky~\cite{Cutkosky:1960sp}, when the masses of $\I_G(p)$ are generic and the Pham locus $\P_{\kappa}$ is principal, the discontinuity of $\I_G(p)$ across the branch cut
associated with the Landau diagram $G^\kappa$
is given by the \textbf{absorption integral}
\begin{equation}
     \cA_{G}^{\kappa} (p) = \int \prod_{c \in \Chat(G)} \rd^d k_c
          \prod_{e \in \Eint(G^\kappa) }  (-2\pi i)\; \theta_\ast(q_e^0)\delta(q_{e}^2 -m_{e}^2)
          \prod_{e^\prime \in E(G) \backslash E(G^\kappa)} \frac{1}{q_{e^\prime}^2-m_{e^\prime}^2 + i\varepsilon}
     \label{eq:absorption_int_2} \, .
\end{equation}
For a particular edge $e$, the surface $q_e^2=m_e^2$ has two branches, given by $q_e^0 = \pm \sqrt{\smash[b]{\vec{q_e}^2+m_e^2}}$, which never intersect for $m_e>0$. 
However, the internal energy flow through each of the edges in $\Eint(G^\kappa)$, including signs, is fixed by the solution to the Landau equations under consideration.
The factors of $\theta_\ast(q_e^0)$ in Eq.~\eqref{eq:absorption_int_2} are defined to evaluate to either $\theta(q_e^0)$ or $\theta(-q_e^0)$, depending on the sign of the energy flowing through the edge $e$.\footnote{
The real points of a given Landau singularity can have multiple parts which are separated by intersections with other Landau singularities.  Therefore, one cannot smoothly connect two such parts without crossing through a non-smooth point.  Simple paths going around different parts of a Landau singularity yield different elements in the first homotopy group of the complement of the Pham loci (see Fig.~\ref{fig:co1homotopy}).  These different regions correspond to different signs for the energies of the internal propagators.  Two regions which differ only by the sign of the energy of an internal propagator are separated by an intersection with a Landau singularity where this propagator is contracted.  This ensures that the signs of the energies can be assigned consistently.}
This factor enforces that the same branch of $q_e^0 =\pm\sqrt{\smash[b]{\vec{q_e}^2+m_e^2}}$ is maintained throughout the absorption integral. In Sec.~\ref{sec:deform_contours} we will discuss how Cutkosky's formula does not actually hold for the discontinuity of Feynman integrals with respect to any Pham locus, but only applies to principal loci.

When studying the singularities arising from a graph, we need to pay close attention to the signs of the energies of internal particles.  As we will discuss in more detail, the signs of energies of the internal particles  determine the region where a singularity occurs and, for massive particles, regions which correspond to different signs for the energies are disconnected.  When computing the discontinuity by using Cutkosky's formula, we need to find the signs of the energies of internal particles which are compatible with the location of the singularity we are encircling.  One can also find singularities where the signs of the energies are such that the corresponding arrows form closed oriented loops, as in the case of the pseudo-threshold singularity of the bubble integral.  This process cannot be realized as a physical scattering in Lorentzian signature space-time, so its corresponding singularity does not appear in the physical region; it does however appear on the next sheet and can in principle be understood by studying the singularities of the discontinuity of the bubble integral around the normal threshold.

The discontinuities of Feynman integrals can also be computed by analytic continuation in the space of external kinematics. In particular, the discontinuity computed by the absorption integral $\cA_{G}^{\kappa} (p)$ corresponds to the difference between the original expression, and the one obtained by computing the monodromy of $\I_G(p)$ around the variety $\P_\kappa$:
\begin{equation}
        \cA_G^{\kappa} (q) = \Big(\bbone - 
   \monM^{\P_\kappa}_{\circlearrowleft} \Big) \I_G(p) 
   \,.
   \label{Mdef}
\end{equation}
where $\smash{\monM^{\P_\kappa}_{\circlearrowleft}}$ is an operator that maps $\I_G(p)$ to its value after being analytically continued around an infinitesimal counterclockwise circle centered on the branch point $\P_\kappa$. We will often abbreviate $\monM^{\P}_{\circlearrowleft}$ by $\monM_{\P}$. For more details on this notation and how monodromies act on Feynman integrals, we refer the reader to~\cite{Bourjaily:2020wvq}.

Importantly, the discontinuity computed in~\eqref{Mdef} is not the same as the \textbf{total discontinuity} of $\I_G(p)$, which is defined
to be the difference between the integral with a $+i \eps$ prescription and a $-i \eps$ prescription. The value of the total discontinuity of $\I_G(p)$ depends on the kinematic region in which it is computed; in particular, in a region where only the edges in $G^\kappa$ can all be put on-shell, the discontinuity would be given by a sum over all possible cuts through the graph, with the amplitude on the right side of the cut conjugated (in particular, with $+i \epsilon$ on propagators to the left of each
cut and $-i \epsilon$ on propagators to the right of each cut). The total discontinuity is equal to the
monodromy of $\I_G(p)$ along a finite contour from this region to the Euclidean region (where
no edges can go on-shell) and back to the original region. Such a monodromy in general will
be the sum of many different absorption integrals which correspond to different subsets of the edges that can go
on shell in the physical region. In particular, individual absorption integrals in Eq.~\eqref{eq:absorption_int_2} can have both real and imaginary parts, and need not correspond to physical absorption. See~\cite{Hannesdottir:2022bmo} for an explicit formula relating the imaginary part of $I_G(p)$ to a sum of absorption integrals from Eq.~\eqref{eq:absorption_int_2}, and conditions under which this imaginary part is equal to the total discontinuity of the amplitude in the case of non-generic masses. See also~\cite{Bourjaily:2020wvq} for an extensive discussion of the relationship between
monodromies and total discontinuities, and on the importance of the choice of analytic continuation contour.

  \section{Discontinuities of Feynman Integrals}
\label{sec:PicardLefschetz}

In Sec.~\ref{sec:review}, we reviewed Landau's equations and Cutkosky's formula for the discontinuities of Feynman integrals.
In this section we explore how Cutkosky's formula can be derived using Picard--Lefschetz theory~\cite{pham, Bloch:2015efx,Muhlbauer:2022ylo}. This will prepare us for subsequent sections, in which we will show how the same techniques can be used to put constraints on the sequential discontinuities of Feynman integrals. 

In the Picard--Lefschetz approach, we start from an integral $I(p) = \int_h \omega(p)$ of a closed differential form $\omega(p)$ over a closed integration contour $h$ (the case where the path $h$ has boundaries can also be described using relative homology).  
Notably, this integral depends only on the homology class $[h]$ of $h$, and not on the further details of this contour. 
As we analytically continue the variables $p$ to compute one of the monodromies of $I(p)$, the singularities of $\omega(p)$ may collide with the contour $h$. To avoid this, the contour must be deformed in conjunction with our analytic continuation, leading to a new contour $h'$. The Picard--Lefschetz formula describes how the corresponding discontinuity of $I(p)$ can be written as an integral over the difference between these contours $\operatorname{Var}[h]=[h']-[h]$. In particular, when applied locally, Picard--Lefschetz gives us an explicit way to compute $\operatorname{Var}[h]$ that eliminates the need to explicitly trace what happens to the integration contour during the analytic continuation. 
The remaining integral over $\operatorname{Var}[h]$ can then be performed using Leray's higher-dimensional analogue of Cauchy's residue formula~\cite{BSMF_1959__87__81_0}.

\subsection{Monodromies in Feynman-Parameter Space}
\label{sec:alphaspace}

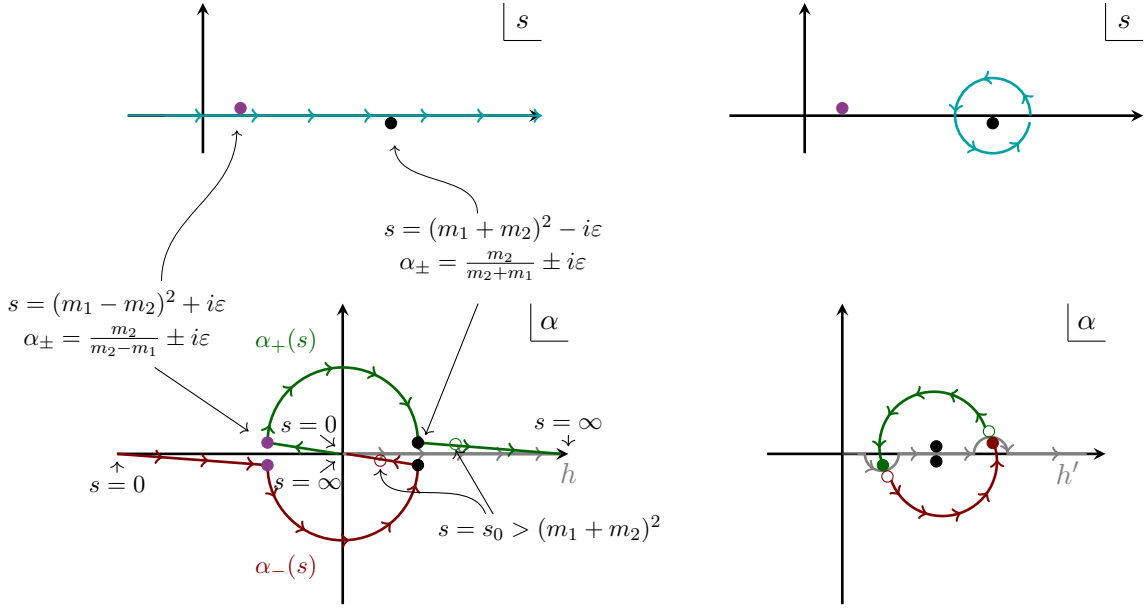
\begin{figure}[t]
 \centering
 \begin{tikzpicture}
\node at (0,0) {
\begin{tikzpicture}[scale=0.5]
\draw[-stealth,line width=1] (-4,0) -- (7,0) coordinate (xaxis);
\draw[-stealth,line width=1] (-2,-1) -- (-2,3) coordinate (yaxis);
\draw[line width=0.7,darkcyan] (-4,0) --(7,0);
\draw[line width=0.7,darkcyan,
 postaction={decorate,decoration={markings, mark=between positions 0.1 and 1 step 0.15 with {\arrow[line width=1]{to}}}}
] (-3,0) --(7,0);
\filldraw[black] (3,-0.2) circle (0.15); 
\filldraw[etacol2] (-1,0.2) circle (0.15); 
\draw [draw=black] (6,2) -- (6,3);
\draw [draw=black] (6,2) -- (7,2);
\node [black,scale=1] at  (6.5,2.5) {$s$};
\end{tikzpicture}
};
\node at (8,0) {
\begin{tikzpicture}[scale=0.5]
\draw[-stealth,line width=1] (-4,0) -- (7,0) coordinate (xaxis);
\draw[-stealth,line width=1] (-2,-1) -- (-2,3) coordinate (yaxis);
\draw[line width=1,darkcyan,postaction={decorate,decoration={markings, mark=between positions 0.1 and 1 step 0.2 with {\arrow[line width=1]{to}}}}
] (4,0)  arc[start angle=0, delta angle=350,radius=1cm];
\filldraw[black] (3,-0.2) circle (0.15); 
\filldraw[etacol2] (-1,0.2) circle (0.15); 
\draw [draw=black] (6,2) -- (6,3);
\draw [draw=black] (6,2) -- (7,2);
\node [black,scale=1] at  (6.5,2.5) {$s$};
\end{tikzpicture}
};
\node at (0,-5) {
\begin{tikzpicture}[scale=0.5]
\draw[white,line width=1] (-8,-7) -- (-8,7);
\draw[-stealth,line width=1] (-6,0) -- (6,0) coordinate (xaxis);
\draw[-stealth,line width=1] (0,-4) -- (0,4) coordinate (yaxis);
\draw [draw=black] (5,3) -- (5,4);
\draw [draw=black] (5,3) -- (6,3);
\node [black,scale=1] at  (5.5,3.5) {$\alpha$};
\draw[line width=1,black!50,
 postaction={decorate,decoration={markings, mark=between positions 0.2 and 1 step 0.18 with {\arrow[line width=1]{to}}}}
] (0,0) -- (5.8,0);
\draw[line width=1,col4, postaction={decorate,decoration={markings,
 mark=between positions 0.1 and 1 step 0.1 with {\arrow[line width=1]{to}}}}] (-6,0) -- (-2,-0.3) arc[start angle=180, delta angle=180,radius=2cm] -- (0.1,0);
\draw[line width=1,col3, postaction={decorate,decoration={markings,
 mark=between positions 0.1 and 1 step 0.1 with {\arrow[line width=1]{to}}}}] (-0.1,0) -- (-2,0.3) arc[start angle=180, delta angle=-180,radius=2cm]  -- (5.8,0);
\filldraw[black] (2,0.3) circle (0.15); 
\filldraw[black] (2,-0.3) circle (0.15); 
\filldraw[etacol2] (-2,0.3) circle (0.15);
\filldraw[etacol2] (-2,-0.3) circle (0.15);
\draw[darkred] (1,-0.18) circle (0.15); 
\draw[darkgreen] (3,0.26) circle (0.15); 
\node[col4,scale=0.85] at (-1.5,-3) {$\alpha_-(s)$};
\node[col3,scale=0.85] at (-1.5,3) {$\alpha_+(s)$};
\node[black,scale=0.85] at (-6,-0.8) {$s=0$};
\draw[->] (-6,-0.5) -- (-6,-0.2);
\node[black,scale=0.85] at (-0.9,0.8) {$s=0$};
\draw[->] (-0.6,0.5) -- (-0.2,0.2);
\node[black,scale=0.85] at (-0.93,-0.8) {$s=\infty$};
\draw[->] (-0.6,-0.5) -- (-0.2,-0.2);
\node[black,scale=0.85] at (6,0.8) {$s=\infty$};
\draw[->] (6,0.5) -- (6,0.2);
\node[black,scale=0.85] at (-6,4) {$s=(m_1-m_2)^2+i\varepsilon$};
\node[black,scale=0.85] at (-6,3) {$\alpha_\pm=\frac{m_2}{m_2-m_1}\pm i \varepsilon$};
\draw[->] (-5,2.3) -- (-2.5,0.6);
\node[black,scale=0.85] at (4,6) {$s=(m_1+m_2)^2-i\varepsilon$};
\node[black,scale=0.85] at (4,5) {$\alpha_\pm=\frac{m_2}{m_2+m_1} \pm i \varepsilon$};
 \draw[->] (3.6,4) -- (2.2,0.5);
\node[black,scale=0.85] at (5.5,-2) {$s=s_0>(m_1+m_2)^2$};
 \draw[->] (4,-1.6) -- (3,0.1);
 \draw[->] (3.8,-1.6) to[out=110,in=-90] (1,-0.4);
\node[black!50,scale=1] at (6,-0.5) {$h$};
\end{tikzpicture}
};
\draw[->,line width=0.07] (-2.3,-2.7) to[out=90,in=-90] (-1.3,-0.7);
\draw[->,line width=0.07] (1.9,-1.8) to[out=90,in=-90] (0.8,-0.9);
\node at (8,-5) {
\begin{tikzpicture}[scale=0.5]
\draw[-stealth,line width=1] (-2,0) -- (7,0) coordinate (xaxis);
\draw[-stealth,line width=1] (0,-4) -- (0,4) coordinate (yaxis);
\draw [draw=black] (6,3) -- (6,4);
\draw [draw=black] (6,3) -- (7,3);
\node [black,scale=1] at  (6.5,3.5) {$\alpha$};
 \draw[line width=1,black!50,
 postaction={decorate,decoration={markings, mark=between positions 0.1 and 1 step 0.2 with {\arrow[line width=1]{to}}}}
 ] (0.4,0) -- (0.6,0) 
 arc[start angle=180, delta angle=180,radius=0.45cm] 
 -- (3.5,0)
arc[start angle=180, delta angle=-180,radius=0.45cm] --
(6.5,0);
\draw [draw=black!50] (0,0) -- (0.5,0);
\draw[col3, line width=1,
 postaction={decorate,decoration={markings, mark=between positions 0.1 and 1 step 0.2 with {\arrow[line width=1]{to}}}}
 ] (3.8,0.7)
    arc[start angle=20, delta angle=180,radius=1.45cm] ;
\draw[col4, line width=1,
 postaction={decorate,decoration={markings, mark=between positions 0.1 and 1 step 0.2 with {\arrow[line width=1]{to}}}}] (1.3,-0.7)
arc[start angle=200, delta angle=180,radius=1.45cm] ;
\filldraw[col3] (1.08,-0.3) circle (0.15);
\filldraw[col4] (4,0.29) circle (0.15);
\draw[col3] (3.9,0.6) circle (0.15);
\draw[col4] (1.2,-0.6) circle (0.15);
\filldraw[black] (2.5,0.2) circle (0.15); 
\filldraw[black] (2.5,-0.2) circle (0.15); 
\node[black!50,scale=1] at (6,-0.5) {$h'$};
\end{tikzpicture}
};
\end{tikzpicture}
       \caption{ The bubble diagram can be written as an integral over the contour $0<\alpha<1$, shown in the lower left panel in grey. There are poles in the integrand at $\alpha_+(s)$ and $\alpha_-(s)$; these poles move along the green and red curves in the lower left panel as $s$ is moved along the real axis in the upper left panel. At $s=(m_1\pm m_2)^2$ these two curves nearly meet at $\alpha_+=\alpha_- = \frac{m_2}{m_2\pm m_1}$, corresponding to the position of the purple or black dot (they don't quite meet due to their displacement by $\pm i\varepsilon$). Only the singularity at the normal threshold $s=(m_1+m_2)^2$ pinches the integration contour. 
       In the right panes, we see that as $s$ is rotated in the complex plane from a point $s_0 > (m_1+m_2)^2$ around the branch point at $(m_1+m_2)^2$, the corresponding $\alpha_\pm(s)$ also rotate, at half the rate. The grey integration contour $h$ must also be deformed as this analytic continuation is carried out, so that the singularities $\alpha_\pm(s)$ never cross the contour.  The monodromy corresponding to this analytic continuation can be computed as integral over the difference between the integration contour before and after this analytic continuation, and the result is the absorption integral $A(s_0)$.
       \label{fig:bubble_cycle}
}
\end{figure}

Let us begin with an example of an integral over a single complex variable, where one can perform the analytic continuation in the external variables and trace what happens to the integration contour directly. Such an example is given by the two-dimensional bubble integral in Feynman-parametrized form.\footnote{An analogous example was worked out in section 3-1 of Ref.~\cite{Hwa:102287}.} In momentum space, this integral is defined to be
\begin{equation}
    I_{\bubrgT}(p) =
\resizebox{!}{1cm}{
\begin{tikzpicture}[baseline=(current bounding box.center),
    line width=1.5,scale=0.7]
    \draw[black] (-3,0) -- node[midway, above,scale=1.2] {$p$} (-1,0);
    \draw[black,-latex] (-3,0) -- (-2,0);
     \draw[darkred] (1,0) arc (0:180:1);
     \draw[olddarkgreen] (1,0) arc (0:-180:1);
    \draw[black] (1,0) -- node[midway, above,scale=1.2] {$p$} (3,0); 
    \draw[black,-latex] (1,0) -- (2,0);
    \draw[-latex,darkred] (0.1,1) -- (0.2,1)  node[above,scale=1.2] {$p-k,m_2$};
    \draw[-latex,olddarkgreen] (0.1,-1) -- (0.2,-1) node[below,scale=1.2] {$k,m_1$};
\end{tikzpicture}
}
=
    \lim_{\varepsilon \to 0^+} \int \rd^2 k \frac{1}{k^2-m_1^2+i\varepsilon}\frac{1}{ (p-k)^2-m_2^2+i\varepsilon} \,,
    \label{eq:bub2D_mom}
\end{equation}
which in terms of Feynman parameters becomes 
\begin{equation}
    I_{\bubrgT}(s) = \lim_{\varepsilon \to 0^+} \int_0^1 \rd \alpha \frac{- i \pi }{s \alpha (1-\alpha)-m_1^2 \alpha - m_2^2 (1-\alpha) + i\varepsilon} \, ,
    \label{eq:bub_alpha}
    \end{equation}
where we have denoted $s=p^2$.

To find the singularities of the integral in Eq.~\eqref{eq:bub_alpha}, we look for kinematic configurations and values of $\alpha$ for which the denominator and its derivative with respect to $\alpha$ are simultaneously zero. The denominator in this integral vanishes at the two roots of the quadratic equation, 
\begin{equation}
    \alpha_{\pm} = \frac{s+m_2^2-m_1^2 \pm \sqrt{s^2-2 s (m_1^2+m_2^2)+(m_1^2-m_2^2)^2+ i s \varepsilon}}{2 s} \,.
    \label{eq:apm}
\end{equation}
If we further impose that its derivatives vanish, we obtain two solutions:
\begin{align} \label{eq:s_alpha_bubble_soln}
    s & = (m_1 + m_2)^2 - i \varepsilon \,, \qquad \alpha_{\pm} = \frac{m_2}{m_2 + m_1} + i \varepsilon \, \text{sgn}(m_2-m_1) \,, \\
    s & = (m_1 - m_2)^2 + i \varepsilon \,, \qquad \alpha_{\pm} = \frac{m_2}{m_2 - m_1} - i \varepsilon \, \text{sgn}(m_2-m_1) \,.
\end{align}
We would thus like to find the monodromies of $I_{\bubrgT}(s)$ as we encircle $s=(m_1\pm m_2)^2\mp i \varepsilon$ in the space of external momenta, which here is simply the space of complex $s$. We first do this by closely following how the integration contour is deformed by this analytic continuation, and then show in the next subsection how the Picard--Lefschetz formula reproduces this result.

As a first step, we trace what happens to the roots $\alpha_{\pm}$ in $\alpha$ space as $s$ is moved along the real axis. These paths are depicted in Fig.~\ref{fig:bubble_cycle}. When $m_2<m_1$, the two roots collide on the $\alpha<0$ part of the real axis as $s \to (m_1-m_2)^2$ and $\varepsilon \to 0$. 
This point, referred to as the \textit{pseudothreshold}, represents a solution to the Landau equations, but does not lead to a pinch of the integration contour over $0<\alpha<1$. 
Thus, if $s$ is analytically continued in a circle around $(m_1 -m_2)^2$ in the space of external momenta as $\varepsilon\to 0$, the two roots $\alpha_+$ and $\alpha_-$ switch places, but there is never any need to modify the integration contour to avoid a singularity.
We therefore have
\begin{equation}
     \left(\bbone-\monM_{s=(m_1-m_2)^2} \right) I_\bubrgT(s) = 0 \,,
\end{equation}
as the integration contour in $I_\bubrgT(s)$ is the same before and after applying the monodromy operator.
If we instead increase the value of $s$ so it approaches $(m_1+m_2)^2$, the two roots $\alpha_\pm$ get closer and closer to the integration contour that runs between $0<\alpha<1$, until finally they pinch the contour when $s=(m_1+m_2)^2$ and $\varepsilon \to 0$. This singularity is referred to as the \textit{threshold}.

Now, let us compute the monodromy around the threshold singularity by analytically continuing $s$ in a circle around $(m_1+m_2)^2-i\varepsilon$ in the complex $s$ plane, starting at a point $s_0 > (m_1+m_2)^2$. This is shown on the right in Fig.~\ref{fig:bubble_cycle}. As we perform this continuation, the two roots will cross the real $\alpha$ axis, so the integration contour must be deformed along with the roots. Once we arrive back at the point $s_0$, the two roots $\alpha_+$ and $\alpha_-$ have switched places, and the integration contour has been modified. 
The corresponding absorption integral, evaluated at $s_0 > (m_1+m_2)^2$, is then (in the $\varepsilon \to 0$ limit) given by
\begin{equation}
    A_\bubrgT^\bubrgT(s_0)\equiv \left(\bbone-\monM_{s=(m_1+m_2)^2} \right) I_\bubrgT(s_0) = \int_{c} \rd I\, , \label{eq:absorption_normal_thresh}
\end{equation}
where 
\begin{equation}
    \rd I = \frac{-i\pi}{s_0}\frac{1}{[\alpha_+(s_0)-\alpha][\alpha-\alpha_-(s_0)]} \rd\alpha
\end{equation}
and $c$ is the difference between the original integration contour $h$ and the one obtained after analytic continuation $h^\prime$:
\begin{equation}
\label{varc1}
c = \text{Var}[h] = [h']-[h]
=
\resizebox{2cm}{!}{
\begin{tikzpicture}[baseline=(current bounding box.center)]
     \draw[line width=2mm,black!50,
 postaction={decorate,decoration={markings, mark=between positions 0.02 and 1 step 0.15 with {\arrow[line width=2mm]{to}}}}
 ] (0,0) -- (0.6,0) 
 arc[start angle=180, delta angle=180,radius=0.7cm] 
 -- (3.5,0)
arc[start angle=180, delta angle=-180,radius=0.7cm] --
(6,0);
\filldraw[olddarkgreen] (1.22,-0.2) circle (0.2);
\filldraw[darkred] (4.2,0.2) circle (0.2);
\node[black!50,scale=3] at (2.8,0.8) {$h'$};
\node[opacity=0] at (2.8,-1.8) {$h$};
\end{tikzpicture}
}
-
\resizebox{2cm}{!}{
\begin{tikzpicture}[baseline=(current bounding box.center)]
     \draw[line width=2mm,black!50,
 postaction={decorate,decoration={markings, mark=between positions 0.02 and 1 step 0.3 with {\arrow[line width=2mm]{to}}}}
 ] (0,0) -- (6,0);
\filldraw[olddarkgreen] (1.08,-0.3) circle (0.2);
\filldraw[darkred] (4,0.29) circle (0.2);
\node[black!50,scale=3] at (3,0.8) {$h$};
\node[opacity=0] at (3,-1.8) {$h$};
\end{tikzpicture}
}
=
\resizebox{2cm}{!}{
\begin{tikzpicture}[baseline=(current bounding box.center)]
 \draw[line width=1mm,black!50,
 postaction={decorate,decoration={markings, mark=between positions 0.02 and 1 step 0.2 with {\arrow[line width=0.5mm,black!50]{>}}}}
 ] (1.4,0) circle (0.45cm);
 \draw[line width=1mm,black!50,
 postaction={decorate,decoration={markings, mark=between positions 0.02 and 1 step 0.2 with {\arrow[line width=0.5mm,black!50]{<}}}}
 ] (3.8,0) circle (0.45cm);
\filldraw[olddarkgreen] (1.4,0) circle (0.15);
\filldraw[darkred] (3.8,0) circle (0.15);
\end{tikzpicture}
}
\end{equation}
Since the singularities of the integrand at $\alpha_+(s_0)$ and $\alpha_-(s_0)$ are simple poles, we can deform the integration contour $c$ into two circles, each of which encircles one of the roots $\alpha_{\pm}(s_0)$. Note that for $\varepsilon=0$ these poles are on the real $\alpha$ line, which is precisely why the $i\varepsilon$ is needed to make the Feynman integral well-defined.

We can now use Cauchy's residue theorem to compute the integral in Eq.~\eqref{eq:absorption_normal_thresh} over $\rd I$. Namely, 
\begin{equation}
    A_\bubrgT^\bubrgT(s_0)= 
    - \frac{2 \pi^2}{s_0} 
    \Bigl(\operatorname{res}_{\alpha = \alpha_+} \frac{\rd \alpha}{(\alpha - \alpha_+) (\alpha - \alpha_-)} - \operatorname{res}_{\alpha = \alpha_-} \frac{\rd \alpha}{(\alpha - \alpha_+) (\alpha - \alpha_-)}\Bigr)  \,,
    \label{eq:mon_bub_delta}
\end{equation}
where the sign of each contribution is determined by the orientation of the contours. 
Evaluating these residues, one finds
\begin{equation}
A_\bubrgT^\bubrgT(s_0)=
    \frac{-4 \pi^2}{ \sqrt{s_0-(m_1-m_2)^2} \sqrt{s_0-(m_1+m_2)^2}} \,.
    \label{eq:mon_bub}
\end{equation}
This value can be compared to what one gets by first evaluating the bubble integral as a logarithm, and then computing the monodromy of the transcendental function directly. Explicitly, performing the integration in Eq.~\eqref{eq:bub_alpha}, one finds
\begin{multline}
    \I_\bubrgT(s) = \frac{-2 \pi}{\sqrt{-[s-(m_1-m_2)^2] [s-(m_1+m_2)^2]}} \times \\ \log \left( \frac{\sqrt{(m_1+m_2)^2-s}-i\sqrt{s-(m_1-m_2)^2}}{\sqrt{(m_1+m_2)^2-s}+i \sqrt{s-(m_1-m_2)^2}} \right)\,,
    \label{eq:bub_full}
\end{multline}
where the square roots are evaluated on their principal branches. Computing the monodromy of this expression around $s=(m_1+m_2)^2$, one recovers Eq.~\eqref{eq:mon_bub}.

Next, let us compute an additional discontinuity 
of the bubble integral with respect to the threshold at $s = (m_1 + m_2)^2$ using the same method of tracking the integration contours. That is, we compute the monodromy of the absorption integral $A_\bubrgT^\bubrgT(s_0)$ around $s = (m_1 + m_2)^2$. Since the integration contour of the absorption integral is simply given by the two small circles around the roots $\alpha_{\pm}$, it is deformed by the analytic continuation in an especially simple way: the two circles just follow the roots $\alpha_{\pm}$ as these roots are analytically continued around the same circular path as before. As the contours around $\alpha_+$ and $\alpha_-$ have the opposite orientation, the difference between the original contour and the new one is just $c' = 2 c$. Thus, we have that
\begin{equation}
\left(\bbone-\monM_{s=(m_1+m_2)^2} \right)^2 \I_\bubrgT(s_0)  =  \left(\bbone-\monM_{s=(m_1+m_2)^2} \right) A_\bubrgT^\bubrgT(s_0) =\int_{c'} \rd I_\bubrgT = 2 A_\bubrgT^\bubrgT(s_0)
    \,,
    \label{eq:abs_normal}
\end{equation}
for $A_\bubrgT^\bubrgT(s_0)$ as given in Eq.~\eqref{eq:mon_bub}.

To recap, we have computed single and double monodromies around the normal and pseudonormal thresholds of the bubble integral, starting at a point $s_0$. By tracking how the integration contour gets dragged in the Schwinger-parameter space, we found
\begin{align}
    \left(\bbone-\monM_{s=(m_1+m_2)^2} \right) \I_\bubrgT(s_0) & = A_\bubrgT^\bubrgT(s_0) \,,
    \label{eq:mon_bub1}
    \\ 
    \left(\bbone-\monM_{s=(m_1-m_2)^2} \right) \I_\bubrgT(s_0) & = 0 \,,
     \label{eq:mon_bub2}
    \\
   \left(\bbone-\monM_{s=(m_1+m_2)^2} \right)^2 \I_\bubrgT(s_0)  & =  2 A_\bubrgT^\bubrgT(s_0) \,,
     \label{eq:mon_bub3}
\end{align}
where $A_\bubrgT^\bubrgT(s_0)$ is the absorption integral where the internal particles of the bubble are put on shell.
All of these results can be compared to what one gets by computing the monodromies directly from Eq.~\eqref{eq:bub_full} around $s=(m_1+m_2)^2$ and $s=(m_1-m_2)^2$. Doing so, one recovers Eqs.~\eqref{eq:mon_bub1}--\eqref{eq:mon_bub3}.

While the above analysis illuminates how monodromies in the space of external kinematics can be computed by tracking how the integration contour must be deformed to avoid singularities in the integrand, it is hard to carry out the corresponding procedure in more than one complex dimension; tracing how to avoid singularities in many variables simultaneously quickly gets out of hand. In more complicated examples, we therefore resort to Picard--Lefschetz theory, which provides us with a general prescription for how to represent monodromies computed in the space of external kinematics as integrals over modified integration contours. 

\subsection{Picard--Lefschetz in Feynman Parameter Space}
Picard--Lefschetz theory states that the difference between an integral and the monodromy of the integral around a branch point $s=s^\ast$ can be written as
\begin{equation}
    \left(\bbone-\monM_{s=s^\ast} \right) \int_h \rd I
    =
    N_0 \int_{c} \rd I\,,
    \label{eq:PLteaser}
\end{equation}
where $N_0$ is an integer and $c$ is a new integration contour. In the case of the bubble integral in Feynman parameter space, we worked out the contour $c$ for the threshold monodromy in Eq.~\eqref{varc1}: the new integration contour became circles around the two roots $\alpha_\pm$, and the resulting integral could be evaluated using Cauchy's residue theorem. The rest of this section, along with Sec.~\ref{sec:iterated}, Sec.~\ref{sec:deform_contours}, and App.~\ref{sec:kronecker_index}, will be devoted to understanding when Eq.~\eqref{eq:PLteaser} applies to Feynman integrals, and to giving prescriptions for how the integer $N_0$ and the modified contour $c$ can be determined. Before diving into the general theory, though, let us introduce the relevant ideas by continuing our analysis of the bubble Feynman integral by showing how its discontinuities can be computed using Picard--Lefschetz theory.

First, we identify the contour $e$ that connects the two the singular surfaces that have the potential to pinch the integration contour when they coincide. In the above example, these singular surfaces are $\alpha_+(s_0)$ and $\alpha_-(s_0)$. The contour that connects them, the {\it{vanishing cell}}, vanishes when $s_0 \to (m_1+m_2)^2$. We conventionally choose the orientation of the vanishing cell to be from $\alpha_+(s_0)$ to $\alpha_-(s_0)$. The boundary of $e$ is given by the roots $\alpha_\pm(s_0)$ and the {\it coboundary} of these boundary points are the circles in Fig.~\ref{fig:bubble_cycle}, which correspond to the difference contour that appears in the absorption integral $A_\bubrgT^\bubrgT(s_0)$:\footnote{In  Picard--Lefschetz theory, the orientation of the coboundary is determined by the orientation of the vanishing cell.}
\begin{equation}
    \begin{tikzpicture}[baseline=(current bounding box.center),scale=0.8]
 \draw[line width=2,black!50,
 postaction={decorate,decoration={markings, mark=between positions 0.02 and 1 step 0.2 with {\arrow[line width=1,black!50]{<}}}}
 ] (1.4,0) circle (0.45);
 \draw[line width=2,black!50,
 postaction={decorate,decoration={markings, mark=between positions 0.02 and 1 step 0.2 with {\arrow[line width=1,black!50]{>}}}}
 ] (3.8,0) circle (0.45);
 \draw[red!50,line width=2] (1.4,0) -- (3.8,0);
\draw[-latex,red!50,line width=2] (2.8,0) -- (2.95,0);
\filldraw[olddarkgreen] (1.4,0) circle (0.15);
\filldraw[darkred] (3.8,0) circle (0.15);
\node[black,scale=1] at (-0.5,1.5) {vanishing cell $e$};
\draw[-latex,red!50,line width=0.8] (1,1.1) -- (2.5,0.1);
\node[black,scale=1] at (1,-1.5) {boundary};
\node[black,scale=1] at (1,-2.2) {(vanishing sphere)};
\draw[-latex,olddarkgreen,line width=0.8] (1,-1) -- (1.3,-0.15);
\draw[-latex,darkred,line width=0.8] (1.5,-1) -- (3.7,-0.15);
\node[black,scale=1] at (5.5,2.2) {coboundary $c$};
\node[black,scale=1] at (5.5,1.5) {(vanishing cycle)};
\draw[-latex,black!50,line width=0.8] (4.7,1.1) -- (3.9,0.6);
\draw[-latex,black!50,line width=0.8] (4.3,1.1) -- (1.8,0.4);
\end{tikzpicture}
\end{equation}
If we abuse our notation by using $h$ to refer to both the original integration contour and the homology class to which it belongs, Picard--Lefschetz theory tells us that the discontinuity of the bubble integral with respect to the threshold singularity can be rewritten as
\begin{equation} 
    \left(\bbone-\monM_{s=(m_1+m_2)^2} \right) I_\bubrgT(s_0) = \langle e,h\rangle \int_{c} \rd I \, , \label{PLformsimple}
\end{equation}
where $\langle e,h\rangle$ is the Kronecker index, which gives the number of times (weighted by orientations) the vanishing cell $e$ intersects a contour in the homology class $h$. In the case of the bubble, we can arrange that the vanishing cell intersects the original integration contour at a single point:

\begin{equation} \label{contour_def}
\resizebox{!}{0.8cm}{
\begin{tikzpicture}
\draw[line width=1.5mm,black!50,
 postaction={decorate,decoration={markings, mark=between positions 0.02 and 1 step 0.3 with {\arrow[line width=1mm]{>}}}}
] (0,0) -- (6,0);
\draw[red!50,line width=1.5mm,
 postaction={decorate,decoration={markings, mark=between positions 0.3 and 1 step 0.3  with {\arrow[line width=1mm]{to}}}}
] (2,-0.5) -- (4,0.5);
\filldraw[olddarkgreen] (2,-0.5) circle (0.15);
\filldraw[darkred] (4,0.5) circle (0.15);
\node[scale=2] at (7,0) {$\approx$};
\end{tikzpicture}
}
\resizebox{!}{0.8cm}{
\begin{tikzpicture}
\draw[line width=1.5mm,black!50,
 postaction={decorate,decoration={markings, mark=between positions 0.02 and 1 step 0.3 with {\arrow[line width=1mm]{>}}}}
] (0,0) to[out=30,in=150] (3,0) to[out=-30,in=210] (6,0);
\draw[red!50,line width=1.5mm,
 postaction={decorate,decoration={markings, mark=between positions 0.3 and 1 step 0.3  with {\arrow[line width=1mm]{to}}}}
] (2,0) -- (4,0);
\filldraw[olddarkgreen] (2,0) circle (0.15);
\filldraw[darkred] (4,0) circle (0.15);
\end{tikzpicture}
}
\end{equation}
On the left, we have kept the poles $\alpha_\pm(s)$ (and the vanishing cell) complex for real $s$, as consistent with the $i\varepsilon$ prescription. On the right we have used the $i\varepsilon$ prescription to deform the integration contour, which allows us to keep the vanishing cell real. Keeping the vanishing cell real is helpful for applying the Picard--Lefschetz theorem and will be our default prescription going forward. With either prescription, we see that in this case the two curves intersect only once, so the Kronecker index is 1.  In App.~\ref{sec:kronecker_index}, we explain how to compute Kronecker indices in more detail.

We can rephrase the discontinuity of the absorption integral with respect to the threshold singularity in a similar way. The vanishing cell will be the same as before, while the integration contour $c$ that appears in the absorption integral is given by the two circles from Eq.~\eqref{varc1}:
\begin{equation}
\resizebox{2.3cm}{!}{
\begin{tikzpicture}
 \draw[line width=1mm,black!50,
 postaction={decorate,decoration={markings, mark=between positions 0.02 and 1 step 0.2 with {\arrow[line width=0.5mm,black!50]{<}}}}
 ] (2,0) circle (0.45cm);
 \draw[line width=1mm,black!50,
 postaction={decorate,decoration={markings, mark=between positions 0.02 and 1 step 0.2 with {\arrow[line width=0.5mm,black!50]{>}}}}
 ] (4,0) circle (0.45cm);
\draw[red!50,line width=1.5mm,
 postaction={decorate,decoration={markings, mark=between positions 0.3 and 1 step 0.3  with {\arrow[line width=1mm]{to}}}}
] (2,0) -- (4,0);
\filldraw[olddarkgreen] (2,0) circle (0.15);
\filldraw[darkred] (4,0) circle (0.15);
\end{tikzpicture}
}
\label{circles}
\end{equation}
This implies that the Kronecker index $\langle e,c\rangle$ is 2, since we can arrange that the vanishing cycle intersects the integration contour twice with the positive orientation. It is then straightforward to check that the Picard--Lefschetz formula, which tells us that
\begin{equation} 
    \left(\bbone-\monM_{s=(m_1+m_2)^2} \right) A_\bubrgT^\bubrgT(s_0)  = \langle e,c \rangle \int_{c} \rd I = 2 A_\bubrgT^\bubrgT(s_0)  \, , \label{PLformsimple2}
\end{equation}
reproduces the previous result in Eq.~\eqref{eq:abs_normal}. That the integration contour on the right side of~\eqref{PLformsimple2} is given by $c$ follows from the fact that we have the same vanishing cell as when we computed the first discontinuity; we will describe how this contour is determined in more detail in Sec.~\ref{sec:momspacePL}.

\begin{figure}[t]
    \centering
 \begin{tikzpicture}
\node at (0,0) {
\begin{tikzpicture}[scale=0.5]
\draw[-stealth,line width=0.7] (-4,0) -- (7,0) coordinate (xaxis);
\draw[-stealth,line width=0.7] (-2,-2) -- (-2,3) coordinate (yaxis);
\draw[line width=1,darkcyan,
 postaction={decorate,decoration={markings, mark=between positions 0.02 and 1 step 0.09 with {\arrow[line width=1]{to}}}}
] (5.5,0) --(-0.5,0.4)
 arc[start angle=20, delta angle=330,radius=0.5cm]
 -- (5.5,0);
\filldraw[black] (3,-0.2) circle (0.15); 
\filldraw[etacol2] (-1,0.2) circle (0.15); 
\draw[black,line width=1.0] (5.5,0) circle (0.2);
\draw [draw=black] (6,2) -- (6,3);
\draw [draw=black] (6,2) -- (7,2);
\node [black] at  (6.5,2.5) {$s$};
\end{tikzpicture}
};
\node at (8,0) {
\begin{tikzpicture}[scale=0.5]
\draw[-stealth,line width=0.7] (-4,0) -- (7,0) coordinate (xaxis);
\draw[-stealth,line width=0.7] (-2,-2) -- (-2,3) coordinate (yaxis);
\draw[line width=1,darkcyan,
 postaction={decorate,decoration={markings, mark=between positions 0.02 and 1 step 0.1 with {\arrow[line width=1]{to}}}}
] (5.5,0) to[out=-120,in=-40] (-0.5,0)
 arc[start angle=-30, delta angle=330,radius=0.5cm]
 to[out=-45,in=-100] (5.5,0);
\filldraw[black] (3,-0.2) circle (0.15); 
\filldraw[etacol2] (-1,0.2) circle (0.15);
\draw[black,line width=1.0] (5.5,0) circle (0.2);
\draw [draw=black] (6,2) -- (6,3);
\draw [draw=black] (6,2) -- (7,2);
\node [black] at  (6.5,2.5) {$s$};
\end{tikzpicture}
};
\node at (0,-4) {
\begin{tikzpicture}[scale=0.5]
\draw[-stealth,line width=0.7] (-4,0) -- (7,0) coordinate (xaxis);
\draw[-stealth,line width=0.7] (0,-2) -- (0,4) coordinate (yaxis);
\draw [draw=black] (6,3) -- (6,4);
\draw [draw=black] (6,3) -- (7,3);
\node [black] at  (6.5,3.5) {$\alpha$};
 \draw[line width=1,black!50,
 postaction={decorate,decoration={markings, mark=between positions 0.02 and 1 step 0.15 with {\arrow[line width=1]{to}}}}
 ] (0,0) -- (5,0);
\draw[darkred, line width=1,
 postaction={decorate,decoration={markings, mark=between positions 0.02 and 1 step 0.15 with {\arrow[line width=1]{to}}}}
 ] (1.3,-0.3) to[out=210,in=-80] (-0.5,0)
.. controls (-0.5,1) and (1,1.5) .. (3.5,0.6);
\draw[olddarkgreen, line width=1,
 postaction={decorate,decoration={markings, mark=between positions 0.02 and 1 step 0.09 with {\arrow[line width=1]{to}}}}]
 (4,0.6) to[out=150,in=30] (-1.5,0.6)
arc[start angle=45, delta angle=280,radius=1cm] 
.. controls (0,-1) and (1,-1) ..
(1.8, -0.3);
\draw[darkred] (1.3,-0.3) circle (0.15);
\filldraw[olddarkgreen] (1.8,-0.3) circle (0.15);
\draw[olddarkgreen] (4,0.6) circle (0.15);
\filldraw[darkred] (3.5,0.6) circle (0.15);
\filldraw[black] (2.5,0.2) circle (0.15); 
\filldraw[black] (2.5,-0.2) circle (0.15); 
\filldraw[etacol2] (-2.5,0.2) circle (0.15); 
\filldraw[etacol2] (-2.5,-0.2) circle (0.15); 
\node[darkred,scale=1] at (1.5,0.5) {$\alpha_-(s)$};
\node[olddarkgreen,scale=1] at (-1,1.5) {$\alpha_+(s)$};
\end{tikzpicture}
};
\node at (8,-4) {
\begin{tikzpicture}[scale=0.5]
\draw[-stealth,line width=0.7] (-4,0) -- (7,0) coordinate (xaxis);
\draw[-stealth,line width=0.7] (0,-2) -- (0,4) coordinate (yaxis);
\draw [draw=black] (6,3) -- (6,4);
\draw [draw=black] (6,3) -- (7,3);
\node [black] at  (6.5,3.5) {$\alpha$};
\draw[olddarkgreen, line width=1] 
(1.5,-0.3) to[out=90,in=30] (-1.5,0.6) 
.. controls (-4,2) and (-4,-2) .. (-1.5,-0.6)
.. controls (1,-2) and (4,-1) .. (5,0.6);
\draw[line width=1,black!50,
 postaction={decorate,decoration={markings, mark=between positions 0.03 and 1 step 0.05 with {\arrow[line width=1]{to}}}}]
(3.9,0) -- (4.2,0) .. controls (3,-1.3) and (0,-1.5) .. 
(-1.5,-0.35) .. controls (-3.7,-2) and (-3.7,2) ..
(-1.5,0.35) to[out=30,in=105] (1.3,-0.2) 
arc[start angle=180, delta angle=180,radius=0.2]
 to[out=90,in=30] (-1.5,0.8)
 .. controls (-4.3,2) and (-4.3,-2) ..(-1.5,-0.8)
 .. controls (1,-2) and (4,-1.2) ..(4.8,0) -- (6,0);
\draw[line width=1,black!50,
 postaction={decorate,decoration={markings, mark=between positions 0.02 and 1 step 0.15 with {\arrow[line width=1]{to}}}}]
 (0,0) -- (0.3,0) .. controls (-0.5,0.5) and (-0.5,-0.1) .. (0,-0.3) .. controls (2,-1) and (3,-0.3) .. (3.6,0.7)
.. controls (3.8,1) and (4.1,1) .. (4.1,0.7) 
.. controls (3,-1.5) and (0,-1.3) .. (-1,0) to[out=90,in=120] (0.9,0) -- (1.1,0);
\draw[line width=1,black!50,
 postaction={decorate,decoration={markings, mark=between positions 0.4 and 1 step 0.6 with {\arrow[line width=1]{to}}}}]
 (1.85,0) -- (2.7,0);
\draw[darkred, line width=1] 
(3.8,0.6) .. controls (3,-1) and (1,-1) .. (0,-0.5)
.. controls (-2,0.5) and (0.7,0.7) ..( 0.8,-0.3);
\filldraw[olddarkgreen] (1.5,-0.3) circle (0.15);
\draw[darkred] (0.8,-0.3) circle (0.15);
\draw[olddarkgreen] (5,0.6) circle (0.15);
\filldraw[darkred] (3.8,0.6) circle (0.15);
\filldraw[black] (2.3,0.2) circle (0.15); 
\filldraw[black] (2.3,-0.2) circle (0.15); 
\filldraw[etacol2] (-2.5,0.2) circle (0.15); 
\filldraw[etacol2] (-2.5,-0.2) circle (0.15); 
\end{tikzpicture}
};
\end{tikzpicture}
       \caption{Whether the monodromy of the bubble integral around the pseudothreshold is nonzero or not depends on the analytic continuation path chosen for $s$. In particular, in the top two plots we depict how one must choose whether to analytically continue $s$ around the pseudothreshold (represented by the purple dot) by either going above or below the normal threshold (represented by the black dot) when starting from a point $s_0>(m_1+m_2)^2$ (represented by the hollow black circle). 
       The bottom two panels depict the corresponding paths that are traversed by the roots $\alpha_{\pm}(s)$, which move from the hollow green and hollow red dots to the solid red and green dots. When $s$ is analytically continued above the normal threshold, we see that the integration contour is not pinched by this continuation. Conversely, when $s$ is analytically continued below the normal threshold, the contour must be deformed and the monodromy can be nonzero.
       \label{fig:bubble_cycle2}
}
\end{figure}
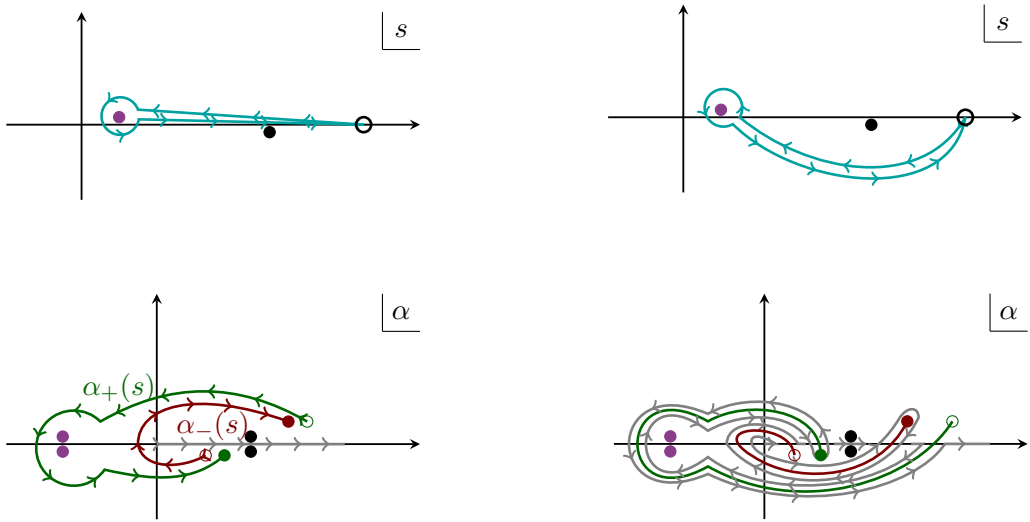
Picard--Lefschetz can also explain what happens when we compute a monodromy around the pseudothreshold $s=(m_1-m_2)^2$.    
Recall that the original integration contour is not pinched by the roots $\alpha_{\pm}$ at the pseudothreshold, so the monodromy around $s=(m_1-m_2)^2$ is zero on the physical sheet. We can easily see this using the Picard--Lefschetz theorem; the vanishing cell between the two roots is a line between them that crosses the negative $\alpha$-axis, away from the integration contour:
\begin{equation} \label{contour_def2}
\resizebox{5cm}{!}{
\begin{tikzpicture}
\draw[line width=1.5mm,black!50,
 postaction={decorate,decoration={markings, mark=between positions 0.02 and 1 step 0.3 with {\arrow[line width=1mm]{>}}}}
] (0,0) -- (6,0);
\draw[red!50,line width=1.5mm,
 postaction={decorate,decoration={markings, mark=between positions 0.3 and 1 step 0.3  with {\arrow[line width=1mm]{to}}}}
] (-2,-1) -- (-2,1);
\filldraw[olddarkgreen] (-2,-1) circle (0.15);
\filldraw[darkred] (-2,1) circle (0.15);
\end{tikzpicture}
}
\end{equation}
Thus, the Kronecker index is $\langle e,h\rangle=0$, since the integration contour and the vanishing cell do not intersect.

The vanishing of the discontinuity around the pseudothreshold is implicitly due to the $i \varepsilon$ prescription, which (implicitly) tells us that we should remain above the normal threshold when analytically continuing $s$ around the pseudothreshold. More generally, though, we can consider discontinuities of $\I_\bubrgT(s)$ on different Riemann sheets.
Consider for example what happens if one crosses the branch cut at $s>(m_1+m_2)^2$ before analytically continuing around the pseudothreshold. This situation is depicted in Fig.~\ref{fig:bubble_cycle2}, where we see there that the deformed contour is homologous to a contour surrounding the singular points:
\begin{equation} \label{eq:pseudothreshold_contour}
h' =
\resizebox{!}{1cm}{
\begin{tikzpicture}[baseline=(current bounding box.center)]
\draw[line width=1mm,black!50,
 postaction={decorate,decoration={markings, mark=between positions 0.03 and 1 step 0.05 with {\arrow[line width=0.8mm]{to}}}}
 ]
(3.9,0) -- (4.2,0) .. controls (3,-1.3) and (0,-1.5) .. 
(-1.5,-0.35) .. controls (-3.7,-2) and (-3.7,2) ..
(-1.5,0.35) to[out=30,in=105] (1.3,-0.2) 
arc[start angle=180, delta angle=180,radius=0.2]
 to[out=90,in=30] (-1.5,0.8)
 .. controls (-4.3,2) and (-4.3,-2)  ..(-1.5,-0.8)
  .. controls (1,-2) and (4,-1.2)  ..(4.8,0) -- (6,0);
\draw[line width=1mm,black!50,
 postaction={decorate,decoration={markings, mark=between positions 0.02 and 1 step 0.15 with {\arrow[line width=0.8mm]{to}}}}
] (0,0) -- (0.3,0) .. controls (-0.5,0.5) and (-0.5,-0.1) .. (0,-0.3) .. controls (2,-1) and (3,-0.3) .. (3.6,0.7)
.. controls (3.8,1) and (4.1,1) .. (4.1,0.7) 
.. controls (3,-1.5) and (0,-1.3) .. (-1,0) to[out=90,in=120] (0.9,0) -- (1.1,0);
\draw[line width=1mm,black!50,,
 postaction={decorate,decoration={markings, mark=between positions 0.4 and 0.6 step 0.5 with {\arrow[line width=0.8mm]{to}}}}
 ] (1.9,0) -- (2.7,0);
\filldraw[olddarkgreen] (1.5,-0.3) circle (0.15);
\filldraw[darkred] (3.8,0.6) circle (0.15);
\end{tikzpicture}
}
=
\resizebox{!}{0.5cm}{
\begin{tikzpicture}[baseline=(current bounding box.center)]
  \draw[line width=1mm,black!50,
  postaction={decorate,decoration={markings, mark=between positions 0.02 and 1 step 0.08 with {\arrow[line width=0.8mm]{to}}}}
 ] (0,0) -- (4,1) arc[start angle=100, delta angle=-260,radius=0.5cm]
-- (1,0) arc[start angle=90, delta angle=270,radius=0.5cm] --
(5,0);
\filldraw[olddarkgreen] (1.1,-0.5) circle (0.15);
\filldraw[darkred] (4,0.6) circle (0.15);
\end{tikzpicture}
}
\end{equation}
As the difference between the original contour and this deformed contour encircles the roots $\alpha_\pm$, it gives rise to a nonzero discontinuity. As before, the value of this discontinuity can be computed using the Picard--Lefschetz formula. However, in this example one must be careful when computing the Kronecker index, as we have changed the original integration contour in order to leave the physical sheet. To avoid this complication below, we will only apply the Picard--Lefschetz formula to discontinuities that can be computed via analytic continuations that are restricted to an infinitesimally-small neighborhood around the singularity.\footnote{More specifically, the part of the analytic continuation that passes through one of the branch cuts that defines the current sheet should be infinitesimal; in general, we will also have to analytically continue a finite distance while staying on the same sheet to reach the singular point of interest.} When we restrict to these small neighborhoods,
no detailed knowledge of the analytic continuation contour is needed, and the Picard--Lefschetz formula can be applied algorithmically. %

When computing a monodromy around a threshold on the physical sheet, we end up on a different sheet and can expose further singularities such as the pseudothreshold. 
We can compute the monodromy of the absorption integral $A_\bubrgT^\bubrgT(s_0)$ around the pseudothreshold by leaving the infinitesimal neighborhood of the normal threshold. As the integration contour $c$ in this integral is just given by two small circles around $\alpha_{\pm}$, the contour will follow these roots as we analytically continue them to the pseudothreshold. Then, the monodromy computation is analogous to the second monodromy around the normal threshold in Eq.~\eqref{eq:abs_normal}; namely, the contribution to the contour we pick up with this analytic continuation is simply twice the contour $c$, so
\begin{equation}
    \left(\bbone-\monM_{s=(m_1-m_2)^2} \right) \left(\bbone-\monM_{s=(m_1+m_2)^2} \right) I_\bubrgT (s_0) = \left(\bbone-\monM_{s=(m_1-  m_2)^2} \right)A_\bubrgT^\bubrgT(s_0) =
    2 A_\bubrgT^\bubrgT(s_0) \, .
    \label{eq:abs_pseudo}
\end{equation}
It is easy to verify that this result matches what one would obtain by analytically continuing the algebraic expression for $A_\bubrgT^\bubrgT(s_0)$ in Eq.~\eqref{eq:mon_bub}, since analytically continuing a square root around its branch point results in the same expression with an overall minus sign, so the monodromy is twice the original expression.  

\subsection{Picard--Lefschetz in Momentum Space}
\label{sec:momspacePL}

Having introduced the basic ingredients that go into the Picard--Lefschetz formula above, we now describe them more formally. We stick to the example of the bubble integral, but move to the loop momentum form of this integral, as our analysis generalizes more easily in momentum space. The upshot will be that the monodromies of this Feynman integral can be computed by modifying the integration contour to the \textit{coboundary} of the \textit{boundary} of the \textit{vanishing cell} in the space of loop momenta, just like we found empirically for the bubble integral in Feynman-parameter space. We will extend this statement to generic Feynman integrals in Sec.~\ref{sec:intro_PL}.

The momentum space version of the two-dimensional bubble integral was given in Eq.~\eqref{eq:bub2D_mom}. When the factors of $i \varepsilon$ are included in the denominator, the integration contour is taken to be the whole real $k^\mu$ plane. An alternative is to instead write the integral as 
\begin{equation}
    \I_\bubrgT(p)= \int_h \frac{\rd^2 k}{[k^2-m_1^2] [(p-k)^2-m_2^2]} \, , \label{eq:bubble_momentum_h}
\end{equation}
where the integration contour $h$ is now taken to have small imaginary parts that are consistent with the $i \varepsilon$ prescription, similar to what was done in~\eqref{contour_def}. We will have more to say about these contours in Sec.~\ref{sec:iepaths}.

We work in the rest frame of the external momentum so that $p^\mu =(Q,0)$. Thus, we have that $s=Q^2$, and we denote the two denominators in Eq.~\eqref{eq:bubble_momentum_h} by  
\begin{align}
    s_1(p,k) &=  (k^0)^2-(k^1)^2-m_1^2 \, ,\\
    s_2(p,k) &=  (Q-k^0)^2-(k^1)^2-m_2^2 \,.
\end{align}
The integrand has singularities where $s_1(p,k)=0$ or $s_2(p,k)=0$. These algebraic conditions define a pair of codimension-one hypersurfaces $S_1$ and $S_2$.

Since the integration contour in Eq.~\eqref{eq:bubble_momentum_h} is deformed away from the real plane by small imaginary parts, these singular surfaces do not lead to singularities in the integral, except when they pinch the integration contour. This will happen where
\begin{equation}
    s = (m_1 \pm m_2)^2\, , \qquad k^\mu = \frac{m_1}{m_1 \pm m_2} p^\mu \, ,
    \label{eq:loop-landau_bubble}
\end{equation}
namely at the solution to the Landau equations in loop momentum space.

In order to study the analytic continuations of $\I_\bubrgT(p)$, we first write the integral in such a way that the Picard--Lefschetz formula can be directly applied. This first requires complexifying and compactifying the integration contour. We thus allow the differential form
\begin{equation} \label{eq:bubble_form}
    \omega = \frac{1}{s_1(p,k)} \frac{1}{s_2(p,k)} \rd k^0 \wedge \rd k^1 \, 
\end{equation}
to be complex, and compactify the space of internal kinematics.
We denote the compactified and complexified space of momentum (that satisfy momentum conservation) by $\Xs$. %
We will not discuss the choice of compactification here in more detail (for a discussion, see Refs.~\cite{Bloch:2015efx,Muhlbauer:2022ylo}); however, we note that, if possible, this compactification should be chosen to be consistent with the symmetries of the problem.\footnote{If the space is not compact then several problems can arise.  For example, if the space has a boundary, then we need to work with relative homology.  It can also happen that pinches arise ``at infinity''. These are known as Landau singularities of second type. If the space is compact, then there is no special infinity anymore so all pinches can be understood in a uniform language.} In general, we should also check that our integral is invariant under small deformations of the contour $h$. Assuming that the integrals at hand are UV and IR finite, and that the space of internal kinematics can be compactified such that the integral is invariant under small deformations of $h$, the Feynman integral can be viewed as a pairing $([\omega],[h])\to {\mathbb C}$ between the homology class of the integration contour and the cohomology class of the form.\footnote{The contour $h$ is closed in the compactified space of kinematics, as it does not have a boundary. To show that $\omega$ is closed, consider $\omega = \omega(z_1, \dotsc, z_n) \rd z_1 \wedge \dotso \wedge \rd z_n$.  Then, 
\begin{equation}
\rd \omega = \sum_k \Bigl((\partial_{z_k} \omega)(z_1, \dotsc, z_n) \rd z_k \wedge \rd z_1 \wedge \dotso \wedge \rd z_n +
(\partial_{\bar{z}_k} \omega)(z_1, \dotsc, z_n) \rd \bar{z}_k \wedge \rd z_1 \wedge \dotso \wedge \rd z_n\Bigr).
\label{eq:omega_closed}
\end{equation}
The first kind of terms are zero by the antisymmetry of the wedge product while the second kind of terms vanish by the holomorphy of $\omega(z_1, \dotsc, z_n)$. Note that if $\omega$ was meromorphic (was allowed to have poles in its domain), then the right-hand side of Eq.~\eqref{eq:omega_closed} could get non-zero contributions since $\bar{\partial} \frac 1 z \propto \delta^2(z)$. But we have removed all the points from the domain for which $\omega$ has poles, and thus conclude that $\rd \omega = 0$.  We emphasize that $\omega$ also depends on external kinematic parameters and it is only closed when the differentials do not act with respect to these external parameters.} In the example at hand, the two-form $\omega$ has singularities along $S_1 \cup S_2$ and is therefore a representative of the second cohomology class
\eq{
    [\omega] \in H^2(\Xs \setminus (S_1 \cup S_2))\,,
}
while the contour $h$ is a representative of the second homology class,
\eq{
    [h] \in H_2(\Xs \setminus (S_1 \cup S_2))\,.
}
meaning it does not intersect $S_1 \cup S_2$ in $\Xs$.\footnote{A generalization of this construction for an integration contour with boundaries involves using relative cohomology and homology (see Ref.~\cite{BSMF_1959__87__81_0}).  The relative version was used in Sec.~\ref{sec:alphaspace}.}

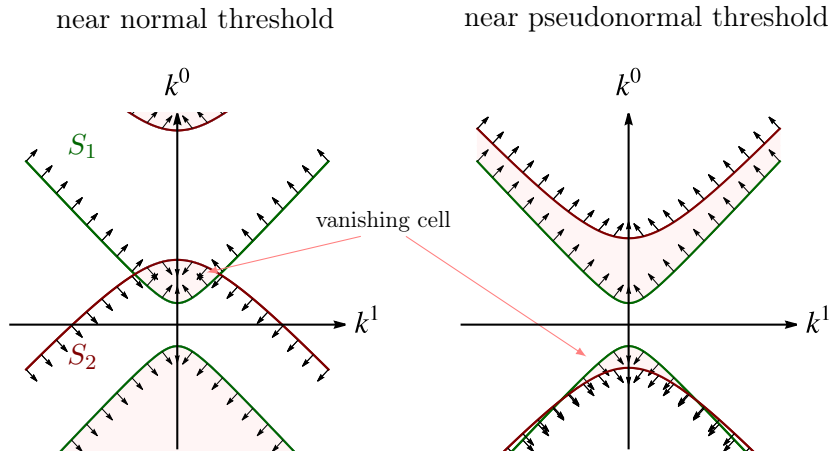
\begin{figure}
    \centering
\begin{tikzpicture}
 \node (image) at (0,0) {
        \includegraphics[width=0.32\textwidth]{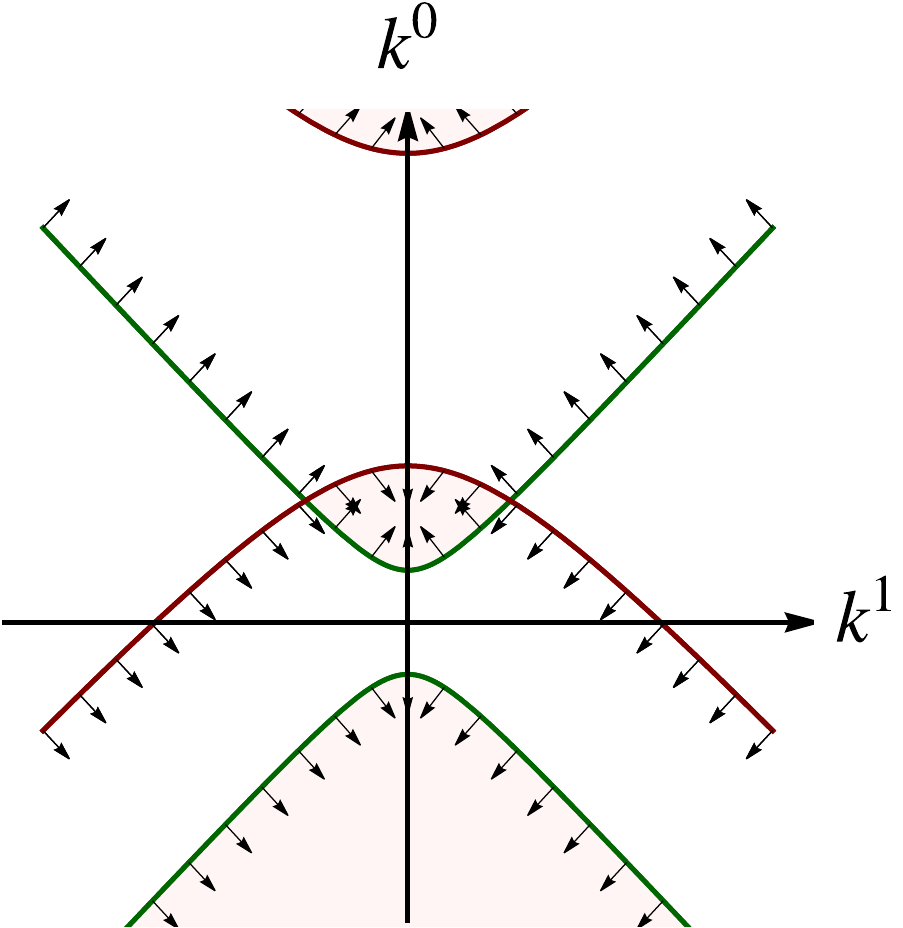}
    };    
    \node[scale=0.8] at (2.5,0.5) {vanishing cell};
    \draw[red!50,-latex] (2.2,0.3) -- (0.15,-0.18);
    \draw[red!50,-latex] (2.8,0.3) -- (5.2,-1.3);
    \node[col3,scale=1] at (-1.5,1.5) {$S_1$};
    \node[col4,scale=1] at (-1.5,-1.3) {$S_2$};
    \node[scale=1] at (0,3.2) {near normal threshold};
 \node (image) at (6,0) {
        \includegraphics[width=0.32\textwidth]{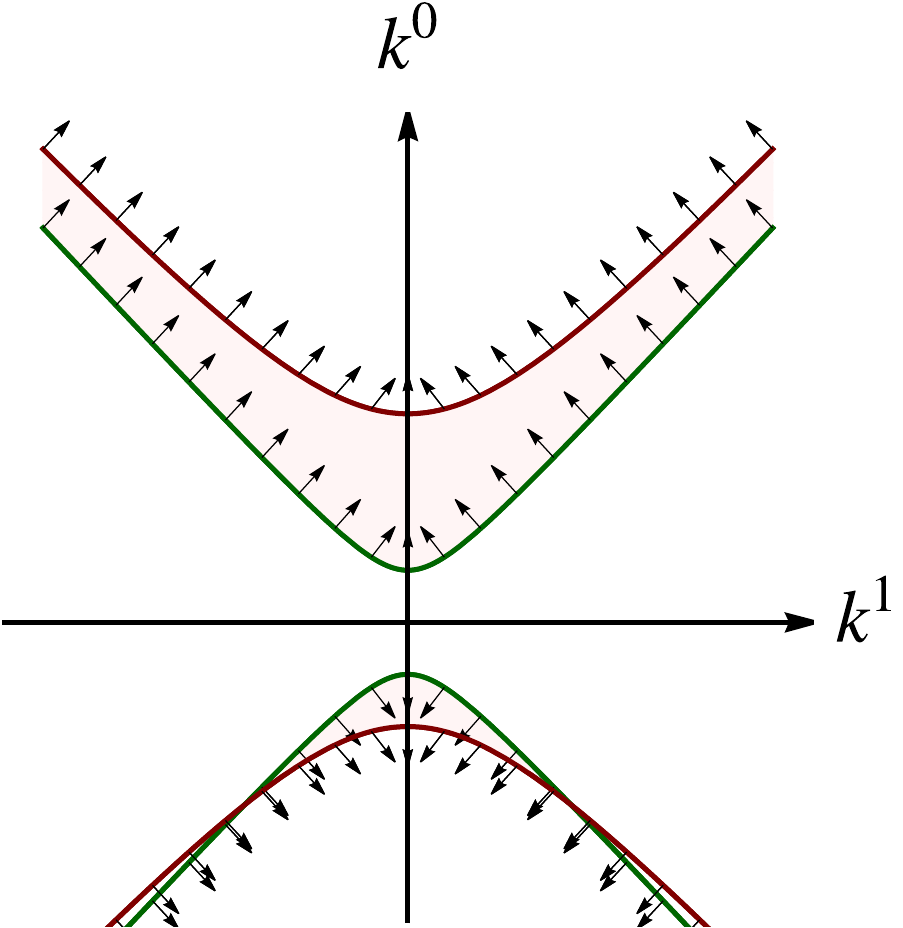}
};    
    \node[scale=1] at (6,3.2) {near pseudonormal threshold};
\end{tikzpicture}
    \caption{The on-shell surfaces $S_1$ and $S_2$ of the two-dimensional bubble (the green and red lines) in the $(k^0, k^1)$ plane. The left plot shows a point where the external momenta is slightly above the normal threshold ($Q \gtrsim m_1+m_2$). The right plot shows a point near the pseudonormal threshold ($Q=m_1-m_2$). The $i\varepsilon$ prescription deforms the on-shell contours out of the real $(k^0, k^1)$ plane, giving them imaginary parts proportional to the arrows. The vanishing cell shown on the left is the region where $s_1\geq 0$ and $s_2\geq 0$. It intersects the integration plane at a point where the vector field vanishes.}
    \label{fig:eyelashes}
\end{figure}

The $i \varepsilon$ prescription for Feynman integrals  dictates which contour in $H_2(\Xs \setminus (S_1 \cup S_2))$ corresponds to the physical amplitude as $\varepsilon \to 0^+$. That is, the Feynman $i \varepsilon$ moves the singularities off the real axis to make the contour well-defined, and as $\varepsilon \to 0^+$ the singularities move back on to the real axis and force us to deform the contour. The deformation that this prescription implies for the bubble integral on the surfaces $S_1$ and on $S_2$ is shown in Fig.~\ref{fig:eyelashes}, where the arrows indicate the direction of the imaginary contour deformation at different points. (For instance, an arrow pointing straight up would indicate a deformation in the $+i k^0$ direction, while an arrow pointing to the left would indicate a deformation in the $-i k^1$ direction.)
In the region around the normal threshold, the integration contour then looks like
\begin{equation}
\resizebox{8cm}{!}{
\begin{tikzpicture}[baseline=(current  bounding  box.center)]
 \node (image) at (0,0) {
        \includegraphics[width=0.255\textwidth]{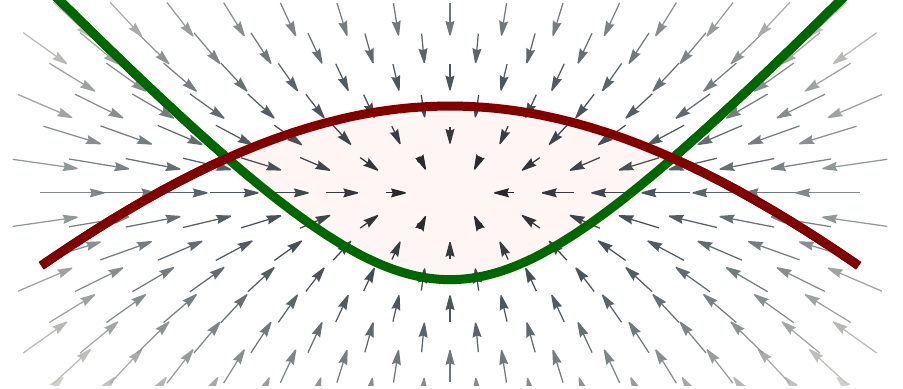}
    };  
\end{tikzpicture}
}
\label{eq:vectorfield}
\end{equation}
\\[-5mm]
By deforming the contour instead of inserting an $i \varepsilon$ into each propagator, one does not need to take the limit as $\varepsilon \to 0$ at the end of the calculation; $\varepsilon$ simply needs to be sufficiently small to determine the homology class of the contour (recall that the integral is invariant under deformations of the contour as long as it remains in the same homology class).  More details of how to deform the integration contour in loop-momentum space will be given in Sec.~\ref{sec:iepaths}, while an analogous contour deformation in Feynman-parameter space is described in~\cite{Hannesdottir:2022bmo}.

For fixed values of the external kinematics, we say that singular surfaces such as $S_1$ and $S_2$ are in a \textit{general position} when the corresponding integral is well-defined.\footnote{Here, well-defined means avoiding singularities at particular values of the external kinematics, like poles or branch points, as well as free of UV or IR divergences.}
When the Landau equations are satisfied, these singular surfaces will not be in a general position and the integral may itself also become singular. (Note that we do not necessarily mean that the value of the integral becomes infinite, but rather that there is a branch point at this value of the external kinematics, where it becomes multivalued.) At this non-generic position of the external kinematics, the integration contour becomes pinched. If we now vary the external kinematics in a small loop around this singularity, the pinched integration contour gets dragged along, leading the integral to pick up a contribution. The Picard--Lefschetz theorem makes this intuitive picture precise and provides a formula for calculating the extra contribution picked up by the integral under an analytic continuation in the space of external kinematics.

Let us denote the value of $p$ at the normal-threshold solution of the Landau equations, where ${p}^2=(m_1+m_2)^2$, by $p^\ast$. At this threshold, the surfaces $S_{1}$ and $S_{2}$ are not in a general position, and there exist nonzero complex numbers $\alpha_1, \alpha_2$ as well as some value $k^\ast$ of the loop momenta such that
\eq{
    \alpha_1 \rd s_1(p^\ast, k^\ast) + \alpha_2 \rd s_2(p^\ast, k^\ast) = 0\, .
}
This implies that $S_{1}$ and $S_{2}$ are tangent to each other at some point $(p^\ast,k^\ast) \in \Xs$. Indeed, we saw in~\eqref{eq:s_alpha_bubble_soln} that $\alpha_1$ and $\alpha_2$ take nonzero values at the normal threshold of the bubble. 
For the purpose of analytically continuing around this singular point in the space of external kinematics, it is useful to consider a small neighborhood $U\subset \Xs$ around the point $(p^\ast, k^\ast)$, so that we avoid complications coming from other singular configurations.  This also allows us to have a simple description of the homotopy path; otherwise we would need to make a choice between several possibilities as at the top of Fig.~\ref{fig:bubble_cycle2}.  We refer to $U$ as the \textbf{Leray bubble}. It should be small enough not to include any branch points other than the one we are interested in. Let us also define $U_p$ to be the slice of $U$ for which the external momentum takes a fixed value $p$.
It is from this point $p$ that we will compute the monodromy around $p^\ast$.\footnote{One can only analytically continue around smooth points on the Landau variety. To see why, recall how a path around a codimension-one variety is constructed: first we pick a transversal space at a point $p \in \LL$.  This space is complex one-dimensional, and chosen such that $p$ sits at the origin.  Then, a path around $p \in \LL$ is a path in this transversal one-dimensional complex space.  If the point $p$ is singular, then the transversal space does not necessarily exist.}$^{,}$\footnote{Note that points where multiple Pham loci intersect have to be treated specially.}

We also associate a \textbf{vanishing cell} $e$ with the singularity at $p^\ast$ by the conditions $s_1, s_2 \ge 0$. This cell vanishes as we approach the singularity $p \to p^\ast$ from above, since at $p^\ast$ we have $s_1 = s_2 =0$. In general, the vanishing cell associated with a singularity does not always have to be real. For now, we will simply assume it is real, postponing technical details of sufficient conditions until Sec.~\ref{sec:principal}.
Since the vanishing cell is bounded by the real singular surfaces $S_1$ and $S_2$, it is an element of the \textit{relative} homology class $[e] \in H_2(\Up, S_1 \cup S_2)$, namely of the second homology class of $\Up$ with boundaries in $S_1 \cup S_2$. Given such a cycle in $H_2(\Up, S_1 \cup S_2)$, we define boundary operators $\partial_i$ that restrict to the boundary of $e$ that intersects $S_i$. Thus, we have
\begin{align}
    \partial_1 \colon H_2(\Up, S_1 \cup S_2) & \to H_1(\Up \cap S_1, S_2)\, , \nonumber \\
    \partial_2 \colon H_2(\Up, S_1 \cup S_2) & \to H_1(\Up \cap S_2, S_1)\, , \label{ddHH} \\
    \partial_2 \circ \partial_1 \colon H_2(\Up, S_1 \cup S_2) & \to H_0(\Up \cap S_1 \cap S_2)\, . \nonumber
\end{align}
That is, the operator $\partial_1$ acts on the vanishing cell to give (the homology class of) its boundary in $\Up \cap S_1$. This boundary is shown as the green curve Fig.~\ref{fig:eyelashes}. If we then apply $\partial_2$ to this green curve, we just get the pair of intersection points that constitute $S_1 \cap S_2$. We call $\partial_2 \circ \partial_1 [e] \in H_0(\Up \cap S_1 \cap S_2)$ the \textbf{vanishing sphere}.

Next, we define the \textbf{vanishing cycle} to be the Leray coboundary of the vanishing sphere.  Leray coboundaries are higher-dimensional analogs of the coboundary we constructed in the example in Sec.~\ref{sec:alphaspace}, in which we used the fact that the coboundary of a point in the complex plane is a small circle around that point. 
Just as a point has real codimension two and its coboundary has real codimension one in the complex plane, the Leray coboundary operator $\delta$ acts on a cycle $\sigma$ of real dimension $k$ to produce a cycle $\delta \sigma$ of real dimension $k+1$. It does this by constructing a circle in the complex space transverse to $s$ for each point along the cycle. 
These circles for each point along the cycle $\sigma$ generate the coboundary $\delta \sigma$, which does not intersect the cycle $s$ itself. One can also define a tubular neighborhood by attaching a small enough disk centered at every point along the cycle.  Then the coboundary can be seen as the boundary of this tubular neighborhood. 
More technically, starting with a homology cycle which belongs to a submanifold $S$ in $\Xs$,\footnote{The condition that $S$ be a manifold, and therefore free of singularities, is essential for this construction.} the coboundary operator associates a homology cycle of one higher dimension in $\Xs \setminus S$. Applying this construction to the subvarieties $S_1$ and $S_2$ we get
\begin{gather}
    \delta_1 \colon H_0(\Up \cap S_1 \cap S_2) \to H_1((\Up \cap S_2) \setminus S_1)\, , \nonumber \\
    \delta_2 \colon H_0(\Up \cap S_1 \cap S_2) \to H_1((\Up \cap S_1) \setminus S_2)\, ,  \label{deltadeltaHH} \\
    \delta_1 \circ \delta_2 \colon H_0(\Up \cap S_1 \cap S_2) \to H_2(\Up \setminus (S_1 \cup S_2))\, . \nonumber 
\end{gather}
The vanishing cycle, given by coboundary of the vanishing sphere, is therefore $\delta_2 \circ \delta_1 \circ \partial_1 \circ \partial_2 [e]$. 

At a point just above the normal threshold, we thus have the following picture:
\begin{equation}
\resizebox{12cm}{!}{
\begin{tikzpicture}[baseline=(current  bounding  box.center)]
 \node (image) at (0,0) {
        \includegraphics[width=0.255\textwidth]{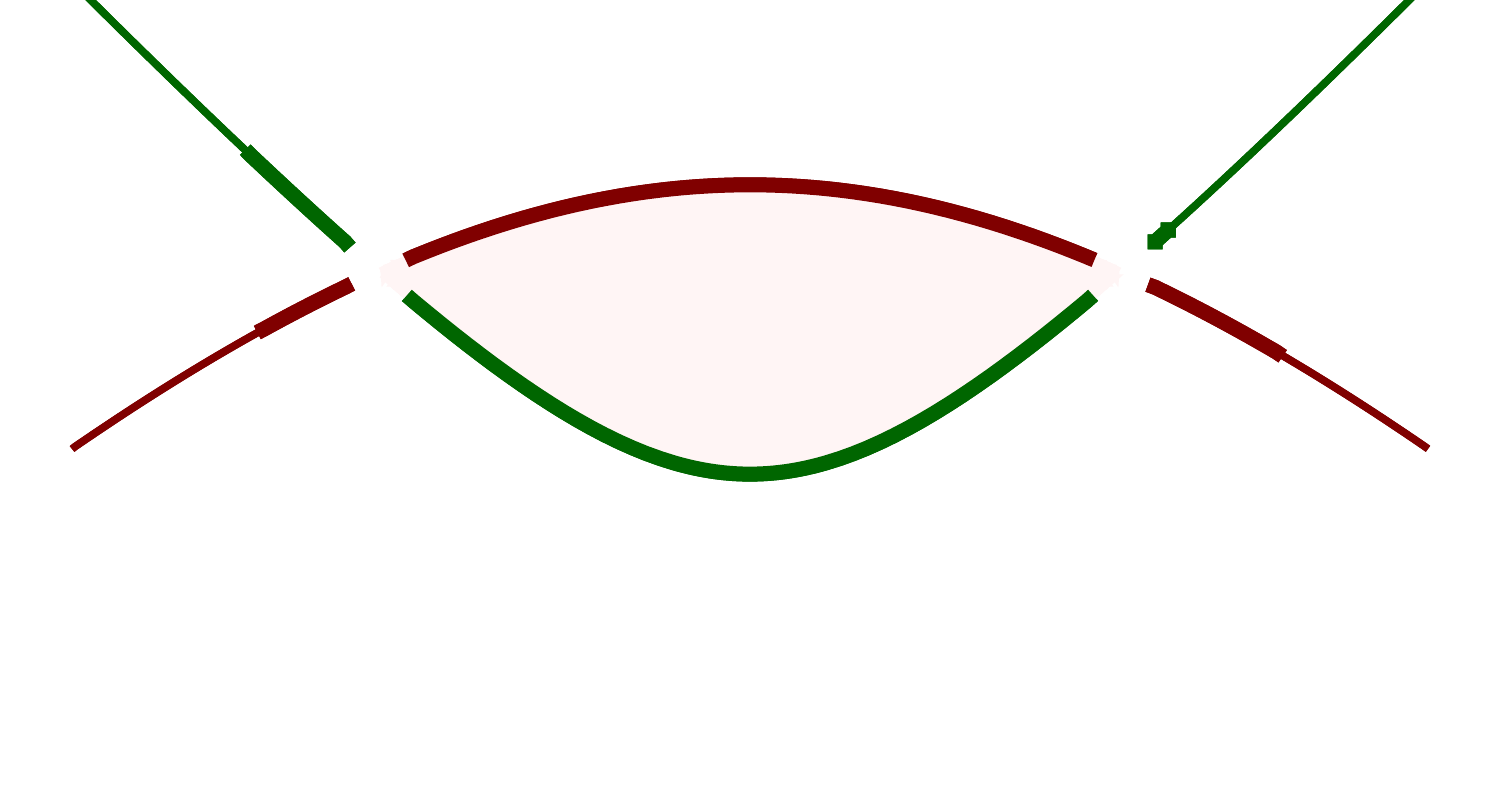}
    };  
\filldraw[black] (1,0.35) circle (0.02);
\filldraw[black] (-1,0.35) circle (0.02);
\draw[rotate=-20,thick,darkcyan!20,pattern=north west lines, pattern color=darkcyan!10] (0,0.2) ellipse (1.5 and 0.8);
\node[darkred,scale=0.5] at (-0.1,0.7) {$\Up \cap S_2 \setminus S_1 $};
\node[black,scale=0.5] at (2.3,0.35) {$~~~~\Up \cap S_1 \cap S_2 $};
\node[black,scale=0.5] at (2.5,0.15) {(vanishing sphere)};
\draw[-latex] (1.65,0.36) to[out=150,in=30] (0.6,0.5) to[out=210,in=-30] (-0.95,0.34);
\draw[-latex] (1.6,0.3) to[out=150,in=30] (1.3,0.4) to[out=210,in=0] (1.05,0.34);
\node[olddarkgreen,scale=0.5] at (0.1,-0.33) {$\Up \cap S_1 \setminus S_2 $};
 \node[darkcyan,scale=0.5,right] at (-0.8,1.3) {$\Up$ (Leray bubble over $p$)};
\node[olddarkgreen,scale=0.5] at (-1.45,1.0) {$S_1$};
\node[darkred,scale=0.5] at (-1.6,0.2) {$S_2$};
\node[darkred!60,scale=0.5] at (-3,0.3) {$U_p \cap (s_1\ge 0)\cap (s_2 \ge 0)$};
\node[darkred!60,scale=0.5] at (-3,0) {(vanishing cell $e_p$)};
\draw[darkred!60,-latex] (-2.5,0.5) to[out=30,in=160] (-0.2,0.3);  
\end{tikzpicture}
}
\label{eyelashU}
\end{equation}
\\[-10mm]
\noindent
Here, the black dots give the vanishing sphere, while the pink shaded area is the vanishing cell.
To construct the vanishing cycle, one first applies $\delta_1$ to construct a tubular neighborhood in the imaginary direction around the two dots, while staying on the surface $S_2$, then one applies $\delta_2$ to take a tubular neighborhood around the resulting object. The boundary of the resulting tubular neighborhood is a vanishing cycle. These coboundary maps are difficult to illustrate in ${\mathbb C}^2$, but are analogous to the complex circles around the endpoints of the vanishing cell depicted in Eq.~\eqref{circles}. Nevertheless, the vanishing cycle is a contour enveloping the vanishing sphere, which is the intersection of the boundary of the vanishing cell with the on-shell surface $S_1 \cap S_2$. 

Having introduced all of the required elements, we can finally state how the Picard--Lefschetz theorem allows us to compute the discontinuity of $\I_\bubrgT(p)$ around the threshold at $p^\ast$. We begin by analytically continuing $p$ in a small circle around the point $p^\ast$ in the space of external kinematics. While the original integration contour was given by an element $h\in [h] \in H_2(\Xs \setminus (S_1 \cup S_2))$, the contour has been modified to a different contour $h' \in [h']\in H_2(\Xs \setminus (S_1 \cup S_2))$ after this analytic continuation. The Picard--Lefschetz theorem tells us this monodromy in the space of external kinematics as an integral over the difference contour $h'-h$ by constructing a mapping
\begin{equation}
    \operatorname{Var} \colon H_2(\Xs \setminus (S_1 \cup S_2)) \to H_2(U \setminus (S_1 \cup S_2)),
\end{equation}
which maps the homology class $[h]$ to the vanishing cycle. In particular, the theorem says that
\begin{equation}
    \operatorname{Var} [h] = \langle e, h\rangle (\delta_1 \delta_2 \partial_2 \partial_1 [e]),
\end{equation}
where we have omitted ``$\circ$'' between operators for conciseness. This formula is valid for two-dimensional cycles; we will present the general formula below.  The integer $\langle e, h\rangle$ is called the \textbf{Kronecker index}, and is the intersection number between the vanishing cell and the original integration contour $h$. Applying this formula to the bubble integral then gives
\begin{equation} \label{eq:bubble_pic_lef}
    \left(\bbone - \monM_{s=(m_1+m_2)^2} \right) \I_{\bubrgT}(p) = -\langle e, h\rangle \int_{\delta_1 \delta_2 \partial_2 \partial_1 e} \frac{\rd k_0 \wedge \rd k_1}{[k^2-m_1^2][(p-k)^2-m_2^2]} \,,
\end{equation}
where $\delta_1 \delta_2 \partial_2 \partial_1 e$ denotes a contour $c$ in the homology class $\delta_1 \delta_2 \partial_2 \partial_1 [e]$.  The extra sign arises from the fact that the left-hand side computes the integral over $h - h'$ while the Picard--Lefschetz theorem computes $\operatorname{Var}[h] = [h'] - [h]$.

\subsection{Multivariate Residues}
\label{sec:multivariate_res}

The final step in obtaining the discontinuity of the bubble integral as an absorption integral involves rewriting the modified integration contour as a set of $\delta$-functions in momentum space. Intuitively, this involves replacing the integral of $\omega$ over the Leray coboundary with a multivariate residue, similar to what we did in our one-dimensional example to get Eq.~\eqref{eq:mon_bub_delta}. Fortunately, the Leray coboundary is precisely the ingredient that enters the theory of multivariate residues, as discussed in the context of Feynman integrals in Ref.~\cite{Abreu:2017ptx}. 

In carrying out this type of analysis, it will be helpful to use notation in which we allow ourselves to divide by forms, for instance writing
\begin{equation}
    \psi = \frac{\omega}{\rd s_1 \wedge \rd s_2}  \quad \Leftrightarrow\quad \omega = \rd s_1 \wedge \rd s_2 \wedge \psi \, ,
\end{equation}
where the expression on the left makes sense whenever the quantity on the right exists. This notation was introduced by Gelfand and Chilov~\cite{MR0132390} and later employed by Leray and Pham.\footnote{More details for this notation can be found in~\cite{pham1968singularities}.} To connect it to the language normally employed by physicists, we can write these divisions by differential forms in terms of Dirac $\delta$-functions. To do this, we use Poincar\'e duality (see App.~\ref{sec:signs}) to extend the integration to the embedding space:
\begin{equation}
    \int_{S_1 \cap S_2} \frac{\omega}{\rd s_1 \wedge \rd s_2} = \int_{U} \delta(s_1) \delta(s_2) \rd s_1 \wedge \rd s_2 \, \frac{\omega}{\rd s_1 \wedge \rd s_2} =
    \int_U \delta(s_1) \delta(s_2) \omega \,, \label{eq:convert_to_deltas}
\end{equation}
where $U$ is a neighborhood of the intersection $S_1 \cap S_2$.
As an example, suppose that $\omega$ takes the form $\omega = \psi(s_1,s_2,s_3) \, \rd s_1 \wedge \rd s_2 \wedge \rd s_3$, while $S_1$ and $S_2$ denote the surfaces defined by $s_1=0$ and $s_2=0$ respectively. Then \begin{align}
    \int_{S_1 \cap S_2} \frac{\omega}{\rd s_1 \wedge \rd s_2}\Big|_{S_1 \cap S_2}
    &= \int_{s_1=s_2=0} \frac{\omega}{\rd s_1 \wedge \rd s_2}\Big|_{s_1=s_2=0} = \int \rd s_3 \psi(0,0,s_3) \\
    &= \int \delta(s_1) \delta(s_2) \psi(s_1,s_2,s_3) \rd s_1 \rd s_2 \rd s_3.
\end{align}
Effectively, we have that
\begin{equation}
    \delta(s_i) \sim \frac{1}{\rd s_i}.
\end{equation}
This relation acts as a useful mnemonic for remembering the transformation properties of $\delta(s_i)$, if nothing else. 

Now, suppose $\omega$ has simple poles at $s_1 = 0$ and $s_2 = 0$, so that we can write
$\omega = \frac{\rd s_1}{s_1} \wedge \frac{\rd s_2}{s_2} \wedge \psi$, where $\psi$ is regular on $S_1\cup S_2$. The \textbf{multivariate residue} of such a form is\footnote{Here we have adopted the convention that $\operatorname{res}_S \frac{\rd s}{s} \wedge \psi = \psi\vert_S$.  This is the convention of Leray (see Ref.~\cite{BSMF_1959__87__81_0}) and of Pham (see Ref.~\cite{pham2011singularities}).  This convention is however not universal; Hwa \& Teplitz (see Ref.~\cite{Hwa:102287}) use the opposite convention where $\operatorname{res}_S \psi \wedge \frac{\rd s}{s} = \psi\vert_S$.}
\begin{equation}
\operatorname{res}_{S_2} \operatorname{res}_{S_1}\omega
 = \psi \big\vert_{S_1 \cap S_2}
=  s_1 s_2 \frac{\omega}{\rd s_1 \wedge \rd s_2}\Big|_{S_1 \cap S_2}.
\end{equation}
Then, if we have a cycle $\sigma \in S_1 \cap S_2$ with coboundary $\delta_1 \delta_2 \sigma$, the integral over the coboundary gives
\begin{equation}
    \int_{\delta_1\delta_2 \sigma} \omega 
    = (2\pi i)^2 \int_\sigma \operatorname{res}_{S_2} \operatorname{res}_{S_1}\omega.
\end{equation}
This is the \textbf{Leray residue formula}, which generalizes Cauchy's theorem to the multivariate case.

Starting with the two-form $\omega$ for the two-dimensional bubble integral (which we gave in Eq.~\eqref{eq:bubble_form}), we can change variables so that $\omega = \frac{\rd s_1}{s_1} \wedge \frac{\rd s_2}{s_2} \wedge \psi$.
In this case, $\psi$ is a zero-form (a function). 
We can therefore replace the coboundaries $\delta_1 \delta_2$ with the corresponding residues around $S_1$ and $S_2$. Combining Leray's residue formula with the Picard--Lefschetz formula relating the monodromy to the integral over the vanishing cycle, we then have for the bubble integral $\I_{\bubrgT}(p)$,
\begin{equation}
    \left(\bbone - \monM_{s=(m_1+m_2)^2} \right) \I_{\bubrgT}(p) = (2 \pi i)^2 \int_{\partial_2 \partial_1 e} \operatorname{res}_{S_2} \operatorname{res}_{S_1} \omega
    = (2 \pi i)^2 \int_{\partial_2 \partial_1 e} s_1 s_2 \frac{\omega}{\rd s_1 \wedge \rd s_2} \,,
\label{eq:cutkosky_bubble0}
\end{equation}
which we can write using the Poincar\'e duality as
\begin{equation}
    \left(\bbone - \monM_{s=(m_1+m_2)^2} \right) \I_{\bubrgT}(p) = (2 \pi i)^2 \int_{U_p} \delta(s_1) \delta(s_2) \psi \,.
\label{eq:cutkosky_bubble}
\end{equation}
Recalling that $U_p$ must be defined to include only the neighborhood around the intersections of $S_1$ and $S_2$, we see that if we restrict the integration contour to be over the positive part of the energies of the two cut particles, the pseudothreshold will not contribute.
We can enforce this by including appropriate theta functions, and so we compute Eq.~\eqref{eq:cutkosky_bubble} by writing
\begin{equation}
    \left(\bbone - \monM_{s=(m_1+m_2)^2} \right) \I_{\bubrgT}(p) =
    (2 \pi i)^2 \int \rd^2 k \, \theta(k^0) \delta(k^2-m_1^2)  \theta(p^0 - k^0) \delta[(p-k)^2-m_2^2]  \,,
\end{equation}
where we have assumed that $p^0 > 0$.  If $p^0 < 0$ then we need to impose negative-energy conditions on the cut momenta.  This integral can be evaluated to give
\begin{equation}
    \left(\bbone -\monM_{s=(m_1+m_2)^2} \right) \I_{\bubrgT}(p) = \frac{-4 \pi^2}{\sqrt{[s-(m_1-m_2)^2][s-(m_1+m_2)^2]}} \,,
\end{equation}
which agrees with Eq.~\eqref{eq:mon_bub}, as well as the direct monodromy computation in Eq.~\eqref{eq:bub_full}.

\subsection{Cutkosky from Picard--Lefschetz}
\label{sec:intro_PL}
Finally, let us move beyond the example of the bubble integral to state the analogous Picard--Lefschetz result for generic scalar Feynman integrals. Consider the Feynman integral associated with a graph $G$:
\begin{equation} \label{eq:feyn_int_PL}
\I_G(p) = \int_h \prod_{c \in \Chat(G)} \rd^d k_c \prod_{e \in \Eint(G)} \frac{1}{\left[q_e(k, p)\right]^2-m_e^2} \,.
\end{equation}
This integral involves the differential form
\begin{equation}
    \omega = \bigwedge_{c \in \Chat(G)} \rd^d k_c \prod_{e \in \Eint(G)} \frac{1}{\left[q_e(k, p)\right]^2-m_e^2} \, ,
\end{equation}
which is a representative of the cohomology class $H^{n}(\Xs \setminus (S_1 \cup S_2 \cup \cdots \cup S_{\nint}))$, and a contour $[h] \in H_n(\Xs \setminus (S_1 \cup S_2 \cup \cdots \cup S_{\nint}))$. As before, we take each of the surfaces $S_e$ to be defined by the on-shell equations $s_e(k,p) = \left[q_e(k, p)\right]^2-m_e^2 = 0$.

Now consider the complexified and compactified space of on-shell external kinematics associated with $I_G(p)$, which we denote by $X$. For every value of the external kinematics $p \in X$, there exists a fiber $\Xp$ of complex dimension $n$, which corresponds to the domain over which $\omega$ is integrated in~\eqref{eq:feyn_int_PL}. 
Each copy of the space $\Xp$ contains a set of codimension-one varieties that are defined by the on-shell conditions. While these codimension-one varieties will be in a general position for generic values of the external kinematics, they will not be in a general position in the fibers over points in the Landau variety. Recall, for example, that for the bubble Feynman integral from Sec.~\ref{sec:momspacePL}, the Landau branch point was the projection of the momentum-space configuration in which the on-shell surfaces $S_1$ and $S_2$ became tangent in loop-momentum space to the space of external variables. The higher-dimensional analog of this condition is when the normals to the surfaces $S_1, S_2, \ldots S_m$ become linearly dependent. 

The Picard--Lefschetz theorem tells us how to obtain the difference in the integration contour before and after analytic continuation around a branch point from knowledge of just the boundary operators and coboundary operators associated with the surfaces $S_1,\dots,S_m$, given by
\begin{equation}
    \partial_1 \colon H_* (U, S_1 \cup S_2 \cup \cdots \cup S_m) \to H_{*-1} (U \cap S_1, S_2 \cup \cdots \cup S_m) \,,
\end{equation}
\begin{equation}
    \delta_m \colon H_*(S_1 \cap S_2 \cap \cdots \cap S_m) \to H_{*+1} (S_1 \cap \cdots S_{m-1} \setminus S_m)\,.
\end{equation}
Specifically, the Picard--Lefschetz theorem can be applied to Feynman integrals as follows:
\begin{theorem*}[Picard--Lefschetz]
We consider a closed form $\omega$ corresponding to the integrand of a Feynman integral, a closed contour $h$, and a codimension-one branch $\mathcal{\P_{\kappa}}$ of its Landau variety where the integration contour may be pinched between some number of surfaces $S_1,S_2,\ldots,S_m$, for $m \leq \nint$.  We impose the constraint that the pinch is \emph{simple}, which means that it fits in a single Leray bubble $U \subset X \setminus\bigcup_{i = 1}^m S_i$.
We then consider a homotopy class $\gamma \in \pi_1(X \setminus \P_{\kappa})$ which fits in the projection of the Leray bubble $U$ to external kinematics and represents the path along which we are interested in analytically continuing $\int \omega$ in the space of external kinematics.
Under the action of this analytic continuation, the homology class $[h]$ is sent to $[h'] = \gamma_* [h]$. The variation 
\begin{equation}
    \operatorname{Var} \colon H_n(\Xs \setminus \bigcup_{i = 1}^m S_i) \to H_n(U \setminus \bigcup_{i = 1}^m S_i)
\end{equation}
associated with this map is defined by $\operatorname{Var}[h] = [h'] - [h]$.
For analytic continuations around a codimension-one branch locus $\P_{\kappa}$ that are confined to an infinitesimal region around $p^\ast \in \P_{\kappa}$, we have %
    \begin{equation}
    \operatorname{Var} h = (-1)^{\frac{(n+1)(n+2)}{2}} \langle e, h\rangle (\delta_1 \dotso \delta_m \partial_m \dotso \partial_1 e),
    \label{eq:PLtheorem}
\end{equation}
where $\langle e, h\rangle$ is the integer-valued intersection number between the vanishing cell $e$ and the contour $h$.
\end{theorem*}
\noindent Namely, using this theorem we are able to express the discontinuities of $\I_G(p)$ as an integral over the \textit{original} form $\omega$, but with a new integration contour given by $(\delta_1 \dotso \delta_m \partial_m \dotso \partial_1 e)$. For Feynman integrals involving generic masses, we therefore obtain that its discontinuities can be written as:
\begin{align}
    \left(\bbone-\monM_{p=p^\ast}\right) I_G(p) &= \int_h \omega - \int_{h'} \omega \nonumber \\
    &= - \int_{\operatorname{Var} h} \omega \, \\
    &= - (-1)^{\frac{(n + 1)(n + 2)}{2}} \langle e, h\rangle \int_{\delta_1 \dotso \delta_m \partial_m \dotso \partial_1 e} \omega. \nonumber 
\end{align}
Since we can apply the Picard--Lefschetz theorem to any closed contour $h$, we can also apply it to compute sequential discontinuities of $I_G(p)$. For example, the Picard--Lefschetz theorem tells us that when encircling the same branch point iteratively, we get
\begin{equation}
    \left(\bbone-\monM_{p=p^\ast}\right)^r I_G(p) = (-1)^r \int_{\operatorname{Var}^r h} \omega \,.
\end{equation}
We will apply the Picard--Lefschetz theorem in this iterative fashion in later sections.

It is important to emphasize that the vanishing cell, the vanishing sphere, and the vanishing cycle are all in one-to-one correspondence. That is, the maps in Eq.~\eqref{ddHH} and Eq.~\eqref{deltadeltaHH} (which respectively construct the vanishing sphere from the vanishing cell, and the vanishing cycle from the vanishing sphere) are isomorphisms. Therefore, all three of these ingredients are uniquely determined by any single one of them.\footnote{This is not true when there are at least as many singular denominators as the dimension of the integration cycle. We will consider an example of this type in Sec.~\ref{sec:critical}.} Thus, when computing sequential monodromies, it is sufficient to track the vanishing sphere, which lives within the on-shell surface $S_1\cap \cdots \cap S_m$, rather than the vanishing cell or vanishing cycle, which live in higher dimensions. 

We can now use the Picard--Lefschetz theorem on the Feynman integral $\I_G(p)$ to obtain
\begin{equation}
    \left(\bbone - \monM_{p=p^\ast} \right) \I_{G}(p) = \int_{\delta_1\cdots \delta_m \partial_m\cdots \partial_1 e} \omega \,. \label{eq:picard_lefschetz_formal}
\end{equation}
Here, we have used the fact (shown in App.~\ref{sec:kronecker_index}) that $\langle e, h\rangle = (-1)^{\frac{n (n - 1)}{2}}$, which (when multiplied by the prefactor $(-1)^{\frac{(n + 1)(n + 2)}{2}}$ in the Picard--Lefschetz theorem) yields an overall factor of $-1$, which is compensated by the fact that here we are computing the difference between the function and its analytic continuation around the singularity, while the Picard--Lefschetz theorem computes the difference between the function \emph{after} and before analytic continuation.
We can rewrite this using the identities from Sec.~\ref{sec:multivariate_res} as,
\begin{multline}
    \left(\bbone - \monM_{p=p^\ast} \right) \I_{G}(p) = (2 \pi i)^m \int_{\partial_m\cdots \partial_1 e} \operatorname{res}_{S_m}\cdots \operatorname{res}_{S_1} \omega  \\ = (2 \pi i)^m \int_{\partial_m\cdots \partial_1 e} s_1 \cdots s_m \frac{\omega}{\rd s_1 \wedge\cdots \wedge \rd s_m} \, .
\end{multline}
We finally apply the Poincar\'e duality from Eq.~\eqref{eq:stokes_with_theta_iter} in App.~\ref{sec:signs} to get
\begin{equation}
    \left(\bbone - \monM_{p=p^\ast} \right) \I_{G}(p)
    = (-2 \pi i)^m \int_{U_p} \delta(s_1)\cdots \delta(s_m)
\prod_{c \in \Chat(G)} \rd^d k_c \prod_{e \not\in \{1,\dots , m\}} \frac{1}{s_e} \,.
\label{IGmform3}
\end{equation}
This expression matches Cutkosky's formula in Eq.~\eqref{eq:absorption_int_2}. Note that the positive-energy flow, which was denoted with a theta function in Eq.~\eqref{eq:absorption_int_2}, is implicit in the restriction to the Leray bubble $U_p$: the signs of the energies of $s_1, \ldots, s_m$ must match the ones of the Landau-equation solution.

\subsection{Complex Integration Contours}
\label{sec:iepaths}

In employing Picard--Lefschetz theory, we have preferred to deform the integration contour in a way that is consistent with the $i \varepsilon$ prescription, rather than keep these infinitesimal imaginary factors in the integrand. We end this section by commenting on how this can be done in general. Namely, we will describe the contour $h$ that allows Feynman integrals to be written in the form
\begin{equation}
    \I_G (p) =  \prod_{c \in \Chat(G)} \int_{h} \rd^d k_c  \prod_{e \in \Eint(G)} \frac{1}{s_e} \,,
\end{equation}
where the singular surfaces in the denominator are given by
\begin{equation}
    s_e \equiv \left[q_e(k, p)\right]^2-m_e^2 \,.
\end{equation}
Since the integrand $\I_G(p)$ can become singular at complex points, we want to make sure that the integration cycle $h$ not only avoids the singularities in the complex space of loop momenta, but is also consistent with the $i \eps$ prescription for momenta in the physical region.

Before we describe how to construct such a contour, let us discuss the advantages of this construction over the standard Feynman $i \varepsilon$ prescription. First, the analytic structure of the integrand is not modified---its branch points are not displaced by an unphysical, dimensionful parameter $\varepsilon$. %
Second, since the value of the integral is determined by the homology class of the integration contour, and not the exact path taken, one does not need to consider infinitesimal deformations from the real contour to get the correct physical value of the Feynman integral. Instead, it is sufficient to take the contour deformation to finite but small, making sure that it leads to the correct homotopy class. It is therefore better-suited for numerical computations than deforming the integrand by $i \varepsilon$.\footnote{For a discussion of a similar contour deformation for numerical evaluations of amplitudes using local unitarity, see Ref.~\cite{Capatti:2019edf}.}

To construct a contour $h$ that is in the correct homology class, we will need to deform slightly away from real values of loop momenta. To visualize this deformation in the complex neighborhood of each real point, we can construct a vector field whose components depict the size of the deformation in each imaginary direction.  Indeed, suppose we have a variety $S=S_1 \cup S_2 \cup \cdots \cup S_m$, where each $S_e$ is given by a real local equation $s_e(k^\mu) = 0$, where $k^\mu$ are also real vectors.
Extending the on-shell equation to complex variables amounts to imposing the condition $s_e(k^\mu + i v^\mu) = 0$, where $v^\mu$ is a real vector. This means that the variety $S$ satisfies $\Re\, s_e(k^\mu + i v^\mu) = 0$ and $\Im\, s_e(k^\mu + i v^\mu) = 0$.  By definition, the points in a small complex neighborhood are such that the components of $v^\mu$ are small, so we can expand the on-shell equation to obtain
\begin{equation}
    \label{eq:singular_surfce_imaginary_expansion}
    s_e(k^\mu) + i v^\nu \partial_{k^\nu} s_e(k^\mu) + \mathcal{O}(v^2) = 0.
\end{equation}
If $v^\mu(k)$ is a small tangent vector at the point of $S$ of coordinates $k^\mu$, then this equation is solved to first order in $v^\mu$, since $s_e(k^\mu)=0$ by assumption.  Therefore, in order for the integration cycle to avoid complex singularities it is sufficient to pick an imaginary part $k^\mu \to k^\mu + i v^\mu(k)$ such that $v^\mu$ is transversal to the real points of $S$. In other words, we see from Eq.~\eqref{eq:singular_surfce_imaginary_expansion} above that it is equivalent to choosing transverse vectors
such that $v \cdot \partial s_e \neq 0$, or equivalently $\rd s_e(v^\mu) \neq 0$.\footnote{In loop-momentum space, the partial derivative is taken with respect to the loop momenta $k^\mu$. The vectors $v$ are then normals to the on-shell surfaces $S_e$ in the space of the loop momenta.} To be consistent with the causal $i \varepsilon$ prescription, we choose the orientation of $v^\mu$ such that $\Im \, s_{e} >0$, which amounts to
\begin{equation}
    v^\nu \partial_{k^\nu} s_e(k^\mu) >0 \,.
\end{equation}
For principal Pham loci, we can choose the signs of $\Im \, s_{e}$ uniformly, and then the vector field will be either inward-pointing or outward-pointing on the boundary of the vanishing cell, which we recall is defined by $s_{e} \geq 0$ for all edges that enter the definition of $S$.

As an example of this construction, let us work out the transverse vector field for the bubble diagram in two dimensions. We take $s_1=k^2-m_1^2$ and $s_2=(p-k)^2-m_2^2$, where $p^\mu=(Q,0)$ in the center-of-mass frame.  Then, choosing a transverse component of the imaginary part amounts to finding a vector field $v^\mu$ that satisfies $\rd s_1 (v^\mu) \equiv 2 k \cdot v > 0$, and $\rd s_2 (v^\mu) \equiv 2 (k-p) \cdot v > 0$. The changes in $s_1$ and $s_2$ as we shift by $i \varepsilon v^\mu$ are obtained as
\begin{equation}
    \Delta s_e \equiv s_e (k^\mu+ i \varepsilon v^\mu) - s_e (k^\mu) \approx i \varepsilon \, \rd s_e (v^\mu) 
    \, .
\end{equation}
Written out in components, this becomes
\begin{equation}
    \rd s_1 = 2 k^0 \rd k^0- 2 k^1 \rd k^1 \,, \qquad \rd s_2 = 2 (k^0-p^0) \rd k^0  - 2 k^1 \rd k^1 \,.
\end{equation}
A \textit{tangent} vector to the curve bounded by $s_1=0$ is therefore given by $v_{\parallel, \pm}^\mu = \pm (k^1,k^0)$ while a \textit{transverse} vector is given by $v_{\perp, \pm}^\mu = \pm (k^0,-k^1)$. We construct the imaginary part at each point to include the \textit{transverse} vector field, which gives $\Delta \Im \,s_1 = 2 \varepsilon \left[ (k^0)^2+(k^1)^2 \right]$.
This choice of transverse vector leads to $\Im \, s_1>0$, and is therefore equivalent to deforming $s_1$ by $+ i\varepsilon$ in the limit as $\varepsilon \to 0^+$.

This type of construction---of a vector field with transversal component for the bubble integral---was already depicted in Sec.~\ref{sec:momspacePL}, where we found that the threshold and pseudothreshold constructions correspond to the following configurations:
\begin{equation}
\hspace{1.2cm}
\resizebox{6cm}{!}{
\begin{tikzpicture}[baseline=(current  bounding  box.center)]
 \node (image) at (0,0) {
        \includegraphics[width=0.255\textwidth]{Figures/normalonshellvin.pdf}
    };  
\end{tikzpicture}
}
\resizebox{6cm}{!}{
\begin{tikzpicture}[baseline=(current  bounding  box.center)]
 \node (image) at (0,0) {
        \includegraphics[width=0.255\textwidth]{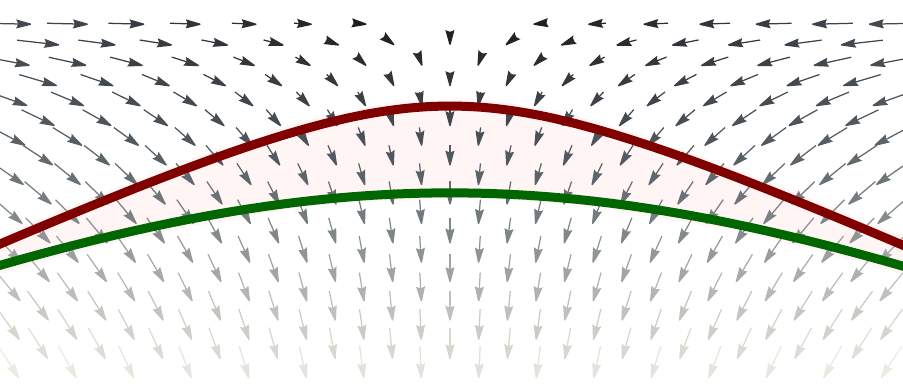}
    };  
\end{tikzpicture}
}
\label{eq:vectorfield_rep}
\end{equation}
We see that for the threshold, there will be at least one point in $k^\mu$ space where this construction of an integration contour that avoids the singular surfaces $S_1$ and $S_2$ will inevitably fail. Intuitively, the vector field $v^\mu$ has inward-pointing components at the boundary of the vanishing cell (defined by $s_1=0$ and $s_2=0$); since $v^\mu$ is taken to be continuous, this implies that there must exist at least one point at which $v^\mu$ vanishes.\footnote{To apply the Picard--Lefschetz theorem, one can keep the vanishing cell finite, and the intersection point can be chosen anywhere within the cell. Note, however, that as the critical value at $s=(m_1+m_2)^2$ is approached, and the vanishing cell shrinks to zero at a certain point in the loop-momentum space, the intersection point necessarily coincides with that point. } Taking the vanishing cell to be small shows that the singularity is unavoidable as the critical value $s=(m_1+m_2)^2$ is approached, leading to a singularity or a branch point of the integral $I_G(p)$. This analysis, supplemented by the arguments for the sign in App.~\ref{sec:kronecker_index}, yields $\langle e, h \rangle=1$ for the intersection number. Conversely, we see in that in the case of the pseudothreshold, there is no obstruction to taking a non-zero vector field that avoids the integration contour at all points. Hence, the integral $I_G(p)$ does not have a singularity at $s=(m_1-m_2)^2$.

We remark that in general, only two choices of the vector field for the vanishing cell bounded by the surfaces $S_e$ lead to a non-vanishing intersection number: either all outward or all inward pointing. For $\alpha$-positive singularities, in which the vanishing cell is bounded by $s_e\geq 0$, we can take the contour deformation that is consistent with causality to be uniformly inward-pointing. If the vanishing cell is bounded by a different set of inequalities, which require $s_e \geq 0$ for some edges and $s_e\leq 0$ for others, the intersection number between the contour and the vanishing cell will vanish.

Finally, let us discuss how to relate a contour deformation in the loop momenta $k^\mu$ to a change in the external variables $p$. Viewing the Feynman integral as $\varepsilon \to 0^+$ as a function of complexified kinematics $p$ with branch cuts in the complex plane, we can ask from which direction in the complexified kinematics one should approach the physical value. This approach must be consistent with causality, so we use the construction analogous to the one in Eq.~\eqref{formvanish}. The detours in external kinematics dictate on which side of the branch cut the physical amplitudes are evaluated. If we take the bubble integral to have internal momenta $k^\mu$ and $p^\mu - k^\mu$, then $k^\mu = \frac{m_1}{m_1 + m_2} p^\mu$ at the threshold where $s = (m_1 + m_2)^2$. But since the causal deformation in internal momentum is according to $k \to k + i \varepsilon v$, then we can equivalently deform the external momentum to $p \to p + i \varepsilon \frac{m_1+m_2}{m_1} v$. Hence, since $p \cdot v > 0$ to leading order in $\varepsilon$, we see that $s \to s + i\varepsilon$ for $s>(m_1+m_2)^2$. In more general examples, taking $+ i\varepsilon$ in the propagators does not necessarily correspond to taking $s \to s + i\varepsilon$ in the physical $s$ channel. For explicit examples where a positive imaginary part for $s$ fails to be equivalent to the causal prescription, see Ref.~\cite{Hannesdottir:2022bmo}.

  \section{Discontinuities of Absorption Integrals}
\label{sec:iterated}
In Sec.~\ref{sec:PicardLefschetz}, we saw how the Picard--Lefschetz theorem can be combined with Leray's theory of residues to reproduce Cutkosky's formula for absorption integrals.  While Cutkosky's formula is already known, 
the same technique can be used to compute the discontinuities of absorption integrals, or equivalently sequential discontinuities of Feynman integrals. 
Anticipating 
that we will want to compute such discontinuities in later sections, we now apply the methods developed in the last two sections to absorption integrals. More specifically, we study where absorption integrals can become singular, and illustrate how discontinuities around these singularities can again be computed by modifying the integration contour to the coboundary of the vanishing sphere corresponding to the relevant Pham locus.

\subsection{Singularities of Absorption Integrals}
We begin our study of absorption integrals by recalling that these integrals can be written as Feynman integrals in which the integration contour has been modified (as in Eq.~\eqref{eq:picard_lefschetz_formal}), rather than in terms of $\delta$-functions as done in Cutkosky's formula.
It follows that the Landau equations also describe the singularities of absorption integrals, since these equations make no reference to the integration contour.
However, stronger conditions than the Landau equations can be placed on the locations of singularities in absorption integrals. Just as we can use the $\alpha$-positivity of the integration contour in Feynman integrals to restrict to singularities on the physical sheet, we can exploit what we know about the integration contour that appears in absorption integrals: this integration contour forces certain propagators to be on-shell. It follows that when staying inside the Leray bubble, absorption integrals can only develop singularities on the branches of the Landau variety where these propagators are also on shell.

Recall from Sec.~\ref{sec:absorption_integrals} that the absorption integral $\cA_G^\kappa$ is defined to be the cut integral corresponding to the discontinuity of $\I_G(p)$ around the Pham locus $\P_{\kappa}$. It is associated with the singularity locus in which all of the propagators in the diagram $G^\kappa$ are on-shell and Landau loop equations for the loops in $G^\kappa$ are satisfied, and can be represented by the sequence
\begin{equation}
 G\xrightarrowdbl{~~~} G^\kappa \xrightarrowdbl{~\kappa~} G_0 \, .
\end{equation}
To study singularities of $\cA_G^\kappa$ we need to consider longer sequences of contractions, like
\begin{equation}
    G \xrightarrowdbl{~~~} G^\kappap \xrightarrowdbl{~\kappab~} G^\kappa \xrightarrowdbl{~\kappa~} G_0 \, .
    \label{longerseq}
\end{equation}
We then need to determine to what extent the on-shell and loop conditions of the first discontinuity may be relaxed. 

Now, the integration contour in  $\cA_G^\kappa$ forces all the propagators in $G^\kappa$ to be on-shell. This is made clear in Cutkosky's formulation of the absorption integral in terms of cut lines and $\delta$-functions. Thus, singularities of $\cA_G^\kappa$ should maintain these on-shell conditions, but can have additional lines on-shell. On the other hand, the loop equations which were imposed to find the singularity $\P_\kappa$ are not necessarily enforced by the absorption integral. Again, writing this integral as over $\delta$-functions imposing on-shell constraints does not restrict to the single phase-space point we get from solving the loop equations. 
With these observations in mind, let us now rewrite the  sequence of contractions in Eq.~\eqref{longerseq} as
\begin{equation}
\label{dominating}
\begin{gathered}
\begin{tikzpicture}
\node[left] at (-1.5,0) {$G$};
\draw [->>] (-1.5,0) -- (-0.7,0) ;
\node[right] at (-0.7,0) {$G^\kappap$};
\draw [->>] (0,0) -- (1,0) ;
\draw [->>] (1.3,-0.3) -- (1.3,-1.2);
\draw [->>] (0,-0.2) -- (1.1,-1.2);
\node [right] at (1,0) {$G^\kappa$};
\node [right] at (1,-1.5) {$G_0$};
\node [above] at (0.5,0) {$\kappab$};
\node [] at (1.6,-0.6) {$\kappa$};
\node [] at (0.3,-0.8) {$\kappap$};
\end{tikzpicture}
\end{gathered} 
\end{equation}
When such a diagram can be formed, with $\kappap = \kappa \circ \kappab$ we say that the contraction $\kappap$ \textbf{dominates} $\kappa$. 
Using this terminology, our observations above suggest that singularities of $\cA_G^\kappa$ should be on Pham loci associated with contractions that dominate $\kappa$. This result can be stated as a theorem~\cite{pham}:
\begin{theorem}[Pham] \label{thm:pham}
The absorption integral $\cA^{\kappa}_{G}(p)$ associated with the contraction $\kappa$ and the sequence $G \twoheadrightarrow G^\kappa \twoheadrightarrow G_0$ is analytic everywhere for $p$ in physical regions, outside of the Pham loci $\P_\kappap$ of contractions $\kappap$ that dominate $\kappa$ through sequences $G \twoheadrightarrow G^\kappap \twoheadrightarrow G^\kappa \twoheadrightarrow G_0$.
\end{theorem}
\noindent
We sketch the proof of this theorem in Appendix~\ref{sec:sing_absorption}.

At this point, it is worth making some historical comments. A similar result can be found in the $S$-matrix literature from the 1960s under the name of the hierarchical principle (see for instance~\cite{Landshoff1966,boyling1968homological}). In that context, one begins with an elementary graph $G_0$ and proceeds to sequentially expand internal vertices by adding internal lines and loops (Pham also took this perspective). The diagram that appears at each step in this expansion is taken to represent a leading singularity---that is, the part of the Landau variety in which all of the Feynman parameters in this diagram are nonzero---with smaller diagrams representing \emph{lower-order} singularities, and larger ones representing \emph{higher-order} singularities. The \textbf{strict hierarchical principle} was then the claim
that 
the singularities of an 
absorption integral $\cA_G^\kappa$ should only be on the part of the Landau variety associated with higher-order leading singularities~\cite{Landshoff1966}. The hope was that this nesting of singularities would simplify the problem of determining the full analytic structure of a given scattering process by allowing one to study the lower-order singularities first. Namely, if a singularity is not 
the leading singularity of some lower-order diagram, it should not appear as a singularity of the diagram with more internal lines inserted---or, stated differently, higher-order diagrams should not suddenly switch on lower-order singularities. However, the strict hierarchical principle was found to have exceptions. One example is the bubble-like singularity of the ice cream cone~\cite{Landshoff1966}, which corresponds to a singularity where all lines of the ice cream cone are on shell, but two of the Feynman parameters are zero:
\begin{equation}
G_{\icet}^{\text{bubble like}}
=
\begin{tikzpicture}[baseline=(current bounding box.center),scale=0.8]
\path [darkred,line width=1.07,out=130,in=230] (0,-0.6,0) edge (0,0.6);
\path [darkred,line width=1.07,out=50,in=-40] (0,-0.6,0) edge (0,0.6);
\node[darkred, scale=0.8] at (3.5,0.8) {$q_3^2=m_3^2$};
\node[darkred, scale=0.8] at (3.5,0.4) {$\alpha_3=0$};
\node[darkred, scale=0.8] at (3.5,-0.8) {$q_4^2=m_4^2$};
\node[darkred, scale=0.8] at (3.5,-1.2) {$\alpha_4=0$};
\draw [newdarkblue2,line width=1.07] (0,0.6) -- (1.5,0);
\draw [newdarkblue2,line width=1.07] (0,-0.6) -- (1.5,0);
\node[newdarkblue2, scale=0.8] at (1.2,1.2) {$q_1^2=m_1^2$};
\node[newdarkblue2, scale=0.8] at (1.2,0.8) {$\alpha_1 \neq 0$};
\node[newdarkblue2, scale=0.8] at (1.2,-1.2) {$q_2^2=m_2^2$};
\node[newdarkblue2, scale=0.8] at (1.2,-1.6) {$\alpha_2 \neq 0$};
\draw[black,line width=1.07] (0,-0.6) -- ++(-60:0.5);  %
\draw[black,line width=1.07] (0,-0.6) -- ++(-120:0.5); %
\draw[black,line width=1.07] (0,0.6) -- ++(120:0.5);%
\draw[black,line width=1.07] (0,0.6) -- ++(60:0.5);%
\draw[black,line width=1.07] (1.5,0) -- ++(30:0.5);%
\draw[black,line width=1.07] (1.5,0) -- ++(-30:0.5);%
\draw[-latex,black!50] (0,0.05) -- (2.5,0.6);
\draw[-latex,black!50] (0.4,-0.2) -- (2.5,-0.8);
\end{tikzpicture}
\end{equation}
We can contrast this with the actual bubble singularity of the ice cream cone, which does not require the on-shell conditions associated with $q_3$ and $q_4$:
\begin{equation}
G_{\icet}^{\text{bubble}}
=
\begin{tikzpicture}[baseline=(current bounding box.center),scale=0.8]
\path [newdarkblue2,line width=1.07,out=60,in=120] (0,0) edge (1,0);
\path [newdarkblue2,line width=1.07,out=-60,in=240] (0,0) edge (1,0);
\draw[black,line width=1.07] (0,0) -- ++(60:0.5);
\draw[black,line width=1.07] (0,0) -- ++(120:0.5);
\draw[black,line width=1.07] (0,0) -- ++(-60:0.5);
\draw[black,line width=1.07] (0,0) -- ++(-120:0.5);
\draw[black,line width=1.07] (1,0) -- ++(30:0.5);
\draw[black,line width=1.07] (1,0) -- ++(-30:0.5);
\node[darkred, scale=0.8] at (2.5,0.8) {$q_3^2\neq m_3^2$};
\node[darkred, scale=0.8] at (2.5,0.4) {$\alpha_3=0$};
\node[darkred, scale=0.8] at (2.5,-0.8) {$q_4^2 \neq m_4^2$};
\node[darkred, scale=0.8] at (2.5,-1.2) {$\alpha_4=0$};
\node[newdarkblue2, scale=0.8] at (0.5,1.2) {$q_1^2=m_1^2$};
\node[newdarkblue2, scale=0.8] at (0.5,0.8) {$\alpha_1 \neq 0$};
\node[newdarkblue2, scale=0.8] at (0.5,-1.2) {$q_2^2=m_2^2$};
\node[newdarkblue2, scale=0.8] at (0.5,-1.6) {$\alpha_2 \neq 0$};
\draw[-latex,black!50] (0,0) -- (1.5,0.6);
\draw[-latex,black!50] (0,0) -- (1.5,-0.8);
\end{tikzpicture}
\end{equation}
We will have more to say about this example in Sec.~\ref{sec:icecreamtobubbleandsun}. The discovery of these exceptions led to the formulation of the so-called weak hierarchical principle, and further refinements~\cite{Landshoff1966,boyling1968homological}.

As Pham clearly articulated,
the flaw in the strict hierarchical principle was the focus on leading singularities, that is, on requiring all of the Feynman parameters in a given Landau diagram to be nonzero. This is why that this constraint is not imposed in the Pham locus; only the on-shell and Landau loop equations are imposed. Thus, in Pham's language, the bubble-like singularity that appears in the ice cream cone Feynman integral is still associated with the ice cream cone Landau diagram --- it is simply the branch of the Pham locus in which the Feynman parameters $\alpha_3$ and $\alpha_4$ vanish.

\subsection{Applying Picard--Lefschetz to Absorption Integrals}
\label{sec:picard_lefschetz_absorption_int_example}
Theorem~\ref{thm:pham} identifies the surfaces in the space of external kinematics where absorption integrals can become singular. Around those singularities, we can again use the Picard--Lefschetz theorem and Leray's theory of residues to compute discontinuities. In general, new subtleties start to appear when computing the discontinuities of absorption integrals, as these integrals can become singular outside of the physical region. For now, we consider an example of a discontinuity that is still accessible in the physical region, where these complications do not arise, deferring a discussion of what new features can arise outside of this region to Sec.~\ref{sec:deform_contours}.

The example we consider involves the triangle Feynman integral, which has branch points associated with both the triangle and the bubble Landau diagrams. In three dimensions, this integral can be written as
\begin{equation}
I_\tritcol(p) = 
\resizebox{!}{1.3cm}{
\begin{tikzpicture}[baseline= {($(current bounding box.base)-(3pt,3pt)$)},
    line width=1.5,scale=0.7]
    \draw[black] (-4,1.5) -- (-2,1)  node[midway,above] {$p_2$};
    \draw[black,-latex] (-4,1.5) -- (-3,1.25);
    \draw[black] (-4,-1.5) -- (-2,-1) node[midway, below] {$p_1$};
    \draw[black,-latex] (-4,-1.5) -- (-3,-1.25);
    \draw[newdarkblue2] (-2,-1) --  (-2,1) node[midway,left] {$q_3,m_3$};
    \draw[newdarkblue2,-latex] (-2,-1) -- (-2,0.2);
    \draw[darkred] (-2,1) -- (0,0) node[midway,above] {$~~~~~~q_1,m_1$};
    \draw[darkred,-latex] (-2,1) -- (-0.75,0.375);
    \draw[olddarkgreen] (0,0) -- (-2,-1) node[midway,below] {$~~~~~q_2,m_2$};
    \draw[olddarkgreen,-latex] (0,0) -- (-1.25,-0.625);
    \draw[black] (0,0) -- (2,0)  node[midway,below] {$p_3$};
    \draw[black,-latex reversed] (0,0) -- (0.98,0);
    \end{tikzpicture}
}
=\int_h \rd^3 k \frac{1}{s_1(p,k) s_2(p,k) s_3(p,k)} \, ,
\end{equation}
where $s_i(p,k) = q_i^2(p,k) - m_i^2$ and the $+i\varepsilon$ has been moved to the integration contour $h$ using the methods discussed in Sec.~\ref{sec:iepaths}. 
This integral only depends on three independent variables. 
These can be chosen to be $y_{12}$, $y_{23}$, and $y_{13}$, where
\begin{equation}
   y_{ij}  = \frac{(p_i+p_j)^2 - m_i^2 - m_j^2}{2 m_i m_j} \,. \label{yijdef}
\end{equation}
In terms of these variables, the three bubble singularities occur where $y_{ij}^2 =1$, while the triangle singularity occurs where the combination
\begin{equation}
   D  =  1 - y_{12}^2 - y_{23}^2 - y_{13}^2  - 2 y_{12} \, y_{23} \, y_{13}
   \label{eq:dtri}
\end{equation}
vanishes.

We begin by computing the absorption integral for the threshold singularity associated with the pair of momenta $q_1$ and $q_2$, which occurs at $y_{12} = 1$ and corresponds to the bubble Landau diagram in which the third line has been contracted:
\begin{equation}
\resizebox{3cm}{!}{
\begin{tikzpicture}[baseline=(current bounding box.center),
    line width=1.5,scale=0.85]
    \path [darkred,thick,out=60,in=120] (0,0) edge (1,0);
    \path [olddarkgreen,thick,out=-60,in=240] (0,0) edge (1,0);
    \draw[black,thick] (0,0) -- ++(135:0.75)  node[left,above,scale=0.7] {$p_2$};
    \draw[black,thick,-latex reversed] (0,0) -- ++(135:0.48);
    \draw[black,thick] (0,0) -- ++(-135:0.75)  node[left,below,scale=0.7] {$p_1$};
    \draw[black,thick,-latex reversed] (0,0) -- ++(-135:0.48);
    \draw[black,thick] (1,0) -- ++(0:0.75)  node[midway,above,scale=0.7] {$p_3$};
    \draw[black,thick,-latex reversed] (1,0) -- ++(0:0.40);
    \node[darkred,scale=0.7] at (0.5,0.5) {$q_1,\,m_1$};
    \draw[darkred,thick,-latex] (0.55,0.25) -- ++(0:0.1);
    \node[olddarkgreen,scale=0.7] at (0.5,-0.5) {$q_2,\,m_2$};
    \draw[olddarkgreen,thick,-latex] (0.55,-0.25) -- ++(0:0.1);
\end{tikzpicture}
}
\end{equation}
For real momenta, the on-shell space $S_1 \cap S_2$ forms a paraboloid. This paraboloid is depicted in Fig.~\ref{fig:bub} and Eq.~\eqref{SSU} below. For complex momenta, $S_1$ and $S_2$ are complex two-spheres and $S_1 \cap S_2$ is a complex circle.\footnote{A complex $n$-sphere is given by the equation $z_1^2 + \dotso + z_{n+1}^2 = 1$ for $z_i = x_i + i y_i \in \mathbb{C}$.  The real and imaginary parts of the equation read $\smash{\sum_{i=1}^{n+1} (x_i^2 - y_i^2) = 1}$ and $\smash{\sum_{i=1}^{n+1} x_i y_i = 0}$.  If we define $\smash{u_i = x_i / (1 + \sum_{j=1}^{n+1} y_j^2)^{\frac{1}{2}}}$ then $\sum_{i=1}^{n+1} u_i^2 = 1$, so the $u_i$ parameterize a real $n$-sphere.  The remaining coordinates $y_i$ form a tangent vector to this real sphere, so we can say that the complex $n$-sphere is the tangent bundle to the real $n$-sphere.  For the complex circle, the tangent bundle is trivial so the complex circle can be thought of as a cylinder.
A complex sphere has a deformation that retracts it to a real sphere, which implies that its homology is the same as that of a real sphere.  In particular, the first homology of a complex circle is $H_1(S^1(\mathbb{C})) \cong \mathbb{Z}$.} However, to apply Picard--Lefschetz, we do not need to understand this full complex space, we only need to known how to make infinitesimal deformations of the integration contour within the Leray bubble that contains the singularity of interest.
In particular, we should work within a Leray bubble $U \subset X$ around $y_{12}=1$ that avoids the triangle singularity and the other two bubble singularities. For example, we can choose $U$ to contain the values of $Q$ in Fig.~\ref{fig:triangleparaboloid} that are above the bubble branch point and below the triangle one.
It is important that one chooses the Leray bubble such that it avoids the other singular points and branch cuts, since contours in different homotopy classes could give different answers for the monodromy around $y_{12} = 1$.

For fixed values of the external kinematics, the on-shell space $U_p\cap S_1\cap S_2$ is a circle (a slice of the $S_1\cap S_2$ paraboloid; see the red circle in Eq.~\eqref{SSU} below). This circle constitutes the vanishing sphere $\partial_2 \partial_1 e_{12}$, where $e_{12}$ is the vanishing cell that is defined by $s_1\geq 0$ and $s_2\geq 0$ (which is to say, by $q_1^2 \geq m_1^2$ and $q_2^2 \geq m_2^2$). These surfaces constitute higher-dimensional analogs of the vanishing sphere and vanishing cell that were depicted in Eq.~\eqref{eyelashU} for the two-dimensional bubble integral.
Since the vanishing cell is contained in a neighborhood $U$ that does not intersect the surface $S_3$ that corresponds to putting the third propagator on-shell, we can ignore the presence of this extra singularity for now.\footnote{The case where the bubble Pham locus and the triangle Pham locus intersect requires a special treatment.}

Given these ingredients, the Picard--Lefschetz theorem tells us that the absorption integral that computes the discontinuity with respect to this bubble singularity is given by
\begin{equation}
    A_{\tritcol}^{\bubrgt}(p) \,=\, \left(1-\monM_{ y_{12}=1} \right) I_\tritcol \,=\, -\langle {e_{12}}, h\rangle \int_{\delta_1 \delta_2 \partial_2 \partial_1 {e_{12}}} \omega \, .
\end{equation}
Applying the Leray formula $\int_{\delta \sigma} \omega = 2 \pi i \int_\sigma \operatorname{res} \omega$,
this becomes
\begin{align}
    \label{eq:PL_triangle_bubble_disc}
    A_{\tritcol}^{\bubrgt}(p) &= -(2 \pi i)^2 \langle {e_{12}}, h\rangle \int_{\partial_2 \partial_1 e_{12}} \operatorname{res}_2 \operatorname{res}_1 \omega \\
    &=
    -(2 \pi i)^2 \langle {e_{12}}, h\rangle \int_{\partial_2 \partial_1 e_{12}}
    \frac{\rd^3 k}{(\rd s_1 \wedge \rd s_2 )s_3}. \label{eq:bubble_absorption_3d}
\end{align}
Since $\partial_1 \partial_2 {e_{12}} = -\partial_2 \partial_1 {e_{12}}$ and $\operatorname{res}_1 \operatorname{res}_2 \omega = -\operatorname{res}_2 \operatorname{res}_1 \omega$, the expression is symmetric under permutation $S_1 \leftrightarrow S_2$, as expected. To find the Kronecker index $\langle {e_{12}}, h\rangle$, we use the fact that when the integration contour is deformed in a manner consistent with the Feynman $i \varepsilon$ prescription in the physical region, it intersects the vanishing cell ${e_{12}}$ at exactly one point (this result was explained in Sec.~\ref{sec:iepaths}). Comparing the orientations of ${e_{12}}$ and $h$, as in App.~\ref{sec:kronecker_index}, we find $\langle {e_{12}}, h\rangle = -1$. Using Eq.~\eqref{eq:convert_to_deltas} to rewrite the integral in~\eqref{eq:bubble_absorption_3d} in terms of delta functions, the full expression can then be evaluated to give
\begin{equation}
     A_{\tritcol}^{\bubrgt}(p) = \frac{\pi^3 i}{\sqrt{D}} \, ,
     \label{eq:mon_bub_tri}
\end{equation}
where $D$ was defined in Eq.~\eqref{eq:dtri}.

Now that we have the absorption integral for the bubble singularity in hand, we can compute a second discontinuity with respect to the triangle singularity at $D=0$. While it would be simple to compute this discontinuity using the explicit algebraic result in Eq.~\eqref{eq:mon_bub_tri}, we instead are interested in seeing how this discontinuity can be computed by applying Picard--Lefschetz to the expression in Eq.~\eqref{eq:bubble_absorption_3d}, which we have left as an integral over the real one-dimensional cycle $\partial_2 \partial_1 e_{12}$. 

Since the integration contour in $A_{\tritcol}^{\bubrgt}(p)$ already forces two of the propagators to be on-shell, the triangle singularity can be reached by restricting to the surface $S_3$. Thus, we are interested in the points along the integration contour $\partial_2 \partial_1 e_{12}$ that satisfy the constraint $s_3 = 0$. The situation can be visualized as follows: %
\begin{equation}
\label{SSU}
\resizebox{12cm}{!}{
\begin{tikzpicture}[baseline=(current  bounding  box.center)]
 \node (image) at (0,-0.2) {
        \includegraphics[width=0.3\textwidth]{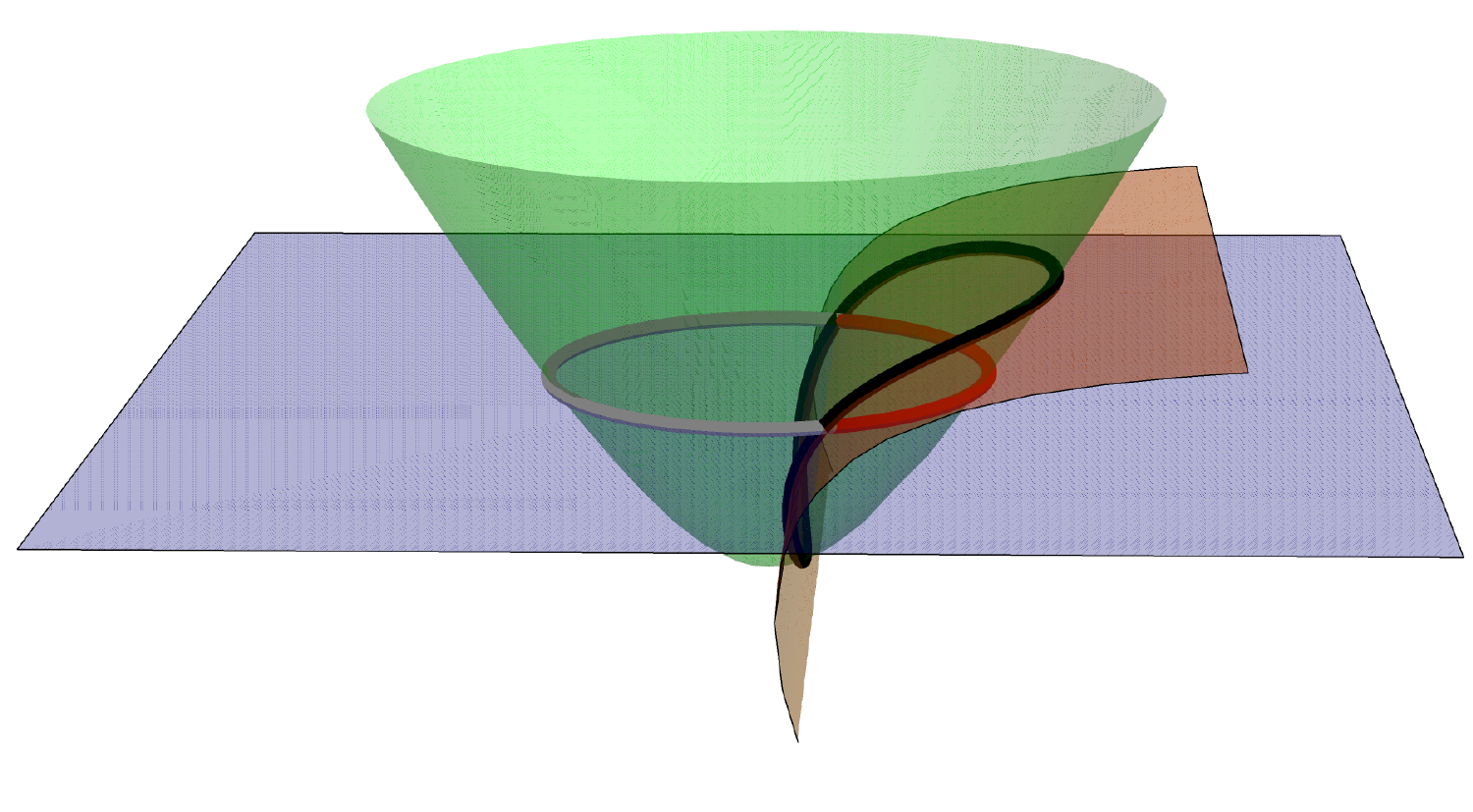}
    };  
\node[olddarkgreen,scale=0.8] at (-3,0.8) {$S_1 \cap S_2$};
\draw[-latex,olddarkgreen] (-2.3,0.8) -- (-1.3,0.7);
\node[orange,scale=0.8] at (3,0.8) {$S_3$};
\draw[-latex,orange] (2.8,0.8) -- (1.5,0.5);
\node[newdarkblue2!80,scale=0.8] at (-3.2,-0.5) {$U_p$};
\draw[newdarkblue2!80,-latex] (-2.8,-0.5) -- (-2.4,-0.4);
\node[black,scale=0.8] at (3.6,-0.5) {$S_1 \cap S_2 \cap S_3$};
\draw[-latex,black] (2.6,-0.5) -- (1.0,0.2);
\node[red,scale=0.8] at (0,-1.4) {$S_1 \cap S_2 \cap U_p$};
\draw[-latex,black!50] (-0.2,-1.2) -- (-0.6,-0.3);
\draw[-latex,red] (0.2,-1.2) -- (0.7,-0.3);
\end{tikzpicture}
}
\end{equation}
\\[-2mm]
In this picture, the new vanishing cell $e_{12 | 3}$ is the part of the circle $S_1 \cap S_2 \cap U_p$ that respects $s_3\geq 0$. This inequality is satisfied below the orange curved surface $S_3$, so $e_{12 | 3}$ corresponds to the bright red semi-circular arc. The vanishing sphere then corresponds to the boundary $\partial_3$ of this vanishing cell. Since all three propagators are now on shell, the vanishing sphere is precisely $\partial_3 e_{12 | 3} = \partial_3 \partial_2 \partial_1 e_{123}$, where $e_{123}$ is the cell defined by $s_1 \ge 0$, $s_2 \ge 0$, and $s_3 \ge 0$. For fixed external kinematics, this corresponds to the two ends of the red arc:
\begin{equation}
\resizebox{12cm}{!}{
\begin{tikzpicture}
 \draw[->,line width=1] (-8.5,-0.5) -- (-7.5,-0.5) node[right,scale=1.5] {$k^{y}$};
 \draw[->,line width=1] (-8.5,-0.5) -- (-8.5,-1.5) node[below,scale=1.5] {$k^{x}$};
\draw[line width=1.2mm,black!50] (2,0) arc[start angle=45, delta angle=90,radius=2];
\draw[line width=2mm,red!50] (2,0) arc[start angle=45, delta angle=-270,radius=2];
\node[black,scale=1.5] at (0.5,-1) {$\partial_3 \partial_2 \partial_1 e_{123}$};
\draw[-latex,black,line width =0.8mm] (0.3,-0.6) -- (-0.8,0);
\draw[-latex,black,line width =0.8mm] (0.7,-0.6) -- (2,0);
\draw[dashed,orange,line width = 0.5mm] (-3,0) -- (4,0);
\node[orange,scale=1.5] at (-4.5,0) {$s_3 = 0$};
\node[orange,scale=1.5] at (-3,-1.3) {$s_3 > 0$};
\draw[-latex,orange,line width =0.8mm] (-2,-0.5) -- (-2,-2);
\node[red!50,scale=1.5] at (7,-1.5) {(vanishing cell)};
\draw[->,red!50,line width=0.8mm] (7,-2) to[out=210,in=-30] (2.8,-2);
\end{tikzpicture}
}
\label{circlecell}
\end{equation}
Finally, the vanishing cycle is given by $\delta_3 \partial_3 \partial_2 \partial_1 e_{123}$, which is a complex neighborhood around the vanishing sphere (which we have not depicted).
Note that in this analysis, we have assumed that $\alpha_3 > 0$ at the critical point.\footnote{If instead $\alpha_3 < 0$, we should choose as our vanishing cell the complementary arc between the points $\partial_1 \partial_2 \partial_3  e_{123}$ in the circle depicted in Eq.~\eqref{circlecell}; this vanishing cell is defined by $s_3\vert_{S_1(\mathbb{R}) \cap S_2(\mathbb{R})} \leq 0$.}

Using Eq.~\eqref{eq:PLtheorem} for $n = 1$, we then have that
\begin{equation}
    \operatorname{Var} (\partial_1 \partial_2 e_{12}) = -\langle e_{12|3}, \partial_2 \partial_1 e_{12} \rangle \delta_3 \partial_3 \partial_2 \partial_1 e_{123} \, .
\end{equation}
To determine the Kronecker index $\langle e_{12|3}, \partial_2 \partial_1 e_{12} \rangle$, we need to know how the integration contour in Eq.~\eqref{eq:bubble_absorption_3d} is detoured around the two points where $s_3=0$ and the integrand becomes singular. Just as for the Feynman integral we started with, this detour is dictated by causality and can be represented as a vector field.  In this case, the Feynman $i \epsilon$ prescription amounts to choosing a detour represented by a vector field $v$ on the real circle such that
$\operatorname{sgn} \bigl(\, \rd s_3\vert_{S_1(\mathbb{R}) \cap S_2(\mathbb{R})}(v)\bigr) = +1$,
where $\rd s_3\vert_{S_1(\mathbb{R}) \cap S_2(\mathbb{R})}$ is the restriction of the differential form $\rd s_3$ to the space
$S_1(\mathbb{R}) \cap S_2(\mathbb{R})$.
This amounts to the vector field having a sink-type zero inside the vanishing cell, which looks like
\begin{equation}
\resizebox{3cm}{!}{
\begin{tikzpicture}
\draw[line width=2mm,red!50] (2,0) arc[start angle=45, delta angle=-270,radius=2];
\draw[line width=2,blue!50,-latex] (2.4,-0.5) -- (3.2,-2);
\draw[line width=2,blue!50,-latex] (-1.2,-0.5) -- (-2.0,-2);
\draw[line width=2,blue!45,-latex] (2.6,-1.5) -- (2.6,-3);
\draw[line width=2,blue!45,-latex] (-1.4,-1.5) -- (-1.4,-3);
\draw[line width=2,blue!40,-latex] (1.9,-2.9) -- (1.6,-3.6);
\draw[line width=2,blue!40,-latex] (-0.7,-2.9) -- (-0.4,-3.6);
\draw[line width=2,blue!30,-latex] (0.6,-3.4) -- (0.6,-3.8);
\draw[line width=1.2mm,black!50] (2,0) arc[start angle=45, delta angle=90,radius=2];
 \end{tikzpicture}
}
\label{circlecell2}
\end{equation}
According to the computation in appendix~\ref{sec:kronecker_index}, we then have that $\langle e_{12|3}, \partial_1 \partial_2 e_{12} \rangle = -1$.

Putting everything together, the variation of $ A_{\tritcol}^{\bubrgt}(p)$ around the singular locus defined by the vanishing of the propagator $s_3$ is given by
\begin{equation}
   \left(1-\monM_{ D=0} \right)  A_{\tritcol}^{\bubrgt}(p)
  =  (2 \pi i)^2 \langle e_{12|3}, \partial_2 \partial_1 e_{12} \rangle \langle {e_{12}}, h\rangle \int_{\delta_3 \partial_3 \partial_2 \partial_1 e_{123}}
    \frac{\rd^3 k}{(\rd s_1 \wedge \rd s_2 )s_3} \, .
\end{equation}
Finally, we can integrate over the coboundary using the Leray residue formula and plug in the value of the Kronecker indices, which leads us to
\begin{equation}
\left(1-\monM_{ D=0} \right) A_{\tritcol}^{\bubrgt}(p)
= (2 \pi i)^3 \int_{\partial_1 \partial_2 \partial_3 e_{123}} \frac{\rd^3 k}{\rd s_1 \wedge \rd s_2 \wedge \rd s_3}\, . \label{eq:double_disc_absorp}
\end{equation}
Evaluating this integral explicitly, we find
\begin{equation}
    \left(1-\monM_{ D=0} \right) A_{\tritcol}^{\bubrgt}(p)
    = \frac{2 \pi^3 i}{\sqrt{D}} \,.
\end{equation}
Quite surprisingly, this is the same result one finds when computing just a single discontinuity with respect to the triangle singularity. The vanishing cell for the triangle singularity is $e_{123}$, so the Picard--Lefschetz theorem implies that
\begin{align}
    (1 - \monM_{D=0}) I_\tritcol &= -\langle  {e_{123}}, h\rangle \int_{\delta_1 \delta_2 \delta_3 \partial_3 \partial_2 \partial_1  {e_{123}}} \frac{\rd^3 k}{s_1 s_2 s_3}  \\
    &=
    (2 \pi i)^3 \int_{\partial_3 \partial_2 \partial_1  {e_{123}}} \frac{\rd^3 k}{\rd s_1 \wedge \rd s_2 \wedge \rd s_3}  \label{I3tri1} %
\end{align}
where we have used the fact that $\langle {e_{123}}, h\rangle = -1$. This is the same expression we just found in Eq.~\eqref{eq:double_disc_absorp}, as claimed above. While one might have been able to anticipate that the integration contour would be the same in these two cases, it is nontrivial that the intersection numbers that enter the Picard--Lefschetz theorem also conspire to give the same result. In fact, in Sec.~\ref{sec:sequential_disc}, we will show that this is just the simplest example of a general class of identities that hold between the discontinuities of Feynman integrals.

  \section{A Geometric Angle on Landau Varieties}
\label{sec:deform_contours}

In the previous sections, we worked out a number of examples to illustrate how the Picard--Lefschetz formula can be used to compute  discontinuities and sequential discontinuities of Feynman integrals. In these calculations, we offered only minimal details on how to compute Kronecker indices, despite the fact that this is generally the most difficult part of the computation (especially if one starts considering discontinuities outside of the physical region). In part, we have been able to do this because we have only computed discontinuities with respect to principal Pham loci, which we show can only give rise to Kronecker indices that evaluate to $\pm 1$ or $0$ in Appendix~\ref{sec:kronecker_index}. 

In this section, we characterize the properties of principal Pham loci, which correspond to a subset of the singularities that  appear on the physical sheet. In addition, we show that all Pham loci can be recast as the critical values of projection maps between on-shell spaces, and illustrate how the codimension of a Landau singularity in the space of external kinematics can be read off of the solution to the Landau equations in Feynman parameter space when all masses are generic. Many of the results in this section will be needed to establish the relations among sequential discontinuities presented in Sec.~\ref{sec:sequential_disc} and Sec.~\ref{sec:codim2}.

When considering Pham loci as the critical values of projections maps, we will make frequent reference to the kernel of graph contractions. The \textbf{kernel of a contraction} is defined to be the graph that is formed by the edges that are contracted.
For example, if $\kappab$ represents the contraction of the bubble out of the ice cream cone diagram,
\begin{equation}
\label{eq:ice_cream_cone_contraction}
G_{\icet}= 
\begin{tikzpicture}[baseline= {($(current bounding box.base)-(2pt,2pt)$)},scale=0.8]
\path [darkred,thick,out=130,in=230] (0,-0.6,0) edge (0,0.6);
\path [darkred,thick,out=50,in=-40] (0,-0.6,0) edge (0,0.6);
\node[darkred] at (0.5,0) {$q_3$};
\node[darkred] at (-0.5,0) {$q_4$};
\draw [newdarkblue2,thick] (0,0.6) -- (1.5,0);
\draw [newdarkblue2,thick] (0,-0.6) -- (1.5,0);
\node[newdarkblue2] at (0.7,0.6) {$q_1$};
\node[newdarkblue2] at (0.7,-0.6) {$q_2$};
\draw[black,thick] (0,-0.6) -- ++(-60:0.5);  %
\draw[black,thick] (0,-0.6) -- ++(-120:0.5); %
\draw[black,thick] (0,0.6) -- ++(120:0.5);%
\draw[black,thick] (0,0.6) -- ++(60:0.5);%
\draw[black,thick] (1.5,0) -- ++(30:0.5);%
\draw[black,thick] (1.5,0) -- ++(-30:0.5);%
\end{tikzpicture}
~~
\xrightarrowdbl{~~\kappab~~}
~~
G_{\bubt} =
\begin{tikzpicture}[baseline= {($(current bounding box.base)-(2pt,2pt)$)},scale=0.8]
\path [newdarkblue2,thick,out=60,in=120] (-1,0) edge (0,0);
\path [newdarkblue2,thick,out=-60,in=240] (-1,0) edge (0,0);
\node[newdarkblue2] at (-0.5,0.6) {$q_1$};
\node[newdarkblue2] at (-0.5,-0.6) {$q_2$};
\draw[black,thick] (-1,0) -- ++(60:0.5);
\draw[black,thick] (-1,0) -- ++(120:0.5);
\draw[black,thick] (-1,0) -- ++(-60:0.5);
\draw[black,thick] (-1,0) -- ++(-120:0.5);
\draw[black,thick] (0,0) -- ++(30:0.5);
\draw[black,thick] (0,0) -- ++(-30:0.5);
\end{tikzpicture} \, ,
\end{equation}
then the kernel of $\kappab$ is the bubble itself:
\begin{equation}
\ker \kappab = 
\begin{tikzpicture}[baseline= {($(current bounding box.base)-(2pt,2pt)$)},scale=0.8]
\path [darkred,thick,out=130,in=230] (0,-0.6,0) edge (0,0.6);
\path [darkred,thick,out=50,in=-40] (0,-0.6,0) edge (0,0.6);
\node[darkred] at (0.5,0) {$q_3$};
\node[darkred] at (-0.5,0) {$q_4$};
\end{tikzpicture} \, .
\end{equation}
Together with $\kappab$, the kernel of $\kappab$ forms a short exact sequence:\footnote{The arrows $\twoheadrightarrow$ and $\rightarrowtail$ signify that these maps are (respectively) epimorphisms or monomorphisms, in a category whose objects are graphs and whose morphisms are maps between graphs that preserve path composition (see Pham's paper~\cite{pham}).  An exact sequence is a sequence of maps $H \rightarrowtail G \twoheadrightarrow K$ where the image of the first map is the kernel of the second.  Then, $G$ is called an extension of $K$ by $H$.  The reader who is unfamiliar with this language can safely skip this remark and think of the maps $\twoheadrightarrow$ as graph contractions and $\rightarrowtail$ as sub-graph embeddings.}
\begin{equation} \label{eq:short_exact_sequence}
\begin{tikzcd}
  \text{ker\,} \kappab \arrow[r, rightarrowtail, ""] & G^\kappap \arrow[r, twoheadrightarrow, "\kappab"] & G^\kappa \, .
  \end{tikzcd}
\end{equation}
We will see many instances of these short exact sequences in later sections. 

\subsection{Pham Loci as Critical Values of Projection Maps}
\label{sec:critical}

The Landau equations constitute algebraic conditions for a pinch singularity to occur along the integration contour of a Feynman integral. However, they also turn out to have an alternative interpretation. Namely, the Landau equations as can be viewed as identifying the critical points of projection maps $\pi_\kappa$ from the on-shell space $\S(G^\kappa)$ of a Landau diagram $G^\kappa$ to the on-shell space  of external momenta $\S(G_0)$:
\begin{equation}
    \pi_\kappa \colon \S(G^\kappa) \to \S(G_0) \, ,
\end{equation}
where $\pi_k$ simply removes the momentum associated with every internal edge $e \in E(G^\kappa)$. That is, if we collectively denote all external momenta by $p$ and all internal momenta by $q$, $\pi_k$ acts as 
\begin{equation}
     \pi_\kappa \colon (p, q) \mapsto p \, .
\end{equation}
What makes the map non-trivial is that all of the on-shell constraints in $\S(G^\kappa)$ are satisfied.

The projection operator $\pi_\kappa$ maps from a space of dimension $d_{G_0}=d (\nex - 1) - \nex$ to a space of dimension $d_{G^\kappa} = d_{G_0} + L d - \nint$, where $d$ is the spacetime dimension and $L$ is the number of loops in $G^{\kappa}$.
A point in this map is called a \textbf{critical point} if the rank of the $d_{G^\kappa} \times d_{G_0}$ Jacobian matrix is less than $d_{G_0}$ at that point.  The image of a critical point is referred to as a \textbf{critical value}. A powerful way to think about singularities of Feynman integrals is provided by the following result:
\begin{lemma}[Pham] The critical values of the projection map $\pi_\kappa$ constitute the Pham locus $\P_\kappa$.
\label{lem:crit}
\end{lemma}
\noindent
While the idea of describing Landau singularities as critical values of differential maps was primarily advocated by Pham~\cite{pham} (see also Refs.~\cite{pham1968singularities, pham2011singularities}),  Pham gives credit to Ren\'e Thom for the idea. Before sketching a proof of this correspondence, we illustrate how it works in a handful of examples.

\paragraph{Landau Singularities as Critical Values: First Example}
As a first example, consider the contraction of the bubble to the elementary four-point graph:
\begin{equation}
\label{projbubble}
\resizebox{!}{1.1cm}{
$
\begin{tikzpicture}[baseline= {($(current bounding box.base)-(2pt,2pt)$)},scale=0.8]
\path [darkred,thick,out=60,in=120] (-1,0) edge (0,0);
\path [olddarkgreen,thick,out=-60,in=240] (-1,0) edge (0,0);
\node[darkred,scale=0.8] at (-0.5,0.5) {$q_1,\,m_1$};
\draw[darkred,thick,-latex] (-0.45,0.25) -- ++(0:0.1);
\node[olddarkgreen,scale=0.8] at (-0.5,-0.5) {$q_2,\,m_2$};
\draw[olddarkgreen,thick,-latex] (-0.45,-0.25) -- ++(0:0.1);
\draw[black,thick] (-1,0) -- ++(150:1);
\draw[black,thick] (-1,0) -- ++(-150:1);
\draw[black,thick] (0,0) -- ++(30:1);
\draw[black,thick] (0,0) -- ++(-30:1);
\node[black,scale=0.8] at (-2,0.8) {$p_2,\, M_2$};
\node[black,scale=0.8] at (-2,-0.8) {$p_1,\, M_1$};
\node[black,scale=0.8] at (1,0.8) {$p_3,\, M_3$};
\node[black,scale=0.8] at (1,-0.8) {$p_4,\, M_4$};
\draw[black,thick,-latex reversed] (-1,0) -- ++(150:0.68);
\draw[black,thick,-latex reversed] (-1,0) -- ++(-150:0.68);
\draw[black,thick,-latex] (0,0) -- ++(30:0.75);
\draw[black,thick,-latex] (0,0) -- ++(-30:0.75);
\end{tikzpicture}
~~~~~~~
\hspace{-0.75cm}
\xrightarrowdbl{~~\kappa~~}
\hspace{-0.75cm}
~~~~~~~
\begin{tikzpicture}[baseline= {($(current bounding box.base)-(2pt,2pt)$)},scale=0.8]
    \draw[black,thick] (0,0) -- ++(150:1);
    \draw[black,thick] (0,0) -- ++(-150:1);
    \draw[black,thick] (0,0) -- ++(30:1);
    \draw[black,thick] (0,0) -- ++(-30:1);
    \draw[black,thick,-latex] (0,0) -- ++(30:0.75);
    \draw[black,thick,-latex] (0,0) -- ++(-30:0.75);
    \draw[black,thick,-latex reversed] (0,0) -- ++(150:0.68);
    \draw[black,thick,-latex reversed] (0,0) -- ++(-150:0.68);
    \node[black,scale=0.8] at (-1,0.8) {$p_2,\, M_2$};
    \node[black,scale=0.8] at (-1,-0.8) {$p_1,\, M_1$};
    \node[black,scale=0.8] at (1,0.8) {$p_3,\, M_3$};
    \node[black,scale=0.8] at (1,-0.8) {$p_4,\, M_4$};
\end{tikzpicture}
$
}
\end{equation}
We label the incoming momenta by $p_1$ and $p_2$ and the outgoing momenta by $p_3$ and $p_4$, and assign each of these external lines a corresponding mass $M_i$. The internal momenta are denoted $q_1$ and $q_2$, and are assigned masses $m_1$ and $m_2$.

For $2\to2$ scattering we can choose coordinates in the center-of-mass frame where $p_1^\mu$ and $p_2^\mu$ are back-to-back in the $x$ direction:
\begin{align}
    p_1^\mu & =(p_1^0,p^x,\vec{0}) \,, \\
    p_2^\mu & =(p_2^0,-p^x,\vec{0}) \,.
\end{align}
The on-shell conditions $p_1^2=M_1^2$ and $p_2^2=M_2^2$ fix the energies $p_1^0$ and $p_2^0$ in terms of $p^x$, and we can trade $p^x$ for the center-of-mass energy $Q=p_1^0+p_2^0$. This allows us to express the incoming momenta entirely in terms of $Q$. 
The outgoing momenta are also fixed by momentum conservation up to a scattering angle $\theta$. 
Let us write the internal momenta as $q_1^\mu =k^\mu$ and $q_2^\mu = p_1^\mu+p_2^\mu-k^\mu$.
When momentum conservation and the on-shell conditions are imposed on the internal momenta, their energies can be determined to be $\smash{q_1^0 = ({m_1^2 + \vert \vec{k} \vert ^2})^{\frac12}}$ and $\smash{q_2^0 = (m_2^2 + \vert \vec{k} \vert ^2)^{\frac12}}$ in terms of a single $(d-1)$-dimensional momentum $\vec{k}$. Energy conservation then requires that
\begin{equation}
    \label{eq:bubble-internal-kinematics}
    \vert \vec{k} \vert ^2 = \frac{(Q + m_1 + m_2)(Q - m_1 - m_2)(Q - m_1 + m_2)(Q + m_1 - m_2)}{4 Q^2},
\end{equation}
which implies that the spatial part of the loop momentum lives on the surface of a $(d\!-\!1)$-sphere whose radius is fixed by the external kinematics.\footnote{This assumes that the two internal particles are not identical.  In the case of identical particles, we have to divide by the permutations of the particles and then the on-shell space is a real projective space instead of a real sphere.}  If we are above the threshold for production of the particles in the loop, namely if $Q > m_1 + m_2$, then the right-hand side in the formula above is positive and we can find a real solution for $\vec{k}$.  Right at the threshold, when $Q = m_1 + m_2$, the radius of the sphere is zero so we have $\vec{k} = \vec{0}$.  Below the threshold, when $Q<m_1+m_2$, there is no real solution for $\vec{k}$.

\begin{figure}[t]
    \centering
\begin{tikzpicture}
 \node (image) at (0,0) {
        \includegraphics[width=0.42\textwidth]{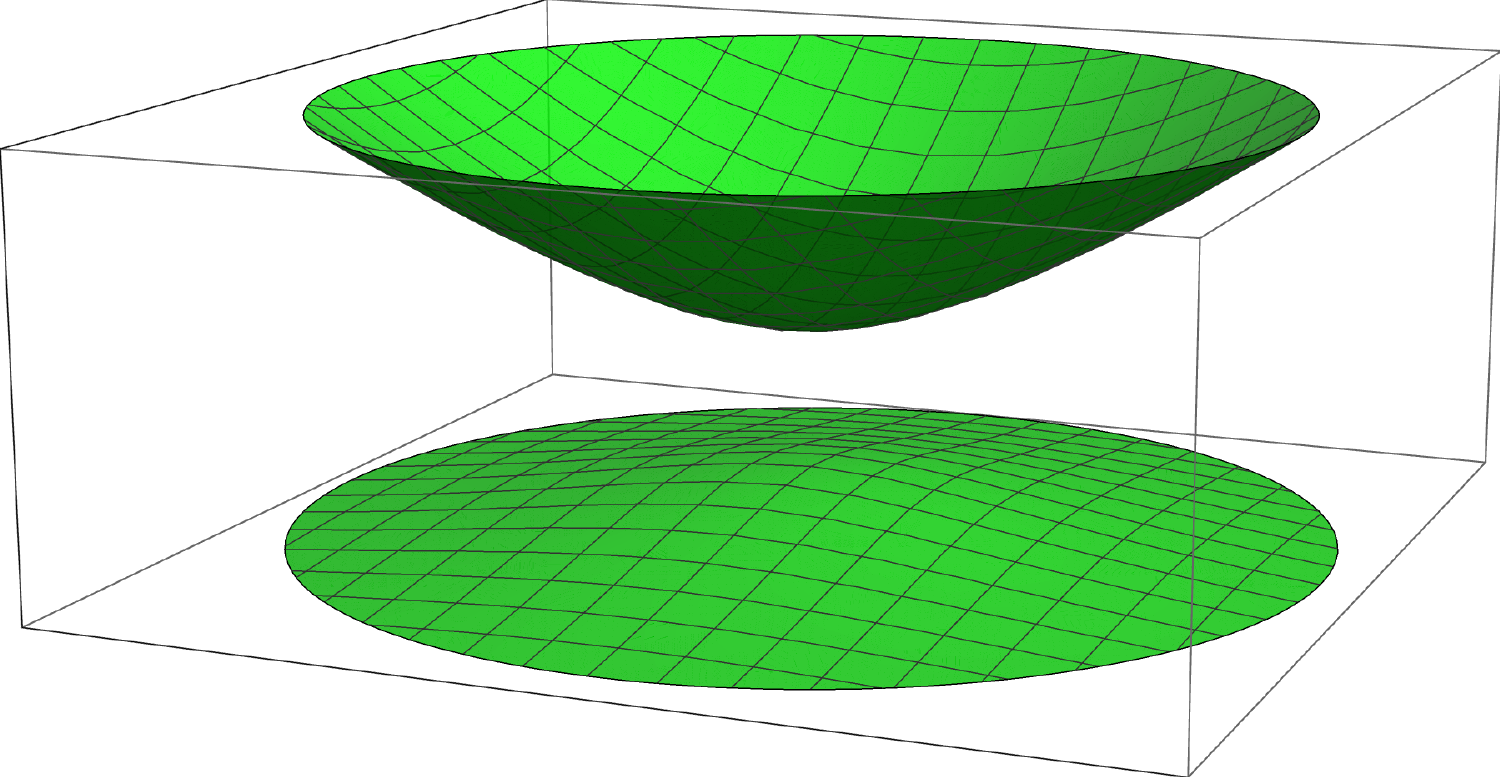}
    };
\node[] at (-1,-1.5) {$k^x$};
\node[] at (3,-1.2) {$k^y$};
\node[] at (-4,0) {$Q$};
\node[] at (4.8,1) {$\xmapsto{~~\pi_\kappa~~}$};
\filldraw[black] (0.2,0.2) circle (0.1); 
\draw[black] (0.2,-0.2) circle (0.1); 
\filldraw[black] (6,0.2) circle (0.1); 
\draw[black] (6,-0.2) circle (0.1); 
\draw[->,line width=0.75mm, olddarkgreen] (6,0.2) -- (6,2);
\draw[-,line width=0.75mm, olddarkgreen] (6,-0.9) -- (6,-0.3);
\node[] at (6.5,1) {$Q$};
\draw[line width=0.25mm, black,dotted] (1.5,0.2) -- (5.75,0.2);
\draw[line width=0.25mm, black,dotted] (1.5,-0.2) -- (5.75,-0.2);
\node[right] at (6.5,0.3) {Threshold};
\node[right] at (6.5,-0.3) {Pseudothreshold};
\end{tikzpicture}
\caption{%
The singular points that are encoded by the contraction of the four-point bubble diagram to the elementary graph in Eq.~\eqref{projbubble} correspond to the critical values of the projection map $\pi_\kappa$ depicted here. The surface on the left depicts the space of on-shell internal and external kinematics in $d=3$, which are constrained to a parabolic surface. When $m_1=1$ and $m_2=3$, this surface is determined by the constraint $(k^x)^2 + (k^y)^2 = \frac{1}{4Q^2}(Q^4-20 Q^2+64)$,
which we have depicted in green. The critical points of the map from this on-shell space $\{k^x,k^y,Q\}$ to the on-shell space $\{Q\}$  of the external momenta constitute the Landau variety of the bubble. The critical point marked by the filled dot occurs at $Q=m_1+m_2=4$ and corresponds to the threshold, while the critical point marked by the empty dot occurs at $Q=m_2-m_1=1$ and corresponds to the pseudothreshold.
}
\label{fig:bub}
\end{figure}
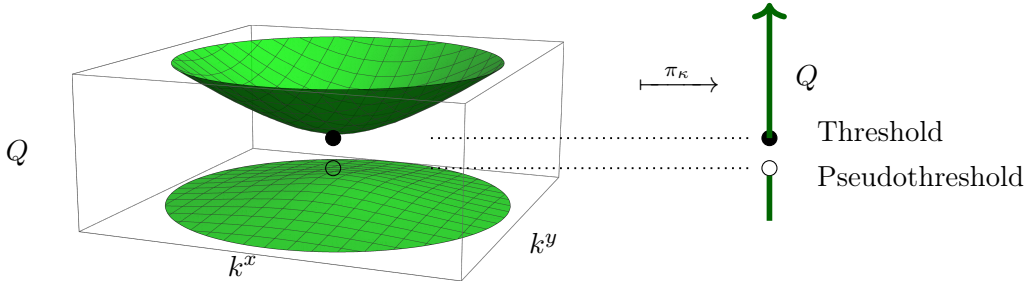

The space $\S(G_0)$ of external momenta is two-dimensional, and can be parameterized by $Q$ and the scattering angle $\theta$. The space $\S(G)$ of on-shell internal and external momenta includes the momentum $\vec{k}$ (which remains subject to the constraint in Eq.~\eqref{eq:bubble-internal-kinematics}) in addition to $Q$ and $\theta$. In $\S(G)$, we can trade $Q$ for $\vert \vec{k} \vert ^2$, so that the $d$-dimensional space $\S(G)$ is parameterized by $\theta$ and $\vec{k}$ (which is now unconstrained). The projection map $\pi_\kappa$ then maps $\theta\to \theta$ and $\vec{k} \to Q$.  
To see how the Landau equations appear as the critical points of this map, it is instructive to solve for $Q$ in terms of $\vec{k}$ to get
\begin{equation}
    Q = \left(\sqrt{\vert \vec{k} \vert^2+m_1^2} \pm \sqrt{\vert \vec{k} \vert^2+m_2^2} \right) \,,
\end{equation}
(assuming without loss of generality that $m_1 \geq m_2$), so the Jacobian for the projection map (ignoring $\theta$) is given by the $1\times(d-1)$ dimensional matrix whose entries are
\begin{equation}
    \frac{\partial Q}{\partial \vec{k}} = \vec{k} \left[\frac{1}{\sqrt{\vert \vec{k} \vert^2+m_1^2}} \pm \frac{1}{\sqrt{\vert \vec{k} \vert^2+m_2^2}} \right] \,.
\end{equation}
The rank of this matrix is generically one, unless $\vec{k}=0$, in which case it drops rank.  This point is therefore a critical point of the map $Q(\vec{k})$. The values $Q(\vec{0}) = m_1 \pm m_2 $ are the critical values of the projection map, which correspond to the threshold and pseudothreshold of the bubble diagram. This situation is depicted in Fig.~\ref{fig:bub}.

\paragraph{Landau Singularities as Critical Values: Second Example}
Another type of critical map arises already at tree level.  Consider the same four-point process as above, but now look at the exchange of a single particle in the $s$-channel with momentum $q_1$, which is subject to the on-shell condition $q_1^2 = m_1^2$:
\begin{equation}
\resizebox{!}{0.8cm}{
\begin{tikzpicture}[baseline=(current bounding box.center),
    line width=1.5,scale=0.8]
\draw[black] (-1,1) -- node[midway, above] {$~~~p_2$} (0,0);
\draw[black,-latex] (-1,1) -- ++(-45:1);
\draw[black] (-1,1) -- node[midway, above] {$~~~p_2$} (0,0);\draw[black] (-1,-1) -- node[midway, below] {$~~~p_1$} (0,0);
\draw[black,-latex] (-1,-1) -- ++(45:1);
\draw[darkred] (0,0) -- node[midway, above] {$q_1,m_1$} (2,0); 
\draw[darkred,-latex] (0,0) -- (1.2,0); 
\draw[black] (2,0) -- node[midway, above] {$p_3~~~$} (3,1);
\draw[black,-latex] (2,0) -- ++(45:1);
\draw[black] (2,0) -- node[midway, below] {$p_4~~~$} (3,-1);
\draw[black,-latex] (2,0) -- ++(-45:1);
\end{tikzpicture}
}
\end{equation}
We can again choose a center-of-mass frame, where $p_1 + p_2 = (Q, \vec{0})$, but we see that this value is compatible with the internal on-shell condition for the internal line only if $Q = m_1$. Moreover, because of momentum conservation, the internal momentum is fixed to be $q_1^\mu = (Q, \vec{0})$. So the space of external kinematics is again two-dimensional, but now the on-shell space of internal and external kinematics is only one-dimensional, since it obeys an additional constraint.
Thus, we are in a special case where the projection from $\S(G^\kappa)$ to $\S(G_0)$ maps to a space of higher dimension than the original space.
In this case we adopt the convention that all the points in the domain of this map are critical points, and all the points in the image are critical values. Correspondingly, in this example, the critical points would be the ones for which $Q=m_1$.\footnote{Recall our definition of critical points: given a differentiable map $f \colon \mathbb{R}^m \to \mathbb{R}^n$ a point $x \in \mathbb{R}^n$ is called a \emph{critical point} if the rank of the Jacobian matrix at $x$ is smaller than $n$.  Since the Jacobian matrix has shape $m \times n$, if $m < n$ then the maximal rank is $m$.  Therefore, in this case all points in $\mathbb{R}^m$ are critical points.  Working by patches there is an obvious extension of these notions to differentiable maps of manifolds $f \colon M \to N$ and one can show that the definition is independent on the choice of local coordinates.}\textsuperscript{,}\footnote{We emphasize that here we are using Pham's definition of Landau variety, which corresponds to taking the on-shell conditions to be satisfied even for the edges with $\alpha = 0$ (see the remark on pages 31--32 of Ref.~\cite{pham1968singularities}).}

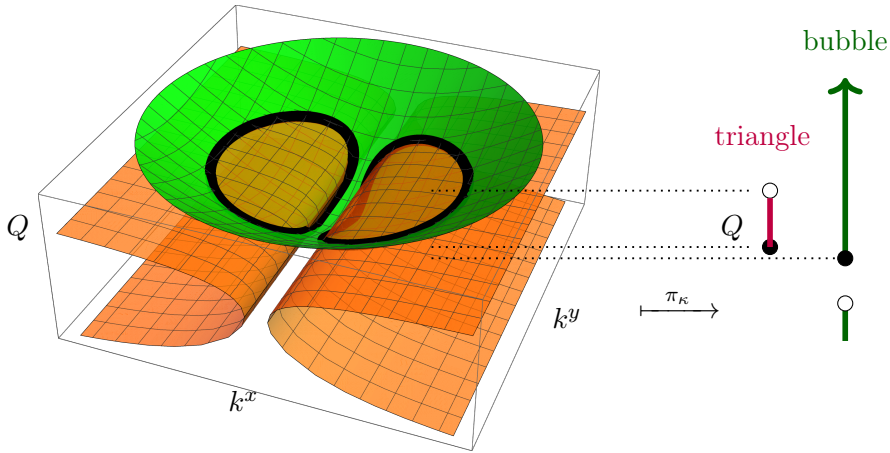
\begin{figure}[t]
    \centering
\begin{tikzpicture}
 \node (image) at (0,0) {
        \includegraphics[width=0.48\textwidth]{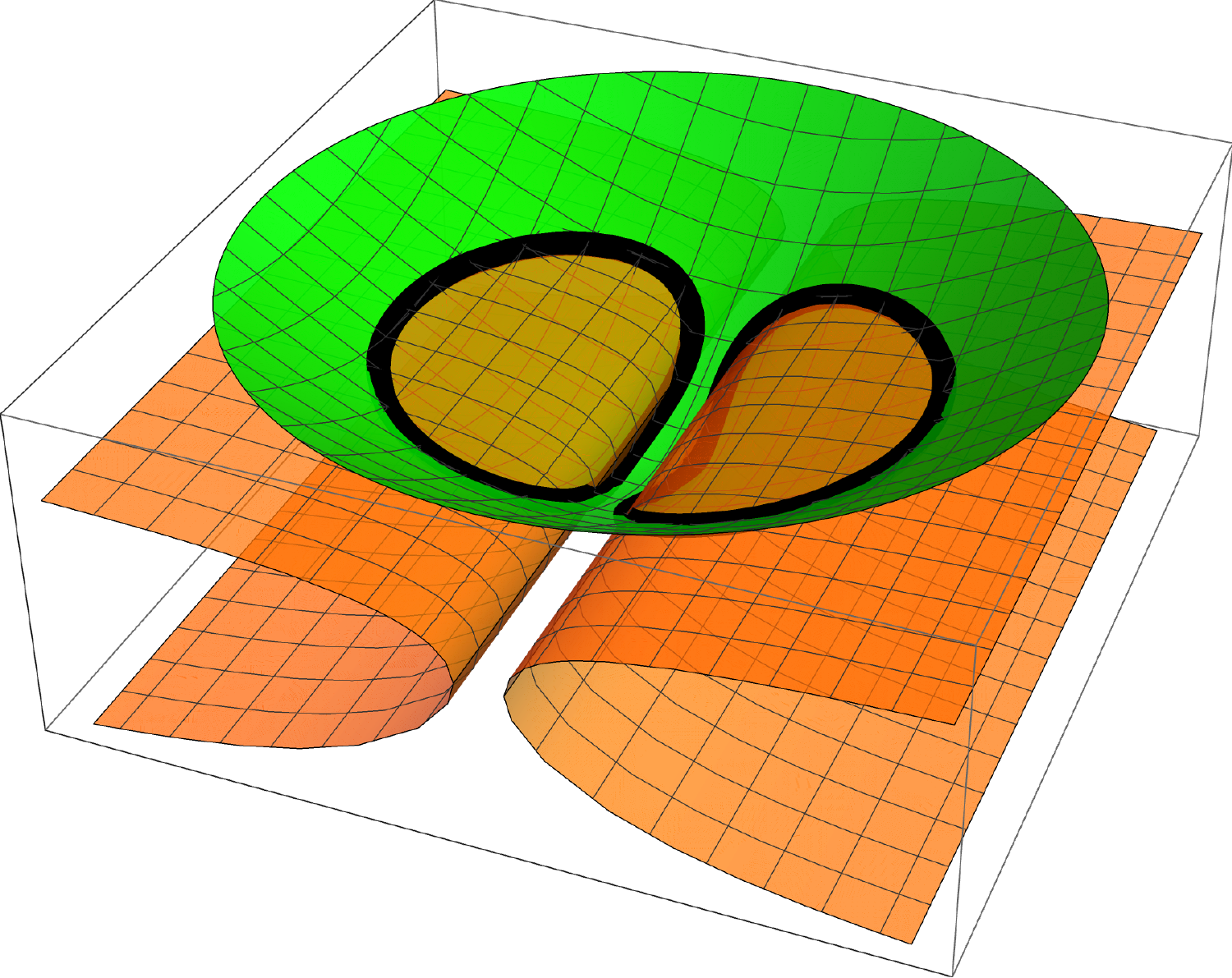}
    };
\node[] at (-1,-2.3) {$k^x$};
\node[] at (3.3,-1.2) {$k^y$};
\node[] at (-4,0) {$Q$};
\node[] at (4.8,-1) {$\xmapsto{~~\pi_\kappa~~}$};
\draw[black] (6,0.5) circle (0.1); 
\filldraw[black] (6,-0.25) circle (0.1); 
\draw[line width=0.75mm, purple] (6,-0.25) -- (6,0.4);
\node[] at (5.5,0) {$Q$};
 \draw[line width=0.25mm, black,dotted] (1.5,0.5) -- (5.75,0.5);
 \draw[line width=0.25mm, black,dotted] (1.5,-0.25) -- (5.75,-0.25);
\draw[->,line width=0.75mm, olddarkgreen] (7,-0.35) -- (7,2);
\draw[-,line width=0.75mm, olddarkgreen] (7,-1.5) -- (7,-1.1);
\draw[black] (7,-1) circle (0.1); 
\filldraw[black] (7,-0.4) circle (0.1); 
\node[purple] at (5.9,1.2) {triangle};
\node[olddarkgreen] at (7,2.5) {bubble};
 \draw[line width=0.25mm, black,dotted] (1.5,-0.4) -- (7,-0.4);
\end{tikzpicture}
\caption{The triangle singularity corresponds to the critical points of the map from the on-shell space of external and internal kinematics that involves three internal propagators to the space of external kinematics. Here, we have included the green paraboloid that depicts the on-shell space for two internal propagators in $d=3$ from Fig.~\ref{fig:bub} (only the upper branch is shown). The orange surface is the on-shell space for the third propagator. Their intersection (the black curve) gives the full on-shell space, which closes off at a maximum and minimum value of $Q$. These extrema (the critical points of the map from $\{k^y,Q\} \to \{Q\}$) give the Landau variety for the triangle singularity. These figures correspond to kinematics in which $m_1=1$, $m_2=3$, $m_3=5$, $M_1=2$, and $M_2=9$. The lower limit on $Q$ is the $\alpha$-positive triangle threshold (which occurs at $Q=\frac{1}{5} \sqrt{767-4 \sqrt{7854}}\approx4.06$), while the upper limit (at $Q=\frac{1}{5} \sqrt{767+4 \sqrt{7854}}\approx6.69$) corresponds to a pseudothreshold.
    \label{fig:triangleparaboloid}}
\end{figure}

\paragraph{Landau Singularities as Critical Values: Third Example}
Now let us consider the triangle diagram that contributes to the same four-point scattering process:
\begin{equation}
\resizebox{!}{1.3cm}{
\begin{tikzpicture}[baseline=(current bounding box.center),
    line width=1.5,scale=0.7]
    \draw[black] (-4,1.5) -- (-2,1)  node[midway,above,yshift=2] {$p_2,\,M_2$};
    \draw[black,-latex] (-4,1.5) -- (-3,1.25);
    \draw[black] (-4,-1.5) -- (-2,-1) node[midway, below] {$p_1,\,M_1$};
    \draw[black,-latex] (-4,-1.5) -- (-3,-1.25);
    \draw[newdarkblue2] (-2,-1) --  (-2,1) node[midway,left] {$q_3,\,m_3$};
    \draw[newdarkblue2,-latex] (-2,-1) -- (-2,0.2);
    \draw[darkred] (-2,1) -- (0,0) node[midway,above] {$~~~~~~q_1,\,m_1$};
    \draw[darkred,-latex] (-2,1) -- (-0.75,0.375);
    \draw[olddarkgreen] (0,0) -- (-2,-1) node[midway,below] {$~~~~~q_2,\,m_2$};
    \draw[olddarkgreen,-latex] (0,0) -- (-1.25,-0.625);
    \draw[black] (0,0) -- ++(30:2)  node[midway,above,yshift=10,xshift=10] {$p_3,\,M_3$};
    \draw[black] (0,0) -- ++(-30:2)  node[midway,below,yshift=-10,xshift=10] {$p_4,\,M_4$};
    \draw[black,-latex] (0,0) -- ++(30:1.2);
    \draw[black,-latex] (0,0) -- ++(-30:1.2);
\end{tikzpicture}
}
\label{tri3diag}
\end{equation}
We again have all of the same on-shell and momentum conservation constraints on $q_1^\mu$ and $q_2^\mu$ as we had in the bubble example. But we now need to impose additional constraints on $q_3^\mu$. The on-shell condition for this extra propagator is
\begin{equation}
    m_3^2 = q_3^2 = (q_1 - p_2)^2 = M_2^2 + m_1^2 - 2 p_2 \cdot q_1.
\end{equation}
Let us denote $q_1^\mu = k^\mu = (k^0,k^x,\vec{k}_\perp)$. 
The energies of $p_2^\mu$ and $q_1^\mu$ are fixed by the on-shell conditions and momentum conservation, as they were for the bubble:
\begin{equation}
    p_2^0 = \frac{Q^2 + M_2^2 - M_1^2}{2 Q}, \quad
    k^0 = \frac{Q^2 + m_1^2 - m_2^2}{2 Q}.
    \label{k0form}
\end{equation}
If we keep $\vec{p}_2$ in the $x$-direction and write $p_2^\mu = \left(p_2^0,p^x,\vec{0}_\perp \right)$, the on-shell condition for $q_3$ becomes
\begin{equation}
    \label{eq:triangle_parallel_component}
    \frac{-M_2^2 - m_1^2 + m_3^2}{2} + \frac{Q^2 + M_2^2 - M_1^2}{2 Q} \frac{Q^2 + m_1^2 - m_2^2}{2 Q} = p^x k^x.
\end{equation}
Recalling that $p^x$ can be expressed in terms of $Q$, this fixes the value of $k^x$. 
The constraint in Eq.~\eqref{eq:bubble-internal-kinematics} also applies in this example:
\begin{equation}
     \frac{(Q + m_1 + m_2)(Q - m_1 - m_2)(Q - m_1 + m_2)(Q + m_1 - m_2)}{4 Q^2} = (k^x)^2 + \vec{k}_\perp^2.
    \label{eq:triangle_sphere_component}
\end{equation}
Plugging Eq.~\eqref{eq:triangle_parallel_component} in to Eq.~\eqref{eq:triangle_sphere_component}, we see that $k_\perp^2$ is fixed in the on-shell space of internal and external kinematics associated with the triangle graph, so it described by the surface of a  $(d\!-\!2)$-sphere. The surfaces described by Eqs.~\eqref{eq:triangle_sphere_component} and \eqref{eq:triangle_parallel_component} are illustrated in Fig.~\ref{fig:triangleparaboloid}.

Recalling that the on-shell space of external kinematics is parametrized by $Q$ and $\theta$, the projection $\pi_\kappa$ maps from a $(d\!-\!1)$-dimensional space to a two-dimensional space. By computing the Jacobian, we find that the critical points in this map occur when $\vec{k}_\perp^2$ vanishes. We can check that this corresponds to the same value for $Q$ as is obtained by solving the Landau equations for the triangle diagram. Thus, we again see that the solutions to the Landau equations are reproduced by the critical points of the projection map $\pi_\kappa$.

It is worth making some further comments about how the nature of the critical values in this example depend on the space-time dimension $d$. Note that the condition in Eq.~\eqref{eq:triangle_parallel_component} is linear in the loop momentum, while the condition in Eq.~\eqref{eq:triangle_sphere_component} is quadratic.  When a quadratic constraint is involved, the critical point is called a \emph{quadratic pinch} (see chapter~V.2 of Ref.~\cite{pham2011singularities}). If we are in two dimensions, however, there is no $\vec{k}_\perp$ component.  Then, the three on-shell conditions for the three internal lines of the triangle can be written, after a change of coordinates, as
\begin{gather}
  s_1(x, t) = t - (x_1 + x_2), \\
  s_2(x, t) = x_1, \qquad
  s_3(x, t) = x_2,
\end{gather}
where $t$ can be expressed in terms of only external kinematic variables, while $x_1$ and $x_2$ are related to the two independent loop variables after solving for momentum conservation (we could take $x_1$ to be $k^0$ minus the value obtained by solving the equations above, and similarly for $x_2$ and $k^x$).  The surfaces defined by these equations intersect in the $(x_1, x_2)$ space only for $t = 0$, which is the location of the Landau singularity.  Note that the normals to these three surfaces are automatically linearly dependent (since there are three normals in a two-dimensional space). This situation is called a \emph{linear pinch} since we only have linear constraints on the loop momentum.  In the case of the triangle integral in two dimensions, the linear pinch singularity encodes a singularity of pole type.

\paragraph{Landau Singularities as Critical Values: Sketch of a Proof}~\\[-10pt]%
\label{sec:critproof}

\noindent Having worked through some examples, we now sketch the proof of Lemma~\ref{lem:crit}.  We start by considering a graph contraction $\kappa \colon G^\kappa \twoheadrightarrow G_0$ and the projection map between the associated on-shell spaces $\pi_\kappa \colon \S(G^\kappa) \to \S(G_0)$.  We would like to show that the critical points of the map $\pi_\kappa$ arise when the Landau equations corresponding to the contraction $\kappa$ are satisfied.

To show this, let us first describe the tangent spaces to $\S(G^\kappa)$ and $\S(G_0)$.  Given a point $p \in \S(G^\kappa)$, the tangent space at $p$ is denoted by $T_p \S(G^\kappa)$.  Since $\S(G^\kappa)$ is given by a set of constraints of the form $s_e = q_e^2 - m_e^2 = 0$, the tangent vector $\mathcal{X} \in T_p \S(G^\kappa)$ has to satisfy the conditions $\mathcal{X} s_e = 0$ for all $e \in E(G^\kappa)$.  Choosing a basis of circuits $\Chat(G^\kappa)$ for the diagram $G^\kappa$ (i.e.\ a basis of independent loop and external momenta, collectively denoted with $k_c^\mu$), any vector $\mathcal{X}$ in the tangent space can be written in components as
\begin{equation}
  \mathcal{X} = \sum_{c \in \Chat(G^\kappa)} X_c^\mu \frac \partial {\partial k_c^\mu}\, , \label{eq:tangent_components}
\end{equation}
where $X_c^\mu$ are the expansion coefficients. In addition to pointing in some tangent direction, we must also guarantee that the vector $\mathcal{X}$ is on $\S(G^\kappa)$. The conditions $\mathcal{X} s_e = 0$ become
\begin{equation}
  \sum_{c \in \Chat(G^\kappa)} b_{c e}(G^\kappa) q_e \cdot X_c = 0, \quad \forall e \in E(G^\kappa), \label{eq:tangent_vec_constraints}
\end{equation}
where we have used the notation $b_{ce}$ from Sec.~\ref{sec:review} for the circuit matrix of $G^\kappa$, in addition to $\frac {\partial q_e^\nu}{\partial k_c^\mu}=b_{c e}(G^\kappa) \delta_\mu^\nu$.

These conditions for $e \in E(G^\kappa)$ are linear constraints on the components $k_c^\mu$ of the tangent vector $\mathcal{X}$.  If these linear conditions are independent, then the space $\S(G^\kappa)$ has (real) dimension $d_{G^\kappa}= d (L+ \nex - 1)   - \nex -\nint$ where $\nint$ and $\nex$ are the number of internal and external edges respectively, and $L$ is the number of loops in $G^\kappa$.  A similar analysis can be done for the on-shell space $\S(G_0)$.  If the linear relations that arise for the graph $G_0$ are also all independent then we call the kinematics \textbf{generic}.  We will discuss exceptional kinematics in Sec.~\ref{sec:exceptional_kinematics}.

The tangent spaces of $\S(G^\kappa)$ and $\S(G_0)$ are thus given by
\begin{gather}
  T_p \S(G^\kappa) = \Bigl\{\mathcal{X} = \sum_{c \in \Chat(G^\kappa)} X_c^\mu \frac \partial {\partial k_c^\mu} \mid \sum_{c \in \Chat(G^\kappa)} b_{c e}(G^\kappa) q_e \cdot X_c = 0, \forall e \in E(G^\kappa)\Bigr\}, \\
  T_{\pi_\kappa(p)} \S(G_0) = \Bigl\{\mathcal{X} = \sum_{c \in \Chat(G_0)} X_c^\mu \frac \partial {\partial k_c^\mu} \mid \sum_{c \in \Chat(G_0)} b_{c e}(G_0) q_e \cdot X_c = 0, \forall e \in E(G_0)\Bigr\}.
\end{gather}
It is convenient to take, with some abuse of language, $\Chat(G^\kappa) = \Chat(\ker \kappa) \cup \Chat(G_0)$ and $E(G^\kappa) = E(G_0) \cup E(\ker \kappa)$ (see App.~\ref{sec:graph_theory} for a more detailed discussion of these decompositions).  The meaning of the first equality is that we can pick a set of fundamental circuits of the graph $G^\kappa$ which consists of the circuits arising by the graph embedding of $\ker \kappa$ in $G^\kappa$ and the circuits which contract to $\Chat(G_0)$. Finally, the tangent map $(\pi_\kappa)_{*, p} \colon T_p \S(G^\kappa) \to T_{\pi_\kappa(p)} \S(G_0)$ is defined by
\begin{equation}
  (X_1, X_2, \dotsc, X_{\nex-1}, X_{k_1}, X_{k_2}, \dotsc, X_{k_L}) \mapsto (X_1, X_2, \dotsc, X_{\nex-1}) \,,
\end{equation}
where the components are assumed to satisfy the linear constraints in the definitions of $T_p \S(G^\kappa)$ and $T_{\pi_\kappa(p)} \S(G_0)$.

Let us assume that the external kinematics are generic.  Then, $\pi_\kappa$ will only have critical points when there exists some linear dependence between the columns of the Jacobian matrix that go beyond the constraints that appear in the definition of $T_{\pi_\kappa(p)} \S(G_0)$. Using the fact that we can decompose $\Chat(G^\kappa) = \Chat(\ker \kappa) \cup \Chat(G_0)$ and $E(G^\kappa) = E(G_0) \cup E(\ker \kappa)$, we can rewrite Eq.~\eqref{eq:tangent_vec_constraints} in the form
\begin{equation}
  \label{eq:kernelvecs}
  \sum_{c \in \Chat(\ker \kappa)} b_{c e}(\ker \kappa) q_e \cdot X_c = 0, \quad \forall e \in E(\ker \kappa),
\end{equation}
These linear constraints are not independent when there exist $\alpha_e$ for $e \in E(\ker \kappa)$, not all zero, such that
\begin{equation}
  \sum_{e \in E(\ker \kappa)} \alpha_e b_{c e}(\ker \kappa) q_e^\mu = 0, \quad \forall c \in \Chat(\ker \kappa).
  \label{eq:projLandau}
\end{equation}
This means that the space of $X_c$ for $c \in \Chat(\ker \kappa)$ is larger than na\"ive dimension counting would suggest, which in turns means that they impose extra constraints on $X_c$ for $c \in \Chat(G_0)$.  Therefore the image of $(\pi_\kappa)_{*, p}$ is smaller than na\"ive dimension counting would suggest.  This is precisely the critical condition for the map $\pi_\kappa$, and Eq.~\eqref{eq:projLandau} is precisely the loop Landau equation.

\subsection{Codimension of Landau Singularities}
\label{sec:codimension}

One of the crucial properties of different solutions to the Landau equations that will enter our definition of principal Pham loci is their (complex) \textbf{codimension} in the space of external kinematics. The most important class of solutions are those that are of codimension one, since contours can get trapped around codimension-one singularities.
For example, a point is codimension one in the complex plane and a closed contour around such a point cannot be unraveled. In contrast, a point has codimension two in ${\mathbb C}^2$. A curve in ${\mathbb C}^2$ can be deformed around any such point. Correspondingly, determining when a solution has codimension one will help us determine the nature of the corresponding singularity.

Unfortunately, determining the codimension of a Pham locus is not usually straightforward; in general, it depends on both the topology of the graph and the number of spacetime dimensions. However, one simple rule exists when all masses are generic: the codimension of the solution to the Landau equations is the same as the dimension of the solution in the $\alpha$ variables, ignoring the $\sum \alpha_e = 1$ constraint. As the Feynman parameters appear linearly and homogeneously in the Landau loop equations, the sets of $\alpha_e$ that satisfy these equations form a vector space, which makes determining this dimension a much easier problem.
In cases that involve the same number of internal edges as Landau loop equations, the solution space in the $\alpha_e$ variables will be one dimensional and the only freedom we have is to rescale all of the Feynman parameters by the same factor.\footnote{In a Feynman integral, the Feynman parameters are constrained by the inhomogeneous equation $\sum \alpha_e = 1$ which fixes the scaling freedom in $\alpha$. Then the minimal solution space is zero-dimensional: all the $\alpha$'s are fixed. We prefer not to impose this constraint at the moment to avoid a plethora of $-1$'s in our formulae.} If there are fewer equations than internal edges, then the solution space for $\alpha_e$ can have higher dimension. This possibility was already recognized by Landau in his original paper~\cite{landau1959}, and proven by Pham~\cite{pham1968singularities}. 

As a first example, consider the bubble with incoming momentum $p^\mu$,
\begin{equation}
\resizebox{!}{1.1cm}{
\begin{tikzpicture}[baseline=(current bounding box.center),
    line width=1.5,scale=0.7]
    \draw[black] (-3,0) -- node[midway, above] {$p,M$} (-1,0);
    \draw[black,-latex] (-3,0) -- ++(0:1.2);
    \draw[darkred] (1,0) arc (0:180:1);
    \draw[olddarkgreen] (1,0) arc (0:-180:1);
    \draw[black] (1,0) -- node[midway, above] {$p,M$} (3,0); 
    \draw[black,-latex] (1,0) -- ++(0:1.2);
    \draw[-latex,darkred] (0.1,1) -- (0.2,1)  node[above] {$q_1,m_1$};
    \draw[-latex,olddarkgreen] (0.1,-1) -- (0.2,-1) node[below] {$q_2,m_2$};
\end{tikzpicture}
} \label{eq:bubble_codim}
\end{equation}
Since there is only one external momentum, the loop and momentum conservation equations force $q_1^\mu, q_2^\mu$ and $p^\mu$ all to be proportional. In the frame where $\vec{p}=0$, this means that the four-vector of the internal momenta only has an energy component, equal to plus or minus its mass (due to the on-shell conditions). In this frame, the loop equation $\alpha_1 q^\mu_1 - \alpha_2 q^\mu_2=0$
thus reduces to $\alpha_1 m_1 \pm \alpha_2 m_2=0$, which leads to a one-dimensional solution for the $\alpha_i$. As a result, we expect the solution to the Landau equations to be of codimension one. Indeed, in the rest frame momentum conservation implies that $\sqrt{p^2} = m_1\pm m_2$, which are the conditions for the normal and pseudonormal thresholds. This constitutes a zero-dimensional subspace of the one-complex dimensional space spanned by $p^2$. When the loop momentum goes to infinity one also finds a second-type singularity, for $\alpha_1=\alpha_2$ and $p^2=0$.

For a higher-codimension example, consider the following two-loop graph, which still only depends on a single external momentum:
\begin{equation}
\resizebox{!}{1cm}{
\begin{tikzpicture}[baseline=(current bounding box.center),
    line width=1.5,scale=0.7]
    \draw[black] (-3,0) -- node[midway, above] {$p$} (-1,0);
    \draw[black,-latex] (-3,0) -- ++(0:1.2);
    \draw[newdarkblue2] (-0.2,0.95) -- node[midway,left] {$q_5$} (0.2,-0.95);
    \draw[black] (1,0) -- node[midway, above] {$p$} (3,0); 
    \draw[black,-latex] (1,0) -- ++(0:1.2);
     \draw[newdarkblue2,-latex] (-0.2,0.95) -- (0.05,-0.2375);
     \draw[darkred,-latex] ({+cos(135)},{sin(135)}) -- ({+cos(135)+0.1},{sin(135)+0.1});
     \draw[darkred,-latex] ({+cos(30)},{sin(30)}) -- ({+cos(30)+0.1},{sin(30)-0.1});
     \draw[darkred,-latex] ({+cos(-135)},{sin(-135)}) -- ({+cos(-135)+0.1},{sin(-135)-0.1});
     \draw[darkred,-latex] ({+cos(-30)},{sin(-30)}) -- ({+cos(-30)+0.1},{sin(-30)+0.1});
     \draw[darkred] (1,0) arc (0:180:1)  node[pos=0.25,above] {$q_2$}
     node[pos=0.75,above] {$q_1$};
     \draw[darkred] (1,0) arc (0:-180:1)
     node[pos=0.25,below] {$q_3$}
     node[pos=0.75,below] {$q_4$};
\end{tikzpicture}
}
\label{eq:2loopbub}
\end{equation}
For simplicity, let us look at a potential solution of the Landau equations with the energy flow indicated with arrows in Eq.~\eqref{eq:2loopbub}, and ignore solutions with infinite loop momenta. We also assume that each propagator with momentum $q_i$ has a nonzero mass $m_i$, which cannot be varied. 
In this case, there are two one-dimensional loop equations: $\alpha_1 m_1 + \alpha_5 m_5 - \alpha_4 m_4=0$ and $\alpha_2 m_2 - \alpha_3 m_3 - \alpha_5 m_5=0$. Since there are five Feynman parameters, the solution space is three-dimensional. This leads us to expect that the Landau equations will give rise to a codimension-three solution. However, since $p^\mu$ is still one-dimensional this means that this configuration has no solution for generic masses. For example, momentum conservation implies $\sqrt{p^2} = m_1+m_2 = m_2+m_3$, which is already in conflict with the generic-mass assumption. This diagram does, however, have solutions to the Landau equations if we take some of the $\alpha_e=0$, thereby lowering the dimension of the space of $\alpha_e$ without (necessarily) reducing the number of equations. For example, taking $\alpha_1=\alpha_4=\alpha_5=0$ reduces the equations to the same ones as for the bubble diagram in Eq.~\eqref{eq:bubble_codim}, which we saw corresponded to a solution of codimension one. Moreover, setting $\alpha_1=\alpha_3=0$ reduces this graph to the two-loop sunrise graph, whose Landau equations admit a one-dimensional solution in the space of Feynman parameters (for the chosen energy flow, the normal-threshold solution would be at $p^2=(m_2+m_4+m_5)^2$), corresponding to a codimension-one Pham locus~\cite{landau1959,OKUN1960261}.

As a third and final example, consider the triangle-box diagram:
\begin{equation}
\resizebox{!}{1.2cm}{
\begin{tikzpicture}[baseline=(current bounding box.center),
    line width=1.5,scale=0.4]
    \draw[black] (-2.5,0) -- (0,0);
    \draw[darkred] (2,-1.5) -- (0,0) -- (2,1.5);
    \draw[darkorange] (2,1.5) -- (2,-1.5);
    \draw[olddarkgreen] (2,1.5) -- (4.5,1.5) -- (4.5,-1.5) -- (2,-1.5);
    \draw[black] (4.5,1.5) -- (6,2.3);
    \draw[black] (4.5,-1.5) -- (6,-2.3);
\end{tikzpicture}
}
\end{equation}
We can work in the center-of-mass frame of the incoming $p_1^\mu$, and see that the outgoing momenta $p_2^\mu$ and $p_3^\mu$ can be taken to have only two components (since their spatial momenta are back-to-back in this frame). The loop momenta $k_1^\mu$ and $k_2^\mu$ therefore have two undetermined components each, leading to four loop Landau equations in total.  Accounting for the $\text{GL}(1)$ covariance and recalling that the on-shell equations do not involve the $\alpha_e$, we note that there are only five linear equations for the six unknown $\alpha_e$. So, if a solution exists, it must have at least one free parameter in the $\alpha_e$-solution, leaving us with a two-dimensional surface of solutions in the space of Feynman parameters. Thus, the leading singularity of the triangle-box has a codimension of at least two, if it exists. The Landau-equation solution of this diagram was indeed found to be of codimension-two in Ref.~\cite{Mizera:2021icv}. One can find more worked examples in Ref.~\cite{OKUN1960261,Mizera:2021icv}.

\subsection{Principal Loci}
\label{sec:principal}

Not all of the singularities of $I_G(p)$ lead to discontinuities that can be computed by Cutkosky's formula, once this integral is analytically continued off the physical sheet. In part, this is because the validity of Cutkosky's formula requires applying the Picard--Lefschetz theorem, which can only be done for singularities of type $S_1$, which is to say singularities that are codimension one and that are associated with a projection map which is locally spherical (taking the form $\pm x_1^2 \pm \cdots \pm x_n^2$).\footnote{$S$-type is also called transversality. We prefer to use the term $S$-type to avoid confusion with transversally intersecting Pham loci, which will play an important role in Sec.~\ref{sec:codim2}.} The former property ensures that the contour cannot be trivially deformed to avoid the singularity, while the latter property ensures that the singularity is of stable topological type---or, more informally, that the topology of the singularity is invariant under small deformations.\footnote{The classification of stable singularities of differentiable maps was studied by Thom~\cite{MR87149}, and in the context of scattering amplitudes by Pham~\cite{pham}.} In addition, the vanishing cell should be real, so that the delta functions in Cutkosky's formula make sense. Luckily, these properties are all satisfied by the \textbf{principal Pham locus}, which is defined as the union of all $\alpha$-positive branches $\P_\kappa$ of codimension one that have at least one non-vanishing $\alpha$ in each loop. Many of the singularities of $I_G(p)$ that are in the physical region are on the principal Pham locus (with the exceptions being singularities for which all of the $\alpha_e$ in a loop vanish). In the remainder of this section, we describe these properties in more detail and prove them for principal Pham loci.

\begin{figure}[t]
    \centering
\begin{tikzpicture}
 \node (image) at (-0.2,0) {
        \includegraphics[width=0.45\textwidth]{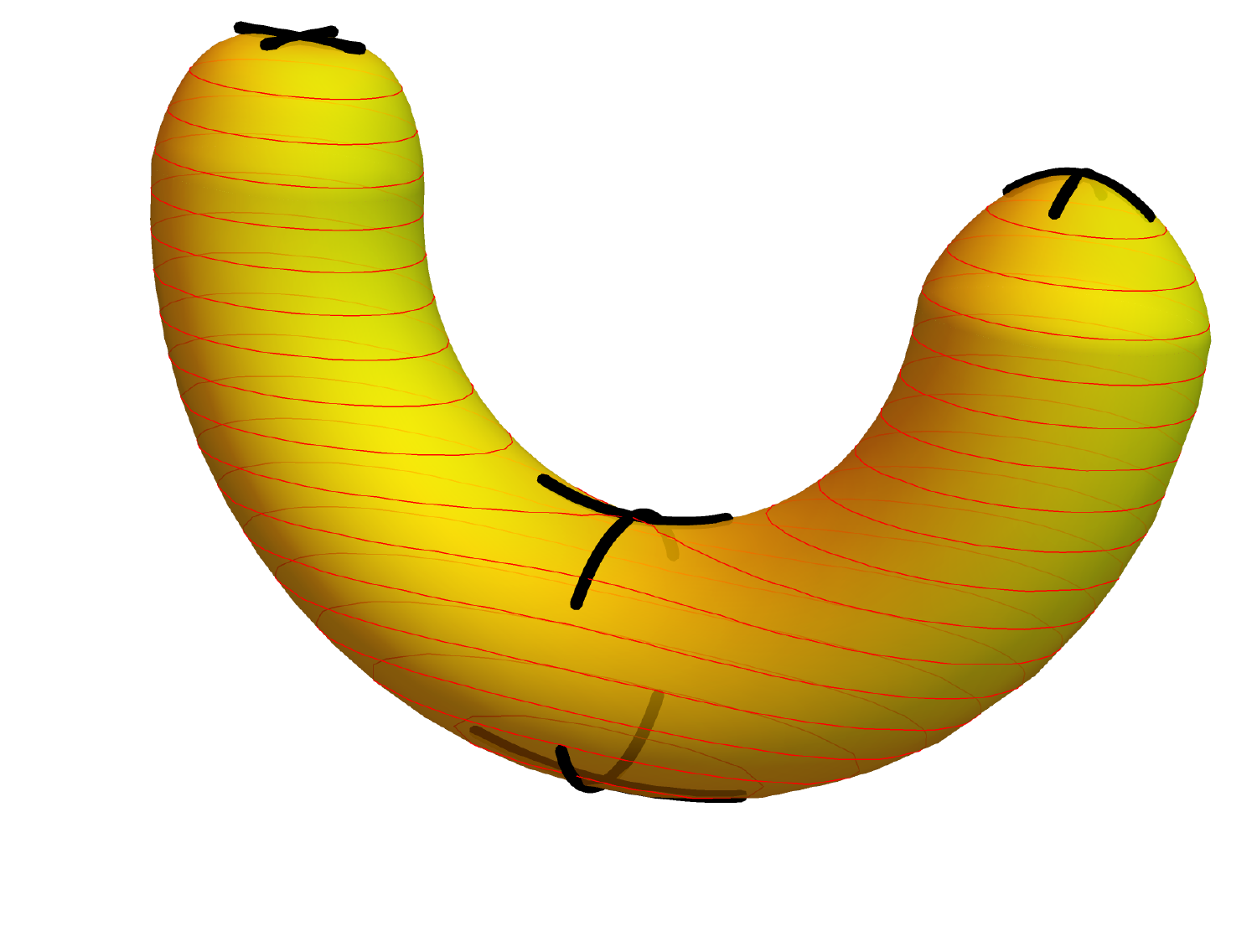}
    };   
\draw[->] (3.5,1.6)  node[black,right,scale=0.8] {$S^-$} -- (2.8,1.6);
\draw[->] (-0.2,0.8)  node[above,above,scale=0.8] {$S$-type} -- (-0.2,0);
\draw[->] (-2.5,-1.2)  node[black,left,scale=0.8] {$S^+$} to[out=-30,in=210] (-0.6,-1.8);
\draw[->] (-3.5,2.4)  node[black,left,scale=0.8] {not $S$-type} -- (-2.8,2.4);
\node[] at (4.8,0) {$\xmapsto{~~\pi_\kappa~~}$};
\node[] at (6.5,0) {$p$};
\draw[->,line width=0.75mm, olddarkgreen] (6,-1.5) -- (6,2);
\end{tikzpicture} 
\caption{Critical points of projection maps can be characterized as $S$-type if all of the eigenvalues of the Hessian are non-zero. They are $S^+$ if the Hessian has all positive eigenvalues, while they are $S^-$ if the eigenvalues are all negative. A saddle point is of $S$-type, but has eigenvalues of both signs. If some eigenvalues are zero, as is the case for the critical point on the top-left of our shape, which scales like $-r^4$,
the topological type is considered unstable. In the physical region all singularities of Feynman integrals are of $S^+$-type.}
    \label{fig:Storus}
\end{figure}
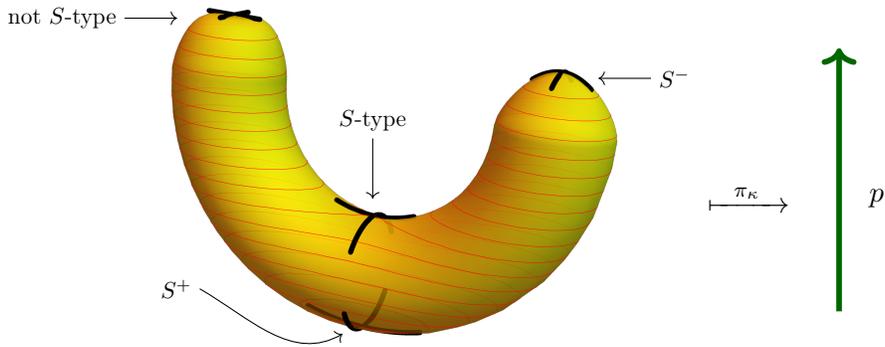

As described in Sec.~\ref{sec:critical}, we can analyze the singularities of Feynman integrals by studying differentiable maps $f \colon \bbR^n \to \bbR^m$, or more generally maps $f \colon M \to N$ between pairs of manifolds $M$, $N$. More precisely, to analyze the properties of the branch $\B_\kappa$ of the Landau variety we can study the singularities of the map $\pi_\kappa \colon \S(G^\kappa) \rightarrow \S(G_0)$. Such projections between on-shell spaces can be put in a canonical form in which at most one of the component functions is quadratic in the coordinates, while the rest are linear. We will give an explicit example of this type of coordinate chart in Sec.~\ref{sec:trangentialmaps}; for now, we merely assume $\pi_\kappa$ can be put in this form.\footnote{A coordinate-independent construction of the transversal Hessian can also be found in Sec.~2.5 and Sec.~2.6 of Ref.~\cite{pham1968difftop}.  In App.~\ref{sec:hessians} we provide a simpler discussion based on Lagrange multipliers.}

A differential map $f \colon \mathbb{R}^n \to \mathbb{R}$ has a singularity of type $S_1$ if it can be written as $f(x_1, \dotsc, x_n) = \pm x_1^2 \pm \dotso \pm x_n^2$ in the neighborhood of a critical point. (Differentiable maps whose image has dimension higher than one can also have $S_1$ singularities; then, these singularities are identified by choosing a parametrization in which all other components are linear, as can be done for $\pi_\kappa$.) An $S_1$ singularity is more specifically considered to be of type $S_1^{+}$ or $S_1^{-}$ if the Hessian matrix for the map $f$ at the critical point is either positive or negative definite, respectively. More generally, a differential map $f \colon \mathbb{R}^n \to \mathbb{R}$ is said to have transversal index $\eta$ if $f(x_1, \dotsc, x_n) = \pm x_1^2 \pm \dotso \pm x_n^2$ with exactly $\eta$ minus signs. 
This corresponds to the Hessian matrix for $f$ having exactly $\eta$ negative eigenvalues. So, the $S_1^+$ and $S_1^-$ singularities discussed here will have a transversal index of 0 and $n$, respectively.

In the context of Feynman integrals, the key property of $S_1^{\pm}$ singularities is that there is a well-defined notion of being on one side or the other of the singularity for real kinematics. This is illustrated in Fig.~\ref{fig:Storus}. There, we see that it is clear whether you are above or below the $S^{\pm}$ critical point in the image of $\pi_\kappa$, while there is no distinct notion of being above or and below the saddle point after this projection. In this way, every $S_1^{\pm}$ singularity splits the real points of $\S(G_0)$ in two. On one side of this division, which we refer to as \textbf{above the threshold}, the on-shell loop momenta are all real. Since the projection map is a positive quadratic form near the singularity, this implies that some of the momenta are necessarily complex on the other side of the singularity. We refer to this region as \textbf{below the threshold}.

To check whether a given branch point $p^\ast \in \B_\kappa$ is of type $S_1^+$, we use the fact from Sec.~\ref{sec:critical} that the Pham locus corresponding to $p^\ast$ is the critical value of the projection map $\pi_\kappa$ from $\S(G^\kappa)$ to $S(G_0)$. Choosing coordinates $X_c^\mu$ on $\S(G^\kappa)$ as in Sec.~\ref{sec:critical}, we must expand around the critical points of $\S(G^\kappa)$ to see how the map behaves in the vicinity of the singularity.
We can define a quadratic form
\begin{equation}
A(X_c^\mu) = \sum_{\substack{c_1, c_2 \in \Chat(\ker \kappa) \\ e \in E(\ker \kappa)}} \alpha_e b_{c_1 e} b_{c_2 e} X_{c_1} \cdot X_{c_2} \, , \label{eq:quad_form}
\end{equation}
where $X_c^\mu$ are the coordinates on $\S(G^\kappa)$ used in Sec.~\ref{sec:critproof}.  We show in App.~\ref{sec:hessians} that the definiteness of this quadratic form implies that the critical point is a local extremum, not a saddle point.
Then the $S$-type condition is that the quadratic form $A(X_c^\mu)$, which by definition is restricted to the kernel of the tangent map (defined in Eq.~\eqref{eq:kernelvecs}), is positive or negative definite when evaluated at $p^\ast$. If, instead, $\eta \neq 0$, then the vanishing cell that enters the monodromy computation using Eq.~\eqref{eq:PLtheorem} could be complex. If some of the eigenvalues of the Hessian vanish, then the singularity is not even of $S$-type (see again an example in Fig.~\ref{fig:Storus}).

In~\cite{pham}, it was shown that all singularities on the principal Pham locus are of type $S_1^+$. The proof goes as follows.  We first define a set of vectors $Y_e^\mu = \sum_{c \in \Chat} b_{ce} X_c^\mu$. We can think of the vectors $X_c^\mu$ as small perturbations around the singularity in the on-shell space $\S(G^\kappa)$, so the $Y_e^\mu$ represent the shift in each on-shell momenta $q_e^\mu$. In terms of these vectors, the quadratic form becomes
\begin{equation}
    A(X_c^\mu) = \sum_e \alpha_e Y_e^2 \,.
\end{equation}
Likewise, we know that the vectors $Y_e^\mu$ must satisfy momentum conservation.
Since all of the $Y_e^\mu$ are orthogonal to time-like vectors $q_e^\mu$, we have that $Y_e^2 \leq 0$ and thus $A(X_c^\mu)$ is negative semidefinite for $\alpha_e \geq 0$. To show under which conditions $A(X_c^\mu)$ is negative \textit{definite}, we assume that a solution exists for which $A(X_c^\mu) = 0$ with $X_c^\mu$ being nonzero. Then, we need $Y^\mu_{e'} = 0$ for all $e'$ such that $\alpha_{e'} \neq 0$. Let us analyze what momentum conservation would imply for such a configuration. Recall that the $Y_e^\mu$ are defined in terms of the $X_c^\mu$, which are in the tangent space of $\pi_\kappa$, so the momentum-conservation constraint can be thought of as one for a graph without any external edges. Moreover, the $Y_e^\mu$ which are zero can be trivially excluded from momentum-conservation constraints. So, the graph that satisfies the momentum conservation consists only of the lines for which $\alpha_{e}=0$ and $Y_{e}^\mu \neq 0$, and none of the external ones. It is impossible to satisfy these constraints for a tree graph.
Therefore, a singularity must be $S$-type unless  there is a loop in which all Feynman parameters vanish.\footnote{Note that a loop with all $\alpha_e=0$ is not the same as the leading singularity of the Pham locus of the diagram in which all these legs have been contracted, since all edges are on shell in the Pham locus, even the ones for which $\alpha_e=0$.}

Since principal Pham loci correspond to singularities of type $S_1^+$, the vanishing cells that they give rise to are real. 
The remaining problem is to show that for $\alpha$-positive singularities, the cell is carved out by $s_e \geq 0$ for all $e$, or equivalently that the cell defined by these inequalities vanishes as the singularity is approached. The argument is as follows. We can expand in the loop momenta around the Pham locus to get
\begin{equation}
    \sum_{e \in E (\ker \kappa)} \alpha_e (q_e^2-m_e^2) = 
    \ell + A(X_c^\mu) + \,\cdots\,,
    \label{eq:Landau_expansion}
    \end{equation}
where $\ell = 0$ at the singularity $\B_\kappa$, and
$A(X_c^\mu)$ is negative definite, as shown above. Our conventions are such that $\ell < 0$ below the threshold.
Since the quadratic form $A(X_c^\mu)$ is negative-definite, the expression in Eq.~\eqref{eq:Landau_expansion} cannot not vanish below the threshold, since it is always negative.  Above the threshold $\ell > 0$, so the equation $-A(X_c^\mu) = \ell$ has solutions. However, since $A(X_c^\mu)$ is negative-definite, these solutions in $X_c^\mu$ form a compact set with the topology of a sphere.  This is the vanishing sphere, since its radius shrinks to zero as the singularity is approached.  We can take the vanishing cell to be the interior of this sphere. In other words, if we approach the singularity in the direction $\ell \to 0^+$, then the hypersurface defined by $q_e^2-m_e^2 \geq 0$ will be small and bounded if all $\alpha_e \geq 0$.

Before concluding this subsection, let us again highlight that it is the fact that principal Landau varieties correspond to singularities of type $S_1^+$ that allows us to replace propagators with delta functions in Cutkosky's formula,  and thus localize to real kinematics. This is also  emphasized in the proofs in Refs.~\cite{Bloch:2015efx,Muhlbauer:2022ylo}. Second, we emphasize that if $A(X_c^\mu)$ is not a definite quadratic form, then the expression $\ell + A(X_c^\mu)$ in the denominator can vanish for either $\ell > 0$ and for $\ell < 0$.  This type of situation can arise when computing sequential discontinuities of Feynman integrals, and significantly complicates the analysis.

  \section{Hierarchical Sequential Discontinuities}
\label{sec:sequential_disc}
At the end of Sec.~\ref{sec:iterated}, we noticed that computing the double discontinuity of the all-mass triangle integral with respect to the bubble singularity at $y_{12}=1$ and then the triangle singularity at $D=0$ gave the same result as just computing a single discontinuity with respect to the triangle singularity. In the notation introduced in that section, this corresponds to the relation:
\begin{equation}
     \big(\bbone-\monM_{ D=0} \big)
     \big(\bbone-\monM_{ y_{12}=1} \big)
         I_\tritcol =
     \big(\bbone-\monM_{D=0} \big) 
     I_\tritcol \, .
     \label{PhamAtri2}
\end{equation}
In this section, we show that this is just the simplest example of a broad class of constraints that take this form. The requirements for a relation such as Eq.~\eqref{PhamAtri2} to exist are easy to spell out using the concepts we have already introduced. We must simply have that both singularities are principal Pham loci, and that the contraction associated with the second singularity dominates the contraction associated with the first. More formally, this result can be stated as follows:

\begin{theorem}[Pham] 
For a series of contractions $G \twoheadrightarrow G^\kappap \twoheadrightarrow \cdots \twoheadrightarrow G^\kappa \twoheadrightarrow G_0$ the relation 
\begin{equation}
\Big(\bbone - 
  \monM_{\P_\kappap} \Big)\cdots \Big(\bbone - \monM_{\P_\kappa} \Big) \I_G(p) 
  = \Big(\bbone -   \monM_{\P_\kappap} \Big) \I_G(p)
  \label{eq:phamcodim1rel}
\end{equation}
\noindent
holds when $\P_\kappa \cdots \P_\kappap$ correspond to principal Pham loci, and $p$ is in the physical region. 
\label{thm:co1}
\end{theorem}
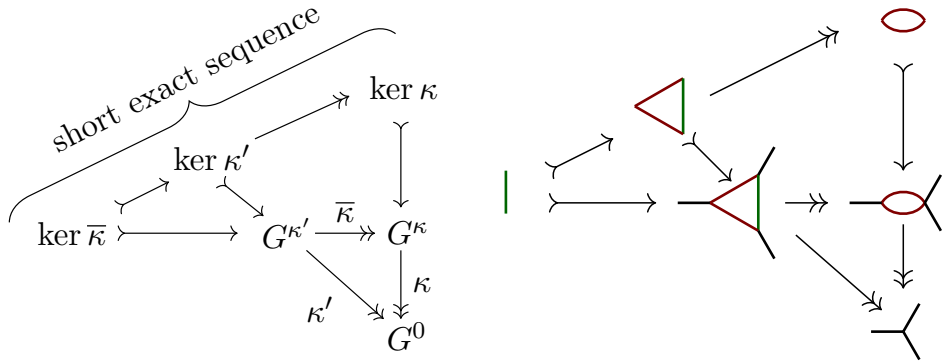
\begin{figure}[t]
\centering
\hspace{-5mm}
\resizebox{6cm}{!}{
\begin{tikzpicture}
\node[left] at (-0.4,0) {$G^\kappap$};
\node [right] at (0.2,0) {$G^\kappa$};
\node [right] at (0.2,-1.2) {$G^0$};
\draw [->>] (-0.5,0) -- (0.2,0) node[midway,above] {$\kappas$};
\draw [->>] (0.5,-0.2) -- (0.5,-1) node[midway,right] {$\kappa$};
\draw [->>] (-0.6,-0.2) -- (0.3,-1) node[midway,below left] {$\kappap$};
\draw [>->] (-1.6,0.6) -- (-1.1,0.2);
\draw [>->] (-2.8,0.3) --(-2.2,0.6);
\node [above] at (-1.7,0.6) {$\ker \kappap$};
\draw [>->] (-2.8,0) -- (-1.4,0);
\node [left] at (-2.8,0) {$\ker \kappas$};
\draw [->>] (-1.2,1.1) -- (-0.1,1.6);
\node [right] at (-0,1.7) {$\ker \kappa$};
\draw [>->] (0.5,1.3) -- (0.5,0.3);
\draw [decorate,decoration={brace,amplitude=10pt,raise=4pt}]
(-4,0) -- (0.5,2.2);
\node[rotate=27] at (-2,1.8) {short exact sequence};
\end{tikzpicture}}
\hspace{3mm}
\resizebox{6cm}{!}{
\begin{tikzpicture}
\begin{scope}[shift={(-1.0,0)}]
    \coordinate (a) at (60:0.3);
    \coordinate (b) at (180:0.3);
    \coordinate (c) at (-60:0.3);
    \draw[darkred, line width=0.7] (a) -- (b);
    \draw[olddarkgreen, line width=0.7] (c) -- (a);
    \draw[darkred, line width=0.7] (b) -- (c);
    \draw[line width=0.7] (a) -- ++(60:0.3);
    \draw[line width=0.7] (b) -- ++(180:0.3);
    \draw[line width=0.7] (c) -- ++(-60:0.3);
\end{scope}
\begin{scope}[shift={(0.5,0)}]
    \coordinate (a) at (-0.2,0);
    \coordinate (b) at (0.2,0);
    \draw[darkred, line width=0.7] (a) to [bend right=60] (b);
    \draw[darkred, line width=0.7] (a) to [bend left=60] (b);
    \draw[line width=0.7] (a) -- ++(180:0.3);
    \draw[line width=0.7] (b) -- ++(60:0.3);
    \draw[line width=0.7] (b) -- ++(-60:0.3);
\end{scope}
\begin{scope}[shift={(0.5,-1.2)}]
    \coordinate (a) at (60:0);
    \coordinate (b) at (180:0);
    \coordinate (c) at (-60:0);
    \draw[line width=0.7] (a) -- ++(60:0.3);
    \draw[line width=0.7] (b) -- ++(180:0.3);
    \draw[line width=0.7] (c) -- ++(-60:0.3);
\end{scope}  
 \begin{scope}[shift={(-1.7,0.9)}]
    \coordinate (a) at (60:0.3);
    \coordinate (b) at (180:0.3);
    \coordinate (c) at (-60:0.3);
    \draw[darkred, line width=0.7] (a) -- (b);
    \draw[olddarkgreen, line width=0.7] (c) -- (a);
    \draw[darkred, line width=0.7] (b) -- (c);
\end{scope}
\begin{scope}[shift={(-3.2,0.1)}]
    \draw[olddarkgreen, line width=0.7] (0,-0.2) -- (0,0.2);
\end{scope}
\begin{scope}[shift={(0.5,1.7)}]
    \coordinate (a) at (-0.2,0);
    \coordinate (b) at (0.2,0);
    \draw[darkred, line width=0.7] (a) to [bend right=60] (b);
    \draw[darkred, line width=0.7] (a) to [bend left=60] (b);
\end{scope}
\draw [->>] (-0.6,0) -- (-0.2,0); 
\draw [->>] (0.5,-0.2) -- (0.5,-0.8);
\draw [->>] (-0.5,-0.3) -- (0.3,-1);
\draw [>->] (-1.5,0.6) -- (-1.1,0.2);
\draw [>->] (-2.8,0.3) --(-2.2,0.6);
\draw [>->] (-2.8,0) -- (-1.8,0);
\draw [->>] (-1.3,1.0) -- (-0.1,1.6);
\draw [>->] (0.5,1.3) -- (0.5,0.3);
\end{tikzpicture}}
\caption{
When one Pham locus dominates another, the kernels of their contraction maps form a short exact sequence. If the loci are principal, the discontinuity of a Feynman integral $\I_G(p)$ with respect to $\P_\kappap$ is unaffected by first computing a discontinuity with respect to $\P_\kappa$. The diagram on the left depicts the generic case, while the one on the right shows the example of the triangle and bubble singularities.}
\label{fig:co1pham}
\end{figure}
\noindent We will focus on the case that involves just two discontinuities on the left, as the generalization to further discontinuities is straightforward.

We first note that both $\P_\kappa$ and $\P_\kappap$ (as well as any additional Pham loci that appear in Eq.~\eqref{eq:phamcodim1rel}) are of codimension one, since these loci are principal. So, the discontinuity around both $\P_\kappa$ and $\P_\kappap$ is in general nonzero. Indeed, the relation would be trivial if either locus was of higher codimension. In addition, the absorption integral from the first discontinuity on the left must be nonzero, which by Theorem~\ref{thm:pham} implies that $\kappap$ dominates $\kappa$. We therefore have the following picture that describes the sequential discontinuity around two loci:
\begin{equation}
\label{eq:dominating}
\begin{gathered}
\begin{tikzpicture}
\node[left] at (-1.5,0) {$G$};
\draw [->>] (-1.5,0) -- (-0.7,0) ;
\node[right] at (-0.7,0) {$G^\kappap$};
\draw [->>] (0,0) -- (1,0) ;
\draw [->>] (1.3,-0.3) -- (1.3,-1.2);
\draw [->>] (0,-0.2) -- (1.1,-1.2);
\node [right] at (1,0) {$G^\kappa$};
\node [right] at (1,-1.5) {$G_0$};
\node [above] at (0.5,0) {$\kappas$};
\node [] at (1.6,-0.6) {$\kappa$};
\node [] at (0.3,-0.8) {$\kappap$};
\end{tikzpicture}
\end{gathered}
\end{equation}
These contractions can be combined with their kernels into a larger diagram, as shown in Fig.~\ref{fig:co1pham}.

As written, Theorem~\ref{thm:co1} places constraints on sequential discontinuities of Feynman integrals. However, we emphasize that we can also apply the theorem iteratively to constrain absorption integrals.
That is, the same constraints must also be satisfied by the absorption integrals that are associated with all diagrams that $G^\kappa$ can be contracted to. In these cases, the bottom graph in~\eqref{eq:dominating} should also be replaced by whatever Landau diagram $G^\kappa$ is contracted to:
\begin{equation}
\label{eq:dominating2}
\begin{gathered}
\begin{tikzpicture}
\node[left] at (-1.5,0) {$G$};
\draw [->>] (-1.5,0) -- (-0.7,0) ;
\node[right] at (-0.7,0) {$G^\kappap$};
\draw [->>] (0,0) -- (1,0) ;
\draw [->>] (1.3,-0.3) -- (1.3,-1.2);
\draw [->>] (0,-0.2) -- (1.1,-1.2);
\node [right] at (1,0) {$G^\kappa$};
\node [right] at (1,-1.5) {$G^{\kappa''}$};
\node [above] at (0.5,0) {$\kappas$};
\node [] at (1.6,-0.6) {$\kappa$};
\node [] at (0.3,-0.8) {$\kappap$};
\end{tikzpicture}
\end{gathered}
\end{equation}
In this way, Theorem~\ref{thm:co1} should be thought of as placing constraints on arbitrarily long sequences of discontinuities of Feynman integrals.  Of course, in many cases, this constraint will be trivially satisfied, as the discontinuities with respect to $\mathcal{P}_\kappa$ and $\mathcal{P}_{\kappa^\prime}$ will both be zero. But as seen at the end of Sec.~\ref{sec:iterated}, and as we will see in further examples in this section, this constraint is not always satisfied in this trivial way.

\subsection{Tangential Maps \label{sec:trangentialmaps}}

One of the requirements for Pham loci to be principal is that they correspond to singularities of type $S_1^+$. This in turn implies that the Hessian of the projection map is positive or negative definite, as discussed in Sec.~\ref{sec:principal}. 
We now take a moment to discuss this requirement in more detail. As we will see, when $\P_\kappa$ and $\P_\kappap$ are principal Pham loci and $\kappap$ dominates $\kappa$,
the definiteness of the Hessian leads naturally to the geometric picture that
$\mathcal{P}_\kappa$ and $\mathcal{P}_{\kappa^\prime}$ intersect  \textbf{tangentially}.

In the neighborhood of $S_1^+$ critical points, Morse's lemma tells us that the projection map $\pi$ can be written as quadratic form of corank one.  More explicitly, if $\pi: \bbR^n\to \bbR^p$ with $n>p$, then one can find coordinates near the critical point such that $p-1$ of the coordinates transform linearly, while the remaining $p^{\text{th}}$ coordinate is a quadratic function of the other $n-p$ variables in $\bbR^n$. To be concrete, let us say that $\S(G^\kappap)$ has dimension $n$,  $\S(G^\kappa)$ has dimension $p$, and $\S(G_0)$ has dimension $q$. 
We pick coordinates $x \in \mathbb{R}^n$, $y \in \mathbb{R}^p$, and $z \in \mathbb{R}^q$, and thus have maps that can be depicted as
\begin{equation}
\begin{tikzcd}
G^\kappap \arrow[r,twoheadrightarrow, "\kappas"] \arrow[dr,twoheadrightarrow,"\kappap"{below}] & G^\kappa 
\arrow[d,twoheadrightarrow, "\kappa"] \\
& G_0
\end{tikzcd}
\hspace{10mm}
\begin{tikzcd}
x\in \mathbb{R}^n \arrow[r, "\pi_\kappas"] \arrow[dr,"\pi_\kappap"{below}] & y \in \mathbb{R}^p \arrow[d, "\pi_\kappa"] \\
&  z \in \mathbb{R}^q
\end{tikzcd}
\end{equation}
For definiteness, we take $n > p > q$; other cases can be treated similarly. 
As both $\P_\kappa$ and $\P_\kappap$ are principal, they are singularities of type $S_1^+$ and the projection maps are corank one. We now explore the forms taken by the projection maps by trial and error.  For a more systematic approach, see Ref.~\cite{pham}.

Since $\pi_{\kappap} = \pi_{\kappa} \circ \pi_{\kappas}$, we begin by parameterizing $\pi_\kappa$ and $\pi_\kappas$, after which
we can deduce $\pi_\kappap$. 
Starting with the projection associated with $\P_\kappa$, by Morse's lemma we can choose coordinates $z = \pi_\kappa(y)$
in the neighborhood of the critical point of the form
\begin{equation}
  \pi_{\kappa} : \quad (z_1, \ldots, z_{q-1}) = (y_1, \ldots y_{q-1}), \quad z_q =  y_q^2 + \cdots + y_p^2\,.
\end{equation}
If we choose the map $\pi_\kappas$ to take a similar form,
\begin{equation}
  \pi_{\kappas} : \quad (y_1, \ldots, y_{p-1}) = (x_1, \ldots x_{p-1}), 
  \quad y_p =  x_p^2 + \cdots + x_n^2 \, ,
\end{equation}
then the concatenated projection is given by
 \begin{equation}
  \pi_{\kappap} : \quad  
  (z_1, \ldots, z_{q-1}) = (x_1, \ldots x_{q-1}),
  \quad z_q = x_q^2 + \cdots + x_{p-1}^2 + 
  (x_p^2 + \cdots + x_n^2)^2 .
\end{equation}
The dependence of each change of coordinates thus ends up looking like:
\begin{equation}
  \underbrace{\underbrace{x_1, x_2, \ldots x_{q - 1}}_{y_1 \ldots y_{q -
  1}}}_{z_1 \ldots z_{q - 1}}, \underbrace{\underbrace{x_q, \ldots, x_{p -
  1}}_{y_q \ldots y_{p - 1}} \underbrace{x_p, \ldots, x_n}_{y_p}}_{z_q}.
\end{equation}
However, due to the quartic dependence of $z_q$ on $x_p$ through $x_n$, the singularity described by $\pi_\kappap$ is not $S_1^+$. Correspondingly, this is not the parametrization we are looking for.

We can instead try keeping $\pi_\kappas$ as it is, while we take $\pi_{\kappa}$ to have the form:
\begin{equation}
  \pi_{\kappa} : \quad (z_1, \ldots, z_{q-1}) = (y_1, \ldots y_{q-1}), \quad z_q =  y_q^2 + \cdots +y_{p-1}^2+ y_p\, .
\end{equation}
With these choices, we have
 \begin{equation}
  \pi_{\kappap} : \quad  
  (z_1, \ldots, z_{q-1}) = (x_1, \ldots x_{q-1}),
  \quad z_q = x_q^2 + \cdots + x_n^2 \, .
\end{equation}
Now the singularity described by $\pi_{\kappap}$ is of type $S_1^+$; however, $\pi_\kappa$ has no critical points. 

A third attempt, in which we again keep the same form of $\pi_\kappas$, is to choose
\begin{equation}
  \pi_{\kappa} : \quad (z_1, \ldots, z_{q-2}) = (y_1, \ldots y_{q-2}), \quad z_{q-1} =y_p,
  \quad z_q =  y_q^2 + \cdots +y_{p-1}^2 \, .
\end{equation}
In this case,
 \begin{equation}
  \pi_{\kappap} : \quad  
  (z_1, \ldots, z_{q-2}) = (x_1, \ldots x_{q-2}),
  \quad z_{q-1} = x_p^2 + \cdots + x_n^2
  \quad z_q = x_q^2 + \cdots + x_{p-1}^2 .
\end{equation}
Now, $\pi_\kappa$ describes a singularity of $S_1^+$, but $\pi_\kappap$ has corank two. Indeed, this describes a case in which two principal Pham loci intersection transversally. We will consider this situation in Sec.~\ref{sec:codim2}. 

Finally, let us consider projection maps
$\pi_{\kappa}$ and $\pi_{\kappap}$ that take the form
\begin{align}
  \pi_{\kappas} &\colon \quad (y_1, \ldots, {\red{y_{p-1}}}) = (x_1, \ldots {\red{x_{p -
  1}}}), \quad y_p =  x_p^2 + \cdots + x_n^2 \, ,\\
  \pi_{\kappa} &\colon \quad     z_{q-1} ={\red{ y_{p - 1}}} + y_p, \quad
    z_{q} = y_{q - 1}^2 + \cdots + {\red{y^2_{p-1}}} \, .
\end{align}
Now both $\pi_{\kappa}$ and $\pi_{\kappap}$ describe $S_1^+$ singularities, and we have
\begin{equation}
  \pi_{\kappap} = \pi_{\kappa} \circ \pi_{\kappas} : \quad 
  z_{q-1} = {\red{x_{p - 1}}} + x_p^2 + \cdots + x_n^2,
  \quad
  z_{q} = {{x_{q-1}^2}} + \cdots + {\red{x^2_{p - 1}}}\, ,
\end{equation}
where the dependence of each change of coordinates looks like:
\begin{equation}
\begin{tikzpicture}[baseline=(current bounding box.center)]
  \node[] at (0,0) {$
  \underbrace{
  \underbrace{x_1, \ldots x_{q - 2}}_{y_1 \ldots y_{q - 2}}}_{z_1\ldots z_{q - 2}} ,
  \underbrace{x_{q - 1}, \ldots, x_{p-2}}_{y_{q-1} \quad\ldots}
  \underbrace{{\red{x_{p-1}}}}_{{{\red{y_{p-1}}}}}
  \underbrace{x_p, \ldots, x_n}_{y_p}$};
  \node[] at (0.2,-0.1) {$\underbrace{\phantom{x_{q-1},\ldots,x_{p-2},x_{p-1}}}$};
  \node[] at (2.1,-0.2) {$\underbrace{\phantom{x_{p-1} x_p,\dots,x_n}}$};
  \node[scale=0.8] at (0.2,-0.5) {$z_{q}$};
  \node[scale=0.8] at (2,-0.7) {$z_{q-1}$};
\end{tikzpicture}
\end{equation}
We can explicitly confirm that the projection $\pi_\kappa$ has a singularity when all the variables that $z_q$ depends on vanish by constructing the (reduced) Jacobian:
\begin{equation}
  J_{\kappa} = \frac{\partial (z_{q}, z_{q-1})}{\partial (y_{q - 1} \ldots
  y_p)}  = \left(\begin{array}{cccc}
    2 y_{q - 1} & \dotso & 2 {\red{y_{p - 1}}} & 0\\
    0 & 0 & 1 & 1
  \end{array}\right) \, .
\label{Jkappa1}
\end{equation}
The determinant of this matrix vanishes, and there is correspondingly a critical point, where $y_{q - 1} = \cdots = y_{p - 1} = 0$. The critical values of this map occur where $z_{q} = 0$, for any value of $z_{q-1}$.
Similarly, the (reduced) Jacobian for $\pi_{\kappap}$ is
\begin{equation}
  J_{\kappap} = \frac{\partial (z_{q}, z_{q-1})}{\partial (x_{q - 1} \dotso
  x_n)}  = \left(\begin{array}{cccccc}
    2 x_{q - 1} & \dotso & 2 {\red{x_{p - 1}}} & 0 & 0 & 0\\
    0 & 0 & 1 & 2 x_p & \cdots & 2 x_n
  \end{array}\right) \, ,
\label{Jkappa2}
\end{equation}
which has two critical points (both of which are $S_1^+$). The first occurs at $x_{q-1} = \dotso = x_{p-1} = 0$, which gives rise to critical values when $z_q=0$. The second occurs where
$x_{q - 1} = \dotso = x_{p - 2} = x_p = \cdots = x_n = 0 $, for any ${\red{x_{p - 1}}}$.
Its critical value is given by $z_q = z_{q - 1}^2$ for any $z_{q-1}$. This encodes a parabola, which is tangent to the first critical surface at $z_{q}=0$. Stated more explicitly, this tells us that Theorem~\ref{thm:co1} describes situations in which pairs of Pham loci intersect tangentially, where this intersection locally looks like the intersection of a line and a parabola. 

The map in Eq.~\eqref{Jkappa2} describes singularities for which the tangent space of $G^\kappa$ is smaller than that of $G^{\kappa'}$, so $n>p$. For some contractions, such as those at one-loop, the tangent space of $G^{\kappa'}$ is more constrained by the on-shell conditions than that of $G^\kappa$, so $n<p$. In those cases, we need a different tangential map, which was worked out in Refs.~\cite{pham,pham2011singularities}. The corresponding Jacobian $J_{\kappa'}$ only has a singularity at $z_q=z_{q-1}^2$, but not at $z_q=0$. For example, the on-shell space of the triangle Landau diagram discussed in Sec.~\ref{sec:iterated} will have a singularity at the triangle Pham locus, but not at the Pham locus corresponding to the bubble Landau diagram.

\subsection{Homotopy Loops}
\label{sec:homotopyloops}
We have now established that when we have two $S_1^+$ contractions where one dominates the other, the relevant coordinates of the Pham loci near their intersection can be parametrized as line and a parabola. We now study how the homotopy group acts near one of these intersection points. Concretely, we consider a pair of Pham loci
\begin{align}
   \P_\kappap & = \{(z_1, z_2) \in \bbC^2 \mid z_1 = z_2^2\} \,, \label{eq:L_tang1-codim1}\\
   \P_\kappa & = \{(z_1, z_2) \in \bbC^2 \mid z_1 = 0\}\,,
    \label{eq:L_tang2-codim1}
\end{align}
and study the homotopy group of the complex space $\bbC^2\setminus (\P_\kappa\cup\P_\kappap)$,  in which these surfaces have been removed. This will teach us how the monodromy group acts on Feynman integrals near this intersection, and which will allow us to establish Theorem~\ref{thm:co1}.

\begin{figure}[t]
    \centering
\resizebox{6cm}{!}{
\begin{tikzpicture}
\draw[-stealth,line width=1.5] (-4,0) -- (4,0) coordinate (xaxis);
\draw[-stealth,line width=1.5] (0,-4) -- (0,4) coordinate (yaxis);
\draw[line width=0.5mm,olddarkgreen,
 postaction={decorate,decoration={markings, mark=between positions 0.1 and 0.9 step 0.25 with {\arrow[line width=0.8mm]{to}}}}
] (1.5,0) 
 arc[start angle=0, delta angle=360,radius=1.5]
;
\node [olddarkgreen,scale=1.7] at  (0.4,1.8) {$\gamma$};
\node[black!50,scale=3] at (1.5,0) {$*$}; 
\draw [draw=black] (3,3) -- (3,4);
\draw [draw=black] (3,3) -- (4,3);
\node [black,scale=1.7] at  (3.5,3.5) {$z_1$};
\filldraw[black] (0,0) circle (0.15); 
\filldraw[polecol] (0.4,0) circle (0.15); 
\node[black,scale=1.5] at (-0.2,0.5) {$0$};
\node[polecol,scale=1.5] at (0.5,0.5) {$z_2^2$};
\end{tikzpicture}
}
\hspace{1cm}
\resizebox{6cm}{!}{
\begin{tikzpicture}
\draw[-stealth,line width=1.5] (-4,0) -- (4,0) coordinate (xaxis);
\draw[-stealth,line width=1.5] (0,-4) -- (0,4) coordinate (yaxis);
\draw[line width=0.5mm,etacol2,
 postaction={decorate,decoration={markings, mark=between positions 0.1 and 0.9 step 0.25 with {\arrow[line width=0.8mm]{to}}}}
] (0,0) arc[start angle=0, delta angle=360,radius=1.8];
\draw[line width=0.5mm,etacol,
 postaction={decorate,decoration={markings, mark=between positions 0.1 and 0.9 step 0.25 with {\arrow[line width=0.8mm]{to}}}}
] (3.6,0) arc[start angle=0, delta angle=360,radius=1.8];
\node [etacol,scale=1.7] at  (2,2.3) {$\eta_+'$};
\node [etacol2,scale=1.7] at  (-2,2.3) {$\eta_-'$};
\draw [draw=black] (3,3) -- (3,4);
\draw [draw=black] (3,3) -- (4,3);
\node [black,scale=1.7] at  (3.5,3.5) {$z_2$};
\draw[line width=0.5mm,olddarkgreen,dashed
, postaction={decorate,decoration={markings, mark=between positions 0.05 and 1.0 step 0.2 with {\arrow[line width=0.8mm]{to}}}}
] (-1.5,0) 
 arc[start angle=180, delta angle=160,radius=1.5]
;
\draw[line width=0.5mm,olddarkgreen,dashed
, postaction={decorate,decoration={markings, mark=between positions 0.05 and 1.0 step 0.2 with {\arrow[line width=0.8mm]{to}}}}
] (1.5,0) 
 arc[start angle=0, delta angle=160,radius=1.5]
;
\filldraw[polecol] (1.5,0) circle (0.15); 
\node[scale=1.3,polecol] at (-2.2,0.3) {$-\sqrt{z_1}$};
\filldraw[polecol] (-1.5,0) circle (0.15); 
\node[scale=1.3,polecol] at (2,0.3) {$\sqrt{z_1}$};
\node[black!50,scale=3] at (0,0) {$*$}; 
\node [olddarkgreen,scale=1.7] at  (1.8,-1) {$\gamma$};
\end{tikzpicture}
}
 \caption{
 The paths ${\color{olddarkgreen} \gamma}$, 
 ${\color{etacol} \eta_+'}$ and ${\color{etacol2} \eta_-'}$  generate the first homotopy group of
 $\pi_1(\bbC^2 \setminus (\P_\kappa \cup \P_\kappap))$. The path ${\color{olddarkgreen} \gamma}$ encircles both $\P_\kappa$ and $\P_\kappap$ while $\eta_\pm'$ only encircle $\P_\kappap$. The base point for the paths is indicated as $\vec{z}_\star=(1,0)$ is marked as $*$. As the path ${\color{olddarkgreen} \gamma}$ is traversed in the complex $z_1$ plane, the singularity at $z_2=\sqrt{z_1}$ moves, as indicated on the right. After the full ${\color{olddarkgreen} \gamma}$ contour, the $\pm\sqrt{z_1}$ branches of $\P_\kappap$ have changed places.}
    \label{fig:gammaeta}
\end{figure}
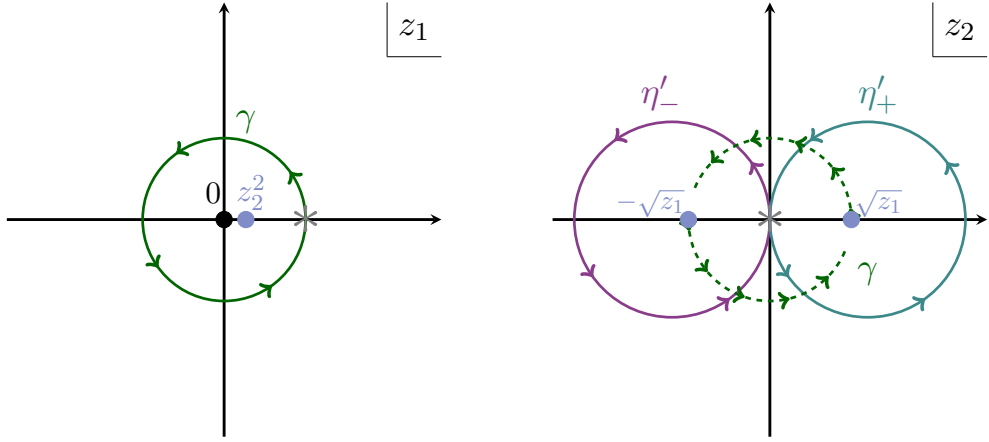

The homotopy group 
$\pi_1 (\bbC^2\setminus (\P_\kappa\cup \P_\kappap))$ is generated by three loops, which can be chosen to be $\MM_+'$, $\MM_-'$, and $\gg$ as depicted in Fig.~\ref{fig:gammaeta}. There, it can be seen that $\MM_+'$ and $\MM_-'$ go around the algebraic branch points that occur at $\pm \sqrt{z_1}$ in the $z_2$ plane, which involves encircling just $\mathcal{P}_{\kappap}$, while $\gamma$ encircles both $\mathcal{P}_{\kappa}$ and $\mathcal{P}_{\kappap}$. In the figure, we have chosen the basepoint of these loops to be 
$\vec{z}_\star=(z_1^\star, z_2^\star) = (1, 0) \in \bbC^2\setminus (\P_\kappa\cup \P_\kappap)$.  

An important pair of relations hold among the loops $\MM_+'$, $\MM_-'$, and $\gg$:
\begin{align}
    \gg \circ \MM_+' \circ \gg^{-1}  &= \MM_-' \, ,
    \label{etapmgammarel1} \\
    \gg \circ \MM_-' \circ \gg^{-1}  &= \MM_+' \, , \label{etapmgammarel2}
\end{align}
where our convention for the composition of loops is\footnote{Some authors use the opposite convention, which is known as the ``topologist's convention''. Our preference for the convention in Eq.~\eqref{eq:composition} stems from the fact that monodromies act on integrals from the left.}
\[
(\alpha \circ \beta)(t) = \begin{cases}
\beta(2 t), &\qquad t \in [0, \frac{1}{2}], \\
\alpha(2 t - 1), &\qquad t \in [\frac{1}{2}, 1].
\end{cases} \label{eq:composition}
\]
The relations in Eqs.~\eqref{etapmgammarel1} and~\eqref{etapmgammarel2} are hard to visualize because they involve the analytic structure of $\sqrt{z}$ in $\bbC$ in an essential way. Intuitively, the idea is that conjugation by $\gamma$ swaps $\eta_+'$ and $\eta_-'$ by interchanging the two roots of $\sqrt{z_1}$. 
As a result, if we want to encircle $\sqrt{z_1}$, we can either travel along the contour $\MM_+'$, or we can first exchange the two algebraic roots using $\gg$, encircle $\sqrt{z_1}$ by travelling along $\MM_-'$, and then swap the algebraic roots back using $\gg^{-1}$. Together, the relations in Eqs.~\eqref{etapmgammarel1} and~\eqref{etapmgammarel2} imply that $\gamma \circ \gamma \circ \eta_\pm' \circ \gamma^{-1} \circ \gamma^{-1} = \eta_\pm'$.

One way to derive Eqs.~\eqref{etapmgammarel1} and~\eqref{etapmgammarel2} is by constructing explicit realizations of the loops $\MM_+'$, $\MM_-'$, and $\gg$. To do that, we first write the loops as maps from $u\in [0,1]$ to $(z_1,z_2) \in  \bbC^2 \setminus (\P_\kappa \cup \P_\kappap)$:
\begin{equation}
    \gamma(u) = \Big(e^{2\pi i u},\, 0\Big),\quad
    \eta_+'(u) = \Big(1,\, 1-e^{2\pi i u}\Big),\quad
    \eta_-'(u) = \Big(1,\, e^{2\pi i u}-1\Big).
\end{equation}
We then construct a map $\psi\colon [0, 1] \times [0, 1] \to  \bbC^2 \setminus (\P_\kappa \cup \P_\kappap)$
between $\eta_+'$ and $\eta_-'$ by interpolating along $\gamma$:
\begin{equation}
    \psi(t,u) 
    =\Big(\gamma(t),\, \sqrt{\gamma(t)}\, \eta_+'(u) \Big)
    =\Big(e^{2\pi i t},\, e^{\pi i t} (1-e^{2\pi i u})\Big).
\end{equation}
This map satisfies $\psi(0,u)=\eta_+'(u)$ and $\psi(1,u) =\eta_-'(u)$. However, since the basepoint changes along the path, this is not a proper homotopy of paths. We can construct such a homotopy, in which the basepoint remains fixed, by conjugating by $\gamma$:
\begin{equation}
\Omega(t, u) = \begin{cases}
\gamma(\frac {2 u} t), &\qquad u \in [0, \frac t 2), \\
\psi(t, 2 u - t), &\qquad u \in [\frac t 2, \frac {1+t} 2], \\
\gamma(\frac {2 u - 1 - t}{1 - t}), &\qquad u \in (\frac {1+t} 2, 1].
\end{cases}
\end{equation}
This satisfies $\Omega(t,0)=\Omega(t,1) = (1,0)$, so the basepoint is fixed for all $t$. It also satisfies $\Omega(0,u) = \gamma \circ \eta_+'$ and $\Omega(1,u) = \eta_-' \circ \gamma$, which proves Eq.~\eqref{etapmgammarel1} and~\eqref{etapmgammarel2}.

\begin{figure}[t]
\centering
\resizebox{9cm}{!}{
\begin{tikzpicture}
\draw[-stealth,line width=4,black!50] (7,-1) -- (10,-1)  node[below,scale=3] {$z_2$};
\draw[-stealth,line width=4,black!50] (-12,2) -- (-12,5)  node[left,scale=3] {$z_1$};
\node[black!50,scale=4] at (0.4,5.6) {$*$};
\begin{knot}[clip width=5]
\strand[etacol,line width =3] (0,5) -- (-7,5) .. controls (-11,4) and (-11,7)  .. (-7,5.2) -- (0.6,5.2);
\strand[darkred,line width =3] (0,5) -- (-7,1) .. controls (-8,-1) and (-6,-1)  .. (-6.4,1) -- (0.4,5);
\strand[olddarkgreen,line width =3] (0.4,5) -- (0.4,1) .. controls (-1,-1.5) and (2,-1.5)  .. (0.6,1) -- (0.6,5.2);
\strand[etacol2,line width =3] (1.2,5) -- (7,5) .. controls (11,4) and (11,7)  .. (7,5.2) -- (1,5.2);
\strand[newdarkblue2,line width =3] (1.2,5) -- (8.2,1) .. controls (9.2,-1) and (7.2,-1)  .. (7.6,1) -- (0.8,5);
\strand[black,line width=3] (-10,0) -- (10,0);
\strand[black,line width=3] (-10,7) parabola bend (0,0) (0,0);
\strand[black,line width=3,dashed] (0,0) parabola bend (0,0) (10,7);
\flipcrossings{1,4,5,6,7,10,13,14,15}
\end{knot}
\node[etacol,scale=3] at (-6,6) {$\eta_+'$};
\node[etacol2,scale=3] at (6,6) {$\eta_-'$};
\node[darkred,scale=3] at (-7.5,2) {$\eta_+$};
\node[newdarkblue2,scale=3] at (8,2) {$\eta_-$};
\node[olddarkgreen,scale=3] at (1,2) {$\gamma$};
\node[above, black,scale=3] at (-10,7) {$\alpha > 0$};
\node[above,black,scale=3] at (10,7) {$\alpha < 0$};
\node[above,black,scale=3] at (3,-2) {$\P_\kappa$};
\node[above,black,scale=3] at (3,1) {$\P_\kappap$};
\end{tikzpicture}}
\caption{Two principal Pham loci $\P_\kappa$ and $\P_\kappap$ with $G^\kappap \supset G^\kappa$
 intersect tangentially. Locally the intersection is described by a parabola $z_1= z_2^2$ and a line $z_1=0$. The intersection is codimension two and it separates the real, non-complexified Pham loci into two regions, corresponding to all $\alpha \geq 0$ and its complement.  The $\alpha$-positive part is a branch hypersurface in the physical region.
 The paths $\eta_+$ and $\eta_+'$ are called simple because they only encircle one branch point, in contrast to $\gamma = \eta_+' \circ \eta_+$, which is not simple.}
\label{fig:co1homotopy}
\end{figure}
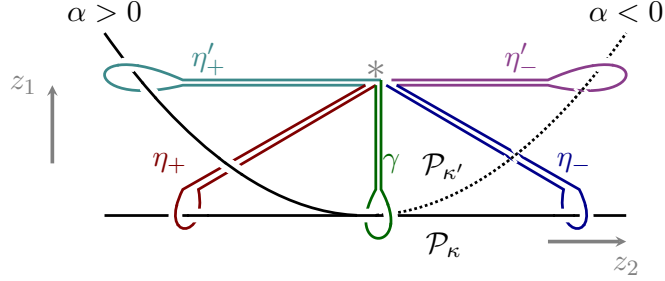

In more general situations, more powerful technology is needed to prove relations like those in Eqs.~\eqref{etapmgammarel1} and~\eqref{etapmgammarel2}. Such relations can be proven algebraically, as described in Ref.~\cite{pham2011singularities}, using the homotopy exact sequence of a fibration. In this approach, one would represent $\bbC^2 \setminus (\P_\kappa \cup \P_\kappap)$ as a fibration of $\bbC \setminus \{-\sqrt{z_1}, \sqrt{z_1}\}$ of coordinate $z_2$ over $\bbC \setminus \{0\}$ of coordinate $z_1$.  An exact sequence for the homotopy of fibrations relates $\pi_1(\bbC^2 \setminus (\P_\kappa \cup \P_\kappap))$ to $\pi_1(\bbC \setminus \{-\sqrt{z_1}, \sqrt{z_1}\})$ and $\pi_1(\bbC \setminus \{0\})$.  The homotopy group $\pi_1(\bbC \setminus \{-\sqrt{z_1}, \sqrt{z_1}\})$ is generated by the loops $\eta_\pm'$, while $\pi_1(\bbC \setminus \{0\})$ is generated by $\gg$.  This approach is described in more detail in appendix~\ref{sec:homotopy}.

Now that we have derived the relations in Eqs.~\eqref{etapmgammarel1} and~\eqref{etapmgammarel2}, we can draw out their implications for Feynman integrals. But first, let us be precise about the relation between the monodromies of Feynman integrals (or more generally on absorption integrals) and the elements of the first homotopy group of $\S(G_0)$. 
The action of the elements in the homotopy group on these integrals is given by the Picard--Lefschetz theorem; that is, the action of the elements $\eta_\pm'$ on the Feynman integral $I_G(p)$ correspond precisely to computing monodromies around $\P_\kappap$. However, because $\eta_+'$ and $\eta_-'$ are not equivalent, we need to be more precise about which path we mean when we talk about the monodromy. For concreteness, we define the monodromy to be the 
`$+$' contour: $\monM_{\P_\kappap} \equiv \monM_{\eta_+'}$.

The element that computes a monodromy around $\P_\kappa$ cannot be any of $\eta_+'$, $\eta_-'$, or $\gamma$, since the first two only encircle $\P_\kappap$, while the last one encircles both $\P_\kappa$ and $\P_\kappap$. In particular, $\gamma$ goes around the intersection point at $z_1 = z_2 = 0$, and therefore, according to the terminology of Ref.~\cite{pham2011singularities}, is not a \emph{simple loop}. 
Conversely, the loops $\eta_+'$ and $\eta_-'$ that encircle the $z_2 > 0$ and $z_2 < 0$ branches of $\P_\kappap$ are simple loops. We can define a similar pair of simple loops $\eta_\pm$ that  encircle the variety $\P_\kappa$ using the relations
\begin{equation}
    \gamma = \eta_+' \circ \eta_+ = \eta_-' \circ \eta_- \, .
    \label{etaetap}
\end{equation}
These loops are depicted in Fig.~\ref{fig:co1homotopy}, where one can see that these relations hold simply by contour composition. This contrasts with the relations in Eqs.~\eqref{etapmgammarel1} and~\eqref{etapmgammarel2}, which required understanding the analytic structure of $\sqrt{z}$ in $\bbC$. Indeed, the requirement that the Pham loci intersect tangentially was essential for Eqs.~\eqref{etapmgammarel1} and~\eqref{etapmgammarel2} to hold, while Eq.~\eqref{etaetap} holds even for a transversal intersection.

Note that we have implicitly chosen a convention in which the loop $\gamma$ is equivalent to traversing the loops $\eta_\pm$ before $\eta_\pm'$ in Eq.~\eqref{etaetap}. In principle, we could have instead adopted the opposite order, and the equalities $\gamma = \eta_+ \circ \eta_+' = \eta_- \circ \eta_-'$. However, a moment's thought shows that this choice would conflict with the definition of the bubble absorption integral above the triangle threshold by analytic continuation through the upper complex plane. We also adopt a convention in which the monodromy around $\mathcal{P}_\kappa$ is given by the action of $\eta_+$, rather than $\eta_-$. That is, $\monM_{\P_\kappa} \equiv \monM_{\eta_+}$.

Now, let us consider the action of the elements of the monodromy group on a generic Feynman integral $\I_G(p)$. Combining Eq.~\eqref{etaetap} with Eqs.~\eqref{etapmgammarel1} and ~\eqref{etapmgammarel2}, we further have that
\begin{equation}
   \gamma =  \MM_+' \circ \MM_+ = \MM_-' \circ \MM_- = \MM_+ \circ \MM_-' = \MM_- \circ \MM_+' \, .
    \label{eq:abpaths-codim1}
\end{equation}
We can therefore write,
\begin{theorem}[Pham]
If two Pham loci $\P_\kappap$ and $\P_\kappapp$ are of codimension one and intersect tangentially, then their sequential monodromies, in a small neighborhood of their intersection and away from other singularities, satisfy:
\begin{equation}
   \monM_{\eta_+'} \circ \monM_{\eta_+} \I_G(p) 
   =
   \monM_{\eta_-'} \circ
   \monM_{\eta_-} \I_G(p)
   =
   \monM_{\eta_+} \circ
   \monM_{\eta_-'} \I_G(p)
   =
   \monM_{\eta_-} \circ
   \monM_{\eta_+'} \I_G(p)\, .
\end{equation}
   \label{thm:codim1gen}
\end{theorem}
\vspace{-6mm}
\noindent Note, in particular, that the monodromies on one side of the intersection do not commute (that is, $\monM_{\eta_+'} \circ \monM_{\eta_+} \neq \monM_{\eta_+} \circ \monM_{\eta_+'}$). Moreover, while we have motivated Theorem~\ref{thm:codim1gen} by studying principal Pham loci, we highlight that this theorem follows just from the properties of the local fundamental group close to the intersection of the two loci $\P_\kappap$ and $\P_\kappapp$. Thus, we do not have to require that these loci are principal or in the physical region for this result to hold. 
However, hierarchically-related singularities of type $S_1^+$ constitute a physically-interesting set of cases where this theorem implies. For instance, all one-loop singularities are of $S_1^+$ type, and the relation in Theorem~\ref{thm:codim1gen} was used to construct the monodromy group for one-loop integrals in Ref.~\cite{Ponzano:1970ch}.

Next, let us look at what further constraints can be derived if we specialize to the physical region. To do this, we take the integration cycle for $\I_G(p)$ to be the physical one---namely, over real loop momenta, except for small deformations when required to avoid singularities, for which we can use the prescription detailed in Sec.~\ref{sec:iepaths}. We then know that the monodromy of $\I_G(p)$ vanishes around the part of $\P_\kappap$ in which $\alpha_e<0$ for any edge $e$. Since the Pham loci $\P_\kappa$ and $\P_\kappap$ meet tangentially, one of the Feynman parameters of $\P_\kappap$ changes sign at the intersection point where $z_1=z_2=0$. 
For example, in the case of the triangle diagram, the line $\P_\kappa$ corresponds to the bubble singularity, in which $\alpha_1,\alpha_2 > 0$ and $\alpha_3=0$, while the parabola $\P_\kappap$ corresponds to the triangle singularity with $\alpha_1,\alpha_2> 0$. However, the remaining Feynman parameter $\alpha_3$ is positive only positive on one branch of the locus $\P_\kappap$, since it flips sign at the point of intersection $z_1=z_2=0$.
Thus, we have that $\monM_{\eta_-'} \I_G(p) = \I_G(p)$. 
This allows us to write
\begin{equation}
    \monM_{\eta_+'} \circ \monM_{\eta_+} \; \I_G(p) = \monM_{\eta_+} \circ \monM_{\eta_-'} \; \I_G(p) = \monM_{\eta_+} \; \I_G(p) \,. 
\end{equation}
Hence, on the physical sheet, in a physical region where the triangle singularity is principal, we have 
\begin{equation}
\Big(\bbone - 
  \monM_{\P_\kappap} \Big) \Big(\bbone - \monM_{\P_\kappa} \Big) \I_G(p) = \Big(\bbone - 
  \monM_{\P_\kappap} \Big) \I_G(p) \, ,
\end{equation}
where we have made use of our adopted conventions, in which $\monM_{\P_\kappa}= \monM_{\eta_+}$ and $\monM_{\P_\kappap}= \monM_{\eta'_+}$. This proves Theorem~\ref{thm:co1}.

\subsection{First Example: the Triangle Integral}
\label{sec:example_triangle}

As a first example that illustrates Theorem~\ref{thm:co1}, we return to the triangle integral $I_\trit(p)$ from Sec.~\ref{sec:picard_lefschetz_absorption_int_example}.
This case is shown on the right side of Fig.~\ref{fig:co1pham}.
While we have already seen that the bubble and triangle singularities satisfy the relation in Eq.~\eqref{PhamAtri2}, additional insight can be gleaned by studying the on-shell surfaces that are relevant to this case, and the $\alpha$-positive condition.

Let us first solve for the leading singularity of the triangle (as is done, for instance, in Ref.~\cite{ELOP}). In the $y_{ij}$ variables in Eq.~\eqref{yijdef}, the Landau equations are given by
\begin{align}
\alpha_1 m_1 - \alpha_2 m_2 y_{12} - \alpha_3 m_3 y_{13} &=0 \, , \label{ltri1}\\
\alpha_1 m_1 y_{23} - \alpha_2 m_2  - \alpha_3 m_3 y_{13} &=0  \, , \label{ltri2}\\
\alpha_1 m_1  y_{23} - \alpha_2 m_2 y_{12} - \alpha_3 m_3  &=0  \, .
\end{align}
These have non-trivial solution when $D =0$, where we recall that $D$ was defined in Eq.~\eqref{eq:dtri}. On the support of this solution, the three Landau equations above become redundant, and we can deduce the relations
\begin{equation}
    \frac{\alpha_1}{\alpha_3} = \frac{m_3}{m_1}\frac{y_{13} + y_{12} y_{23}}{1-y_{12}^2},\qquad
      \frac{\alpha_2}{\alpha_3} =\frac{m_3}{m_2} \frac{y_{23} + y_{12} y_{13}}{1-y_{12}^2} 
      \, , \label{eq:triangle_singularity_ratios}
\end{equation}
from the first two equations.

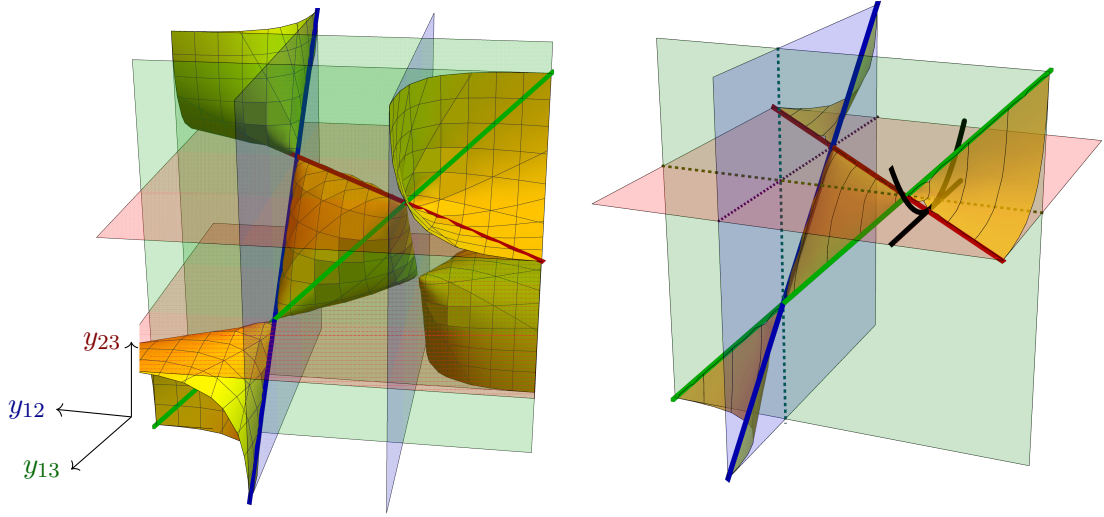
\begin{figure}
    \centering
    \hspace{-2cm}
\begin{tikzpicture}
 \node (image) at (-4,0) {\includegraphics[width=0.4\textwidth]{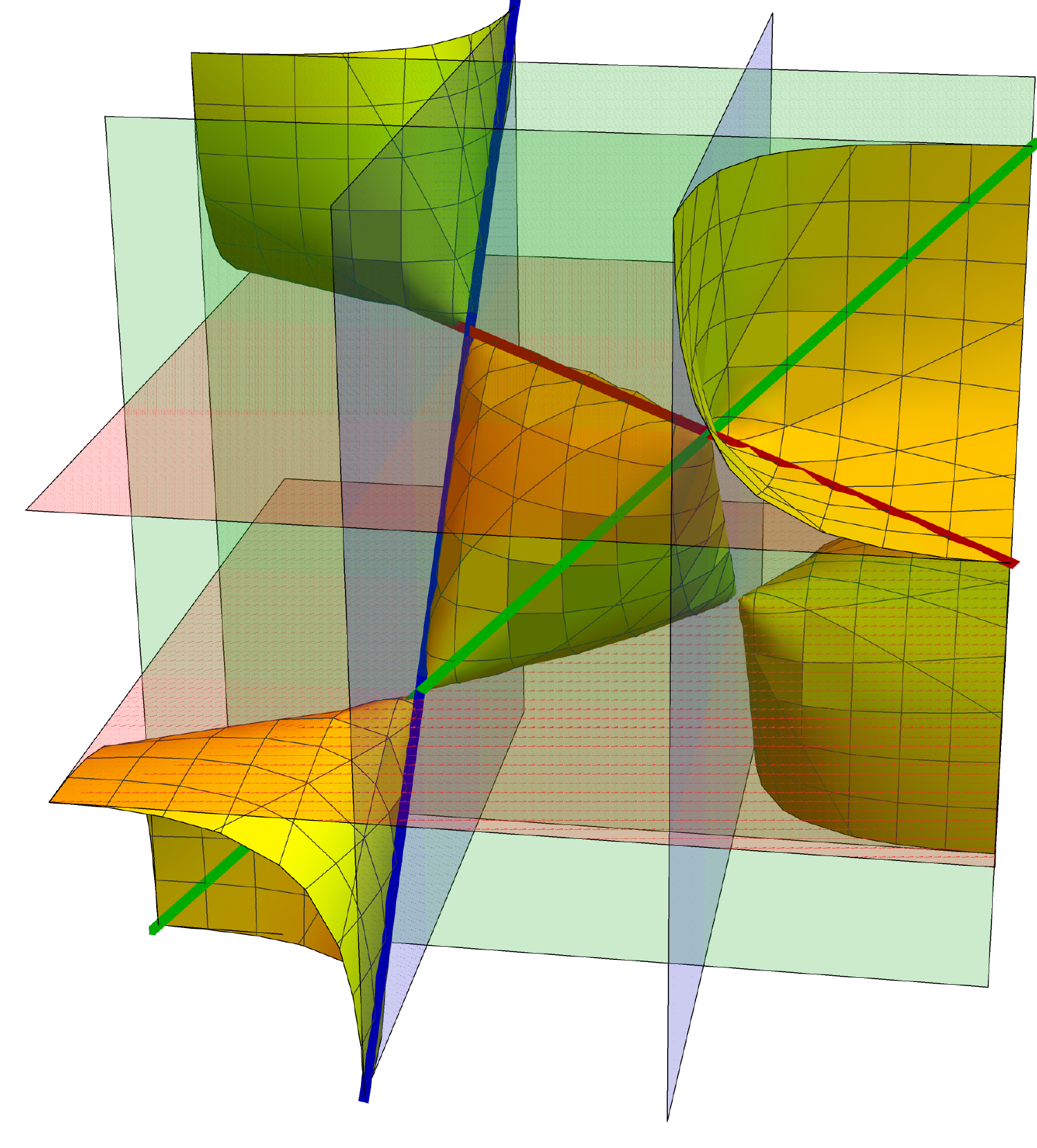}};
 \draw[fill=white,white] (-7,-1.5) rectangle (-6.4,-0.5);
 \draw[->] (-6.5,-2.0) -- (-7.5,-1.9) node[newdarkblue2, left] {$y_{12}$};
 \draw[->] (-6.5,-2.0) -- (-6.5,-1.0) node[darkred, left] {$y_{23}$};
 \draw[->] (-6.5,-2.0) -- (-7.3,-2.7) node[olddarkgreen, left] {$y_{13}$};
  \node (image) at (3,0) {\includegraphics[width=0.45\textwidth]{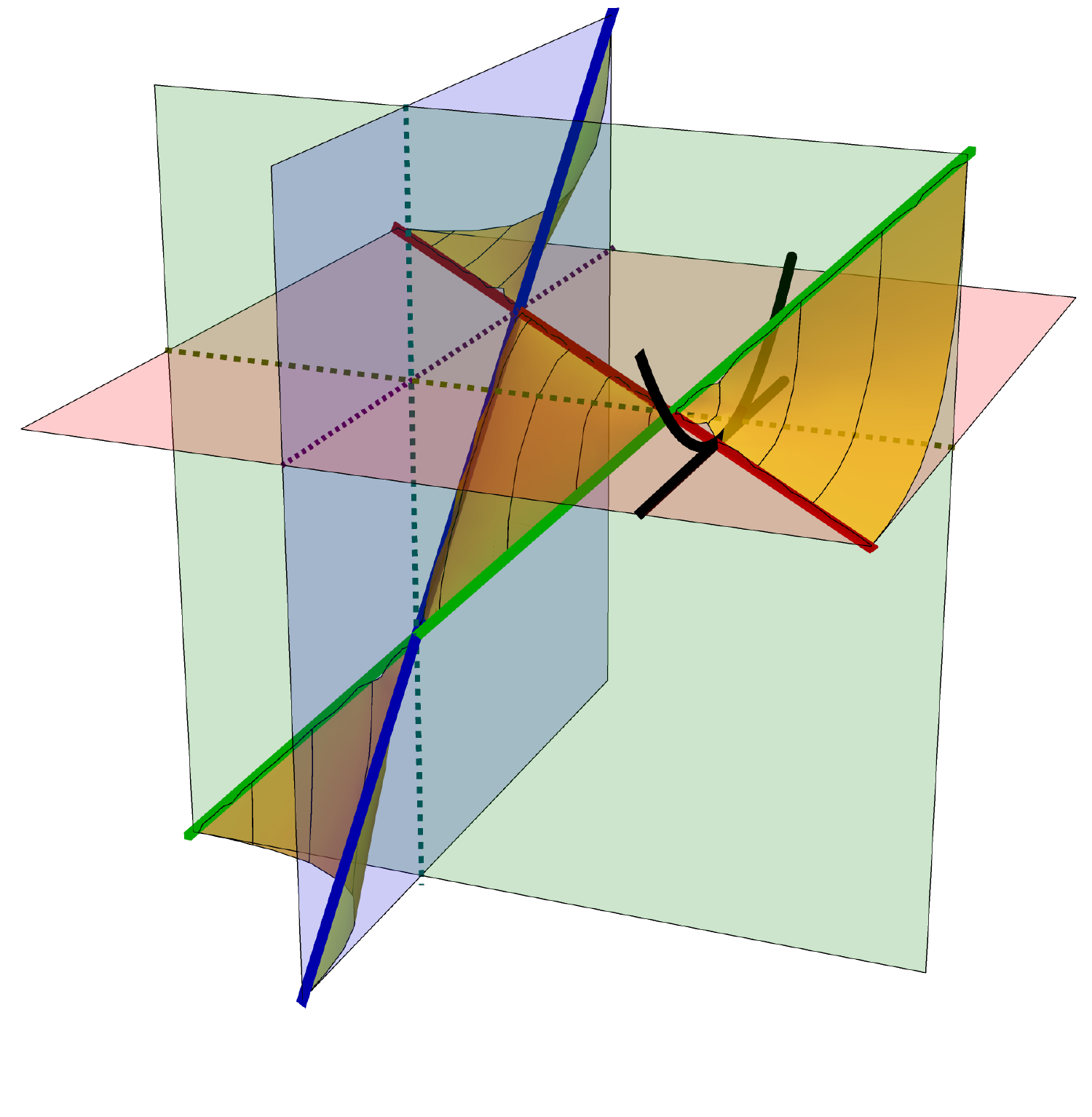}};
\end{tikzpicture}
\caption{The Pham loci for the triangle diagram in three dimensions. The yellow surface depicts the $D=0$ surface where the triangle singularity occurs, while the blue, green, and red planes depict the locations of the bubble singularities at $y_{ij}=\pm 1$. On the right, we show in yellow the $\alpha>0$ region of the triangle singularity, which is 1/4 of the middle tetrahedral ``pillow'' region and 1/3 of the four corner cone-like regions. The straight colored lines represent the intersections between the triangle and the bubble singularities. The black curves on the right depict the $z_1=0$ and $z_1=z_2^2$ curves that approximate the bubble and triangle Pham loci near their point of intersection.
The dotted lines in the right figure show where pairs of bubble singularities intersect; these intersections are transversal rather than tangential.}
\label{fig:triangletangent}
\end{figure}
The location of the bubble singularities can be determined by setting one of the Feynman parameters to zero. For
example, setting $\alpha_3 = 0$ contracts the $q_3$ line to a point. The remaining
Landau equations are then
\begin{align}
  m_1 \alpha_1 - m_2 \alpha_2 y_{12} &= 0 \, ,\\
  - m_1 \alpha_1 y_{12} + m_2 \alpha_2 &= 0 \, .
\end{align}
The only nontrivial solution to this pair of equations is given by $y_{12}^2 = 1$.
On this locus, $m_1 \alpha_1 = m_2 \alpha_2 y_{12}$, which implies that the
$\alpha$-positive region corresponds to the $y_{12} = 1$ branch.

In Fig.~\ref{fig:triangletangent}, we show the on-shell surface for the triangle $(D =
0)$ and the bubble contractions of the triangle ($y_{i j}^2 = 1$). 
There, we see that there are four regions where one can reach the $D = 0$ surface. The first corresponds to the region in which $- 1 < y_{12}, y_{23}, y_{13} < 1$, while the remaining three regions are determined by one of the three variables being less than $-1$, and the remaining two to be greater than 1. Let us specialize to the region in which $y_{13}, y_{23} > 1$ and $y_{12} < - 1$. In this region, $1 - y_{12}^2 <
0$, so from Eq.~\eqref{eq:triangle_singularity_ratios} we can read off the $\alpha$-positivity conditions to be $y_{13} + y_{12}
y_{23} < 0$ and $y_{23} + y_{12} y_{13} < 0$. On the surface $D = 0$, these
constraints reduce to $y_{12} + y_{13} < 0$ and $y_{12} + y_{23} < 0$. The bubble singularity at $y_{23} = 1$
intersects with the triangle singularity on the line $y_{12} + y_{13} = 0$. Since we are interested
in the two directions that are transverse to this line, let us fix $y_{12}$ to
some fiducial value, and introduce a coordinate
$z_1$ that points in the $y_{23}$ direction and a coordinate $z_2$ that points in the $y_{13}$ direction. More precisely, we define 
\begin{align}
    z_1 = 2 (y_{12}^2 - 1) (y_{23} - 1) \, , \qquad z_2 = y_{12} + y_{13} \, . \label{eq:z1_z2_def}
\end{align}
These definitions are chosen so that the intersection of the bubble and triangle singularities occurs at $z_1 = z_2 = 0$, and so the leading term in the expansion of $z_2$ around this intersection point has unit coefficient (as we will see momentarily). In these variables, the line $y_{23} = 1$ is given by $z_1 =
0$, while the $D = 0$ curve becomes 
\begin{align}
  z_1 &= 2 (1 - y_{12}^2) \left( 1 + \sqrt{(1 - y_{12}^2) (1 - (y_{12} -
  z_2)^2)} + y_{12} (z_2 - y_{12}) \right)\\
  &= z_2^2 + \frac{y_{12}}{y_{12}^2 - 1} z_2^3 + \mathcal{O}(z_2^4) \, . \label{z1z22}
\end{align}
This is form we expect this curve to take (at leading order) from our analysis in Sec.~\ref{sec:trangentialmaps}. In terms of $z_2$, the $\alpha$-positive condition becomes
$z_2 = y_{12} + y_{13} < 0$. These regions and surfaces are shown on the right side of Fig.~\ref{fig:triangletangent}.

\subsection{Second Example: the Ice Cream Cone}
\label{sec:icecreamtobubbleandsun}

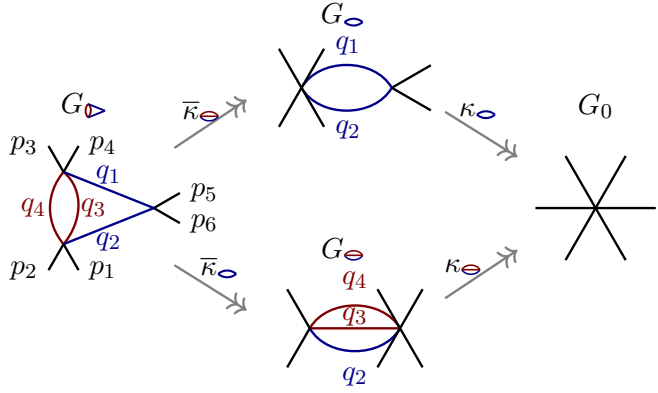
\begin{figure}[t!]
\centering
\begin{tikzpicture}[scale=0.8,line width=0.9]
\node at (0,0) {
\begin{tikzpicture}[baseline= {($(current bounding box.base)-(2pt,2pt)$)},line width=0.9,scale=0.8]
\path [darkred,out=130,in=230] (0,-0.6,0) edge (0,0.6);
\path [darkred,out=50,in=-40] (0,-0.6,0) edge (0,0.6);
\node[darkred] at (0.5,0) {$q_3$};
\node[darkred] at (-0.5,0) {$q_4$};
\draw [newdarkblue2] (0,0.6) -- (1.5,0);
\draw [newdarkblue2] (0,-0.6) -- (1.5,0);
\node[newdarkblue2] at (0.75,0.55) {$q_1$};
\node[newdarkblue2] at (0.75,-0.55) {$q_2$};
\draw[black] (0,-0.6) -- ++(-60:0.5)  node[right] {$p_1$};
\draw[black] (0,-0.6) -- ++(-120:0.5)  node[left] {$p_2$};
\draw[black] (0,0.6) -- ++(120:0.5) node[left] {$p_3$};
\draw[black] (0,0.6) -- ++(60:0.5) node[right] {$p_4$};
\draw[black] (1.5,0) -- ++(30:0.5) node[right] {$p_5$};
\draw[black] (1.5,0) -- ++(-30:0.5) node[right] {$p_6$};
\end{tikzpicture}};
\node at (4,2) {
\begin{tikzpicture}[baseline= {($(current bounding box.base)-(2pt,2pt)$)},line width=0.9,scale=0.8]
\path [newdarkblue2,out=60,in=120] (-0.75,0) edge (0.75,0);
\path [newdarkblue2,out=-60,in=240] (-0.75,0) edge (0.75,0);
\node[newdarkblue2,xshift=-4] at (0.2,0.7) {$q_1$};
\node[newdarkblue2,xshift=-4] at (0.2,-0.7) {$q_2$};
\draw[black] (-0.75,0) -- ++(60:0.75);
\draw[black] (-0.75,0) -- ++(120:0.75);
\draw[black] (-0.75,0) -- ++(-60:0.75);
\draw[black] (-0.75,0) -- ++(-120:0.75);
\draw[black] (0.75,0) -- ++(30:0.75);
\draw[black] (0.75,0) -- ++(-30:0.75);
\end{tikzpicture}};
\node at (4,-2) {
\begin{tikzpicture}[baseline= {($(current bounding box.base)-(2pt,2pt)$)},line width=0.9,scale=0.8]
\path [darkred,out=60,in=120] (-0.75,0) edge (0.75,0);
\draw [darkred] (-0.75,0) -- (0.75,0);
\path [newdarkblue2,out=-60,in=240] (-0.75,0) edge (0.75,0);
\draw[black] (0.75,0) -- ++(60:0.75);
\draw[black] (0.75,0) -- ++(120:0.75);
\draw[black] (0.75,0) -- ++(-60:0.75);
\draw[black] (0.75,0) -- ++(-120:0.75);
\draw[black] (-0.75,0) -- ++(-120:0.75);
\draw[black] (-0.75,0) -- ++(120:0.75);
\node[darkred] at (0,0.2) {$q_3$};
\node[darkred] at (0,0.8) {$q_4$};
\node[newdarkblue2] at (0,-0.8) {$q_2$};
\end{tikzpicture}};
\node at (8,0) {
\begin{tikzpicture}[baseline= {($(current bounding box.base)-(2pt,2pt)$)},line width=0.9,scale=0.8]
\draw[black] (60:1) -- (240:1);
\draw[black] (0:1) -- (180:1);
\draw[black] (120:1) -- (300:1);
\end{tikzpicture}};
\node[black,scale=1] at (-0.5,1.7) {$\Gice$};
\node[black,scale=1] at (3.8,3.2) {$\Gbub$};
\node[black,scale=1] at (3.8,-0.7) {$\Gsun$};
\node[black,scale=1] at (8,1.7) {$G_{0}$};
\draw[->>,black!50] (1,1) -- (2.2,1.8) ;
\node[black,scale=1] at (1.45,1.6) {$\kappasunp$};
\draw[->>,black!50] (5.5,1.6) -- (6.7,0.85);
\node[black,scale=1] at (6.05,1.6) {$\kappabub$};
\draw[->>,black!50] (1,-0.95) -- (2.2,-1.7);
 \node[black,scale=1] at (1.75,-1) {$\kappabubp$};
\draw[->>,black!50] (5.5,-1.5) -- (6.7,-0.7) ;
\node[black,scale=1] at (5.8,-1) {$\kappasun$};
\end{tikzpicture}
\caption{Contractions of the ice cream cone diagram to the bubble and the sunrise diagrams. 
\label{fig:icecreamtobubbleandsun}}
\end{figure}

As a second example, we consider the ice cream cone diagram and its contractions to the bubble and the two-loop sunrise graphs, as shown in Fig.~\ref{fig:icecreamtobubbleandsun}. The ice cream cone diagram involves six external momenta \(p_1, \dotsc, p_6\) of mass \(p_i^2 = M_i^2\) and four internal momenta \(q_1^\mu, \dotsc, q_4^\mu\) of mass \(q_j^2 = m_j^2\):
\begin{equation}
G_\icet = 
\begin{tikzpicture}[baseline= {($(current bounding box.base)-(2pt,2pt)$)},scale=0.8,line width=0.9]
\path [darkred,out=130,in=230] (0,-0.6,0) edge (0,0.6);
\path [darkred,out=50,in=-40] (0,-0.6,0) edge (0,0.6);
\node[darkred] at (0.6,0) {$q_3$};
\node[darkred] at (-0.6,0) {$q_4$};
\draw [newdarkblue2] (0,0.6) -- (1.5,0);
\draw [newdarkblue2] (0,-0.6) -- (1.5,0);
\node[newdarkblue2] at (0.7,0.6) {$q_1$};
\node[newdarkblue2] at (0.7,-0.6) {$q_2$};
\draw[black] (0,-0.6) -- ++(-60:0.5)  node[right] {$p_1$};
\draw[black] (0,-0.6) -- ++(-120:0.5)  node[left] {$p_2$};
\draw[black] (0,0.6) -- ++(120:0.5) node[left] {$p_3$};
\draw[black] (0,0.6) -- ++(60:0.5) node[right] {$p_4$};
\draw[black] (1.5,0) -- ++(30:0.5) node[right] {$p_5$};
\draw[black] (1.5,0) -- ++(-30:0.5) node[right] {$p_6$};
\draw[black,-latex reversed] (0,-0.6) -- ++(-60:0.35);
\draw[black,-latex reversed] (0,-0.6) -- ++(-120:0.35);
\draw[black,-latex reversed] (0,0.6) -- ++(120:0.35);
\draw[black,-latex] (0,0.6) -- ++(60:0.48);
\draw[black,-latex] (1.5,0) -- ++(30:0.48);
\draw[black,-latex] (1.5,0) -- ++(-30:0.48);
\draw[newdarkblue2,-latex reversed] (1.5,0) -- ++(158:0.9);
\draw[newdarkblue2,-latex reversed] (1.5,0) -- ++(-158:0.9);
\draw[darkred,-latex] (0.25,0) -- ++(90:0.2);
\draw[darkred,-latex] (-0.23,0) -- ++(90:0.2);
\end{tikzpicture}
\label{eq:icecreamcone_momentum_labels}
\end{equation}
Momentum conservation reads
\begin{align}
  p_1^\mu + p_2^\mu &= q_2^\mu + q_3^\mu + q_4^\mu, \\
  p_4^\mu - p_3^\mu &= -q_1^\mu + q_3^\mu + q_4^\mu, \\
  p_5^\mu + p_6^\mu &= q_1^\mu + q_2^\mu.
\end{align}
\paragraph{Landau singularities}
The Landau loop equations for the ice cream cone diagram read
\begin{align}
  \alpha_3 q_3^\mu - \alpha_4 q_4^\mu &= 0, \\
  \alpha_1 q_1^\mu - \alpha_2 q_2^\mu + \alpha_4 q_4^\mu &= 0.
\end{align}
Squaring the first equation, we find $\alpha_3^2 q_3^2 = \alpha_4^2 q_4^2$.  Using the on-shell conditions and the constraint that $\alpha_3, \alpha_4 \geq 0$, we have $\alpha_3 m_3 = \alpha_4 m_4$.  This implies that $q_3^\mu = \frac {m_3}{m_3 + m_4} (q_3^\mu + q_4^\mu)$ and $q_4^\mu = \frac {m_4}{m_3 + m_4} (q_3^\mu + q_4^\mu)$.  The second Landau loop equation then becomes
\begin{equation}
    \alpha_1 q_1^\mu - \alpha_2 q_2^\mu + \frac{\alpha_4 m_4}{m_3 + m_4} (q_3^\mu + q_4^\mu) = 0.
\end{equation}
In the generic case, this is a two-dimensional equation and the Pham locus $\P_{\icet}$ is of codimension one, while the corresponding solution in $\alpha$ space is one-dimensional. If all internal masses are taken to be equal for simplicity, the locus $\P_{\icet}$ includes (as can be checked using the techniques from Ref.~\cite{Mizera:2021icv})
\begin{equation}
    p_{12}^2 p_{34}^2 p_{56}^2 + m^2 \left[9 m^2 p_{56}^2 +(p_{12}^2-p_{34}^2)^2 - 5 (p_{12}^2+p_{34}^2) p_{56}^2 +4 p_{56}^4\right] = 0\,,
    \label{eq:leadingsunr1}
\end{equation}
where we have written $p_{12}^2=(p_1+p_2)^2$, $p_{34}^2=(p_3-p_4)^2$ and $p_{56}^2=(p_5+p_6)^2$ for notational convenience.

Next, let us consider the bubble and sunrise subleading singularities of the ice cream cone diagram, as depicted in Fig.~\ref{fig:icecreamtobubbleandsun}. While we can describe these singular loci by setting some of the $\alpha_i$ to zero, we instead show how the same answers can be obtained from an analysis of critical points of the appropriate differentiable maps.  Consider the graph contractions
\vspace{3mm}
\begin{equation}
\begin{tikzcd}[remember picture]
|[alias=A]|
    \arrow[r, twoheadrightarrow]
    \arrow[dr, twoheadrightarrow]
    \arrow[d, twoheadrightarrow]
    &
|[alias=B]|
    \arrow[d, twoheadrightarrow]
\\
|[alias=C]|
    \arrow[r, twoheadrightarrow]
    &
|[alias=D]|
{G_0}
\end{tikzcd}
\begin{tikzpicture}[overlay,remember picture]
\node[] at (A) {$G_{\icet}~~~~$};
\node[] at (B) {$~~~G_{\bubt}$};
\node[]at (C) {$G_{\sunt}~~~~$};
\node[right] at ($(B)!0.5!(D)$)  {$\kappa_{\bubt}$};
\node[above] at ($(A)!0.5!(B)$)  {$\kappas_{\sunt}$};
\node[below] at ($(C)!0.5!(D)$)  {$\kappa_{\sunt}$};
\node[left] at ($(A)!0.5!(C)$)  {$\kappas_{\bubt}$};
\node[above] at ($(A)!0.5!(D)$)  {$\kappa_{\icet}$};
\end{tikzpicture}
\end{equation}
where $\Gice$ is the ice cream cone graph, $\Gbub$ is a bubble graph, $\Gsun$ is a two-loop sunrise graph, and $G_0$ is the elementary graph.
In the case of the contraction $\kappabub$, the Pham locus $\P_{\bubt}$ is given by the condition that the differential $\rd \ell(\kappabub) = \alpha_1 q_1 \cdot d q_1 + \alpha_2 q_2 \cdot d q_2$ vanishes when restricted to the space of external kinematics.  Using momentum conservation, we have $q_2^\mu = p_5^\mu + p_6^\mu - q_1^\mu$, and so
\begin{equation}
    \rd \ell(\kappabub) = (\alpha_1 q_1 - \alpha_2 q_2) \cdot d q_1 + \alpha_2 q_2 \cdot d (p_5 + p_6).
\end{equation}
The condition for having a critical point is thus $\alpha_1 q_1^\mu - \alpha_2 q_2^\mu = 0$, as expected for the bubble threshold.  This condition defines the set of critical points $\Gamma(\kappabub) \subset \S(G)$.

For the contraction $\kappasun$, the Pham locus $\P(\kappasun)$ can be read off of the differential $\rd \ell(\kappasun) = \alpha_2 q_2 \cdot d q_2 + \alpha_3 q_3 d q_3 + \alpha_4 q_4 \cdot d q_4$.  Using momentum conservation, we have $q_2^\mu = p_5^\mu + p_6^\mu - q_1^\mu$ and $q_4^\mu = -p_3^\mu + p_4^\mu + q_1^\mu - q_3^\mu$; hence,
\begin{equation}
    \rd \ell(\kappasun) = (-\alpha_2 q_2 + \alpha_4 q_4) \cdot d q_1 + (\alpha_3 q_3 - \alpha_4 q_4) \cdot d q_3 + \alpha_2 q_2 \cdot d (p_5 + p_6) + \alpha_4 q_4 \cdot d (-p_3 + p_4).
\end{equation}
This vanishes when the external kinematics are kept constant, provided that $-\alpha_2 q_2^\mu + \alpha_4 q_4^\mu = 0$ and $\alpha_3 q_3^\mu - \alpha_4 q_4^\mu = 0$.  As expected, these are the usual Landau equations for the two-loop sunrise singularity.  As with the bubble locus, they also define the set of critical points $\Gamma(\kappasun) \subset \S(G)$.

Let us now analyze the two contractions labeled by $\kappas$ in Fig.~\ref{fig:icecreamtobubbleandsun}, which are contractions where the target graph is not the trivial one. We start with the contraction $\kappasunp \colon \Gice \twoheadrightarrow \Gbub$.  Its Pham locus $\P_{\kappasunp}$ is defined by the condition that the differential $\rd \ell(\kappasunp) = \alpha_3 q_3 \cdot d q_3 + \alpha_4 q_4 \cdot d q_4$ vanishes when the kinematics in the target graph (internal and external) are kept fixed.  Using momentum conservation we have
\begin{equation}
    \rd \ell(\kappasunp) = (\alpha_3 q_3 - \alpha_4 q_4) \cdot d q_3 + \alpha_4 q_4 \cdot d (-p_3 + p_4 + q_1).
\end{equation}
This imposes the condition $\alpha_3 q_3^\mu - \alpha_4 q_4^\mu = 0$, which defines the set of critical points $\Gamma(\kappasunp) \subset \S(\Gice)$. Finally, the contraction $\kappabubp \colon \Gice \twoheadrightarrow \Gsun$ has a Pham locus defined by $\rd \ell(\kappabubp) = \alpha_1 p_1 \cdot d p_1$.  Since we take the masses to be fixed, $p_1 \cdot d p_1 = m_1 d m_1 = 0$.  This imposes no Landau loop equations so the set of critical points $\Gamma(\kappabubp)$ is the entirety of $\S(\Gice)$.  We take $\P(\kappabubp)$ to be the image of $\S(\Gice)$ in
$\S(\kappabubp)$, which imposes the constraint $(p_5 + p_6 - q_2)^2 = m_1^2$ in $\S(\Gice)$.

For the chosen momentum routing in Eq.~\eqref{eq:icecreamcone_momentum_labels}, the only $\alpha$-positive Pham loci are given by the normal two-particle threshold $\P_{\bubt}$ at $(p_5 + p_6)^2 = (m_1 + m_2)^2$, or a normal three-particle threshold $\P_{\sunt}$ at $(p_1 + p_2)^2 = (m_2 + m_3 + m_4)^2$. Since we have taken the internal propagators $q$ to have positive energy, we can set $\alpha_1 = 0$ or $\alpha_4 = 0$ but not $\alpha_2 = 0$.

\paragraph{Codimension-two momentum configuration}
For special values of the external kinematics, the leading ice cream cone singularity becomes codimension-two (the analysis of this codimension-two situation is proposed as an exercise in Sec.~I.3.2 of Ref.~\cite{pham}).  Let us take the external kinematics to be such that the external linear combinations of momenta $p_1^\mu + p_2^\mu$ and $p_5^\mu + p_6^\mu$ (and by momentum conservation, $p_3^\mu - p_4^\mu$) are collinear.  In this configuration, the space of external kinematics becomes one-dimensional and each of the Landau loop equations contribute one constraint.  Hence, only two of the four $\alpha$ are determined and we end up with a codimension-two singularity. 

When the three momenta $p_1^\mu + p_2^\mu$, $p_5^\mu + p_6^\mu$, and $p_3^\mu - p_4^\mu$ are collinear, the map between the components of these momenta and the Mandelstam invariants that describe this process is not regular.  To see this, consider the differential $\beta_1 d (p_1 + p_2)^2 + \beta_2 d (p_3 - p_4)^2 + \beta_3 d (p_5 + p_6)^2$.  One can find values for $\beta_i$ such that it vanishes when the three momenta $p_1^\mu + p_2^\mu$, $p_5^\mu + p_6^\mu$ and $p_3^\mu - p_4^\mu$ are collinear.  As described in Ref.~\cite{pham}, the inverse map from Mandelstam invariants to components of momenta will then not be well-defined.  As a consequence, the form taken by the Pham locus in the neighborhood of special kinematic points where the space of kinematics is lower-dimensional depends on the coordinates one works in terms of.  When described in Mandelstam invariants, we find that these lower-dimensional configurations correspond to points.  In terms of components of momenta, they correspond to curves. 

\paragraph{Ice cream cone variety in terms of momentum components} To illustrate how this works in detail, we take all the masses to be equal to $m$, and work in units where $m=1$. We also work in light-cone coordinates, in which $q_i^\mu = (q_i^+, q_i^-)$, with $q_i^- = 1/{q_i^+}$ and $q_i^2=q_i^+q_i^-$ (since we have three independent external Mandelstam variables for the ice cream cone, we can go to a frame in which the transverse components vanish in light-cone coordinates, so we omit them for simplicity).  We can choose our external kinematics to be such that $p_1^\mu + p_2^\mu = (u, v)$ and $p_5^\mu + p_6^\mu = (w, w)$.
The positivity of energy implies $u + v > 0$ and $w > 0$.  The reason for this choice of parametrization of the $p_5^\mu + p_6^\mu$ components is that the on-shell and momentum-conservation conditions at that vertex can be solved by $q_1 = (x, x^{-1})$ and $q_2 = (x^{-1}, x)$, with $w = x + x^{-1}$.  Since the masses have been chosen to be equal, the bubble Landau loop equation implies that $\alpha_3 = \alpha_4$ and $q_3^\mu = q_4^\mu$.  Hence, a simple parametrization of the internal momenta is provided by
\begin{gather}
    q_3^\mu = q_4^\mu = \frac 1 2 (u - x^{-1}, v - x).
\end{gather}
The on-shell condition for $q_3$ is then $(v - x)(u - x^{-1}) = 4$ and can be parameterized by $v = x + 2 y^{-1}$ and $u = x^{-1} + 2 y$.  Hence, we have found a parametrization in terms of $x$ and $y$ which solve the on-shell conditions as
\begin{gather}
    q_1^\mu = (x, x^{-1}), \qquad
    q_2^\mu = (x^{-1}, x), \qquad
    q_3^\mu = q_4^\mu = (y, y^{-1}).
\end{gather}
The two components of the remaining Landau loop equation are
\begin{gather}
    \alpha_1 x - \alpha_2 x^{-1} + \alpha_4 y = 0, \\
    \alpha_1 x^{-1} - \alpha_2 x + \alpha_4 y^{-1} = 0.
\end{gather}
These always admit a solution, so they do not impose additional constraints on the external kinematics.

We have now obtained a rational parameterization of the codimension-one Pham locus associated with the ice cream cone Landau singularity, in which all Feynman parameters are non-vanishing.  This parameterization is
\begin{gather}
    f \colon \mathbb{R}^2 \to \mathbb{R}^3, \\
    (x, y) \mapsto (u, v, w) = (x^{-1} + 2 y, x + 2 y^{-1}, x + x^{-1}).
\end{gather}
We can eliminate the variables $x$ and $y$ to express the Pham locus by the equation
\begin{equation}
    \label{eq:icecream_cone_ll_equation}
u^2 v^2 - u^2 v w - u v^2 w + u v w^2 + u^2 - 8 u v + v^2 + 3 u w + 
  3 v w + 9 = 0 \,.
\end{equation}
Note that this Pham locus has a self-intersection singularity: there is a one-dimensional locus where one obtains the same values $(u, v, w)$ for \emph{two} different values of $(x, y)$.  Indeed, given values of $x$ and $y$ such that $y - y^{-1} = 2 (x - x^{-1})$, we have that both $(x, y)$ and $(x^{-1}, y^{-1})$ map to the same value of $(u, v, w)$.

For unit masses, the two-loop sunrise Pham locus $\P_{\sunt}$ corresponding to $\alpha_1 = 0$  is given by $(p_1 + p_2)^2 = 9$, or $u v = 9$, and it can be parameterized by $(u, w) \mapsto (u, \frac{9}{u}, w)$.  The bubble Pham locus corresponds to $\alpha_3 = \alpha_4 = 0$, and is given by $(p_5 + p_6)^2 = 4$, or $w^2 = 4$.  It can be parameterized by $(u, v) \mapsto (u, v, 2)$. 
The ice cream cone singularity intersects the two-loop sunrise singularity in a curve parameterized by $x \mapsto (3 x^{-1}, 3 x, x + x^{-1})$.  The ice cream cone intersects the bubble singularity in a curve parameterized by $y \mapsto (1 + 2 y, 1 + 2 y^{-1}, 2)$.  The bubble and two-loop sunrise intersect in a curve defined by $w = 2$ and $u v = 9$.  All three Pham loci intersect at the point $(u, v, w) = (3, 3, 2)$, which corresponds to $x = y = 1$.

\begin{figure}
    \centering
\begin{tikzpicture}
 \node (image) at (-4,0) {    
 \includegraphics[scale=0.6]{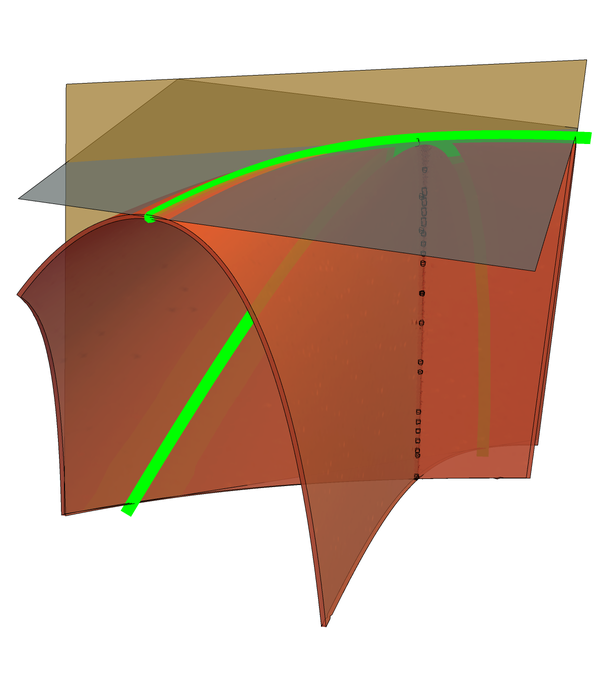}
 };
 \node[] at (-8,1.5) {bubble};
 \node[] at (-8,1) {$\P_{\bubt}$};
 \draw[->] (-7.8,1.8) to [out=90,in=130] (-6.7,1.7);
 \node[] at (-3,3.5) {sunrise~$\P_{\sunt}$};
 \draw[->] (-1.9,3.5) to [out=0,in=90] (-1.3,3);
 \node[] at (-6,-2.5) {ice cream cone};
 \node[] at (-6,-3) {$\P_{\icet}$};
 \draw[->] (-5,-2.3) to [out=90,in=-210] (-4,-2);
 \node (image) at (3,0) {    
    \includegraphics[scale=0.7]{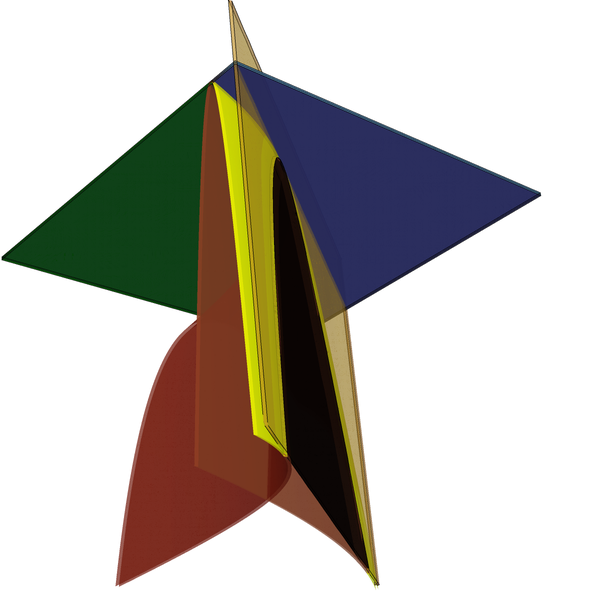}
 };
 \end{tikzpicture}
    \caption{Topology of the singularities of the ice cream cone integral, shown from two perspectives.  Its leading singularity
    is a self-intersecting surface with a Whitney umbrella singularity. The bubble and the two-loop sunrise are subleading singularities. Their Pham loci are given by two-planes.   The ice cream cone intersects these planes along the two light green curves.  These curves correspond to the boundaries of the $\alpha$-positive regions, which are shown as the black, dark green, and yellow regions in the plot on the right. The ice cream cone surface intersects the sunrise and bubble surfaces tangentially, which the bubble and sunrise surfaces intersect transversally.}
    \label{fig:whitney_umbrella}
\end{figure}
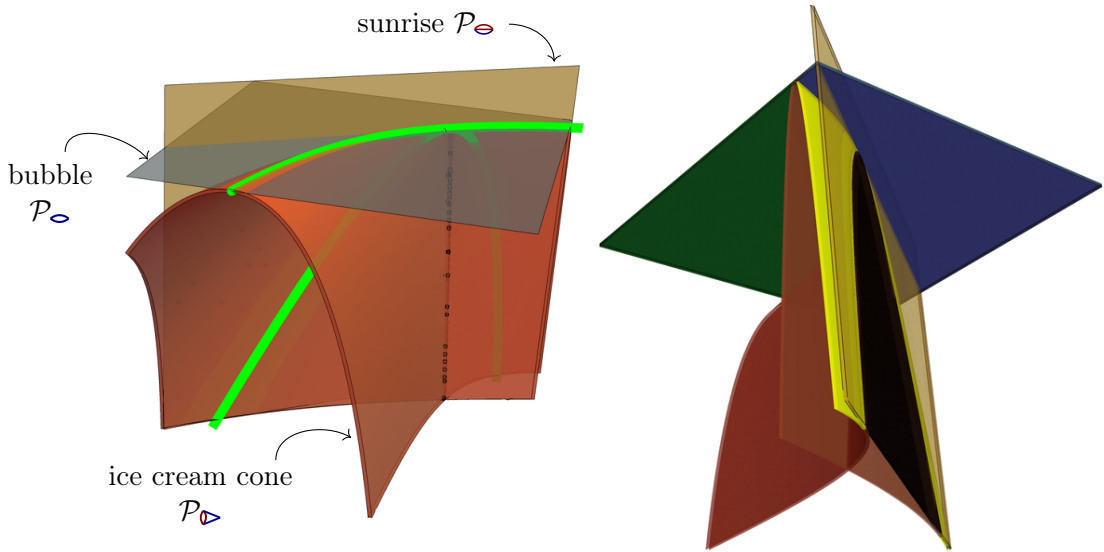

Fig.~\ref{fig:whitney_umbrella} shows the ice cream cone Pham locus (given by the parabolic surface), and the bubble and sunrise Pham loci (which correspond to planes). For visualization purposes, we have changed variables from $v$ to $\vt = \frac{1}{v}$. This makes the sunrise Pham locus the plane $u=9\tilde{v}$. The ice cream cone surface intersects both the sunrise and bubble surfaces tangentially, while the sunrise and bubble surfaces intersect each other transversally.

Not all the points of this locus correspond to positive $\alpha$.  Solving for these variables, we find that, projectively,
\begin{align}
    (\alpha_1 : \alpha_2 : \alpha_4) &= ((1 - x^2 y^2) x : (x^2 - y^2) x : (x^4 - 1) y), \label{eq:alpha_soln_1}\\
    (\alpha_3 : \alpha_4) &= (1 : 1)\,. \label{eq:alpha_soln_2}
\end{align}
To have positive energies flowing through all internal edges, we must have $x > 0$ and $y > 0$.  The positivity of the Feynman parameters in Eqs.~\eqref{eq:alpha_soln_1} and~\eqref{eq:alpha_soln_2} further requires that $x > 1$ and $x y < 1$. The boundary $x = 1$ occurs when to $\alpha_4 = 0$, and similarly $x y = 1$ occurs when $\alpha_1 = 0$.  We can check that these conditions are incompatible with $\alpha_2 = 0$. This can also be read off of the Landau diagram, from the incompatibility of the corresponding energy flow. If we set $\alpha_1 = 0$, Eqs.~\eqref{eq:alpha_soln_1} and~\eqref{eq:alpha_soln_2} describe part of the Landau variety associated with the diagram $\Gsun$. If we instead set $\alpha_3 = \alpha_4 = 0$, these equations describe part of the Landau variety associated with the diagram $\Gbub$. The $\alpha$-positive parts of these surfaces are shown as shaded on the right side of Fig.~\ref{fig:whitney_umbrella}. There, we see that half of the parabola corresponding to the ice cream cone Landau variety is $\alpha$-positive, while the other half is not.

The parameterization of the Pham locus from Eq.~\eqref{eq:icecream_cone_ll_equation} is given in terms of components of the external momenta, in a special frame parametrized by $(u,v,w)$.  As we saw in Eq.~\eqref{eq:leadingsunr1}, the Pham locus looks different in terms of the Mandelstam variables $(p_1 + p_2)^2 = u v$, $(p_3 - p_4)^2 = (w - u)(w - v)$, and $(p_5 + p_6)^2 = w^2$.  In the $(u,v,w)$ parameterization the locus has a singularity of Whitney umbrella type,\footnote{A Whitney umbrella is a singularity of a surface in the neighborhood of which the surface can be represented either by an equation $x^2 - y^2 z = 0$ or, equivalently, by a parameterization $(u, v) \mapsto (x, y, z) = (u v, u, v^2)$.  In this parameterization the surface is ruled by a family of lines; for any fixed value $v$ we have a line in the $(x, y)$ plane at $z = v^2$, defined by $u \mapsto (x, y) = (u v, u)$.  In the same plane, we have a line corresponding to $-v$. These lines intersect at $(x, y) = (0, 0)$.  It follows that the surface has a self-intersection along the $z$ axis; at $z = 0$ the two lines become coincident.  The singularity at $(x, y, z) = (0, 0, 0)$ is called a \emph{pinch point} singularity.  The Whitney umbrella also has a ``handle'' along the $z$ axis defined by $x = y = 0$.  See Fig.~\ref{fig:whitney_umbrella} for a sketch of this singularity, but note that in that figure the surface far away from the singularity looks different (in particular it is not ruled by lines anymore), due to the higher-order polynomial terms that go beyond the ones described above.} while in the Mandelstam parameterization it has a conical singularity at a point.

\paragraph{Bubble-like singularity}
One reason the ice cream cone diagram is important is because 
\begin{equation}
     \left(\bbone-\monM_{\bubt} \right) \left(\bbone-\monM_{\icet} \right) I_\icet(p) \neq 0 \,.
\end{equation}
This nonzero sequential discontinuity contradicts the strict hierarchical principle discussed in Sec.~\ref{sec:iterated} (see also Refs.~\cite{Landshoff1966,boyling1968homological}), which states that one should only be able to take sequential discontinuities of $A_G^\kappa$ around leading singularities of graphs $G^\kappap$ that dominate $G^\kappa$. However, as we alluded in Section~\ref{sec:iterated}, this does not contradict the hierarchical principal when stated in terms of Pham loci, as done in Theorem~\ref{thm:pham}. The reason is that the second monodromy written as $\monM_{\bubt}$ is not exactly around the bubble singularity $\P_\bubt$ but rather around the bubble-like singularity that one gets by setting $\alpha_3=\alpha_4=0$ in $\P_\icet$. So, in our notation, we can also write
\begin{equation}
     \left(\bbone-\monM_{\icet} \right)     \left(\bbone-\monM_{\icet} \right) I_\icet(p)\neq 0 \,,
\end{equation}
which is not surprising.
The Landau loop equations are the same for the bubble-like singularity as for the bubble, but additional on-shell conditions are imposed: $q_4^2=m_4^2$ and $q_3^2=m_3^2$.

Note that while the bubble-like singularity has codimension one, it is not $S$-type and therefore not principal. To see this, it's sufficient to notice that the projection map from the on-shell space of the bubble-like singularity within $\S(G^\icet)$ to $\S(G_0)$ has vanishing eigenvalues. Let us recall some notation from Sec.~\ref{sec:principal} so that we can check this explicitly:\footnote{Here we follow Ref.~\cite{pham1968singularities}.}
\begin{itemize}
\item We associate a $d$-dimensional vector $X_c$ to every fundamental loop $c \in \Chat(\ker \kappa)$, where $d$ is the dimension of the space-time in which the integral is defined.
\item These vectors have to satisfy the constraints
\begin{equation}
\sum_{c \in \Chat(\ker \kappa)} b_{c e} q_e \cdot X_c = 0,
\end{equation}
for all $e \in E(\ker \kappa)$.  This is the constraint that the $X_c^\mu$ belong to the \emph{kernel of the tangent map}.
\item We define a quadratic form
\begin{equation}
A(X) = \sum_{\substack{c_1, c_2 \in \Chat(\ker \kappa)\\ e \in E(\ker \kappa)}} \alpha_e b_{c_1 e} b_{c_2 e} X_{c_1} \cdot X_{c_2}.
\end{equation}
\end{itemize}
The singularity is $S$-type if the quadratic form $A(X)$, restricted to the kernel of the tangent map (defined above), is positive or negative definite for positive values of the $\alpha_e$.

We first construct the quadratic form for the full ice cream cone Pham locus. In $S(G_\icet)$, we have a vector $X_1$ associated with the loop that involves  momenta $q_1^\mu$, $q_2^\mu$, and $q_3^\mu$, and a vector $X_2$ associated with the loop that involves momenta $q_3$ and $q_4$ (see Eq.~\eqref{eq:icecreamcone_momentum_labels} to be reminded of the labeling of momenta).
The kernel of the tangent map conditions read:
\begin{gather}
q_1 \cdot X_1 = 0, \qquad
q_2 \cdot X_2 = 0, \qquad
q_3 \cdot X_1 = 0, \qquad
q_4 \cdot (X_1 - X_2) = 0,
\end{gather}
which can be obtained from the fundamental circuit matrix $b_{ce}$ (given also in App.~\ref{sec:graph_theory})
\begin{equation}
b_{c e} = \begin{pmatrix}
1 & -1 & 0 & 1 \\
0 & 0 & 1 & -1
\end{pmatrix}
\end{equation}
with the rows corresponding to $q_1, q_3$ and the columns to $q_1, q_2, q_3, q_4$.
The quadratic form is thus
\begin{equation}
A(X) = \alpha_1 X_1^2 + \alpha_2 X_1^2 + \alpha_3 X_2^2 + \alpha_4 (X_1 - X_2)^2.
\end{equation}
This quadratic form encodes information about all singularities within $\P_\icet$.

The bubble-like singularity of the ice cream cone corresponds to the subspace of $\P_\icet$ in which $\alpha_3=\alpha_4=0$; accordingly, the relevant contraction projects from this subspace of $\S(G^\icet)$ to $\S(G_0)$. When we have equal masses, we also have $\alpha_1 = \alpha_2 = \frac 1 2$.  In this case, the kernel conditions become $(p_1 + p_2) \cdot X_1 = (p_1 + p_2) \cdot X_2 = 0$ and $q_3 \cdot X_2 = q_4 \cdot (X_1 - X_2) = 0$.  The first two constraints imply that $X_1$ and $X_2$ can be chosen to be space-like.\footnote{Here we take the ice cream cone integral in more than two dimensions so the vectors $X_1$ and $X_2$ can exist.  In four dimensions the integral contains a logarithmic ultraviolet divergence from a sub-bubble and this complicates the analysis.}  Then the quadratic form becomes
\begin{equation}
A(X) = X_1^2.
\end{equation}
In particular, this does not depend on $X_2$ anymore so not only is the quadratic form not definite, but it is in fact \emph{degenerate}.  Therefore, this singularity cannot be understood in terms of quadratic pinches anymore. In this case a generalization of the Picard--Lefschetz formula has to be applied which instead of vanishing spheres involves generalizations called Pham--Brieskorn spheres (see Refs.~\cite{pham1965formules, MR198497}). For further discussions on the bubble-like singularity, see Refs.~\cite{Kawai1976DiscontinuityFA,Kawai:1981fs,Honda2014OnTG}.

  \section{Non-hierarchical Sequential Discontinuities}
\label{sec:codim2}
In Sec.~\ref{sec:sequential_disc} we analyzed situations in which we had two principal Pham loci $\P_\kappa$ and $\P_\kappap$ that were hierarchically related, so that $G \twoheadrightarrow G^\kappap \twoheadrightarrow G^\kappa \twoheadrightarrow G_0$. 
In such cases, we found that the value of the discontinuity around $\P_\kappap$ was unaffected by whether or not one computed a discontinuity around $\P_\kappa$ first. 
In this section, we consider situations in which different types of relations hold for pairs of Pham loci $\P_\kappa$ and $\P_\kappap$. These relations will take the form of double discontinuities that are forced to vanish,
\begin{equation}
    \Big(\bbone - \monM_{\P_\kappap} \Big) 
   \Big(\bbone - \monM_{\P_\kappa}\Big) \I_G(p) 
=   0  \, ,
\label{phamcodim2a}
\end{equation}
or double discontinuities that commute,
\begin{equation}
    \Big(\bbone - 
   \monM_{\P_\kappap} \Big) 
   \Big(\bbone - \monM_{\P_\kappa}\Big) \I_G(p) 
=
   \Big(\bbone - 
   \monM_{\P_\kappa}\Big) \Big(\bbone - 
   \monM_{\P_\kappap} \Big) \I_G(p) \, .
   \label{phamcodim2b}
\end{equation}
We will also be able to formulate a graphical condition for when these relations hold.

Actually, we already know situations in which the relation in Eq.~\eqref{phamcodim2a} holds. 
By Theorem~\ref{thm:pham}, the absorption integral $A_G^\kappa(p) = \Big(\bbone - \monM_{\P_\kappa} \Big) I_G(p)$ can only have singularities on Pham loci $\P_\kappap$ when $\kappap$ dominates $\kappa$. Thus, when $\kappap$ does not dominate $\kappa$, the sequential relation in Eq.~\eqref{phamcodim2a} immediately follows. Conversely, we know that when $\kappap$  dominates $\kappa$, the sequential discontinuity does \emph{not} generically vanish when $\mathcal{P}_\kappa$ and $\mathcal{P}_\kappap$ are principal loci; it is equal to the discontinuity around $\P_\kappap$, as shown in Sec.~\ref{sec:sequential_disc}. %
However, in this section, we will go beyond these cases by considering pairs of Pham loci that are not hierarchically related.

It is worth first commenting on why double discontinuities with respect to non-hierarchical Pham loci can be nontrivial, as this seems to contradict Theorem~\ref{thm:pham}. For instance, in the case of relations taking the form of Eq.~\eqref{phamcodim2b}, Theorem~\ref{thm:pham} must be satisfied on both sides of the equation. This simultaneously requires that $\kappap$ dominates $\kappa$ on the left side while $\kappa$ dominates $\kappap$ on the right side, which can only be the case if $\kappa = \kappap$. This would seem to restrict Eq.~\eqref{phamcodim2b} to cases in which it is tautological.
All is not lost, however, because Pham loci can have different branches; recall that their definition requires all the lines in $\ker \kappa$ to be on shell, but puts no restrictions on the values of the associated $\alpha_e$ parameters, except that they cannot all be zero. 
Thus, like in the example in Sec.~\ref{sec:icecreamtobubbleandsun}, we can consider subspaces of these loci where some $\alpha_e$ are zero, even though the associated edges are on-shell. For example, if we take a one-vertex reducible diagram
\begin{equation}
G =    \begin{tikzpicture}[baseline= {($(current bounding box.base)-(2pt,2pt)$)},
    line width=1.2,scale=0.5]
    \node[black] at (-2.5,0.2) {\vdots};
    \draw[black] ({-1+cos(150)} ,{sin(150)} ) -- ++(150:1);
    \draw[black] ({-1+cos(130)} ,{sin(130)} ) -- ++(130:1);
    \draw[black] ({-1+cos(-150)} ,{sin(-150)} ) -- ++(-150:1);
    \draw[black] ({-1+cos(-130)} ,{sin(-130)} ) -- ++(-130:1);
    \draw[color=etacol,pattern=north west lines, pattern color=etacol] (-1,0) circle (1);
    \draw[color=olddarkgreen,pattern=north west lines, pattern color=olddarkgreen] (1,0) circle (1);
    \node[color=etacol] at (-1.1,0) {$A$};
    \node[color=olddarkgreen] at (0.9,0) {$B$};
    \draw[color=black,fill=black] (0,0) circle (0.1);
    \node[black] at (2.5,0.2) {\vdots};
    \draw[black] ({1+cos(50)} ,{sin(50)} ) -- ++(50:1);
    \draw[black] ({1+cos(30)} ,{sin(30)} ) -- ++(30:1);
    \draw[black] ({1+cos(-50)} ,{sin(-50)} ) -- ++(-50:1);
    \draw[black] ({1+cos(-30)} ,{sin(-30)} ) -- ++(-30:1);
    \end{tikzpicture}
    \label{eq:factorized}
\end{equation}
then the Pham locus for the combined diagram $\P_{\color{etacol} A \color{olddarkgreen} B}$ includes both $\P_{\color{etacol} A}$ and $\P_{ \color{olddarkgreen} B}$. In this section, we label monodromies by the branch of the full locus $\P_{\color{etacol} A \color{olddarkgreen} B}$ that is being encircled. Thus, expressions such as $\big(\bbone - \monM_{\P_{\color{etacol} A}} \big) \big(\bbone -  \monM_{\P_{\color{olddarkgreen} B}} \big)$ can also be thought of as $\big(\bbone - \monM_{\P_{\color{etacol} A\color{olddarkgreen} B}} \big) \big(\bbone -  \monM_{\P_{\color{olddarkgreen} B}} \big)$, which shows how there is no contradiction with the hierarchical principle. In the generic-mass case, we will in fact see that such factorized diagrams are the only ones that can lead to non-zero sequential discontinuities.

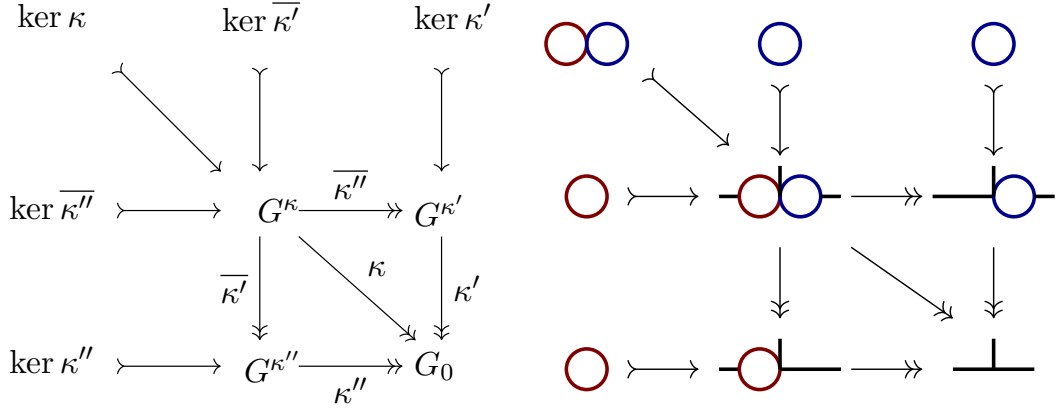
\begin{figure}[t]
    \centering
    \hspace{-5mm}
\resizebox{7cm}{!}{
\begin{tikzpicture}[scale=1.5]
\node[left] at (-0.5,0) {$G^\kappa$};
\node [right] at (0.2,0) {$G^\kappap$};
\node [right] at (0.2,-1.2) {$G_0$};
\draw [->>] (-0.6,0) -- (0.2,0)  
node[midway,above] {$\overline{\kappapp}$};
 \draw [->>] (0.5,-0.2) -- (0.5,-1)
 node[midway,right] {$\kappap$};
\draw [->>] (-0.6,-0.2) -- (0.3,-1);
\node[] at (0,-0.45) {$\kappa$};
\draw [->>] (-0.9,-0.2) -- (-0.9,-1)
node[midway, left] {$\overline{\kappap}$};
\node [] at (-0.8,-1.2) {$G^\kappapp$};
\draw [->>] (-0.6,-1.2) -- (0.2,-1.2)
node[midway, below] {$\kappapp$};
\draw [>->]  (0.5,1.1) -- (0.5,0.3);
\draw [>->]  (-0.9,1.1) -- (-0.9,0.3);
\node [left] at (-0.5,1.5) {$\ker \overline{\kappap}$};
\draw [>->] (-2,0) -- (-1.2,0);
\node [] at (-2.5,0.05) {$\ker \overline{\kappapp}$};
\node [right] at (0.2,1.5) {$\ker \kappap$};
\draw [>->] (-2,-1.2) -- (-1.2,-1.2);
\node [] at (-2.5,-1.15) {$\ker {\kappapp}$};
\draw [>->]  (-2,1.1) -- (-1.2,0.3);
\node [] at (-2.5,1.5) {$\ker \kappa$};
\end{tikzpicture}
}
\hspace{-2mm}
\vspace{-1mm}
\resizebox{7.5cm}{!}{
\begin{tikzpicture}
\node[] at (-0.9,0.5) {\tikz{
    \draw[black,line width=1] (-0.6,0) -- (-0.4,0);
    \draw[darkred,line width=1] (-0.2,0) circle (0.2);
    \draw[newdarkblue2,line width=1] (0.2,0) circle (0.2);
    \draw[black,line width=1] (0,0) -- (0,0.3);
    \draw[black,line width=1] (0.4,0) -- (0.6,0);
    \draw[white,line width=1] (0,-0.2) -- (0,-0.3);
}};
\node[] at (1.2,0.5) {\tikz{
    \draw[black,line width=1] (-0.6,0) -- (0,0);
    \draw[newdarkblue2,line width=1] (0.2,0) circle (0.2);
    \draw[black,line width=1] (0.4,0) -- (0.6,0);
    \draw[black,line width=1] (0,0) -- (0,0.3);
    \draw[white,line width=1] (0,-0.2) -- (0,-0.3);
}};
\node[] at (-0.9,-1.2) {\tikz{
    \draw[black,line width=1] (-0.6,0) -- (-0.4,0);
    \draw[darkred,line width=1] (-0.2,0) circle (0.2);
    \draw[black,line width=1] (0,0) -- (0.6,0);
    \draw[black,line width=1] (0,0) -- (0,0.3);
    \draw[white,line width=1] (0,-0.2) -- (0,-0.3);
}};
\node[] at (1.2,-1.2) {\tikz{
    \draw[black,line width=1] (-0.4,0) -- (0.4,0);
    \draw[black,line width=1] (0,0) -- (0,0.3);
    \draw[white,line width=1] (0,-0.2) -- (0,-0.3);
}};
\node[] at (-2.8,-1.2) {\tikz{
    \draw[darkred,line width=1] (-0.2,0) circle (0.2);
}};
\node[] at (-2.8,0.5) {\tikz{
    \draw[darkred,line width=1] (-0.2,0) circle (0.2);
}};
\node[] at (-0.9,2.0) {\tikz{
    \draw[newdarkblue2,line width=1] (-0.2,0) circle (0.2);
}};
\node[] at (1.2,2.0) {\tikz{
    \draw[newdarkblue2,line width=1] (-0.2,0) circle (0.2);
}};
\node[] at (-2.8,2.0) {\tikz{
    \draw[darkred,line width=1] (-0.2,0) circle (0.2);
    \draw[newdarkblue2,line width=1] (0.2,0) circle (0.2);
}};
\draw [->>] (-0.2,0.5) -- (0.5,0.5);
\draw [->>] (1.2,0) -- (1.2,-0.7);
\draw [->>] (-0.9,0) -- (-0.9,-0.7);
\draw [->>] (-0.2,-1.2) -- (0.5,-1.2);
\draw [->>] (-0.2,0) -- (0.8,-0.7);
\draw [>->]  (1.2,1.6) -- (1.2,0.9);
\draw [>->]  (-0.9,1.6) -- (-0.9,0.9);
\draw [>->] (-2.4,0.5) -- (-1.7,0.5);
\draw [>->] (-2.4,-1.2) -- (-1.7,-1.2);
\draw [>->] (-2.2,1.7) -- (-1.4,1.0);
\end{tikzpicture}
}
\caption{Generic Pham diagram (left) and double bubble example (right). 
The two bubble Pham singularities of the double bubble integral have a fiber product which is a double bubble.}
\label{fig:co2pham2}
\end{figure}

As a first example, consider the double-bubble graph that involves three external legs:
\begin{equation}
G_{\dubbubt} =    \begin{tikzpicture}[baseline= {($(current bounding box.base)-(2pt,2pt)$)},
    line width=1.2,scale=0.5]
    \draw[black] (-4,0) -- (-2,0)  node[midway,above,scale=1] {$p_1$};
    \draw[darkred,scale=1] (-1,0) circle (1);
    \node[darkred,scale=1] at (-1.2,1.4) {$m_1$};
    \node[darkred,scale=1] at (-1.2,-1.4) {$m_2$};
    \draw[black] (0,0) -- (0,1.5)   node[above,scale=1] {$p_3$};
    \draw[newdarkblue2] (1,0) circle (1);
    \node[newdarkblue2,scale=1] at (1.2,1.4) {$m_3$};
    \node[newdarkblue2,scale=1] at (1.2,-1.4) {$m_4$};
    \draw[black] (2,0) -- (4,0) node[midway,above,scale=1] {$p_2$};
    \end{tikzpicture}
    \label{eq:doublebub}
\end{equation}
This is just the product of two bubble integrals. The left bubble has an $\alpha$-positive codimension-one branch point at the Pham locus $\P_{\bubredt}$ where $p_1^2= (m_1 + m_2)^2$, while the right bubble has a codimension-one branch point at the $\P_{\bubbluet}$ where  $p_2^2 = (m_3 + m_4)^2$. The Pham locus $\P_\dubbubt$ for the double bubble includes the intersection of these two surfaces, which has codimension two. 
The contractions that correspond to these three Pham loci can be embedded in a larger diagram, as in Fig.~\ref{fig:co2pham2}. 
We call diagrams like this Pham diagrams. 

While the two bubble contractions do not dominate each other, Theorem~\ref{thm:pham} does not forbid their sequential discontinuity being nonzero due to the fact that $\P_{\bubredt} \subset \P_{\dubbubt}$ and $\P_{\bubbluet} \subset \P_{\dubbubt}$. In fact, because $I_\dubbubt(p)$ is the product of the two bubbles, it is simple to compute the double discontinuity, and the answer is the same independent of the order. That is,
\begin{equation}
    \Big(\bbone - 
   \monM_{\bubredt} \Big) 
   \Big(\bbone - \monM_{\bubbluet} \Big) \I_{G}(p) 
   = \Big(\bbone -   \monM_{\bubbluet} \Big) 
   \Big(\bbone -  \monM_{\bubredt} \Big) \I_{G}(p) \, .
  \label{co2dubbub}
\end{equation}
Although this example may seem trivial since the double bubble factorizes, the same singularities can be present in any Feynman diagram that can be contracted to the diagram in Eq~\eqref{eq:doublebub}. For instance, the double bubble is contraction of genuine two-loop graphs such as:
\begin{equation}
\begin{tikzpicture}[baseline= {($(current bounding box.base)-(2pt,2pt)$)},
    line width=1.2,scale=0.4]
    \draw[black] (-4,0) -- (-2,0);
    \path [darkred,out=60,in=180] (-2,0) edge (0,1);
    \path [newdarkblue2,out=0,in=120] (0,1) edge (2,0);
    \path [darkred,out=-60,in=180] (-2,0) edge (0,-1);
    \path [newdarkblue2,out=0,in=-120] (0,-1) edge (2,0);
    \draw[olddarkgreen] (0,1) -- (0,-1);
    \draw[black] (2,0) -- (4,0);
    \draw[black] (0,1) -- (0.5,2);
\end{tikzpicture}
~~~~
\xrightarrowdbl{~~~~~}
~~~~~
\begin{tikzpicture}[baseline= {($(current bounding box.base)-(2pt,2pt)$)},
    line width=1.2,scale=0.4]
    \draw[black] (-4,0) -- (-2,0);
    \draw[darkred] (-1,0) circle (1);
    \draw[newdarkblue2] (1,0) circle (1);
    \draw[black] (2,0) -- (4,0); 
    \draw[black] (0,0) -- (0,1.5);
\end{tikzpicture}
\end{equation}
Eq.~\eqref{co2dubbub} will also hold for these more complicated Feynman integrals.

\subsection{Transversal Intersections}

Just as the situation we considered in Sec.~\ref{sec:sequential_disc} had a geometric interpretation as the Pham loci intersecting tangentially, there is a geometric interpretation of having pairs of loci that intersect transversally: when the intersection of Pham loci is transversal, the discontinuities commute. 
The general statement was proven by Pham:
\begin{theorem}[Pham]
If two Pham loci $\P_\kappap$ and $\P_\kappapp$ intersect transversally, then their sequential discontinuities, in a small neighborhood of their intersection and away from other singularities, commute:
\begin{equation}
    \Big(\bbone - 
   \monM_{\P_\kappap} \Big) 
   \Big(\bbone - \monM_{\P_\kappapp}\Big) \I_G(p) 
   = \Big(\bbone - 
   \monM_{\P_\kappapp} \Big) \Big(\bbone - 
   \monM_{\P_\kappap} \Big) \I_G(p) \, .
\end{equation}
   \label{thm:commuting}
\end{theorem}
\vspace{-6mm}
\noindent Note that this theorem does not put any conditions on the Pham loci involved in this intersection: they can be of any codimension, and do not have to be principal, $\alpha$-positive, or in the physical region. Of course, in many situations both sides of this equation will vanish, but the fact that the discontinuities commute only requires transversality.

\begin{figure}[t]
    \centering
\resizebox{9cm}{!}{
\begin{tikzpicture}
\draw[-stealth,line width=2.5,black!50] (7,-1) -- (10,-1)  node[below,scale=3] {$z_2$};
\draw[-stealth,line width=2.5,black!50] (-12,2) -- (-12,5)  node[left,scale=3] {$z_1$};
\node[black!50,scale=3] at (0.4,5.6) {$*$};
\begin{knot}[
  clip width=5
]
\strand[etacol,line width=2.5] (0,5) -- (-7,1) .. controls (-8,-1) and (-6,-1)  .. (-6.4,1) -- (0.4,5);
\strand[newdarkblue2,line width=2.5] (1.2,5) -- (8.2,1) .. controls (9.2,-1) and (7.2,-1)  .. (7.6,1) -- (0.8,5);
\strand[black,line width=2.5] (-10,0) -- (10,0);
\strand[black,line width=2.5] (-4,-2)  -- (-4,7);
\strand[darkred,line width=2.5] (-0.3,5.7) -- (-3.3,4.5) .. controls (-6,5) and (-6,3)  .. (-3.3,4.2) -- (-0.3,5.4) node[above left,scale=3] {$\eta'$};
\flipcrossings{1,5,9}
\end{knot}
\node[etacol,scale=3] at (-7.5,2) {$\eta_+$};
\node[newdarkblue2,scale=3] at (8,2) {$\eta_-$};
\node[above,black,scale=3] at (3,0) {$\P_\kappapp$};
\node[above,black,scale=3] at (-5,5.5) {$\P_\kappap$};
\end{tikzpicture}
}
\caption{Two Pham loci $\P_\kappap$ and $\P_\kappapp$ that intersect
 transversally. Locally, the intersection is described by the surfaces $z_1=0$ and $z_2=0$. The two cycles around $\P_\kappapp$ labeled ${\color{etacol}{\eta_+}}$ and ${\color{newdarkblue2}{\eta_-}}$ are homotopically equivalent: one can simply slip the loop over the intersection by giving $z_2$ a small imaginary part at the crossing. Thus, at a transversal intersection, the local homotopy group is Abelian: ${\color{darkred} \eta'}$, the monodromy around $\P_\kappap$, commutes with ${\color{etacol}{\eta_+}} \approx {\color{newdarkblue2}{\eta_-}}$.
 }
\label{fig:co2homotopy}
\end{figure}
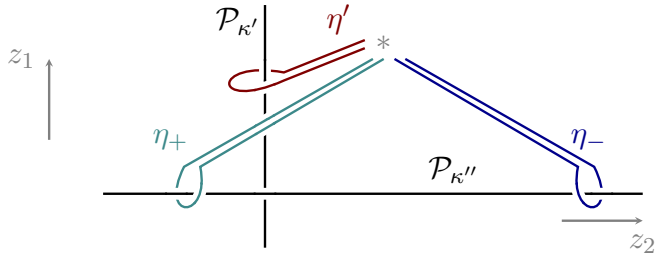

It is easy to see why the transversality condition is sufficient for concluding that discontinuities commute. Let us assume both Pham loci are codimension one (otherwise the corresponding discontinuities are zero and their commutation is trivial).  When these surfaces intersect transversally, we can always choose coordinates for $\bbC^2\setminus (\P_\kappap \cup P_\kappapp)$ in the neighborhood of the intersection in which $\P_\kappap$ corresponds to the surface $z_1=0$ and $\P_\kappapp$ corresponds to the surface $z_2=0$. Then the space itself factorizes, so $\bbC^2\setminus (\P_\kappap \cup P_\kappapp) = \bbC^* \times \bbC^*$. Thus, the monodromies around each singularity are totally independent of the other. We visualize this in Fig.~\ref{fig:co2homotopy}, in which the two monodromies $\eta_+$ and $\eta_-$ are homotopically equivalent. The monodromy $\eta''$ around $\P_\kappapp$ thus commutes with the monodromy $\eta'$ around $\P_\kappap$:
\begin{equation}
    \eta' \circ \eta'' = \eta'' \circ \eta'.
\end{equation}
As a result, computing the corresponding discontinuities in either order gives the same result.

\subsection{Compatibility of Contractions}
\label{sec:sketchco2vanish}

While Theorem~\ref{thm:commuting} does not place any requirements on the pair of Pham loci $\mathcal{P}_\kappa$ and $\mathcal{P}_\kappap$ beyond their transversal intersection, we now turn our attention to a stronger result that can be established for principal Pham loci in physical regions.\footnote{Note that this implies that we only consider cases in which the kernels of $\kappa$ and $\kappap$ are neither disconnected graphs or bouquets of graphs (one-vertex reducible). In Sec.~\ref{sec:exceptional_kinematics} we will consider kernels which are disconnected graphs.} To do so, we first note that, given any such pair of principal loci, one can formulate a Pham diagram of the type shown on the left side of Fig.~\ref{fig:co2pham2}; one just needs to find a graph $G^\kappa$ which contracts to both $G^\kappap$ and $G^\kappapp$. However, in the generic case, there will be no special relation between the kernels of the four contractions $\kappap$, $\kappapp$, $\overline{\kappap}$, and $\overline{\kappapp}$.

The Pham diagram for the double-bubble example on the right side of Fig.~\ref{fig:co2pham2} has the special property that $\ker \overline{\kappap}= \ker \kappap$ and $\ker \overline{\kappapp} = \ker \kappapp$. When both of these relations are satisfied, the contractions $\kappa$ and $\kappap$ are said to be \textbf{compatible}~\cite{pham}.  If either is not satisfied, we say that these contractions are \textbf{incompatible}. 
When a pair of contractions are compatible, one can take $\ker \kappa = \ker \kappap \oplus \ker \kappapp$, meaning $\kappa$ is a fiber product of $\kappap$ and $\kappapp$. The general result that holds in this situation, again proven by Pham, is:
\begin{theorem}[Pham]
   \label{thm:co2vanish}
Assume we have two principal Pham loci $\P_\kappap$ and $\P_\kappapp$ that intersect transversally. Then, for generic masses,  
\begin{equation}
    \Big(\bbone - 
   \monM_{\P_\kappap} \Big) 
   \Big(\bbone - \monM_{\P_\kappapp} \Big) \I_G(p) 
   = 0 \,,
\end{equation}
if $\ker \kappap$ and $\ker \kappapp$ are incompatible, and $p$ is in the physical region. 
\end{theorem}
\noindent 
We first explore this notion of compatibility in an example, then outline a general derivation of this result.

\paragraph{Triangle Diagram Example}
To see an example involving a pair of transversally-intersecting Pham loci that have a vanishing sequential discontinuity, we return to the triangle integral considered in Sec.~\ref{sec:picard_lefschetz_absorption_int_example}. We recall that the three bubble contractions of this diagram correspond to thresholds, which occur at $p_k^2 = (m_i\pm m_j)^2$, or equivalently $y_{ij} = \pm 1$. The surfaces corresponding to each of these bubble contractions intersect transversally, as illustrated in Fig.~\ref{fig:triangletangent}.  

Let us consider the Landau equations that are associated with computing discontinuities around the $p_3^2$ and $p_2^2$ thresholds in more detail. The $p_3^2$ threshold occurs at $(m_1+m_2)^2$, and corresponds to the sequence of contractions
\begin{equation}
\begin{tikzpicture}[baseline={($(current bounding box.base)-(2pt,2pt)$)},
    line width=1,scale=0.5]
    \draw[black] (-4,1.5) -- (-2,1)  node[midway,above,yshift=2,scale=0.8] {$p_2$};
    \draw[black,-latex] (-4,1.5) -- (-3,1.25);
    \draw[black] (-4,-1.5) -- (-2,-1) node[midway, below,scale=0.8] {$p_1$};
    \draw[black,-latex] (-4,-1.5) -- (-3,-1.25);
    \draw[newdarkblue2] (-2,-1) --  (-2,1) node[midway,left,scale=0.8] {$q_3,\,m_3$};
    \draw[newdarkblue2,-latex reversed] (-2,-1) -- (-2,0);
    \draw[darkred] (-2,1) -- (0,0) node[midway,above,scale=0.8] {$~~~~~~q_1,\,m_1$};
    \draw[darkred,-latex] (-2,1) -- (-0.75,0.375);
    \draw[olddarkgreen] (0,0) -- (-2,-1) node[midway,below,scale=0.8] {$~~~~~q_2,\,m_2$};
    \draw[olddarkgreen,-latex] (-2,-1) -- (-0.75,-0.375);
    \draw[black] (0,0) -- ++(0:2);
    \node[black,scale=0.8] at (2,0.5) {$p_3$};
    \draw[black,-latex] (0,0) -- ++(0:1.2);
\end{tikzpicture}
\xrightarrowdbl{~~\overline{\kappapp}~~}
\begin{tikzpicture}[baseline= {($(current bounding box.base)-(2pt,2pt)$)},line width=1,scale=0.5]
\path [darkred,out=60,in=120] (-2,0) edge (0,0);
\path [olddarkgreen,out=-60,in=240] (-2,0) edge (0,0);
\node[darkred,scale=0.8] at (-1,1) {$q_1,\,m_1$};
\draw[darkred,-latex] (-0.9,0.5) -- ++(0:0.2);
\node[olddarkgreen,scale=0.8] at (-1,-1) {$q_2,\,m_2$};
\draw[olddarkgreen,-latex] (-0.9,-0.5) -- ++(0:0.2);
\draw[black] (-2,0) -- ++(150:2);
\draw[black] (-2,0) -- ++(-150:2);
\draw[black] (0,0) -- ++(0:2);
\node[black,scale=0.8] at (-4,1.6) {$p_2$};
\node[black,scale=0.8] at (-4,-1.6) {$p_1$};
\node[black,scale=0.8] at (2,0.5) {$p_3$};
\draw[black,-latex reversed] (-2,0) -- ++(150:1.36);
\draw[black,-latex reversed] (-2,0) -- ++(-150:1.36);
\draw[black,-latex] (0,0) -- ++(0:1.5);
\end{tikzpicture}
 \xrightarrowdbl{~~\kappap~~}
\begin{tikzpicture}[baseline= {($(current bounding box.base)-(2pt,2pt)$)},line width=1,scale=0.5]
\draw[black] (-2,0) -- ++(150:2);
\draw[black] (-2,0) -- ++(-150:2);
\draw[black] (-2,0) -- ++(0:2);
\node[black,scale=0.8] at (-4,1.6) {$p_2$};
\node[black,scale=0.8] at (-4,-1.6) {$p_1$};
\node[black,scale=0.8] at (0,0.5) {$p_3$};
\draw[black,-latex reversed] (-2,0) -- ++(150:1.36);
\draw[black,-latex reversed] (-2,0) -- ++(-150:1.36);
\draw[black,-latex] (-2,0) -- ++(0:1.5);
\end{tikzpicture}
\label{triangley12}
\end{equation}
Writing $q_1^\mu = k^\mu$ and $q_2^\mu = p_3^\mu - k^\mu$, the Landau equations require
\begin{equation}
    k^\mu =  \frac{\alpha_2}{\alpha_1+ \alpha_2} p_3^\mu, \quad k^2 = m_1^2,\quad (k-p_3)^2 = m_2^2 ,
    \label{landaup3m}
\end{equation}
which have the solution $p_3^2 = (m_1+m_2)^2$. In the rest frame of $p_3^\mu$, Eq.~\eqref{landaup3m} imposes $k^\mu=(m_1,0)$ and $p_3^\mu = (m_1+m_2,0)$, which corresponds to a two-parameter, and therefore codimension-one, solution in the space of external momenta.
The discontinuity around the threshold at $p_2^2=(m_1+m_3)^2$ corresponds to the contractions
\begin{equation}
\begin{tikzpicture}[baseline={($(current bounding box.base)-(2pt,2pt)$)},
    line width=1,scale=0.5]
    \draw[black] (-4,1.5) -- (-2,1)  node[midway,above,yshift=2,scale=0.8] {$p_2$};
    \draw[black,-latex] (-4,1.5) -- (-3,1.25);
    \draw[black] (-4,-1.5) -- (-2,-1) node[midway, below,scale=0.8] {$p_1$};
    \draw[black,-latex] (-4,-1.5) -- (-3,-1.25);
    \draw[newdarkblue2] (-2,-1) --  (-2,1) node[midway,left,scale=0.8] {$q_3,\,m_3$};
    \draw[newdarkblue2,-latex reversed] (-2,-1) -- (-2,0);
    \draw[darkred] (-2,1) -- (0,0) node[midway,above,scale=0.8] {$~~~~~~q_1,\,m_1$};
    \draw[darkred,-latex] (-2,1) -- (-0.75,0.375);
    \draw[olddarkgreen] (0,0) -- (-2,-1) node[midway,below,scale=0.8] {$~~~~~q_2,\,m_2$};
    \draw[olddarkgreen,-latex] (-2,-1) -- (-0.75,-0.375);
    \draw[black] (0,0) -- ++(0:2);
    \node[black,scale=0.8] at (2,0.5) {$p_3$};
    \draw[black,-latex] (0,0) -- ++(0:1.2);
\end{tikzpicture}
 \xrightarrowdbl{~~\overline{\kappap}~}
\begin{tikzpicture}[baseline= {($(current bounding box.base)-(2pt,2pt)$)},line width=1,scale=0.5]
\path [darkred,out=60,in=120] (-2,0) edge (0,0);
\draw [newdarkblue2] (-2,0) -- (0,0);
\node[darkred,scale=0.8] at (-1,1) {$q_1,\,m_1$};
\draw[darkred,-latex] (-0.9,0.5) -- ++(0:0.2);
\node[newdarkblue2,scale=0.8] at (-1.8,-0.4) {$q_3,\,m_3$};
\draw[newdarkblue2,-latex] (-0.9,-0.0) -- ++(0:0.2);
\draw[black] (-2,0) -- ++(150:2);
\draw[black] (0,0) -- ++(0:2);
\draw[black] (-3.7,-1.2) to[out=10,in=-140] (0,0);
\node[black,scale=0.8] at (-4,1.6) {$p_2$};
\node[black,scale=0.8] at (-4,-1.6) {$p_1$};
\node[black,scale=0.8] at (2,0.5) {$p_3$};
\draw[black,-latex reversed] (-2,0) -- ++(150:1.36);
\draw[black,-latex] (-3.7,-1.2) -- ++(10:1.36);
\draw[black,-latex] (0,0) -- ++(0:1.5);
\end{tikzpicture}
 \xrightarrowdbl{~~\kappapp~~}
\begin{tikzpicture}[baseline= {($(current bounding box.base)-(2pt,2pt)$)},line width=1,scale=0.5]
\draw[black] (-2,0) -- ++(150:2);
\draw[black] (-2,0) -- ++(-150:2);
\draw[black] (-2,0) -- ++(0:2);
\node[black,scale=0.8] at (-4,1.6) {$p_2$};
\node[black,scale=0.8] at (-4,-1.6) {$p_1$};
\node[black,scale=0.8] at (0,0.5) {$p_3$};
\draw[black,-latex reversed] (-2,0) -- ++(150:1.36);
\draw[black,-latex reversed] (-2,0) -- ++(-150:1.36);
\draw[black,-latex] (-2,0) -- ++(0:1.5);
\end{tikzpicture}
\label{triangley13}
\end{equation}
The Landau equations for this contraction are
\begin{equation}
    k^\mu = \frac{\alpha_3'}{\alpha_1' + \alpha_3'} p_2^\mu, \quad k^2 = m_1^2,\quad (k-p_2)^2 = m_3^2.
    \label{landaup2m}
\end{equation}
In the rest frame of $p_2$, Eq.~\eqref{landaup2m} imposes $k^\mu=(m_1,0)$ and $p_2^\mu = (m_1+m_3,0)$, which is another codimension-one solution.

By constructing the appropriate Pham diagram, one can check that the pair of contractions corresponding to $p_3^2=(m_1+m_2)^2$ and $p_2^2=(m_1+m_3)^2$ is incompatible, and thus by Theorem~\ref{thm:co2vanish} we expect the corresponding double discontinuity of the triangle integral to vanish. To see that this is indeed what happens, we recall that computing a discontinuity with respect to the $p_2^2$ threshold after computing a discontinuity with respect to the $p_3^2$ threshold requires maintaining the on-shell condition $(k-p_3)^2 - m_2^2=0$ when the second discontinuity is computed. This provides an extra constraint on the momentum $p_3^\mu$ that must be imposed in addition to the Landau equations in Eq~\eqref{landaup2m}, which leaves just a one-parameter family of solutions. This encodes a codimension-two surface in $\S(G_0)$, and therefore the monodromy around this surface will vanish. Stated differently, there is no $p_2^2$ normal threshold of the absorption integral $A_G^\kappap$, in accordance with Theorem~\ref{thm:co2vanish}. A similar result holds when the discontinuities are computed in the opposite order.

In contrast, we know from Sec.~\ref{sec:sequential_disc} that the absorption integral $A_G^\kappap$ has a nonzero discontinuity with respect to the triangle singularity at $D=0$. In this case, the Landau equations for the triangle also give a codimension-one surface, but again we must impose the on-shell conditions associated with the bubble diagram in~\eqref{triangley12}. However, because all the edges in the bubble are also in the triangle, they are already on shell; there is no new constraint. Thus, the solution space will remain codimension-one and the monodromy can be nonzero. Indeed, the bubble absorption integral for the three-dimensional triangle diagram in Eq.~\eqref{eq:bubble_absorption_3d} only has a singularity at $D=0$ and not at $y_{ij}^2=1$, in accordance with the sequential-discontinuity analysis.

Another way to understand this result is to recall that the absorption integral that gives the discontinuity with respect to $p_3^2$ has the form given in Eq.~\eqref{eq:PL_triangle_bubble_disc}:
\begin{equation}
   A_G^\kappap
   =
    (2 \pi i)^2  \int_{\partial_1 \partial_2 e_{12}}
    \frac{\rd^3 k}{(\rd s_1 \wedge \rd s_2 )s_3},
\end{equation}
where $s_i = q_i^2-m_i^2$. 
Here we have a function $s_3$ defined on the vanishing sphere $\partial_1 \partial_2 e_{12}$ where $s_1 = s_2 = 0$.  We want to know when the restriction to $s_3 = 0$ of the projection to external kinematics has a critical point.  In other words, we want to know when $\rd s_3\vert_{s_1 = s_2 = 0} = 0$ when restricted to fixed external kinematics.  This is a constrained extremization problem.  To make it unconstrained we introduce Lagrange multipliers and we end up with the condition
\begin{equation}
\rd s_3 + \alpha_1 \rd s_1 + \alpha_2 \rd s_2 = 0,
\end{equation}
which is equivalent to the triangle Landau equation. In other words, the only codimension-one singularity of the bubble absorption integral is on the triangle singularity.

\paragraph{Sketch of a Proof of Theorem~\ref{thm:co2vanish}}
Let us now outline a derivation of Theorem~\ref{thm:co2vanish}. We start by considering an integral of the form
\begin{equation}
    \I_{G} = \int_h \omega \, ,
\end{equation}
where the integration domain $h$ lives in $\EE(G)$, which we recall denotes the space of momentum-conserving internal and external momentum of the graph $G$.
The discontinuity of $\I_G$ with respect to a principal Pham loci $\P_\kappap$ gives an absorption integral
\begin{equation}
A_G^\kappap =  \Big(\bbone - 
   \monM_{\P_\kappap} \Big)  I_{G} =  \int_{\chi'} 
\omega' \, ,
\label{Agkp}
\end{equation}
where $\omega'$ is an integrand of the form seen in Eq.~\eqref{IGmform3}, and $\chi'$ is the vanishing sphere whose construction was described in Sec.~\ref{sec:momspacePL}.

The contour $\chi'$ forms a fiber over $\S(G^\kappap)$ in $\EE(G)$. In particular, all that is really relevant for computing this monodromy is the space $\EX(G^\kappap)$, since no edges other than those of $\ker \kappap$ participate in the Landau equations. Thus, we can define $\chi_\kappap \subset \S(G^\kappap) \subset \EX(G^\kappap)$ as the restriction of the vanishing sphere to this smaller space. Equivalently, it is the fiber over the basepoint $p$ in this space, which can be defined as the inverse image of the projection map $\pi_\kappap \colon \S(G^\kappap) \to \S(G_0)$. That is, $\chi_\kappap = \pi_\kappap^{-1}(p)$. With this definition we can rewrite
\begin{equation}
    A_G^\kappap = \int_{\chi_\kappap}  \int_{h'} \omega' \, ,
    \label{Agkp2}
\end{equation}
where now $h'$ includes whatever additional integrations are in Eq.~\eqref{Agkp} but not in the new contour ${\chi_\kappap}$.

Now let us consider computing the monodromy of $A_G^\kappap$ around $\P_\kappapp$. Following the Picard--Lefschetz approach, this monodromy pushes the integration contour out of the way of this singularity. But one can ask: does it deform the $\chi_\kappap$ contour, the $h'$ contour, or both? Because $\P_\kappap$ and $\P_\kappapp$ intersect transversally, near the intersection the monodromies commute. This implies that $\chi_\kappap$ is unaffected by the second monodromy. 
Thus the monodromy operator can be passed through the outer integral, and we have
\begin{equation}
\Big(\bbone - 
  \monM_{\P_\kappapp} \Big)    \Big(\bbone -
  \monM_{\P_\kappap} \Big) I_{G} = \int_{\chi_\kappap}  \Big(\bbone - 
  \monM_{\P_\kappapp} \Big)\int_{h'} \omega' \, .
  \label{mongoesthrough}
\end{equation}
On the other hand, we can ask more generally for which Pham loci $\P$ the integral
\begin{equation}
     \int_{\chi_\kappap}  \Big(\bbone - 
  \monM_{\P} \Big)\int_{h'} \omega'
\end{equation}
will be nonzero. As with Lemma~\ref{lem:crit}, these loci are critical points of the projection map from one on-shell space to another. In this case, since the first monodromy takes place on a fiber over $\S(G^\kappap)$, the edges in $\ker \kappap$ are on-shell on both sides of the map. That is, the singularities are on Pham loci $\P_{\kappapps}$ for any $\kappapps: G^\kappa \twoheadrightarrow G^\kappap$. 

We are almost done. We are still interested in the second discontinuity for $\kappapp$ in Eq.~\eqref{mongoesthrough}, where the loop equations for the loops in $\ker \kappapp$ are satisfied. But as stated above, the only possible nonzero discontinuities will be for loci $\P_\kappapps$ in which the loop equations for the loops in $\ker \kappapps$ are satisfied. Therefore, we must be able to find a $\kappapps$ for which $\ker \kappapps = \ker \kappapp$. 
Since the discontinuities commute, by Theorem~\ref{thm:commuting}, we must also be able to find a $\kappaps$ for which $\ker \kappaps = \ker \kappap$.
Therefore, the contractions must be compatible for the sequential discontinuity to be nonzero. That completes the sketch of the proof. More details can be found in Sec.~II.3.4 and Sec.~I.3.3 of Ref.~\cite{pham}.

\paragraph{Further Examples} 
Let us reconsider our examples from earlier in this section using the way of thinking about absorption integrals introduced in the above proof. First, returning to the double bubble, we have 
\begin{equation}
    \Big(\bbone - \monM_{\bubredt} \Big)  I_{\dubbubt} =
     \Big(\bbone - \monM_{\bubredt} \Big)  \int \frac {\rd^d k_1 \rd^d k_2}{s_1 s_2 s_3 s_4} = (2 \pi i)^2 \int_{\chi_\bubt} \frac {\rd^d k_1}{\rd s_1 \wedge \rd s_2} \int_{h'} \frac {\rd^d k_2}{s_3 s_4}
\end{equation}
where $\chi_\bubt$ is the surface of constant $k_1^0$ and $|\vec{k_1}_\perp|$, as in Eqs.~\eqref{k0form} and~\eqref{eq:triangle_sphere_component}, while $h'$ is all of $\bbR^d$ for the $k_2$ integration. The monodromy around $\P_\bubbluet$ acts on the second integral, and only on the $\frac{1}{s_3 s_4}$ propagators, since the rest is independent of $k_2$. The contractions are compatible in this case.

For the bubble discontinuity of the triangle in three dimensions, we have
\begin{equation}
     \Big(\bbone - \monM_{\bubrgt} \Big)  I_\tritcol = \Big(\bbone - \monM_{\bubrgt} \Big) \int \frac {\rd^3 k_1}{s_1 s_2 s_3} = (2 \pi i)^2 \int_{\chi_{\bubrgt}} \frac {\rd^3 k_1}{\rd s_1 \wedge \rd s_2} \frac{1}{s_3}
\end{equation}
In this case, there is no second contour $h'$ because the spaces $\EX(G^\tritcol)$ and $\EX(G^\bubt)$ are the same.
Accordingly, the monodromy around a different bubble, such as the one in Eq.~\eqref{triangley13}, vanishes. 

As a third example, we can consider the
bubble and sunrise singularities of the ice cream cone diagram, shown in Fig.~\ref{fig:icecreamtobubbleandsun}. We saw that these singular surfaces intersected transversally in Fig.~\ref{fig:whitney_umbrella}. First, the discontinuity with respect to the bubble singularity is
\begin{equation}
     \Big(\bbone - \monM_{\bubt} \Big) I_{\icet} =  \Big(\bbone - \monM_{\bubt} \Big) \int \frac{\rd^4 k_1 \rd^4 k_2}{s_1 s_2 s_3 s_4} = (2 \pi i)^2 \int_{\chi_\bubt} \frac{\rd^4 k_1}{\rd s_1 \wedge \rd s_2} \int_{h'} \frac{\rd^4 k_2}{s_3 s_4}.
\end{equation}
Here $h'$ is all of $\bbR^4$, as the $k_2$ loop momentum can be chosen not to pass through $q_1$ or $q_2$. This fact also lets us pull $\frac{1}{\rd s_1 \wedge \rd s_2}$ outside of the second integral. There is no sunrise discontinuity of this absorption integral, as the sunrise involves $q_2,q_3$ and $q_4$, but only $q_3$ and $q_4$ participate in the inner integral. The sunrise singularity of the ice cream cone is similarly given by
 \begin{align}
     \Big(\bbone - \monM_{\sunt} \Big)  I_{\icet} 
     &=  \Big(\bbone - \monM_{\sunt} \Big)  \int \rd^4 k_1 \rd^4k_2 \frac{1}{s_1 s_2 s_3 s_4} \\ 
     &= (2 \pi i)^3 \int_{\chi_{\sunt}} \rd^4 k_1 \rd^4 k_2 \frac{1}{s_1} \frac{1}{\rd s_2 \wedge \rd s_3 \wedge \rd s_4}.
\end{align}
In this case, as with the bubble singularity of the triangle, there is no second contour $h'$; rather, $\EX(G^{\icet}) = \EX(G^{\sunt})$ since both loop momenta pass through the sunrise.
So the bubble followed by sunrise or sunrise followed by bubble discontinuities of the ice cream cone vanish. 

\subsection{Compatibility of Landau Equations}
\label{sec:landaucompatibility}
\begin{figure}[t]
\centering
        \tdplotsetmaincoords{70}{110}
\resizebox{7cm}{!}{
\begin{tikzpicture}[tdplot_main_coords]
\begin{scope}[tdplot_main_coords]
 \draw[line width=1,newdarkblue2] plot[variable=\x,domain=-1:4,samples=73,smooth] 
 (\x,2,{1-(1/10)*\x*\x+(1/20)*\x*\x*\x});
  \draw[line width=1,newdarkblue2,dashed] plot[variable=\x,domain=-1:4,samples=73,smooth] 
 (\x,2,0);
  \draw[line width=1,darkred] plot[variable=\x,domain=-1:4,samples=73,smooth] 
 (2,\x,{2.5-(1/10)*\x*\x+(1/20)*\x*\x*\x});
 \draw[line width=1,darkred,dashed] plot[variable=\x,domain=-1:4,samples=73,smooth] 
 (2,\x,0);
\end{scope} 
\node[above,black,scale=1,newdarkblue2] at (5.3,2.2,0) {$\P_{\kappap}$};
\node[above,black,scale=1,darkred] at (3,4.8) {$\P_{\kappapp}$};
\foreach \x in {-1,-0.9,...,4}
 {\draw[newdarkblue2!10] plot[variable=\Z,domain=0:1,samples=73,smooth] 
 (\x,2,{(1-(1/10)*\x*\x+(1/20)*\x*\x*\x)*\Z});}
 \foreach \x in {-1,-0.9,...,4}
 {\draw[darkred!10] plot[variable=\Z,domain=0:1,samples=73,smooth] 
 (2,\x,{(2.5-(1/10)*\x*\x+(1/20)*\x*\x*\x)*\Z});}
  \draw[line width=1,-stealth] (-1,0,0) -- (4,0,0) node[anchor=north east]{$p_1$};
\draw[line width=1,-stealth] (0,-1,0) -- (0,4,0) node[anchor=north west]{$p_2$};
\draw[line width=1,-stealth] (0,0,-1) -- (0,0,3) node[anchor=south]{$k$};
  \draw[line width=1,newdarkblue2] plot[variable=\x,domain=-1:4,samples=73,smooth] 
 (\x,2,{1-(1/10)*\x*\x+(1/20)*\x*\x*\x});
  \draw[line width=1,newdarkblue2,dashed] plot[variable=\x,domain=-1:4,samples=73,smooth] 
 (\x,2,0);
  \draw[line width=1,darkred] plot[variable=\x,domain=-1:4,samples=73,smooth] 
 (2,\x,{2.5-(1/10)*\x*\x+(1/20)*\x*\x*\x});
 \draw[line width=1,darkred,dashed] plot[variable=\x,domain=-1:4,samples=73,smooth] 
 (2,\x,0);
\end{tikzpicture}
}
    \caption{Although two Pham loci $\P_{\kappap}$ and $\P_{\kappapp}$ may intersect transversally in the space formed by some external momenta, labeled as $p_1$ and $p_2$, they do not necessarily intersect in the internal-variable space, schematically shown as $k$. When the discontinuity around $\P_{\kappap}$ is given by Cutkosky's formula, its integration contour $h'$ localizes close to the Landau-singularity solution in the space of loop momenta $k$. If the Landau-equation solution $\P_{\kappapp}$ also puts a constraint on the loop momentum $k$, the corresponding vanishing cell entering the Picard--Lefschetz theorem will generally not intersect the integration contour $h'$.}
    \label{fig:Landauintersection}
\end{figure}
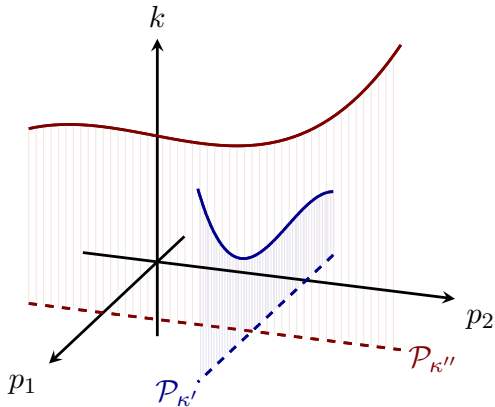

We have defined a pair of contractions to be compatible when the horizontal and vertical kernels are equal in a Pham diagram. There is also a more direct algebraic way to understand the compatibility of pairs of discontinuities through the Landau equations. The claim is that:
\begin{lemma}
Suppose we have two contractions $\kappap$ and $\kappapp$
corresponding to a pair of codimension-one Pham loci $\P_\kappap$ and $\P_\kappapp$ such that neither contraction dominates the other. 
Then, for generic masses, they are compatible if and only if one can solve both the Landau equations for $\P_\kappap$ and $\P_\kappapp$ simultaneously. \label{lemma2}
\end{lemma}
\noindent
When we can solve both sets of Landau equations simultaneously, we say that the \textbf{Landau equations are compatible}. A schematic diagram is shown in Fig.~\ref{fig:Landauintersection}.

Rather than prove this lemma directly, let us first state a related lemma from which Lemma~\ref{lemma2} follows:
\begin{lemma}
Suppose we have a pair of compatible contractions $\kappap$ and $\kappapp$ that describe two codimension-one Pham loci $\P_\kappap$ and $\P_\kappapp$. Then, for generic masses, the Landau diagram that describes the codimension-two Pham locus associated with the combined contraction $\kappap \circ \kappapp = \kappapp \circ \kappap$ factorizes into a pair of diagrams that separately describe the two loci $\P_\kappap$ and $\P_\kappapp$.
\label{lemma3}
\end{lemma}
\noindent
From this lemma, it follows that the only allowed sequential discontinuities that do not vanish according to Theorem~\ref{thm:co2vanish} are those for which the Landau diagram for the codimension-two variety factorizes into a one-vertex reducible diagram, as depicted in Eq.~\eqref{eq:doublebub}.
Below, we will prove this lemma and consider an example in which this factorization property holds; for now, we just sketch how Lemma~\ref{lemma2} is implied by Lemma~\ref{lemma3}. 

To show this, we first assume the pair of contractions $\kappap$ and $\kappapp$ in Lemma~\ref{lemma2} are compatible. Then, by Lemma~\ref{lemma3}, the Landau diagram that describes the combined contraction $\kappap \circ \kappapp$ factorizes, and the kernels of $\kappap$ and $\kappapp$ do not share any loop momenta. As a result, the two sets of Landau equations can be solved simultaneously. To establish the converse, let us now assume that the two contractions are incompatible, and therefore both involve some loop momentum $k_c$. Since by assumption neither contraction dominates the other, both $\ker \kappap$ and $\ker \kappapp$ must also involve at least one edge the other lacks. However, since $\P_\kappap$ and $\P_\kappapp$ are both codimension-one, each of the critical points $p$ these contractions describe corresponds to a unique solution for all the momenta $q_e^\mu$ and $k_c^\mu$ that appear in their kernel. For each point $p^\star$ that exists within the intersection of $\P_\kappap$ and $\P_\kappapp$ in the space of external momenta, the solutions for the loop momenta $k_c^\mu(p^\star)$ will generically be different, since the masses are assumed to be generic. Thus, when contractions are incompatible it is not generically possible to solve both sets of Landau equations simultaneously. This completes our proof.

We can use the fact that the Landau equations are compatible to gain a different, perhaps more physical, perspective on Theorem~\ref{thm:co2vanish}.
Recall that the Picard--Lefschetz analysis from Sec.~\ref{sec:PicardLefschetz} takes place in an arbitrarily small region near the singular point we call the Leray bubble (as illustrated in Eq.~\eqref{eyelashU}). The vanishing cell and the relevant part of the integration contour for computing the discontinuity are both within the Leray bubble. For $\kappap$ this Leray bubble lives in $\EE(G^\kappap)$, while for  $\kappapp$ this Leray bubble lives in $\EE(G^\kappapp)$. Both bubbles are subspaces of the space of internal and external momenta associated with the original graph $\EE(G^\kappa)$.
When two Landau equations are incompatible, the solution for $k_c(p^\star)$ at an intersection point $p^\star$ are different functions of external momenta. Thus the Leray bubbles corresponding to $\kappap$ and $\kappapp$ for $p$ arbitrarily close to $p^\star$ will not intersect in $\EE(G^\kappa)$. This means that the integration contour for $A_G^\kappap$ will not intersect the vanishing cell for $\P_\kappapp$, and vice versa. Therefore, in neither case is the contour of the absorption integral pinched by the second singularity, and therefore the sequential monodromy is zero. This provides another interpretation of Eq.~\eqref{mongoesthrough}.

For an even more direct proof of Theorem~\ref{thm:co2vanish}, we can simply try to solve the Landau equations for $\P_\kappapp$ when the additional on-shell constraints for the edges in $\ker \kappap$ are imposed, and the Landau diagram formed out of both contractions does not factorize. Since $\P_\kappapp$ is a codimension-one solution to the loop equations and the on-shell conditions encoded just by the contraction $\kappapp$, imposing additional on-shell constraints will make it codimension-two or higher in the space of external momenta. Thus, there can be no discontinuity.

\subsection{Allowed Sequential Discontinuities}
\label{sec:allowed}
\begin{figure}[t]
    \centering
\resizebox{8cm}{!}{
\tikzset{internal/.style={line width=0.8}}
\begin{tikzpicture}[scale=1]
\node[] at (-0.9,0.5) {\tikz{
    \draw[black,internal] (-0.7,-0.4) -- (-0.4,-0.3);
    \draw[black,internal] (0.7,-0.4) -- (0.4,-0.3);
    \draw[black,internal] (-0.7,0.4) -- (-0.4,0.3);
    \draw[black,internal] (0.7,0.4) -- (0.4,0.3);
    \draw[etacol,internal] (0.4,0.3) -- (0,0) -- (0.4,-0.3) -- cycle;
    \draw[etacol,internal] (0.4,0.3) -- (0.4,-0.3);
    \draw[etacol,internal] (0.4,-0.3) -- (0,0);
    \draw[darkred,internal] (-0.4,0.3) -- (0,0) -- (-0.4,-0.3) -- cycle;
    \draw[darkred,internal] (-0.4,-0.3) -- (0,0);
    \draw[darkred,internal] (-0.4,-0.3) -- (-0.4,0.3);
}};
\node[] at (1.2,0.5) {\tikz{
    \draw[black,internal] (-0.4,-0.3) -- (0,0);
    \draw[black,internal] (0.7,-0.4) -- (0.4,-0.3);
    \draw[black,internal] (-0.4,0.3) -- (0,0);
    \draw[black,internal] (0.7,0.4) -- (0.4,0.3);
    \draw[etacol,internal] (0.4,0.3) -- (0,0) -- (0.4,-0.3) -- cycle;
    \draw[etacol,internal] (0.4,0.3) -- (0.4,-0.3);
    \draw[etacol,internal] (0.4,-0.3) -- (0,0);
}};
\node[] at (-0.9,-1.2) {\tikz{
    \draw[black,internal] (-0.7,-0.4) -- (-0.4,-0.3);
    \draw[black,internal] (0.4,-0.3) -- (0,0);
    \draw[black,internal] (-0.7,0.4) -- (-0.4,0.3);
    \draw[black,internal] (0.4,0.3) -- (0,0);
    \draw[darkred,internal] (-0.4,0.3) -- (0,0) -- (-0.4,-0.3) -- cycle;
    \draw[darkred,internal] (-0.4,-0.3) -- (0,0);
    \draw[darkred,internal] (-0.4,-0.3) -- (-0.4,0.3);
}};
\node[] at (1.2,-1.2) {\tikz{
    \draw[black,internal] (0.4,-0.3) -- (0,0);
    \draw[black,internal] (0.4,0.3) -- (0,0);
    \draw[black,internal] (-0.4,0.3) -- (0,0);
    \draw[black,internal] (-0.4,-0.3) -- (0,0);
}};
\node[] at (-2.8,-1.2) {\tikz{
    \draw[darkred,internal] (-0.4,0.3) -- (0,0) -- (-0.4,-0.3) -- cycle;
    \draw[darkred,internal] (-0.4,-0.3) -- (0,0);
    \draw[darkred,internal] (-0.4,-0.3) -- (-0.4,0.3);
}};
\node[] at (-2.8,0.5) {\tikz{
    \draw[darkred,internal] (-0.4,0.3) -- (0,0) -- (-0.4,-0.3) -- cycle;
    \draw[darkred,internal] (-0.4,-0.3) -- (0,0);
    \draw[darkred,internal] (-0.4,-0.3) -- (-0.4,0.3);
}};
\node[] at (-0.9,2.0) {\tikz{
    \draw[etacol,internal] (0.4,0.3) -- (0,0) -- (0.4,-0.3) -- cycle;
    \draw[etacol,internal] (0.4,0.3) -- (0.4,-0.3);
    \draw[etacol,internal] (0.4,-0.3) -- (0,0);
}};
\node[] at (1.2,2.0) {\tikz{
    \draw[etacol,internal] (0.4,0.3) -- (0,0) -- (0.4,-0.3) -- cycle;
    \draw[etacol,internal] (0.4,0.3) -- (0.4,-0.3);
    \draw[etacol,internal] (0.4,-0.3) -- (0,0);
}};
\node[] at (-2.8,2.0) {\tikz{
    \draw[etacol,internal] (0.4,0.3) -- (0,0) -- (0.4,-0.3) -- cycle;
    \draw[etacol,internal] (0.4,0.3) -- (0.4,-0.3);
    \draw[etacol,internal] (0.4,-0.3) -- (0,0);
    \draw[darkred,internal] (-0.4,0.3) -- (0,0) -- (-0.4,-0.3) -- cycle;
    \draw[darkred,internal] (-0.4,-0.3) -- (0,0);
    \draw[darkred,internal] (-0.4,-0.3) -- (-0.4,0.3);
}};
\draw [->>] (-0.2,0.5) -- (0.4,0.5)
node[midway,above,scale=0.8] {$\kappapps$};
\draw [->>] (1.2,0) -- (1.2,-0.8)
node[midway,right,scale=0.8] {$\kappap$};
\draw [->>] (-0.9,0) -- (-0.9,-0.8)
node[midway,right,scale=0.8] {$\kappaps$};
\draw [->>] (-0.2,-1.2) -- (0.4,-1.2)
node[midway,above,scale=0.8] {$\kappapp$};
\draw [->>] (-0.2,0) -- (0.8,-0.8);
\draw [>->]  (1.2,1.6) -- (1.2,0.8);
\draw [>->]  (-0.9,1.6) -- (-0.9,0.8);
\draw [>->] (-2.4,0.5) -- (-1.8,0.5);
\draw [>->] (-2.4,-1.2) -- (-1.8,-1.2);
\draw [>->] (-2.2,1.7) -- (-1.4,1.0);
\end{tikzpicture}
}
\caption{Pham diagram for the bow-tie integral, which describes one of the pairs of sequential discontinuities that can appear in the double-box integral.
\label{fig:bowtiephamdiagram}}
\end{figure}

Lemma~\ref{lemma3} puts strong constraints on what sequences of discontinuities can occur in Feynman integrals that depend on generic masses. Namely, all possible sequential discontinuities are described by Landau diagrams that are \textbf{one-vertex reducible}, meaning they factorize into two or more diagrams if one vertex is removed. This property has the effect of factorizing the loop momentum dependence, so that each edge only depends on the loop momenta on the left or right side of the diagram, but not both. As a result, the corresponding Landau equations and Pham loci also trivially factorize. We now consider an example that illustrates this phenomenon, after which we will prove Lemma~\ref{lemma3}.

Consider the double box Feynman integral with generic masses:
\begin{equation}
\I_{\squaresquare} =
\resizebox{!}{1cm}{
\tikzset{internal/.style={line width=1.2}}
\begin{tikzpicture}[baseline= {($(current bounding box.base)-(2pt,2pt)$)},scale=1.5] {
    \draw[black,internal] (-0.9,-0.6) -- (-0.6,-0.3);
    \draw[black,internal] (0.9,-0.6) -- (0.6,-0.3);
    \draw[black,internal] (-0.9,0.6) -- (-0.6,0.3);
    \draw[black,internal] (0.9,0.6) -- (0.6,0.3);
    \draw[black,internal] (0.6,0.3) 
     -- (0.6,-0.3) node[midway,right,etacol] {$q_5$}
     -- (0,-0.3)  node[midway,below,etacol] {$q_6$}
     -- (-0.6,-0.3)  node[midway,below,darkred] {$q_1$}
     -- (-0.6,0.3)  node[midway,left,darkred] {$q_2$}
     -- (0,0.3) node[midway,above,darkred] {$q_3$}
     -- (0.6,0.3) node[midway,above,etacol] {$q_4$};
    \draw[black,internal] (0,0.3) -- (0,-0.3) node[midway,right] {$q_0$} ;
    \draw[darkred,internal] (0,-0.3) 
     -- (-0.6,-0.3)
     -- (-0.6,0.3) 
     -- (0,0.3);
     \draw[etacol,internal]
     (0,0.3)
     -- (0.6,0.3) 
     -- (0.6,-0.3)
     -- (0,-0.3);
    \node[] at (-1.1,0.6) {$p_2$};
    \node[] at (-1.1,-0.6) {$p_1$};
    \node[] at (1.1,0.6) {$p_3$};
    \node[] at (1.1,-0.6) {$p_4$};
    }
    \end{tikzpicture}} \label{eq:dbl_box}
\end{equation}
While this graph itself does not factorize, we can still learn about the discontinuities that can be accessed in this diagram by studying the Landau equations after contracting the middle leg to a point. This gives rise to the bow-tie Landau diagram, whose Pham diagram is depicted in Fig.~\ref{fig:bowtiephamdiagram}. The bow-tie diagram factorizes, so by Lemma~\ref{lemma3} it should describe a possible sequence of discontinuities in the original double box integral. 

To confirm that this is the case, we write the double box integral as
\begin{equation}
    \I_{\squaresquare} = \int_{h_\triangleright} \int_{h_\triangleleft}
    \frac{\rd^d k_\triangleright \wedge \rd^d k_\triangleleft}{s_0 s_1 \cdots s_6}\,,
\end{equation}
where the propagator labels match the momentum labels in Eq.~\eqref{eq:dbl_box}, $k_\triangleright$ denotes the loop momentum going through the left loop, and $k_\triangleleft$ denotes the loop momentum going through the right loop.  The integration contours $h_{\triangleright}$ and $h_{\triangleleft}$ are deformed to have imaginary parts consistent with the Feynman $i \varepsilon$ prescription, as described in Sec.~\ref{sec:iepaths}. Using the Picard--Lefschetz theorem, we find that the discontinuity with respect to the Pham locus $\P_{\triright}$, namely the singularity that is obtained after contracting the right loop to a point, is
\begin{equation}
\label{dbox2}
    \left(\bbone- \monM_{\triright}\right)
    \I_{\squaresquare} = -\langle  e_{123}, h_\triangleright\rangle (2 \pi i)^3 \int_{\partial_1 \partial_2 \partial_3 e_{123}} \frac{\rd^d k_\triangleright}{\rd s_1 \wedge \rd s_2 \wedge \rd s_3} \int_{h_\triangleleft} \frac{\rd^d k_\triangleleft}{s_0 s_4 s_5 s_6},
\end{equation}
where $e_{123}$ is the vanishing cell defined by $s_1 \geq 0$, $s_2 \geq 0$ and $s_3 \geq 0$.  Let us now take the monodromy around the Pham locus $\P_\trileft$, which is obtained by contracting the left loop.  We note that the inner integral over $h_\triangleleft$ has a singularity at $\P_\triangleleft$ and, importantly, its location does not depend on the point in the integration domain for the outer integral.  
We can therefore compute the double discontinuity as
\begin{multline}
    (\bbone - \monM_\trileft) (\bbone - \monM_\triright) \I_{\squaresquare} = (2 \pi i)^6 \langle e_{123}, h_\triangleright\rangle \langle e_{456}, h_\triangleleft\rangle  \\ \times \int_{\partial_1 \partial_2 \partial_3 e_{123}} \frac{\rd^d k_\triangleright}{\rd s_1 \wedge \rd s_2 \wedge \rd s_3} \int_{\partial_4 \partial_5 \partial_6 e_{456}} \frac{\rd^d k_\triangleleft}{s_0 \, \rd s_4 \wedge \rd s_5 \wedge \rd s_6},
    \label{eq:doubleboxseqdisc}
\end{multline}
where $e_{456}$ is defined by $s_4 \geq 0$, $s_5 \geq 0$ and $s_6 \geq 0$.
Since the two Landau singularities $\P_{\triangleleft}$ and $\P_{\triangleright}$ only put constraints on one of the loop momenta, one can easily see that the monodromies commute; that is, $(1 - \monM_\trileft) (1 - \monM_\triright) \I_{\squaresquare}=(1 - \monM_\triright) (1 - \monM_\trileft)\I_{\squaresquare}$.

Note that this construction implies that the inner integral, around which the second monodromy is taken, can be thought of as the product of connected $S$-matrices at the vertices of $G^\kappap$. Let us pause to comment on how this observation connects to the $S$-matrix bootstrap. For the double-box example, we would write Eq.~\eqref{dbox2} as
\begin{equation}
    A_G^{\triright} =
    \int  \cM_1(p_2 \to q_2,q_3)
    \times  \cM_2(p_1, q_1 \to q_2)
    \times  \cM_3(q_1, q_3 \to p_3,p_4)\,,
    \label{eq:Atriright}
\end{equation}
where the integral is with respect to the measure in Cutkosky's formula, and the specific absorption integral in Eq.~\eqref{dbox2} is the one for which $\cM_3$ is itself a box diagram. We can represent Eq.~\eqref{eq:Atriright} diagrammatically as,
\begin{equation}
\big(\bbone-\monM_{\triright} \big) I_G =
    \begin{tikzpicture}[scale=1.8, line width=1.3,baseline= {($(current bounding box.base)-(2pt,2pt)$)}]
    \draw[black] (-0.7,-0.4) -- (-0.4,-0.3);
    \draw[black] (0.4,-0.3) -- (0.2,0);
    \draw[black] (-0.7,0.4) -- (-0.4,0.3);
    \draw[black] (0.4,0.3) -- (0.2,0);
    \draw[darkred] (-0.4,0.3) -- (0.2,0);
    \draw[darkred] (-0.4,-0.3) -- (0.2,0);
    \draw[darkred] (-0.4,-0.3) -- (-0.4,0.3);
    \node[scale=0.9] at (-0.9,0.5) {$p_2$};
    \node[scale=0.9] at (-0.9,-0.5) {$p_1$};
    \node[scale=0.9] at (0.6,0.4) {$p_3$};
    \node[scale=0.9] at (0.6,-0.4) {$p_4$};
    \node[darkred,scale=0.9] at (0,-0.3) {$q_1$};
    \node[darkred,scale=0.9] at (-0.55,0) {$q_2$};
    \node[darkred,scale=0.9] at (0,0.3) {$q_3$};
    \draw[color=white,fill=white] (-0.4,0.3) circle (0.15);
    \draw[color=white,fill=white] (-0.4,-0.3) circle (0.15);
    \draw[color=white,fill=white] (0.2,0) circle (0.15);
    \draw[color=gray,pattern=north west lines, pattern color=gray] (-0.4,0.3) circle (0.15);
    \node[scale=0.7] at (-0.4,0.3) {$\cM_1$};
    \draw[color=gray,pattern=north west lines, pattern color=gray] (-0.4,-0.3) circle (0.15);
    \node[scale=0.7] at (-0.4,-0.3) {$\cM_2$};
    \draw[color=gray,pattern=north west lines, pattern color=gray] (0.2,0) circle (0.15);
    \node[scale=0.7] at (0.2,0) {$\cM_3$};
    \end{tikzpicture}
\end{equation}
where each of the gray blobs labeled with $\cM$ are themselves $S$-matrix elements. Here, $G$ can be any graph that contracts to the triangle. We can go further and represent the absorption integral corresponding to the singularity in Eq.~\eqref{eq:doubleboxseqdisc} as
\begin{equation}
\big(\bbone-\monM_{\trileft}\big)\big(\bbone-\monM_{\triright}\big) I_G =
    \begin{tikzpicture}[scale=1.8, line width=1.3,baseline= {($(current bounding box.base)-(2pt,2pt)$)}]
    \draw[black] (-0.7,-0.4) -- (-0.4,-0.3);
    \draw[black] (-0.7,0.4) -- (-0.4,0.3);
    \draw[darkred] (-0.4,0.3) -- (0.2,0);
    \draw[darkred] (-0.4,-0.3) -- (0.2,0);
    \draw[darkred] (-0.4,-0.3) -- (-0.4,0.3);
    \node[scale=0.9] at (-0.9,0.5) {$p_2$};
    \node[scale=0.9] at (-0.9,-0.5) {$p_1$};
    \node[darkred,scale=0.9] at (0,-0.3) {$q_1$};
    \node[darkred,scale=0.9] at (-0.55,0) {$q_2$};
    \node[darkred,scale=0.9] at (0,0.3) {$q_3$};
    \draw[color=white,fill=white] (-0.4,0.3) circle (0.15);
    \draw[color=white,fill=white] (-0.4,-0.3) circle (0.15);
    \draw[color=white,fill=white] (0.2,0) circle (0.15);
    \draw[color=gray,pattern=north west lines, pattern color=gray] (-0.4,0.3) circle (0.15);
    \node[scale=0.7] at (-0.4,0.3) {$\cM_1$};
    \draw[color=gray,pattern=north west lines, pattern color=gray] (-0.4,-0.3) circle (0.15);
    \node[scale=0.7] at (-0.4,-0.3) {$\cM_2$};
    \begin{scope}[xscale=-1,yscale=1,xshift=-11]
    \draw[black] (-0.7,-0.4) -- (-0.4,-0.3);
    \draw[black] (-0.7,0.4) -- (-0.4,0.3);
    \draw[etacol] (-0.4,0.3) -- (0.2,0);
    \draw[etacol] (-0.4,-0.3) -- (0.2,0);
    \draw[etacol] (-0.4,-0.3) -- (-0.4,0.3);
    \node[scale=0.9] at (-0.9,0.5) {$p_3$};
    \node[scale=0.9] at (-0.9,-0.5) {$p_4$};
    \node[etacol,scale=0.9] at (0,-0.3) {$q_6$};
    \node[etacol,scale=0.9] at (-0.55,0) {$q_5$};
    \node[etacol,scale=0.9] at (0,0.3) {$q_4$};
    \draw[color=white,fill=white] (-0.4,0.3) circle (0.15);
    \draw[color=white,fill=white] (-0.4,-0.3) circle (0.15);
    \draw[color=white,fill=white] (0.2,0) circle (0.15);
    \draw[color=gray,pattern=north west lines, pattern color=gray] (-0.4,0.3) circle (0.15);
    \node[scale=0.7] at (-0.4,0.3) {$\cM_4$};
    \draw[color=gray,pattern=north west lines, pattern color=gray] (-0.4,-0.3) circle (0.15);
    \node[scale=0.7] at (-0.4,-0.3) {$\cM_5$};
    \draw[color=gray,pattern=north west lines, pattern color=gray] (0.2,0) circle (0.15);
    \node[scale=0.7] at (0.2,0) {$\cM_3$};
    \end{scope}
    \end{tikzpicture},
\end{equation}
subject to the usual caveats that the loci $\P_\trileft$ and $\P_\triright$ intersect transversally and do not overlap with other Pham loci.
This picture holds to all orders in perturbation theory, although for any particular perturbative contribution, these matrix elements are sums of Feynman integrals. In this way, the transversally-intersecting case fits naturally into the $S$-matrix bootstrap picture where singularities of higher-order diagrams are built up from lower-order ones.
This Landau diagram has been studied in integrable theories in Ref.~\cite{Coleman:1978kk} where the blobs were exact $S$-matrix elements.

Now let us show that all possible pairs of sequential discontinuities of Feynman integrals with generic masses come from one-vertex reducible Landau diagrams, by proving Lemma~\ref{lemma3}. We begin by considering the commutative diagram
\begin{equation}
\begin{tikzcd}
    G^\kappa \arrow[r, "\kappapps", two heads] \arrow[d, "\kappaps", two heads] \arrow[dr, "\kappa", two heads] & G^\kappap \arrow[d, "\kappap", two heads] \\
    G^\kappapp \arrow[r, "\kappapp", two heads] & G_0
\end{tikzcd}
\end{equation}
that can be constructed for any pair of compatible contractions. The Landau loop equations associated with these contractions can be written as
\begin{equation}
    \sum_{e \in E(\ker \kappap)} \alpha_e b_{c e} q_e^\mu = 0, \qquad \forall c \in \Chat(\ker \kappap) 
    \label{eq:kernels11}
\end{equation}
and 
\begin{equation}
    \sum_{e \in E(\ker \kappapp)} \alpha_e b_{c e} q_e^\mu = 0, \qquad \forall c \in \Chat(\ker \kappapp) \, ,
    \label{eq:kernels11second}
\end{equation}
respectively. Notably, a loop momentum in $\Chat(\ker \kappapp)$ cannot have any edges in $E(\ker \kappaps)$, but a cycle which \emph{contracts} to a cycle in $\Chat(\ker \kappaps)$ may have an edge in $E(\ker \kappapp)$.
Thus, the circuit matrix for $\P_\kappap$ must take the form
\begin{equation}
    \label{eq:bmatrix_lower}
    b_{c e}(\ker \kappa) = \begin{blockarray}{ccc}
     & E(\ker \kappa'') & E(\ker \kappaps) \\
    \begin{block}{c(cc)}
        \Chat(\ker \kappa'') & \ast & 0 \\
        \Chat(\ker \kappaps) & \ast & \ast\\
    \end{block}
\end{blockarray}\, ,
\end{equation}
where the blocks which do not necessarily vanish are marked by $\ast$.  
Moreover, because the intersection is transversal and the monodromies commute, we  can use the same argument with $\kappap$ and $\kappapp$ swapped conclude that the matrix $b_{c e}(\ker \kappa)$ can also be written in the form
\begin{equation}
    \label{eq:bmatrix_upper}
    b_{c e}(\ker \kappa) = \begin{blockarray}{ccc}
     & E(\ker \kappapps) & E(\ker \kappap) \\
    \begin{block}{c(cc)}
        \Chat(\ker \kappapps) & \ast & \ast \\
        \Chat(\ker \kappap) & 0 & \ast\\
    \end{block}
\end{blockarray}\, .
\end{equation}
Using the fact that $\ker \kappa' = \ker \bar{\kappa}'$ and $\ker \kappa'' = \ker \bar{\kappa}''$, we conclude that the circuit matrix for the kernel of the combined contraction $\kappa$ must be block diagonal:
\begin{equation}
    \label{eq:bmatrix-block-diagonal-structure}
    b_{c e}(\ker \kappa) = \begin{blockarray}{ccc}
     & E(\ker \kappa'') & E(\ker \kappa') \\
    \begin{block}{c(cc)}
        \Chat(\ker \kappa'') & \ast & 0 \\
        \Chat(\ker \kappa') & 0 & \ast\\
    \end{block}
\end{blockarray} \, .
\end{equation}
Thus, $\ker \kappa$ can be written as a direct sum of $\ker \kappa'$ and $\ker \kappa''$, namely $b(\ker \kappa) = b(\ker \kappa'') \oplus b(\ker \kappa')$.  This also gives us a \emph{canonical} decomposition of the loop momenta $\Chat(\ker \kappa) = \Chat(\ker \kappa'') \oplus \Chat(\ker \kappa')$, showing that for general kinematics, the two singularities $\P_\kappap$ and $\P_\kappapp$ are incompatible if the fundamental circuits of their kernels are not disjoint.  In other words, the Landau diagram corresponding to the contraction $\kappa$ must factorize into two diagrams, and it is therefore one-vertex reducible. This proves Lemma~\ref{lemma3}. 

Note that the factorization property of Lemma~\ref{lemma3} implies that the kernels of compatible contractions can be put in short exact sequences with the kernel of the combined contraction:
\begin{equation}
    \ker \kappa'' \rightarrowtail \ker \kappa \twoheadrightarrow \ker \kappa'\,, \qquad  \ker \kappa' \rightarrowtail \ker \kappa \twoheadrightarrow \ker \kappa''\,.
\end{equation} 
For the bow-tie diagram in Fig.~\ref{fig:bowtiephamdiagram} the sequences are:
\begin{equation}
\resizebox{!}{0.5cm}{
\begin{tikzpicture}[baseline=(current bounding box.center),
    line width=1.5,scale=2]
\draw[darkred] (-0.4,0.3) -- (0,0) -- (-0.4,-0.3) -- cycle;
\draw[darkred] (-0.4,-0.3) -- (0,0);
\draw[darkred] (-0.4,-0.3) -- (-0.4,0.3);
\end{tikzpicture}
}
\hspace{5mm}
\resizebox{2cm}{!}{
\begin{tikzpicture}[baseline=(current bounding box.center),
    line width=1.5,scale=1.2]
    \draw[>->] (0,0) -- (3,0);
    \draw[<<-] (0,-0.5) -- (3,-0.5);
\end{tikzpicture}
}        
\hspace{5mm}
\resizebox{!}{0.5cm}{
\begin{tikzpicture}[baseline=(current bounding box.center),
    line width=1.5,scale=2]
    \draw[etacol] (0.4,0.3) -- (0,0) -- (0.4,-0.3) -- cycle;
    \draw[etacol] (0.4,0.3) -- (0.4,-0.3);
    \draw[etacol] (0.4,-0.3) -- (0,0);
    \draw[darkred] (-0.4,0.3) -- (0,0) -- (-0.4,-0.3) -- cycle;
    \draw[darkred] (-0.4,-0.3) -- (0,0);
    \draw[darkred] (-0.4,-0.3) -- (-0.4,0.3);
    \end{tikzpicture}
}
\hspace{5mm}
\resizebox{2cm}{!}{
\begin{tikzpicture}[baseline=(current bounding box.center),
    line width=1.5,scale=1.2]
    \draw[<-<] (0,0) -- (3,0);
    \draw[->>] (0,-0.5) -- (3,-0.5);
\end{tikzpicture}
}        
\hspace{5mm}
\resizebox{!}{0.5cm}{
\begin{tikzpicture}[baseline=(current bounding box.center),
    line width=1.5,scale=2]
    \draw[etacol] (0.4,0.3) -- (0,0) -- (0.4,-0.3) -- cycle;
    \draw[etacol] (0.4,0.3) -- (0.4,-0.3);
    \draw[etacol] (0.4,-0.3) -- (0,0);
\end{tikzpicture}
}.
\end{equation}
Such exact sequences are characteristic of compatible Pham diagrams.

\subsection{Disallowed Sequential Discontinuities Beyond Steinmann}
\label{sec:disallowed_discontinuities}
\begin{figure}[t]
    \centering
\resizebox{8cm}{!}{
\tikzset{internal/.style={line width=0.8}}
\begin{tikzpicture}[scale=1]
\node[] at (-0.9,0.5) {\tikz{
    \draw[black,internal] (-0.4,-0.4) -- (-0.2,-0.2);
    \draw[black,internal] (0.4,-0.4) -- (0.2,-0.2);
    \draw[black,internal] (-0.4,0.4) -- (-0.2,0.2);
    \draw[black,internal] (0.4,0.4) -- (0.2,0.2);
    \draw[darkred,internal] (0.2,0.2) -- (0.2,-0.2);
    \draw[darkorange,internal] (0.2,-0.2) -- (-0.2,-0.2);
    \draw[newdarkblue2,internal] (-0.2,-0.2) -- (-0.2,0.2);
    \draw[olddarkgreen,internal] (-0.2,0.2) -- (0.2,0.2);
}};
\node[] at (1.2,0.5) {\tikz{
    \draw[black,internal] (-0.4,-0.3) -- (-0.2,0);
    \draw[black,internal] (0.4,-0.3) -- (0.2,0);
    \draw[black,internal] (-0.4,0.3) -- (-0.2,0);
    \draw[black,internal] (0.4,0.3) -- (0.2,0);
    \draw[olddarkgreen,internal] (-0.2,0) to[out=60,in=120] (0.2,0);
    \draw[darkorange,internal] (-0.2,0) to[out=-60,in=-120] (0.2,0);
}};
\node[] at (-0.9,-1.2) {\tikz{
    \draw[black,internal] (-0.4,-0.3) -- (0,-0.2);
    \draw[black,internal] (0.4,-0.3) -- (0,-0.2);
    \draw[black,internal] (-0.4,0.3) -- (0,0.2);
    \draw[black,internal] (0.4,0.3) -- (0,0.2);
    \draw[newdarkblue2,internal] (0,-0.2) to[out=150,in=210] (0,0.2);
    \draw[darkred,internal] (0,-0.2) to[out=30,in=-30] (0,0.2);
}};
\node[] at (1.2,-1.2) {\tikz{
    \draw[black,internal] (-0.3,-0.3) -- (0,0);
    \draw[black,internal] (0.3,-0.3) -- (0,0);
    \draw[black,internal] (-0.3,0.3) -- (0,0);
    \draw[black,internal] (0.3,0.3) -- (0,0);
}};
\node[] at (-2.8,-1.2) {\tikz{
    \draw[newdarkblue2,internal] (0,-0.2) to[out=150,in=210] (0,0.2);
    \draw[darkred,internal] (0,-0.2) to[out=30,in=-30] (0,0.2);
}};
\node[] at (-2.8,0.5) {\tikz{
    \draw[newdarkblue2,internal] (-0.2,-0.2) -- (-0.2,0.2);
    \draw[darkred,internal] (0.2,-0.2) -- (0.2,0.2);
}};
\node[] at (-0.9,2.0) {\tikz{
    \draw[darkorange,internal] (-0.2,-0.2) -- (0.2,-0.2);
    \draw[olddarkgreen,internal] (-0.2,0.2) -- (0.2,0.2);
}};
\node[] at (1.2,2.0) {\tikz{
    \draw[olddarkgreen,internal] (-0.2,0) to[out=60,in=120] (0.2,0);
    \draw[darkorange,internal] (-0.2,0) to[out=-60,in=-120] (0.2,0);
}};
\node[] at (-2.8,2.0) {\tikz{
    \draw[darkred,internal] (0.2,0.2) -- (0.2,-0.2);
    \draw[darkorange,internal] (0.2,-0.2) -- (-0.2,-0.2);
    \draw[newdarkblue2,internal] (-0.2,-0.2) -- (-0.2,0.2);
    \draw[olddarkgreen,internal] (-0.2,0.2) -- (0.2,0.2);
}};
\draw [->>] (-0.2,0.5) -- (0.4,0.5)
node[midway,above,scale=0.7] {$\overline{\kappa^t}$};
\draw [->>] (1.2,0) -- (1.2,-0.8) 
node[midway,right,scale=0.7] {$\kappa^s$};
\draw [->>] (-0.9,0) -- (-0.9,-0.8)
node[midway,right,scale=0.7] {$\overline{\kappa^s}$};
\draw [->>] (-0.2,-1.2) -- (0.4,-1.2) node[midway,below,scale=0.7] {$\kappa^t$};
\draw [->>] (-0.2,0) -- (0.8,-0.8);
\draw [>->]  (1.2,1.6) -- (1.2,0.8);
\draw [>->]  (-0.9,1.6) -- (-0.9,0.8);
\draw [>->] (-2.4,0.5) -- (-1.8,0.5);
\draw [>->] (-2.4,-1.2) -- (-1.8,-1.2);
\draw [>->] (-2.2,1.7) -- (-1.4,1.0);
\node[black,scale=0.7] at (1.8,1) {$G^{2s}$};
\node[black,scale=0.7] at (-1.3,-0.7) {$G^{2t}$};
\end{tikzpicture}
}
\caption{Contractions of a box singularity to the bubble $s$ and $t$-channel singularities.  In this case the kernels of the horizontal (or vertical) contractions are not the same as graphs; one is connected and the other is disconnected.  However, for the special kinematics the kernels impose the same constraints.
\label{fig:box_s_t_channels}
}
\end{figure}
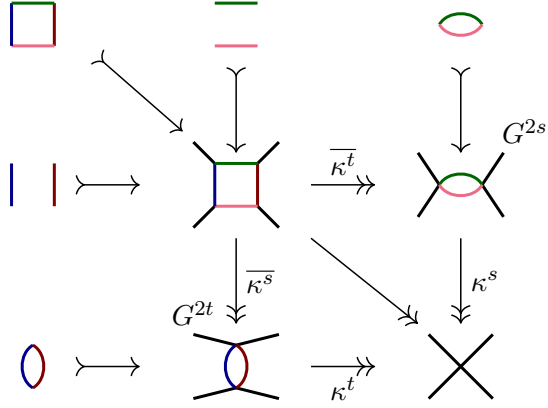

The Steinmann relations famously state that amplitudes cannot have double discontinuities in partially-overlapping momentum channels~\cite{Steinmann,Steinmann2,Cahill:1973qp}. For instance, the double discontinuity of the box diagram with respect to the $s$ and $t$ channels must vanish in the physical region. These types of restrictions have been found to provide extremely powerful constraints for bootstrap approaches to computing amplitudes in recent years~\cite{Caron-Huot:2016owq}, and thus it is interesting to see how these relations also follow from Theorem~\ref{thm:co2vanish}. At the same time, we will be able to see that Theorem~\ref{thm:co2vanish} implies constraints on the sequential discontinuities of Feynman integrals that also go beyond the Steinmann relations. 

We begin by drawing the Pham diagram for the $s$ and $t$ channel discontinuities of the box integral in Fig.~\ref{fig:box_s_t_channels}. We can see from this construction that $\ker \kappa^t \neq \ker \overline{\kappa^t}$ and $\ker \kappa^{s} \neq \ker \overline{\kappa^s}$, which implies that $\kappa^s$ and $\kappa^t$ are incompatible.
Thus, setting all the internal masses in the box to be equal for simplicity (it is easy to check that the Pham loci remain separated in $(s,t)$ space), we conclude that
\begin{equation}
    \big(\bbone - \monM_{s - 4 m^2 } \big) \big(\bbone - \monM_{t - 4 m^2 } \big) I_\squaretcol = 0 \,.
    \label{eq:steinmann_box}
\end{equation}
This relation will hold for the box diagram in the physical region, or similarly for any other diagram that can be contracted to the box diagram. It is not hard to convince oneself that pairs of partially-overlapping momentum channels always give rise to incompatible contractions in the same way, implying that these pairs of sequential discontinuities must always vanish according to Theorem~\ref{thm:co2vanish}.

\begin{figure}[t]
    \centering
\resizebox{8cm}{!}{
\tikzset{internal/.style={line width=0.8}}
\begin{tikzpicture}[scale=1]
\node[] at (-0.9,0.5) {\tikz{
    \draw[black,internal] (-0.4,-0.35) -- (-0.2,-0.2);
    \draw[black,internal] (-0.35,-0.4) -- (-0.2,-0.2);
    \draw[black,internal] (0.35,-0.4) -- (0.2,-0.2);
    \draw[black,internal] (0.4,-0.35) -- (0.2,-0.2);
    \draw[black,internal] (-0.35,0.4) -- (-0.2,0.2);
    \draw[black,internal] (-0.4,0.35) -- (-0.2,0.2);
    \draw[black,internal] (0.35,0.4) -- (0.2,0.2);
    \draw[black,internal] (0.4,0.35) -- (0.2,0.2);
    \draw[darkred,internal] (0.2,0.2) -- (0.2,-0.2);
    \draw[darkorange,internal] (0.2,-0.2) -- (-0.2,-0.2);
    \draw[newdarkblue2,internal] (-0.2,-0.2) -- (-0.2,0.2);
    \draw[olddarkgreen,internal] (-0.2,0.2) -- (0.2,0.2);
}};
\node[] at (1.2,0.5) {\tikz{
    \draw[black,internal] (-0.4,0.05) -- (-0.2,0);
    \draw[black,internal] (-0.4,-0.05) -- (-0.2,0);
    \draw[black,internal] (0.35,-0.3) -- (0.2,0);
    \draw[black,internal] (0.4,-0.25) -- (0.2,0);
    \draw[black,internal] (0.5,0.04) -- (0.2,0);
    \draw[black,internal] (0.5,-0.04) -- (0.2,0);
    \draw[black,internal] (0.4,0.25) -- (0.2,0);
    \draw[black,internal] (0.35,0.3) -- (0.2,0);
    \draw[newdarkblue2,internal] (-0.2,0) to[out=60,in=120] (0.2,0);
    \draw[darkorange,internal] (-0.2,0) to[out=-60,in=-120] (0.2,0);
}};
\node[] at (-0.9,-1.4) {\tikz{
    \draw[black,internal] (0.05,-0.4) -- (0,-0.2);
    \draw[black,internal] (-0.05,-0.4) -- (0,-0.2);
    \draw[black,internal] (0.05,0.4) -- (0,0.2);
    \draw[black,internal] (-0.05,0.4) -- (0,0.2);
    \draw[black,internal] (-0.4,0.35) -- (0,0.2);
    \draw[black,internal] (-0.4,0.25) -- (0,0.2);
    \draw[black,internal] (0.4,0.35) -- (0,0.2);
    \draw[black,internal] (0.4,0.25) -- (0,0.2);
    \draw[darkred,internal] (0,-0.2) to[out=150,in=210] (0,0.2);
    \draw[olddarkgreen,internal] (0,-0.2) to[out=30,in=-30] (0,0.2);
}};
\node[] at (1.2,-1.2) {\tikz{
    \draw[black,internal] (-0.25,-0.2) -- (0,0);
    \draw[black,internal] (-0.2,-0.25) -- (0,0);
    \draw[black,internal] (0.25,-0.2) -- (0,0);
    \draw[black,internal] (0.2,-0.25) -- (0,0);
    \draw[black,internal] (-0.25,0.2) -- (0,0);
    \draw[black,internal] (-0.2,0.25) -- (0,0);
    \draw[black,internal] (0.25,0.2) -- (0,0);
    \draw[black,internal] (0.2,0.25) -- (0,0);
}};
\node[] at (-2.8,-1.2) {\tikz{
    \draw[darkred,internal] (0,-0.2) to[out=150,in=210] (0,0.2);
    \draw[olddarkgreen,internal] (0,-0.2) to[out=30,in=-30] (0,0.2);
}};
\node[] at (-2.8,0.5) {\tikz{
    \draw[darkred,internal] (-0.2,0.2) -- (0.2,0.2);
    \draw[olddarkgreen,internal] (0.2,-0.2) -- (0.2,0.2);
}};
\node[] at (-0.9,2.0) {\tikz{
    \draw[darkorange,internal] (-0.2,-0.2) -- (0.2,-0.2);
    \draw[newdarkblue2,internal] (-0.2,-0.2) -- (-0.2,0.2);
}};
\node[] at (1.2,2.0) {\tikz{
    \draw[newdarkblue2,internal] (-0.2,0) to[out=60,in=120] (0.2,0);
    \draw[darkorange,internal] (-0.2,0) to[out=-60,in=-120] (0.2,0);
}};
\node[] at (-2.8,2.0) {\tikz{
    \draw[darkred,internal] (0.2,0.2) -- (0.2,-0.2);
    \draw[darkorange,internal] (0.2,-0.2) -- (-0.2,-0.2);
    \draw[newdarkblue2,internal] (-0.2,-0.2) -- (-0.2,0.2);
    \draw[olddarkgreen,internal] (-0.2,0.2) -- (0.2,0.2);
}};
\draw [->>] (-0.2,0.5) -- (0.4,0.5);
\draw [->>] (1.2,0) -- (1.2,-0.8);
\draw [->>] (-0.9,0) -- (-0.9,-0.8);
\draw [->>] (-0.2,-1.2) -- (0.4,-1.2);
\draw [->>] (-0.2,0) -- (0.8,-0.8);
\draw [>->]  (1.2,1.6) -- (1.2,0.8);
\draw [>->]  (-0.9,1.6) -- (-0.9,0.8);
\draw [>->] (-2.4,0.5) -- (-1.8,0.5);
\draw [>->] (-2.4,-1.2) -- (-1.8,-1.2);
\draw [>->] (-2.2,1.7) -- (-1.4,1.0);
\end{tikzpicture}
}
\caption{Pham diagram showing how incompatibility of the contractions gives stronger constraints than the Steinmann relations. In this case, the two contractions are not partially overlapping, the sequential discontinuity still vanishes by Theorem~\ref{thm:co2vanish}.
\label{fig:p1p2phamdiagram}
}
\end{figure}
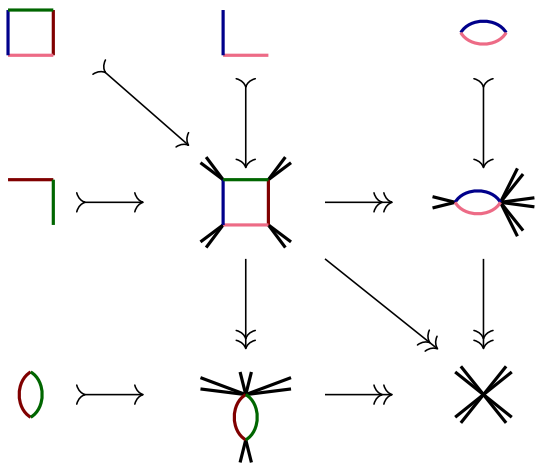

To see an example in which Theorem~\ref{thm:co2vanish} goes beyond the Steinmann relations, let us again consider the box diagram. In general, this diagram can be taken to describe a high-point scattering process with many different particles incoming or outgoing at each vertex, which implies that the kinematic variables $s$, $t$, $p_1^2$, $p_2^2$, $p_3^2$, and $p_4^2$ are all independent, where $p_i^2$ is the momentum flowing in to vertex $i$. In Fig.~\ref{fig:p1p2phamdiagram}, we draw the Pham diagram that describes the computation of sequential discontinuities around $p_1^2=4 m^2$ and $p_3^2=4 m^2$. If we take the process being described to be $4\to4$, the quadruple cut through the box diagram can be realized in the physical region, which is to say with physical on-shell momenta, if two of the corner Mandelstam invariants, say $p_2^2$ and $p_4^2$ are negative. Since the kernels in this Pham diagram are not equal, we get a new restriction
\begin{equation}
    \big(\bbone - \monM_{ p_3^2 - 4 m^2 } \big) \big(\bbone - \monM_{ p_1^2 - 4 m^2 } \big) I_\squaretcol = 0 \,.
    \label{eq:beyond_steinmann_box}
\end{equation}
The Pham relations therefore show that the monodromy around $p_1^2-4m^2=0$ followed by the one around $p_3^2-4m^2=0$ is zero, despite these channels not being overlapping.

It is instructive to compare these results with the formula for the Mandelstam double dispersion relation in Ref.~\cite{mandelstam1959analytic}, in which it is shown that the sequential discontinuity in $s$ and $t$ of the scalar box diagram $\I_\square$ is nonzero. This is not in any contradiction with the result presented here for two reasons. First, as discussed in Ref.~\cite{mandelstam1959analytic}, the support of the double-spectral density $\rho(s,t) \equiv \disc_t \disc_s \I_\square$ is outside of the physical region, so that analytic continuation of $\disc_s \I_\square$ is required to get to the non-vanishing region of $\rho(s,t)$. Second, the second discontinuity in $t$ has support only above the threshold for the box singularity. So the sequential monodromy corresponding to $\rho(s,t)$ is really the one taken first around the bubble singularity at $s=4 m^2$, followed by the one around the box singularity at $4 s t\big[s t-4 m^2(s+t)+12 m^2\big]=0$, where we have assumed that both internal and external masses are equal to $m$, for simplicity.
This sequential monodromy around the box threshold is in agreement with the predictions by Pham, and Eq.~\eqref{eq:steinmann_box} is not violated, even outside of its proven validity in the physical region. A recent discussion about the validity and extension of the double spectral function can be found in  Refs.~\cite{Correia:2020xtr,Correia:2021etg}.

\subsection{Discussion}
\label{sec:co2discussion}
In this section (and in Sec.~\ref{sec:sequential_disc}), we have made a number of technical assumptions in order to be able to make rigorous statements about when sequential discontinuities vanish. However, one can certainly consider situations in which Pham loci are not principal, masses are not generic, singularities are not in the physical region, or singular loci overlap. Indeed, in most theories of physical or mathematical interest such as QCD or supersymmetric Yang-Mills theory, the assumptions we have made for Theorems~\ref{thm:co1},~\ref{thm:commuting}, and~\ref{thm:co2vanish} are violated. However, one can still extract some general lessons from these results, that will allow us to approach examples in which no general theorems yet exist. 

In this respect, the main takeaway from this section is that when two contractions are compatible the associated singularities factorize, meaning that both sets of Landau equations can be simultaneously satisfied without increasing the codimension of the individual solutions. This property is key to understanding the nested singularity structure of Feynman integrals, or, more ambitiously, the $S$-matrix. Note that it is possible for the Landau equations to be compatible in this way even if the contractions associated with a pair of singularities are not compatible, when considering non-generic external kinematics. In such situations, sequential discontinuities can still be nonzero. We next turn to some more detailed considerations of such situations that go beyond the theorems discussed so far.

  \section{Generalizations}
\label{sec:exceptional_kinematics}

In this paper, we have mostly focused on singularities that occur in the physical region, and on studying integrals in generic kinematics.  In this section, we peek beyond these restrictions. To do this, we leverage the intuition we have gained by thinking about Landau singularities as critical values in order to understand how sequential discontinuities can newly appear for restricted kinematics. For example, the projection along the $z$ axis from a two-sphere has critical points at the North and South poles.  But if we restrict the domain by cutting the sphere with a plane, we can see that it is easy to obtain new critical points.  This information is contained implicitly in the Landau equations, but it is not geometrically obvious.

\begin{figure}[t]
\centering
\resizebox{7cm}{!}{
\tdplotsetmaincoords{70}{110}
\begin{tikzpicture}[tdplot_main_coords]
\begin{scope}[tdplot_main_coords]
\node[above,black,scale=1,newdarkblue2] at (5.3,2.2,0) {$\P_{\kappap}$};
\node[above,black,scale=1,darkred] at (3,4.8) {$\P_{\kappapp}$};
\foreach \x in {-1,-0.9,...,4}
 {\draw[newdarkblue2!10] plot[variable=\Z,domain=0:1,samples=73,smooth] 
 (\x,2,{(1-(1/20)*\x*\x+(1/20)*\x*\x*\x)*\Z});}
 \foreach \x in {-1,-0.9,...,4}
 {\draw[darkred!10] plot[variable=\Z,domain=0:1,samples=73,smooth] 
 (1.5,\x,{(1-(1/20)*\x*\x+(1/20)*\x*\x*\x)*\Z});}
  \draw[line width=1,-stealth] (-1,0,0) -- (4,0,0) node[anchor=north east]{$p_1$};
\draw[line width=1,-stealth] (0,-1,0) -- (0,4,0) node[anchor=north west]{$p_2$};
\draw[line width=1,-stealth] (0,0,-1) -- (0,0,3) node[anchor=south]{$k$};
\draw[line width=1,darkred] plot[variable=\x,domain=-1:4,samples=73,smooth] 
 (1.5,\x,{1-(1/20)*\x*\x+(1/20)*\x*\x*\x});
 \draw[line width=1,darkred,dashed] plot[variable=\x,domain=-1:4,samples=73,smooth] 
 (1.5,\x,0);
  \draw[line width=1,newdarkblue2] plot[variable=\x,domain=-1:4,samples=73,smooth] 
 (\x,2,{1-(1/20)*\x*\x+(1/20)*\x*\x*\x});
  \draw[line width=1,newdarkblue2,dashed] plot[variable=\x,domain=-1:4,samples=73,smooth] 
 (\x,2,0);
\end{scope} 
\end{tikzpicture}
}
\caption{While the transversally-intersecting Pham loci $\P_{\kappap}$ and $\P_{\kappapp}$ do not lead to compatible solutions of the Landau equations for generic masses (as illustrated in Fig.~\ref{fig:Landauintersection}), these configurations can become compatible in special kinematic configurations.}
    \label{fig:Landauintersection2}
\end{figure}
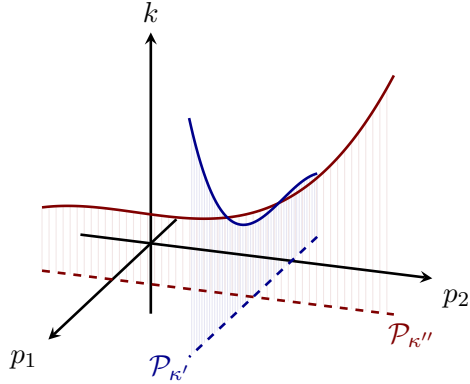

There are many motivations for studying special kinematics. These include:
\begin{itemize}
\item Calculations often simplify in special kinematics, and the value of Feynman integrals and amplitudes in special kinematics can sometimes be used to bootstrap these quantities~\cite{Caron-Huot:2020bkp}.
\item Mathematical features that don't arise in generic kinematics until higher loop orders, such as effective (transversal) intersections of singularity loci, can be simulated and studied in special kinematic configurations already at one loop.
\item  One can explore regions which are not physical, such as $(2,2)$ spacetime signature, which has been useful for understanding twistor constructions and in which massless three-point vertices can be naturally accommodated. 
\item Feynman integrals that evaluate to special functions that go beyond multiple polylogarithms often simplify to polylogarithmic expressions in special kinematics. The techniques developed in this paper for studying Landau singularities may make it possible to systematically find kinematic limits in which these simplifications occur.  For example, the two-loop sunrise integral, which is generically elliptic, reduces to multiple polylogarithms at the pseudo-threshold~\cite{Bloch:2013tra}.
\end{itemize}
We will not explore examples of all these possibilities here. Rather, we focus on exploring the key idea from Sec.~\ref{sec:landaucompatibility}, that sequential discontinuities can only occur when the Landau equations that encode two singularities are compatible. In particular, we will consider two examples in which new double discontinuities appear in special kinematics where the Landau equations that encode these singularities become compatible, as shown in Fig.~\ref{fig:Landauintersection2}, despite the fact that the incompatibility of the corresponding contractions restrict these discontinuities from appearing in generic kinematics.

\subsection{Special Mass Configurations}
\label{sec:special_internal_masses}

In Sec.~\ref{sec:disallowed_discontinuities}, we saw that the double discontinuity of the box integral \begin{equation}
 I_{\squaretcol} =
\resizebox{!}{1.3cm}{
\begin{tikzpicture}[baseline= {($(current bounding box.base)-(2pt,2pt)$)},
    line width=1.5,scale=0.3]
    \draw[black] (-4,-4)   -- (-2,-2)  node[midway,below] {$~~~p_1$};
    \draw[black] (-4,4) -- (-2,2) node[midway,above] {$~~~p_2$}; 
    \draw[black] (4,4) -- (2,2) node[midway,above] {$p_3~~$};
    \draw[black] (4,-4) -- (2,-2)  node[midway,below] {$p_4~~$};
    \draw[black,-latex reversed] (-2,-2) -- ++(-135:1.5);
    \draw[black,-latex reversed] (-2,2) -- ++(135:1.5);
    \draw[black,-latex reversed] (2,-2) -- ++(-45:1.5);
    \draw[black,-latex reversed] (2,2) -- ++(45:1.5);
    \draw[darkred] (2,2) -- (2,-2)  node[midway,right] {$q_4$};
    \draw[darkorange] (2,-2) -- (-2,-2)   node[midway,below] {$q_1$};
    \draw[newdarkblue2] (-2,-2) -- (-2,2)   node[midway,left] {$q_2$};
    \draw[olddarkgreen] (-2,2) -- (2,2)   node[midway,above] {$q_3$};
    \draw[darkred,-latex] (2,2) -- ++(-90:2.5);
    \draw[darkorange,-latex] (2,-2) -- ++(-180:2.5);
    \draw[newdarkblue2,-latex] (-2,-2) -- ++(90:2.5);
    \draw[olddarkgreen,-latex] (-2,2) -- ++(0:2.5);
\end{tikzpicture}
}
\label{Ibox1}
\end{equation}
with respect to the $s$-channel and $t$-channel singularities vanishes. In terms of the momentum labels in Eq.~\eqref{Ibox1}, the $s$-channel singularity occurs when
\begin{equation}
    s=(m_2+m_4)^2, \qquad q_2^\mu = - \frac{\alpha_4}{\alpha_2} q_4^\mu = - \alpha_4 (p_2+p_3)^\mu \, \qquad \alpha_2 = \frac{m_4}{m_2} \alpha_4 = \frac{m_4}{m_2+m_4} \,,
    \label{eq:sbub}
\end{equation}
while the $t$-channel singularity occurs when
\begin{equation}    t=(m_1+m_3)^2, \qquad q_1^\mu = - \frac{\alpha_3}{\alpha_1} q_3^\mu = - \alpha_3 (p_1+p_2)^\mu \, \qquad \alpha_1 = \frac{m_3}{m_1} \alpha_3 = \frac{m_3}{m_1+m_3} \,.
    \label{eq:tbub}
\end{equation}
Although these two loci intersect in the space of \textit{external} momenta, they do not intersect in the larger space of internal and external momenta---the vanishing cell for one does not intersect the integration contour for the other, implying that there is no pinch singularity corresponding to the second discontinuity. However, while this situation holds for generic masses, we can ask if there exists a choice of internal masses for which these solutions to the Landau equations become compatible and the double discontinuity becomes nonzero.

Momentum conservation tells us that $q_1^\mu = -p_1^\mu + q_2^\mu$. Thus, if we are to be on the support the $s$-channel and $t$-channel singularities simultaneously, Eqns.~\eqref{eq:sbub} and~\eqref{eq:tbub} tell us that we must have
\begin{equation}
    - \alpha_3 (p_1+p_2)^\mu = - p_1^\mu - \alpha_4 (p_2+p_3)^\mu \, .
    \label{eq:boxconstr}
\end{equation}
When this equation is satisfied, the $s$-channel and $t$-channel loci 
lead to the same values for the loop momenta, and thus become compatible. 

Although Eq.~\eqref{eq:boxconstr} is strongly constraining, it can be solved for special kinematics. For example, consider the following assignment of masses:
\begin{equation}
\I_{\square} =    \begin{tikzpicture}[baseline= {($(current bounding box.base)-(2pt,2pt)$)},
    line width=1.3,scale=0.5]
    \draw[etacol] (0,0) -- (-1.7,0)  node[left] {$p_1$}
    node[midway,above] {$4m$};
    \draw[darkred] (2,-2) -- (0,0) node[midway,below] {$q_1~~~$} node[midway,above] {$~~~2m$}-- (2,2) node[midway,above]  {$q_2~~~$} node[midway,below] {$~~~2m$};
    \draw[newdarkblue2] (2,2) -- (4,0) node[midway,above] {$~~~q_3$} node[midway,below] {$m~~~$}  -- (2,-2) node[midway,below] {$~~~q_4$} node[midway,above] {$m~~~$};
    \draw[darkred] (4,0) -- (5,0) node[midway,above] {$2m$} node[right] {$p_3$};
    \draw[newdarkblue2] (2,2) -- (4,2.5) node[midway,above] {$~~~m$} node[right] {$p_2$};
    \draw[newdarkblue2] (2,-2) -- (4,-2.5) node[midway,below] {$~~~m$} node[right] {$p_4$};
 \draw [decorate,decoration={brace,amplitude=10pt,raise=4pt}]
 (4.5,3) -- (6,0);
 \draw [decorate,decoration={brace,amplitude=10pt,raise=4pt}]
 (6,0) -- (4.5,-3);
 \node[] at (6.5,2.1) {$s$};
 \node[] at (6.5,-2.1) {$t$};
    \draw[etacol,-latex] (-1.7,0) -- ++(0:1.2);
    \draw[darkred,-latex] (0,0) -- ++(45:1.8);
    \draw[darkred,-latex reversed] (0,0) -- ++(-45:1.5);
    \draw[newdarkblue2,-latex] (2,2) -- ++(-45:1.8);
    \draw[newdarkblue2,-latex reversed] (2,-2) -- ++(45:1.5);
    \draw[newdarkblue2,-latex] (2,2) -- ++(15:1.5);
    \draw[newdarkblue2,-latex] (2,-2) -- ++(-15:1.5);
    \draw[darkred,-latex reversed] (4,0) -- ++(0:0.5);
    \end{tikzpicture}
    \label{steinmanbox}
\end{equation}
We have kept the arrows in this diagram the same as in Eq.~\eqref{Ibox1} to be consistent, but keep in mind that any left-pointing arrow should have negative energy to be physical. This choice of masses corresponds to a singular configuration where all the vertices are allowed on shell, as can be seen in the center-of-mass frame of $p_1^\mu$. In particular, the bubble singularity in the $s$-channel has
the following $\alpha$-positive solution to its Landau equations:
\begin{equation}
    s  = 9m^2, \qquad \alpha_2 =\frac{1}{3},\qquad \alpha_4  = \frac{2}{3}.
    \label{eq:landau_Steinmann_s}
\end{equation}
In the rest frame of $p_1^\mu$, this is the normal threshold for producing the particles associated with the momenta $q_2^\mu$, $q_4^\mu$, and $p_4^\mu$, all at rest. The $t$-channel bubble is similar:
\begin{equation}
    t  = 9 m^2, \qquad \alpha_1 =\frac{1}{3},\qquad \alpha_3 = \frac{2}{3}.
    \label{eq:landau_Steinmann_t}
\end{equation}
Again, in the $p_1^\mu$ rest frame this is a normal threshold, now for producing the particles associated with the momenta $p_2^\mu$, $q_1^\mu$, and $q_3^\mu$, all at rest. At the intersection of these solutions, all momenta in the diagram are at rest.  To check that Eq.~\eqref{eq:boxconstr} is satisfied when both Eq.~\eqref{eq:landau_Steinmann_s} and~\eqref{eq:landau_Steinmann_t} are imposed, we note that in the rest frame it becomes simply
\begin{equation}
    -\frac{2}{3} (4m-m) = - 4m - \frac{2}{3}(-m-2m)\, ,
\end{equation}
which is true. Indeed, Eq.~\eqref{eq:boxconstr} simply came from demanding a consistent assignment of momenta to the two bubbles, and now we have assigned momenta consistently to the whole diagram.

At this point, we have found a special kinematic point where the Landau equations for the $s$-channel and $t$-channel bubble diagrams can be simultaneously satisfied. As a result, sequential discontinuities in these channels are not forbidden, despite the fact that the kernels of the corresponding contractions are incompatible. %
To see how these bubble singularities are related to the box and triangle singularities that are also part of the Pham locus for the contraction of the box to the elementary graph, we can compute the Gram matrix $q_i\cdot q_j$:
\begin{equation}
q_i \cdot q_j  =
\begin{pmatrix}
 4 m^2 & -4 m^2 & \frac{1}{2} \left(5 m^2-t\right) & 2 m^2 \\
 -4 m^2 & 4 m^2 & 2 m^2 & \frac{1}{2} \left(5 m^2-s\right) \\
 \frac{1}{2} \left(5 m^2-t\right) & 2 m^2 & m^2 & -m^2 \\
 2 m^2 & \frac{1}{2} \left(5 m^2-s\right) & -m^2 & m^2 \\
\end{pmatrix}
\end{equation}
At the threshold $s=t=9m^2$, we see that $\det q_i\cdot q_j=0$, which implies that the Landau equations for the box are satisfied. In addition, all the first minors of this matrix (determinants with one row and column removed) vanish at $s=t=9m^2$, indicating that the triangle singularities are also all coincident. So in this singular kinematic configuration, the box, triangles, and bubble singularities all occur at the same locus.

\subsection{Factorization in \texorpdfstring{$(2,2)$}{(2,2)} Signature}
\label{sec:22signature}

Another way that we can get around the constraints implied by Theorem~\ref{thm:co2vanish} is by going outside the physical region, for instance to kinematic regions that correspond to different spacetime signatures. 
While in general, the interpretation of discontinuities as absorption integrals does not hold in outside of the physical region, the monodromy is still defined and can be computed directly using the polylogarithmic form of this integral, or its symbol.

The symbol of $\I_{\squaretcol}$ for generic masses can be found in~\cite{Bourjaily:2019exo} (see also Refs.~\cite{aomoto1977,Davydychev:1997wa}). It can be written in a parsimonious form by adopting variables similar to those we used for the triangle integral in Eq.~\eqref{yijdef}, 
\begin{equation}
    z_{ij} = \frac{p_{ij}^2-m_i^2-m_j^2}{2 m_i m_j} \,,
    \label{eq:yijvars}
\end{equation}
where in $p_{ij}^2=(p_{i}+p_{i+1}+\cdots+p_{j-1})^2$ the cyclic ordering of external momenta is implicit. In this notation,
\begin{equation}
    z_{ii} = -1,\quad 
    z_{i,i+1} = \frac{p_i^2 - m_i^2 -m_j^2}{ 2m_i m_j},\quad 
    z_{13} = \frac{t-m_1^2-m_3^2}{2 m_1 m_3},\quad
    z_{24} =  \frac{s-m_2^2-m_4^2}{2 m_2 m_4},
\end{equation}
so the $s$-channel singularity occurs at $z_{24} = 1$ while the $t$-channel singularity occurs at $z_{13} = 1$.

While keeping all internal masses generic, we now consider values of the external momenta that satisfy $z_{i,i+1}=0$, which implies that $p_i^2 = m_i^2 - m_{i+1}^2$ for all four external momenta.
In such a kinematic configuration, the symbol of the box integral takes an especially simple form, which contains the following pair of terms:
\begin{equation}
  \frac {z_{13} + \sqrt{z_{13}^2 - 1}}{z_{13} - \sqrt{z_{13}^2 - 1}} \otimes \frac {z_{24} + \sqrt{z_{24}^2 - 1}}{z_{24} - \sqrt{z_{24}^2 - 1}} + (z_{13} \leftrightarrow z_{24})\, .
\end{equation}
These terms give precisely the (symbol of the) product of the $s$ and $t$ bubble integrals. 
It is therefore possible to compute a double discontinuity around these bubble thresholds, by analytically continuing around $z_{13}=1$ and $z_{24}=1$. However, while it is clear that one will get a nonzero result from this calculation, it is not clear what the correct interpretation of this double discontinuity ought to be; similar to what we observed in the example in Sec.~\ref{sec:special_internal_masses}, the two bubble thresholds, the four triangle thresholds, and the box threshold all coincide when $z_{13}=z_{24}=1$. This can again be deduced by computing the determinant of the Gram matrix
\begin{equation}
-\frac{q_i \cdot q_j}{m_i m_j} =
    \begin{pmatrix}
    -1 & 0 & z_{13} & 0 \\
    0 & -1 & 0 & z_{24} \\
    z_{13} & 0 & -1 & 0 \\
    0 & z_{24} & 0 & -1
    \end{pmatrix},
\end{equation}
as well as its minors that encode the bubble and triangle singularities. 

\begin{figure}
    \centering
    \resizebox{12cm}{!}{
\begin{tikzpicture}
 \node (image) at (0,0.5) {    
 \includegraphics[scale=1]{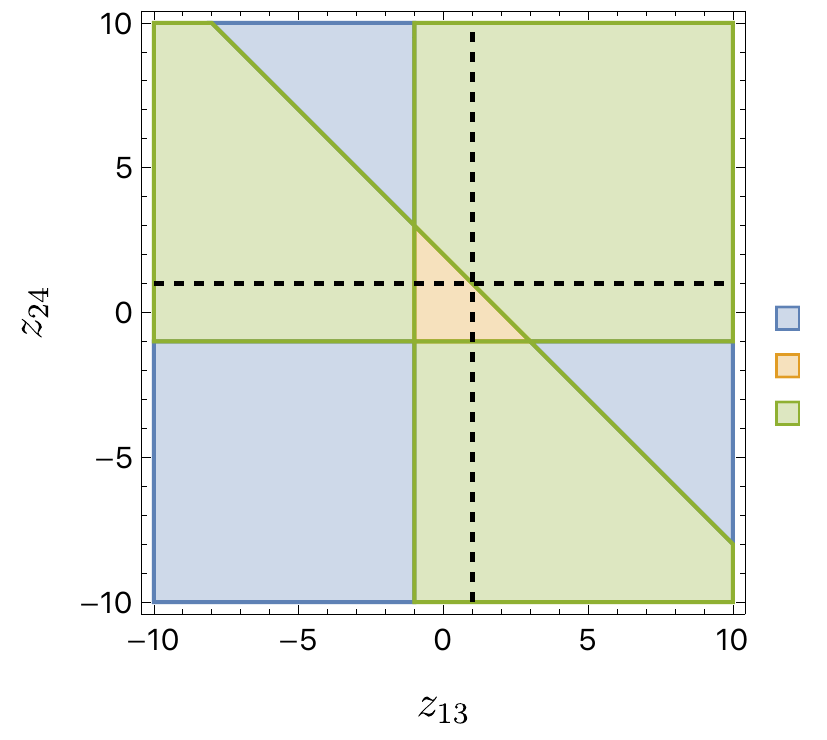}
 };
 \node[black,right] at (4,1) {$(1,3)$ signature};
 \node[black,right] at (4,0.5) {$(4,0)$ signature};
 \node[black,right] at (4,0) {$(2,2)$ signature};
 \end{tikzpicture}
   } 
   \caption{Parameter space for the box with restricted kinematics $z_{i,i+1}=0$ and $m_i=1$. Dashed lines are the $\alpha$-positive branch points at $z_{13}=1$ and $z_{24}=1$.
    The physical region (real momenta) is shown for Lorentzian (1,3) signature 
    as the blue region, which does not include the dashed lines. The intersection of the two singularities at $z_{24}=z_{13}=1$ is on the boundary of the Euclidean region and the $(2,2)$-signature region. The region above the threshold ($z_{13}>1,z_{24}>1$) is in the $(2,2)$ region.
    }
    \label{fig:zregion}
\end{figure}
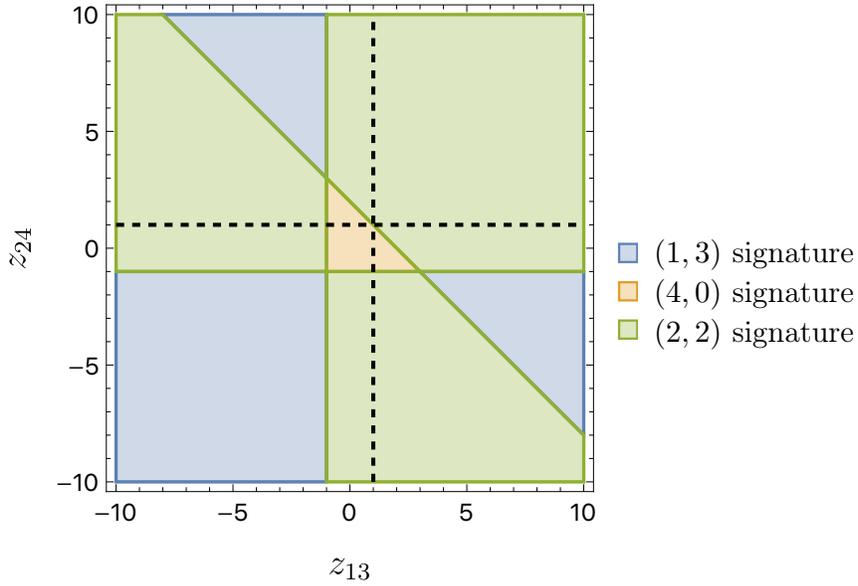

It is also worth asking what spacetime signature the double discontinuity around $z_{13} =1$ and $z_{24} = 1$ can be accessed. Recall that in the physical region, all external momenta are real in (1,3) signature. These reality conditions can be written in terms of Gram matrices constructed out of external momenta. When $z_{i,i+1}=0$, the matrix $p_i \cdot p_j$ takes the form
\begin{equation}
\left(\begin{smallmatrix}
 m_1^2+m_2^2 & m_1 m_3 z_{13}-m_2^2 & -m_1 m_3 z_{13}-m_2 m_4 z_{24} & m_2 m_4 z_{24}-m_1^2 \\
 m_1 m_3 z_{13}-m_2^2 & m_2^2+m_3^2 & m_2 m_4 z_{24}-m_3^2 & -m_1 m_3 z_{13}-m_2 m_4 z_{24} \\
 -m_1 m_3 z_{13}-m_2 m_4 z_{24} & m_2 m_4 z_{24}-m_3^2 & m_3^2+m_4^2 & m_1 m_3 z_{13}-m_4^2 \\
 m_2 m_4 z_{24}-m_1^2 & -m_1 m_3 z_{13}-m_2 m_4 z_{24} & m_1 m_3 z_{13}-m_4^2 & m_1^2+m_4^2
 \end{smallmatrix}\right),
\label{eq:external_gram}
\end{equation} 
and being in the physical region requires that the determinant of the first $2\times 2$ block of this matrix is negative, while the determinant of the first $3\times 3$ is positive~\cite{ELOP}. The physical region is shown in Fig.~\ref{fig:zregion} for the simplified case in which all masses $m_i=1$. There, we see that there is no physical solution for either $z_{13} = 1$ or $z_{24}=1$ in Lorentzian signature, much less at their intersection.

Instead, we can consider this integral in (2,2) signature, which we can also think of as (1,3) signature for a configuration of external momenta that all have a purely imaginary $x$ component. Such a configuration can be achieved when the determinant of the first $2 \times 2$ block of Eq.~\eqref{eq:external_gram} is positive, while the determinant of its first $3 \times 3$ block is negative~\cite{ELOP}. As seen in Fig.~\ref{fig:zregion}, the region above the $s$-channel and $t$-channel thresholds, where $z_{13}>1$ and $z_{24}>1$, corresponds to (2,2) signature, while the intersection of the thresholds $z_{13}=1$ and $z_{24}=1$ occurs at the boundary of this region.

To work out the kinematics explicitly, we pick pairs of light-cone coordinates $X = (X^+, X^-, X^\oplus, X^\ominus)$ with the scalar product
\begin{equation}
    X \cdot Y = \frac 1 2 (X^+ Y^- + X^- Y^+ + X^\oplus Y^\ominus + X^\ominus Y^\oplus).
\end{equation}
Then, on support of $z_{i,i+1}=0$ for all $i$, we can take
\begin{align}
    x_1 &= m_1 (1, 1, 0, 0), \\
    x_3 &= m_3 (x_3^+, (x_3^+)^{-1}, 0, 0), \\
    x_2 &= m_2 (0, 0, 1, 1), \\
    x_4 &= m_4 (0, 0, x_4^\oplus, (x_4^\oplus)^{-1}),
\end{align}
where $x_i$ are dual coordinates to $p_i$, such that $p_{ij}^\mu=(x_j-x_i)^\mu$, and the loop-momentum components are given by $q_i^\mu=(x_i-x_0)^\mu$.
 For these dual coordinates to be consistent with the definition of $z_{ij}$ in Eq.~\eqref{eq:yijvars} %
we need
\begin{gather}
    x_3^+ + \frac{1}{x_3^+} = -2 z_{13}, \\
    x_4^\oplus + \frac{1}{x_4^\oplus} = -2 z_{24}.
\end{gather}
In fact, we can see that we associate the \emph{same} coordinates $z_{13}$ and $z_{24}$ to the configurations in which $x_3^+ \leftrightarrow (x_3^+)^{-1}$ or $x_4^\oplus \leftrightarrow (x_4^\oplus)^{-1}$, which are related by parity.  Which one of them is chosen depends on the choice of square root for $\sqrt{z_{13}^2 - 1}$ and $\sqrt{z_{24}^2 - 1}$.  In this sense, working with the \emph{components} of the momenta naturally realizes the double-cover of all the square roots. We also see that the $\alpha$-positive bubble singularities correspond to taking $x_3^+ = -1$ and $x_4^\oplus = -1$.

To solve for the on-shell space for the box, we need to solve the on-shell equations $(x_i-x_0)^2 - m_i^2 = 0$ for the coordinates $x_i$ described above.  These equations can be rewritten as
\begin{gather}
    (x_1-x_0)^2 = m_1^2, \qquad
    x_0 \cdot (x_2 - x_1) = x_0 \cdot (x_3 - x_1) = x_0 \cdot (x_4 - x_1) = 0.
\end{gather}
One of the solutions is $x_0 = (0, 0, 0, 0)$ while the other is more complicated.  At the intersection of the two bubble Pham loci, where $x_3^+ = -1$ and $x_4^\oplus = -1$, one can check that the two solutions for $x_0$ coincide.

For these values of the $x_i$ variables, we get
\begin{align}
    p_1^\mu & = (-m_1,-m_1,m_2,m_2) \,, \qquad 
    p_2^\mu = (-m_3,-m_3,-m_2,-m_2) \,, \\ 
    p_3^\mu & = (m_3,m_3,-m_4,-m_4) \,,  \qquad 
    p_4^\mu = (m_1,m_1,m_4,m_4) \,.
\end{align}
Plugging these external momenta into Eq.~\eqref{eq:boxconstr}, we see
that 
the two Pham loci indeed give rise to the same values of the loop momentum. So, at the codimension-two point where $z_{13} = 1$ and $z_{24} = 1$, none of the $\alpha$ need vanish; any linear combination of the values of $\alpha_i$ determined by each bubble singularity solves the Landau equations.

  \section{Summary of Results}
\label{sec:summary}

This is a long paper, and the results are interleaved with explanations and examples. We therefore include here a more concise summary of the definitions and main results.

We began by considering generic Feynman integrals, which take the form
\begin{equation}
    \I_G (p) = \int_h \prod_{c \in \Chat(G)} \rd^d k_c \prod_{e \in \Eint(G)}
     \frac{1}{s_e} \,,
\end{equation}
where $s_e = q_e^2 -m_e^2$ and the $+i\varepsilon$'s in the propagators have been transferred to the integration contour $h$. To study the singularities of $I_G$, we considered graphs $G^\kappa$ to which the original graph $G$ could be contracted, via
sequences such as
\begin{equation}
 G\xrightarrowdbl{~\kappas~} G^\kappa \xrightarrowdbl{~\kappa~} G_0 \, .
\end{equation}
When one can write a sequence like this, we say that the composed contraction $\kappa \circ \kappas$ \textbf{dominates} the contraction $\kappa$. To each such contraction, we define a Pham locus $\P_\kappa$ that corresponds to the set of external momenta that satisfy the equations for some $\alpha_e$ that are not all vanishing:
\begin{equation}
\P_\kappa  : \qquad  p^\mu \in \S(G_0) \, \Big|
    \bigcap_{c \in \Chat(G^\kappa)} 
\!\!\Big\{ \sum_{e \in \Eint(G^\kappa)} b_{c e} \, \alpha_e q_e^\mu =0 \Big\}
    \bigcap_{e\in \Eint(G^\kappa)} \!\!
\Big\{ q_e^2 = m_e^2\Big\} .
\end{equation}
Note that these Pham loci correspond to Landau diagrams in which all lines are on shell, but the $\alpha_e$ in $G^\kappa$ are not all required to be non-zero. For instance, while in most of the examples we considered the $\alpha_e$ for all the edges in $G^\kappa$ are non-zero, we also worked through some examples in which some of these $\alpha_e$ are set to zero, such as the bubble-like singularity of the ice cream cone.

Landau showed (in Pham's language) that the singularities of $I_G(p)$ occur on the support of the Pham loci associated with diagrams to which $G$ can be contracted:
\begin{theorem*}[Landau] 
The Feynman integral $I_G(p)$ is analytic outside of the Pham loci $\P_\kappa$ of contractions $\kappa: G^\kappa \twoheadrightarrow G_0$ of diagrams for which there exists a contraction $\kappas: G \twoheadrightarrow G^\kappa$.
\end{theorem*}
\noindent
One can compute the discontinuities of $\I_G(p)$ with respect to the singularities associated with these Pham loci. These discontinuities are especially well understood in the case of principal Pham loci:
\begin{definition*}
A \textbf{principal} Pham locus is (complex) codimension one in the space of external kinematics, $\alpha$-positive, and of type $S_1^+$. 
\end{definition*}
\noindent
Here, being $\alpha$-positive means that the values of all the $\alpha_e$ that occur in the relevant solution to the Landau equations are non-negative. If they are all strictly positive, this is called the \textbf{leading} singularity of the Pham locus. Being a singularity of type $S_1^+$ means that all of the eigenvalues of the Hessian of the projection map $\pi_\kappa$ are positive (see Fig.~\ref{fig:Storus}). 

In the cases where the Pham loci are principal (which is true of many of the singularities of $I_G(p)$ in the physical region on the physical sheet), we can apply the Picard--Lefschetz theorem to derive Cutkosky's form of the absorption integral:
\begin{theorem*}[Cutkosky] 
The discontinuity 
$\cA_G^{\kappa} (q) = \big(\bbone - 
\mathscr{M}_{\P_\kappa} \big) \I_G(p)$  around a principal Pham locus $\P_\kappa$ is given by the absorption integral in Eq.~\eqref{eq:absorption_int_2}:
\begin{equation}
       \cA_{G}^{\kappa} (p)
       =
       \Big(\bbone - 
  \monM_{\P_\kappa} \Big) I_G(p)
  = \int_h \prod_{c \in \Chat(G)} \rd^d k_c
          \prod_{e \in E(\ker \kappa)}  (-2\pi i)\; \theta_\ast(q_e^0)\delta(s_e)
          \prod_{e^\prime \not\in E(\ker \kappa)}
         \frac{1}{s_{e'}}
\end{equation}
\end{theorem*}
\noindent where $\theta_\ast(q_e^0)$ picks out the sign of the energies of the cut lines determined by the solution of the Landau equations at the singular point.
In Sec.~\ref{sec:PicardLefschetz}, we explained how
Cutkosky's formula can be derived using the Picard--Lefschetz theorem and Leray's  multivariate residue calculus. Doing so gives a more flexible form of the absorption integral.
Labeling the edges in $\ker \kappa$ by $e=1,\ldots, m$, the formula is
\begin{equation}
    \cA_{G}^{\kappa} (p) = \int_{\partial_1 \cdots \partial_m e_\kappa}
          \frac{\bigwedge_{c \in \Chat(G)} \rd^d k_c}{\rd s_1 \wedge \cdots \wedge \rd s_m} 
          \prod_{e^\prime \not\in E(\ker \kappa)}
         \frac{1}{s_{e'}}.
\end{equation}
\noindent
Here, $e_\kappa$ is the \textbf{vanishing cell} defined by the conditions $s_e >0$ for all $e \in E(\ker \kappa)$ and $\partial_1 \cdots \partial_m e_\kappa$ is the \textbf{vanishing sphere}. The vanishing sphere comprises the singularities that get pinched at the branch point. The singularities of absorption integrals can also be characterized in terms of Pham loci associated with dominating contractions:
\begin{theorem*}[Pham; Thm.~\ref{thm:pham}] 
The absorption integral $\cA^{\kappa}_{G}(p)$ associated with the contraction $\kappa$ and the sequence $G \twoheadrightarrow G^\kappa \twoheadrightarrow G_0$ is analytic everywhere for $p$ in physical regions, outside of the Pham loci $\P_\kappap$ of contractions $\kappap$ that dominate $\kappa$ through sequences $G \twoheadrightarrow G^\kappap \twoheadrightarrow G^\kappa \twoheadrightarrow G_0$.
\end{theorem*}
\noindent Note that this does not exclude the case in which $\kappap$ is the trivial contraction. The discontinuities of absorption integrals can also be computed using Picard--Lefschetz theorem, as we illustrated in Sec.~\ref{sec:iterated}.

A powerful way of understanding Pham loci is in terms of the critical points of maps between on-shell spaces:
\begin{lemma*}[Pham; Lemma~\ref{lem:crit}] 
Critical values of the projection map $\pi_\kappa \colon \S(G^\kappa) \to \S(G_0)$ constitute the Pham locus $\P_\kappa$.
\end{lemma*}
\noindent
Here $\S(G^\kappa)$ is the space of on-shell internal and external momenta of $G^\kappa$, and $\S(G_0)$ is the space of on-shell external momenta. Thus, by studying the geometric properties of Pham loci in the space of on-shell momenta, we can derive constraints on the discontinuities of Feynman integrals.

We have presented a number of important relations between the sequential discontinuities of Feynman integrals. 
The first puts constraints on hierarchical pairs of singularities:
\begin{theorem*}[Pham; Thm.~\ref{thm:co1}]
For a series of contractions $G \twoheadrightarrow G^\kappap \twoheadrightarrow \cdots \twoheadrightarrow G^\kappa \twoheadrightarrow G_0$ the relation 
\begin{equation}
\Big(\bbone - 
  \monM_{\P_\kappap} \Big)\cdots \Big(\bbone - \monM_{\P_\kappa} \Big) \I_G(p) 
  = \Big(\bbone -   \monM_{\P_\kappap} \Big) \I_G(p)
\end{equation}
\noindent
holds when $\P_\kappa \cdots \P_\kappap$ correspond to principal Pham loci, and $p$ is in the physical region. 
\end{theorem*}
\noindent A more general statement also holds for tangentially-intersection Pham loci, which holds even if these loci are not principal or related hierarchically:
\begin{theorem*}[Pham; Thm.~\ref{thm:codim1gen}]
If two Pham loci $\P_\kappap$ and $\P_\kappapp$ are of codimension one and intersect tangentially, then their sequential monodromies, in a small neighborhood of their intersection and away from other singularities, satisfy:
\begin{equation}
   \monM_{\eta_+'} \circ \monM_{\eta_+} \I_G(p) 
   =
   \monM_{\eta_-'} \circ
   \monM_{\eta_-} \I_G(p)
   =
   \monM_{\eta_+} \circ
   \monM_{\eta_-'} \I_G(p)
   =
   \monM_{\eta_-} \circ
   \monM_{\eta_+'} \I_G(p)\, ,
\end{equation}
\end{theorem*}
\noindent
where the paths $\eta'_{\pm}$ and $\eta_{\pm}$ were defined in Sec.~\ref{sec:homotopyloops}.

If $\kappapp$ does not dominate $\kappap$, then there are also conditions on when double discontinuities with respect to $\P_\kappapp$ and $\P_\kappap$  can be nonzero. Namely, this is only possible if $\P_\kappapp$ is a subspace (not the leading singularity) of some larger Pham locus $\P_\kappa$ that itself dominates $\P_\kappap$. In this case, $\P_\kappa$ must be a codimension-two Pham locus so that the subspace corresponding to $\P_\kappapp$ can be codimension-one. Then
\begin{theorem*}[Pham; Thm.~\ref{thm:commuting}]
If two Pham loci $\P_\kappap$ and $\P_\kappapp$ intersect transversally, then their sequential discontinuities, in a small neighborhood of their intersection and away from other singularities, commute:
\begin{equation}
    \Big(\bbone - 
   \monM_{\P_\kappap} \Big) 
   \Big(\bbone - \monM_{\P_\kappapp}\Big) \I_G(p) 
   = \Big(\bbone - 
   \monM_{\P_\kappapp} \Big) \Big(\bbone - 
   \monM_{\P_\kappap} \Big) \I_G(p) \, .
\end{equation}
\end{theorem*}
\noindent
In this case we can draw the diagram of contractions
\begin{equation}
    \begin{tikzcd}
    G^\kappa \arrow[r, "\kappapps", two heads] \arrow[dr, "\kappa", two heads] \arrow[d, "\kappaps", two heads] & G^\kappap \arrow[d, "\kappap", two heads] \\
    G^{\kappapp} \arrow[r, "\kappapp", two heads] & G_0
    \end{tikzcd}.
\end{equation}
In addition, we defined two notions of compatibility. The first applies to contractions:
\begin{definition*}
The contractions $\kappap$ and $\kappapp$ are \textbf{compatible contractions} if
$\ker \kappap=\ker \kappaps$ and $\ker \kappapp=\ker \kappapps$, otherwise they are \textbf{incompatible contractions}. 
\end{definition*}
\noindent
In terms of this notion of compatibility, the final theorem we discussed is
\begin{theorem*}[Pham; Thm.~\ref{thm:co2vanish}]
Assume we have two principal Pham loci $\P_\kappap$ and $\P_\kappapp$ that intersect transversally. Then, for generic masses,  
\begin{equation}
    \Big(\bbone - 
   \monM_{\P_\kappap} \Big) 
   \Big(\bbone - \monM_{\P_\kappapp} \Big) \I_G(p) 
   = 0 \,,
\end{equation}
if $\ker \kappap$ and $\ker \kappapp$ are incompatible contractions, and $p$ is in the physical region. 
\end{theorem*}
\noindent
The Steinmann relations constitute an important subset of the consequences of this last theorem, but do not exhaust its content. The second notion of compatibility we defined applies to the Landau equations:
\begin{definition*}
A pair of codimension-one solutions $\P_\kappap$ and $\P_\kappapp$ to the Landau equations that correspond to contractions $\kappap$ and $\kappapp$, neither of which dominates the other, are considered to be {\bf compatible} when both solutions can be imposed simultaneously. 
\end{definition*}
\noindent
This notion of compatibility agrees with the compatibility of contractions for Feynman integrals with generic masses:
\begin{lemma*}[Lemma~\ref{lemma2}]
Suppose we have two contractions $\kappap$ and $\kappapp$
corresponding to a pair of codimension-one Pham loci $\P_\kappap$ and $\P_\kappapp$ such that neither contraction dominates the other. 
Then, for generic masses, the contractions are compatible if and only if the solutions to the Landau equations are compatible.
\end{lemma*}
\noindent
Geometrically, in the generic mass case, two principal Pham loci are compatible if they intersect transversely, as in Theorem~\ref{thm:commuting}. 

Lemma~\ref{lemma2} follows from another important result, that only one-vertex reducible Landau diagrams lead to potentially non-vanishing sequential discontinuities:
\begin{lemma*}[Lemma~\ref{lemma3}]
Suppose we have a pair of compatible contractions $\kappap$ and $\kappapp$ that describe two codimension-one Pham loci $\P_\kappap$ and $\P_\kappapp$. Then, for generic masses, the Landau diagram that describes the codimension-two Pham locus associated with the combined contraction $\kappap \circ \kappapp = \kappapp \circ \kappap$ factorizes into a pair of diagrams that separately describe the two loci $\P_\kappap$ and $\P_\kappapp$.
\end{lemma*}
\noindent
Beyond the generic-mass case, sequential discontinuities can be nonzero in situations where pairs of solutions to the Landau equations are compatible even when the corresponding contractions are not compatible. We explored this possibility in Sec.~\ref{sec:exceptional_kinematics}, where we found, among other things, that the symbol for the box diagram factorizes in a special-kinematic configuration where the sequential discontinuity becomes allowed. When internal masses are equal to $m$, and external masses to $M$, this configuration occurs when $M^2=2 m^2$. Since such factorization is characteristic of Landau diagrams for allowed sequential discontinuities according to Lemma~\ref{lemma3}, it would be interesting to explore the connection further.

  \section{Conclusion}
\label{sec:conclusion}

In this paper, we have presented a detailed analysis of the locations and properties of singularities in Feynman integrals, in particular focusing on when these singularities can give rise to sequential discontinuities.
In doing so, we have identified individual branches of the Landau variety with what we call Pham loci, which are specified by the set of propagators that are put on shell in the Landau equations while the Feynman parameters in these equations are allowed to take any value. This definition can be contrasted with what are sometimes called Landau loci, where branches are classified according to which Feynman parameters are required to be nonzero. Pham loci prove to be more useful for extending the application of Landau analysis from the study of the singularities of Feynman integrals to the study of the singularities of the discontinuities of Feynman integrals.

Our analysis has closely followed the work of Pham~\cite{pham}, who studied the analyticity properties of scattering amplitudes using homological methods. However, while Pham studied non-perturbative amplitudes, we have here applied these methods to individual Feynman integrals that involve generic kinematics (including generic masses) in perturbation theory. Our primary focus has been the study of Pham loci that are codimension-one in the space of external kinematics, as these identify the singularities that give rise to nontrivial monodromies. For Feynman integrals with generic masses, these codimension-one surfaces are described by solutions to the Landau equations in which the value of all Feynman parameters are determined up to an overall rescaling.

In studying the properties of Pham loci, we have emphasized the importance of understanding their geometric features in the space of on-shell kinematics. For instance, when we have two codimension-one Pham loci that are associated with the same elementary graph, it is valuable to know whether these loci intersect tangentially or transversally. This distinction helps unravel what types of relations hold among the discontinuities of Feynman integrals with respect to these singularities. More specifically, we can probe the properties of sequential discontinuities by analyzing the local fundamental group in the neighborhood of such intersections: the generators of these homotopy groups in general satisfy various relations, which imply similar relations among the monodromies of Feynman integrals. For instance, in the case of transversally-intersecting Pham loci, the monodromies around the two singularities commute (Theorem~\ref{thm:commuting}). In the case of tangential intersections, one can find relations among discontinuities that correspond to analytically continuing around different sides of the intersection point of the two singular surfaces (Theorem~\ref{thm:codim1gen}). We emphasize that these types of topological constraints follow from relations among generators of the local fundamental group, independently of other properties of Feynman integrals.

We have also derived additional relations that hold among the discontinuities of Feynman integrals in physical kinematics with respect to principal singularities, which are singularities of type $S_1^+$. Singularities that are encoded by one-loop Landau diagrams are of this type, as are all singularities that correspond to branches of Pham loci that involve only non-negative Feynman parameters and in which there are no on-shell loops that involve Feynman parameters that are all zero. Thus, while this requirement puts precise mathematical conditions on many of the results we have presented, these conditions are satisfied by many of the singularities of highest interest of Feynman integrals, for instance in the physical region.

One of the main results we have presented is a general set of restrictions on when pairs of principal singularities can give rise to sequential discontinuities in Feynman integrals. Using the Picard--Lefschetz theorem and the multivariate calculus of Leray and Pham~\cite{BSMF_1959__87__81_0,pham2011singularities}, one can compute a discontinuity by trading one's original integration contour for an infinitesimal contour that encircles the first Pham locus. According to Cutkosky's theorem, this localizes the propagators on loop-momentum values that are close to those of the corresponding solution to the Landau equations. Computing a second discontinuity then involves altering the contour again to encircle the second Pham locus, which tells us we must also localize on the other solution to the Landau equations. If the solutions for the two different branches lead to different constraints on the \textit{internal} loop momenta, the sequential discontinuity must vanish. Intuitively, this corresponds to the observation that while Pham loci may intersect in the space of external momenta, they might not intersect in the larger space of internal and external momenta. The constraints that follow from these considerations include---but also go well beyond---the Steinmann relations~\cite{Steinmann,Steinmann2}, which state that sequential discontinuities of amplitudes in partially-overlapping momentum channels must vanish in physical regions.

Non-principal singularities also appear in Feynman integrals. An example is given by the bubble-like singularity that arises in the ice-cream cone integral, where the two Feynman parameters in the bubble of the ice cream cone Landau diagram both vanish. An immediate difficulty that arises when encountering singularities of this type is the question of how one should construct the vanishing cells that enter the Picard--Lefschetz theorem; since the transverse Hessian is not definite, we cannot build them using the conditions that $s_e \geq 0$. It would be interesting to describe a general method for constructing these vanishing cells. More speculatively, it would also be interesting to study whether the presence of non-principal singularities can serve as some kind of diagnostic for the appearance of integrals that go beyond multiple polylogarithms, such as integrals over elliptic curves or Calabi--Yau manifolds~\cite{Bourjaily:2022bwx}.

While we have briefly illustrated how the ideas we have used in this paper can be applied to Feynman integrals in special kinematics or that do not involve generic masses, a great deal of further investigation is warranted to understand these generalizations. The study of these non-generic situations is what is required for understanding real-world theories, in which scattering generally involves particles that have equal masses as well as particles whose mass is zero. One of the complications that arises in massless theories is the presence of infrared divergences, whose homological study requires a careful analysis of the interplay between these divergences and the further singularities that arise in special kinematic limits. Notably, progress has already been made in this direction, as many of the relations and theorems presented here were already established at one loop in the context of dimensional regularization~\cite{Abreu:2017ptx}. However, we already know that further generalizations are likely to be nontrivial. For example, the Steinmann relations do not seem to hold for partially-overlapping two-particle momentum channels in massless integrals, and the so-called extended Steinmann relations~\cite{Caron-Huot:2019bsq} for massless integrals have been observed to generalize to the non-planar sector for some momentum channels and not others~\cite{Abreu:2021smk}. Developing a better understanding of these special cases is sure to provide us with new insight into the analytic structure of perturbative scattering amplitudes. 

We have mostly made use of the loop-momentum space formulation of Feynman integrals. Alternative formulations can be derived from the Symanzik form of Feynman integrals, in which the loop momenta have already been integrated out~\cite{Symanzik:1958}. The Landau equations for the Symanzik form of Feynman integrals are useful for algorithmically and/or numerically solving the Landau equations in terms of the Feynman parameters $\alpha_e$ (see for instance Refs.~\cite{Klausen:2021yrt,Mizera:2021icv,Correia:2021etg}). However, it is harder to apply Picard--Lefschetz theory to this Symanzik form of Feynman integrals, as the denominator is no longer a product of simple poles. 

By constraining the analytic structure of scattering amplitudes, we get incrementally closer to realizing the goals of the original $S$-matrix program. Indeed, much of the motivation for applying homological methods to scattering amplitudes in the nineteen-sixties came from the study of non-perturbative amplitudes for the strong interaction. The results we have presented in this paper also conjecturally apply to the nonperturbative $S$-matrix, given generous assumptions about its analyticity and singularity properties. For instance, the Landau equations can be interpreted non-perturbatively as identifying singularities that correspond to on-shell classical configurations of scattering particles, where each of the vertices in the corresponding Landau diagram should be understood to represent a non-perturbative $S$-matrix. Singularities that correspond to increasingly large Landau diagrams then arise from iteratively blowing up the $S$-matrix elements at the vertices of smaller Landau diagrams. This approach can be contrasted with the Feynman-diagrammatic perspective presented in this work, where propagators of diagrams are collapsed to find subleading Landau singularities. In the view of recent progress in bootstrapping the non-perturbative $S$-matrix (see e.g.~\cite{Paulos:2016but,Paulos:2017fhb,He:2018uxa,Cordova:2018uop,Guerrieri:2018uew,Bercini:2019vme,Cordova:2019lot,Correia:2020xtr,Guerrieri:2020kcs,Guerrieri:2020bto,Hebbar:2020ukp,Tourkine:2021fqh,He:2021eqn,Guerrieri:2021tak,Albert:2022oes,Miro:2022cbk}), it would be interesting to revisit whether the methods presented here can be extended to apply to the $S$-matrix bootstrap program.

\section*{Acknowledgments}

We thank Nima Arkani-Hamed, Ruth Britto, Simon Caron-Huot, Einan Gardi, Aidan Herderschee, Thomas Lam, Erik Panzer, Andrzej Pokraka and Sebastian Mizera for useful discussions.
We also want to thank Jacob Bourjaily and Lance Dixon for discussions and collaboration at the beginning of the project. HSH gratefully acknowledges support from the Simons Foundation (816048, HSH). MDS is supported in part by the U.S. Department of Energy under contract DE-SC0013607.

 \appendix
  
   \section{Graph Theory}
\label{sec:graph_theory}

In this appendix, we review some useful notions from graph theory that we make use of in the main body of the paper.  A good reference for this material is~\cite{nakanishi1971graph}.

A \emph{directed graph} $G$ is defined by a set of vertices $V(G)$ and a set of oriented edges $E(G)$ that begin and end on vertices drawn from $V(G)$. We denote the number of vertices and edges by $|V(G)|$ and $|E(G)|$. 
Usually only a single edge is permitted between any pair of vertices, but this restriction is lifted when considering \emph{multigraphs}, which we hereafter just refer to as graphs.
In order to avoid treating the external lines in Feynman and Landau diagrams in a special way, we compactify graphs by adding a vertex at infinity and joining the external lines at this vertex, which we denote by $v_\infty$. However, when it proves convenient, we denote the set of internal edges by $\Eint(G)$ (it should always be clear which vertex is at infinity). The edges of these compactified graphs will be oriented according to the flow of energy.

\begin{figure}
  \centering

\tikzset{every picture/.style={line width=0.75pt}} %
\begin{tikzpicture}[x=0.75pt,y=0.75pt,yscale=-1,xscale=1]
\draw [line width=2.25]    (360,142) -- (480,142) ;
\draw [shift={(420,142)}, rotate = 180] [fill={rgb, 255:red, 0; green, 0; blue, 0 }  ][line width=0.08]  [draw opacity=0] (14.29,-6.86) -- (0,0) -- (14.29,6.86) -- cycle    ;
\draw  [dash pattern={on 4.5pt off 4.5pt}]  (411,72) -- (481,142) ;
\draw [shift={(446,107)}, rotate = 225] [fill={rgb, 255:red, 0; green, 0; blue, 0 }  ][line width=0.08]  [draw opacity=0] (8.93,-4.29) -- (0,0) -- (8.93,4.29) -- cycle    ;
\draw [line width=2.25]    (361,142) .. controls (355.5,114) and (388.5,69) .. (411,72) ;
\draw [shift={(373.97,97.56)}, rotate = 482.37] [fill={rgb, 255:red, 0; green, 0; blue, 0 }  ][line width=0.08]  [draw opacity=0] (14.29,-6.86) -- (0,0) -- (14.29,6.86) -- cycle    ;
\draw  [dash pattern={on 4.5pt off 4.5pt}]  (361,142) .. controls (385.5,134) and (410.5,92) .. (411,72) ;
\draw [shift={(394.15,112.76)}, rotate = 485.88] [fill={rgb, 255:red, 0; green, 0; blue, 0 }  ][line width=0.08]  [draw opacity=0] (8.93,-4.29) -- (0,0) -- (8.93,4.29) -- cycle    ;
\draw  [dash pattern={on 4.5pt off 4.5pt}]  (331,162) -- (361,142) ;
\draw [shift={(346,152)}, rotate = 506.31] [fill={rgb, 255:red, 0; green, 0; blue, 0 }  ][line width=0.08]  [draw opacity=0] (8.93,-4.29) -- (0,0) -- (8.93,4.29) -- cycle    ;
\draw  [dash pattern={on 4.5pt off 4.5pt}]  (321,142) -- (361,142) ;
\draw [shift={(341,142)}, rotate = 180] [fill={rgb, 255:red, 0; green, 0; blue, 0 }  ][line width=0.08]  [draw opacity=0] (8.93,-4.29) -- (0,0) -- (8.93,4.29) -- cycle    ;
\draw  [dash pattern={on 4.5pt off 4.5pt}]  (361,62) -- (411,72) ;
\draw [shift={(386,67)}, rotate = 191.31] [fill={rgb, 255:red, 0; green, 0; blue, 0 }  ][line width=0.08]  [draw opacity=0] (8.93,-4.29) -- (0,0) -- (8.93,4.29) -- cycle    ;
\draw  [dash pattern={on 4.5pt off 4.5pt}]  (411,72) -- (451,62) ;
\draw [shift={(431,67)}, rotate = 525.96] [fill={rgb, 255:red, 0; green, 0; blue, 0 }  ][line width=0.08]  [draw opacity=0] (8.93,-4.29) -- (0,0) -- (8.93,4.29) -- cycle    ;
\draw  [dash pattern={on 4.5pt off 4.5pt}]  (481,142) -- (521,142) ;
\draw [shift={(501,142)}, rotate = 180] [fill={rgb, 255:red, 0; green, 0; blue, 0 }  ][line width=0.08]  [draw opacity=0] (8.93,-4.29) -- (0,0) -- (8.93,4.29) -- cycle    ;
\draw [line width=2.25]    (481,142) -- (511,162) ;
\draw [shift={(496,152)}, rotate = 213.69] [fill={rgb, 255:red, 0; green, 0; blue, 0 }  ][line width=0.08]  [draw opacity=0] (14.29,-6.86) -- (0,0) -- (14.29,6.86) -- cycle    ;

\draw (313,164.4) node [anchor=north west][inner sep=0.75pt]    {$\infty $};
\draw (303,124.4) node [anchor=north west][inner sep=0.75pt]    {$\infty $};
\draw (342,54.4) node [anchor=north west][inner sep=0.75pt]    {$\infty $};
\draw (452,54.4) node [anchor=north west][inner sep=0.75pt]    {$\infty $};
\draw (522,124.4) node [anchor=north west][inner sep=0.75pt]    {$\infty $};
\draw (513,154.4) node [anchor=north west][inner sep=0.75pt]    {$\infty $};
\draw (362,145.4) node [anchor=north west][inner sep=0.75pt]    {$1$};
\draw (481,122.4) node [anchor=north west][inner sep=0.75pt]    {$2$};
\draw (407,52.4) node [anchor=north west][inner sep=0.75pt]    {$3$};
\end{tikzpicture}
\caption{A choice of spanning tree (in thick black lines) for the ice cream cone diagram, containing the edges through which momenta $q_2$, $q_4$ and $p_6$ flow. Through the complementary edges flow momenta $p_1$, $p_2$, $p_3$, $p_4$, $p_5$, $q_3$ and $q_1$.  These momenta can be varied independently.  The internal momenta $q_3$ and $q_1$ provide a labeling for the loop momenta of the corresponding Feynman integral.  See also the figure in Eq.~\eqref{eq:icecreamcone_momentum_labels}.}
\label{fig:spanning_tree}
\end{figure}
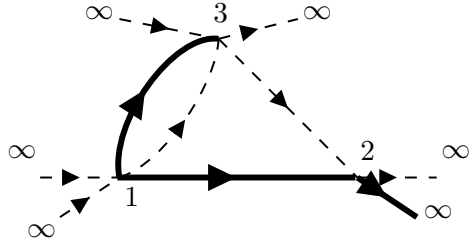

Given a graph $G$, we define its \emph{incidence matrix} to be the $|V(G)| \times |E(G)|$ matrix $a$ whose entries are given by 
\begin{equation}
  \label{eq:incidence_matrix}
a_{v e} = \begin{cases} \ 1 \quad \text{if the edge $e$ starts on the vertex $v$,} \\ 
 -1 \quad \text{if the edge $e$ ends on the vertex $v$,} \\
\  0 \quad \text{otherwise.} \end{cases}
\end{equation}
For example, the incidence matrix of the ice cream cone integral, depicted in Figure~\ref{fig:spanning_tree}, can be written as
\begin{equation}
\begin{blockarray}{ccccccccccc}
& p_1 & p_2 & p_3 & p_4 & p_5 & p_6 & q_2 & q_4 & q_3 & q_1 \\[.1cm]
\begin{block}{c(cccccccccc)}
v_\infty & 1 & 1 & 1 & -1 & -1 & -1 & 0 & 0 & 0 & 0 \\
v_1 & -1 & -1 & 0 & 0 & 0 & 0 & 1 & 1 & 1 & 0 \\
v_2 & 0 & 0 & 0 & 0 & 1 & 1 & -1 & 0 & 0 & -1 \\
v_3 & 0 & 0 & -1 & 1 & 0 & 0 & 0 & -1 & -1 & 1 \\
\end{block}
\end{blockarray} \ \ ,
\end{equation}
where we have labeled the columns and rows of this matrix to make clear how the edges and vertices are indexed.

A \emph{circuit} $c$ in $G$ is a sequence of edges in which each pair of sequential edges share a vertex, and whose initial and final edges also have a common vertex.  A circuit can be assigned two possible orientations. Note that a circuit that is oriented in the same way as all the edges it contains must pass through the vertex at infinity.
A \emph{spanning tree} $T$ of a graph $G$ is a tree that contains all the vertices of $G$. For example, one choice of spanning tree is shown for the ice cream cone diagram in Fig.~\ref{fig:spanning_tree}.
We similarly denote by $\bar{T}$ the \emph{cospanning tree} that is formed by all the edges that are not in $T$.\footnote{Note that if $G$ is a multigraph, cospanning trees can include circuits.} Since adding any of the edges from $\bar{T}$ to the spanning tree $T$ forms a circuit, the edges in $\bar{T}$ are each associated with a distinct circuit. These circuits, called \emph{fundamental circuits}, form a basis for the set of circuits in $G$.
We will denote the set of all circuits associated with a graph $G$ by $C(G)$, and the set of fundamental circuits by $\Chat(G)$.  Since the fundamental circuits are in correspondence with edges in a chosen cospanning tree, we will often abuse notation by parameterizing an independent set of loop momenta by considering all $k_c$ for $c \in \bar{T}$.

The \emph{circuit matrix} $b$ of a graph $G$ is defined as the matrix with entries
\eq{
  \label{eq:circuit_matrix}
    b_{c e} = \begin{cases} \ 1 \quad \text{if the edge $e$ is in the circuit $c$ and is oriented in the same way as $c$,} \\ 
     -1 \quad \text{if the edge $e$ is in the circuit $c$ and is oriented in the opposite way to $c$,} \\
    \  0 \quad \text{otherwise.} \end{cases}
}
The circuit matrix obtained by restricting to fundamental circuits is called the \emph{fundamental circuit matrix}. The fundamental circuit matrix of the ice cream cone integral is given by
\begin{equation}
\begin{blockarray}{ccccccccccc}
 & p_1 & p_2 & p_3 & p_4 & p_5 & p_6 & q_2 & q_4 & q_3 & q_1 \\[.1cm]
\begin{block}{c(cccccccccc)}
      c_{q_3} & 0 & 0 & 0 & 0 & 0 & 0 & 0 & -1 & 1 & 0 \\
      c_{q_1} & 0 & 0 & 0 & 0 & 0 & 0 & -1 & 1 & 0 & 1 \\
      c_{p_1} & 1 & 0 & 0 & 0 & 0 & 1 & 1 & 0 & 0 & 0 \\
      c_{p_2} & 0 & 1 & 0 & 0 & 0 & 1 & 1 & 0 & 0 & 0 \\
      c_{p_3} & 0 & 0 & 1 & 0 & 0 & 1 & 1 & -1 & 0 & 0 \\
      c_{p_4} & 0 & 0 & 0 & 1 & 0 & -1 & -1 & 1 & 0 & 0 \\
      c_{p_5} & 0 & 0 & 0 & 0 & 1 & -1 & 0 & 0 & 0 & 0 \\
\end{block}
\end{blockarray} \ \ ,
\end{equation}
where we have labeled the fundamental circuits by the momenta associated with edges of the cospanning tree, for the choice of spanning tree in Figure~\ref{fig:spanning_tree}.

The incidence matrix and fundamental circuit matrix satisfy an orthogonality relation for each vertex and each circuit in a graph. These orthogonality relations are summarized by the equation
\eq{
\sum_{e \in E(G)} a_{v e} b_{c e} = 0 \, ,
}
which can be proven as follows. First, we note that the sum vanishes if $c$ does not involve the vertex $v$; we can thus take the sum to be over the edges in the circuit $c$ that are incident with the vertex $v$. Tracing $c$ as it passes through $v$, we have an incoming edge and an outgoing edge. If both these edges are oriented towards $v$ or away from $v$, then the $a_{v e}$ values have the same sign but the values for $b_{c e}$ have opposite signs, so $a_{v e} b_{c e}$ cancels when summed over these two edges. If one edge is oriented towards $v$ and the other away from $v$, then the $b_{c e}$ have the same sign but the $a_{v e}$ have opposite signs, so $a_{v e} b_{c e}$ again cancels when summed over the two edges. The orthogonality relation follows from the cancellation obtained from all such pairs in the sum.

To parameterize the independent momenta in our graphs, we will in general choose a spanning tree $T$ that contains only a single external edge. In terms of the fundamental circuits defined by $T$, the momenta associated with each edge $e$ can then be written as
\begin{equation}
  \label{eq:fundamental_momenta}
  p_e = \sum_{c \in \Chat(G)} b_{c e} p_c \, ,
\end{equation}
where $p_c$ denotes the momenta flowing through the edge in the cospanning tree $\bar{T}$ that is associated with the cycle $c$.

Given a graph contraction $\ker \kappa \rightarrowtail G^\kappa \twoheadrightarrow G_0$, we can pick a labeling for the internal momenta so that the fundamental circuits $\Chat(G^\kappa)$ of $G^\kappa$ are such that $\Chat(G^\kappa) = \Chat(G_0) \cup \Chat(\ker \kappa)$ (with some abuse of notation). That is, we decompose the fundamental circuits in $\Chat(G^\kappa)$ into two sets of independent variables: the ones corresponding to the $\nex-1$ external momenta in $\Chat(G_0)$, and the $L$ loop momenta in $\Chat(\ker \kappa)$. This can always be achieved by first picking a set of fundamental circuits for $\Chat(\ker \kappa)$, which---since $\ker \kappa$ embeds in $G^\kappa$---maps to a subset of $\Chat(G^\kappa)$.  We can also pick a set of fundamental circuits of $\Chat(G_0)$ and for each element of this set we can pick an element of $\Chat(G^\kappa)$ which contracts to it.  This last choice is neither unique nor canonical in general, but neither is a choice of fundamental circuits. In practice we label the fundamental circuits by the loop-momenta $k_c$ and the external momenta by $p_c$. 

   \section{Kronecker Indices}
\label{sec:kronecker_index}
In this appendix, we set up the formalism needed to find the integer-valued Kronecker indices that appear in the Picard--Lefschetz theorem.

\begin{definition}[orientation]
Given a manifold $X$ of real dimension $n$ with an open cover $U_\alpha$ and charts $\phi_\alpha \colon U_\alpha \to V_\alpha \subset \mathbb{R}^n$, a choice of \emph{orientation} is an assignment of $o_\alpha \in \{-1, +1\}$ for each patch such that under changes of coordinates they transform as $o_\alpha = \bigl(\operatorname{sgn} \det J_{\alpha \beta}\bigr) o_\beta$, where $J_{\alpha \beta}$ is the Jacobian of changes of coordinates on $U_\alpha \cap U_\beta$ from $U_\alpha$ to $U_\beta$.
\end{definition}
\noindent A choice of orientation can be given by a \emph{system of indicators} (see Ref.~\cite{pham2011singularities}):
\begin{definition}[system of indicators]
Given a manifold $X$ of real dimension $n$ and a point $x \in X$, a \emph{system of indicators} at $x$ is a choice of vectors $v_1, \dotsc, v_n \in T_x(X)$ such that the orientation $o_\alpha$ associated to the chart $\phi_\alpha \colon U_\alpha \to V_\alpha$ is determined by $o_\alpha \operatorname{sgn} \det (v_i)_j^{\alpha} = 1$, where $(v_i)_j^{\alpha}$ is the coordinate $j$ of the vector $v_i$ in the coordinate chart $\alpha$.  This is well-defined since $o_\alpha$ transforms correctly under changes of coordinates.
\end{definition}
\noindent Let $Y_1$ and $Y_2$ be two submanifolds in $X$ whose dimensions add up to $\dim_{\mathbb{R}} X = n$, which intersect transversally at a point $x \in X$.  We take $Y_1$ and $Y_2$ to be oriented and their orientation to be specified by a system of indicators $(v_1, \dotsc, v_p)$ for $Y_1$ and $(v_{p+1}, \dotsc, v_n)$ for $Y_2$.  We say that the orientations of $Y_1$ and $Y_2$ match at $x$ if the system of indicators $(v_1, \dotsc, v_n)$ yields the orientation of $X$ at $x$.

\begin{definition}[Kronecker index]
Given two homology cycles $\gamma$, $\sigma$ in $X$ of complementary codimension, represented by manifolds and which intersect transversally at a finite number of points, the \emph{Kronecker index} $\langle \gamma, \sigma\rangle$ is the sum over the intersection points of contributions $\pm$ according to whether the orientations of $\sigma$ and $\gamma$ at that point match or not.  \\ Warning!  In comparing the orientations, $\sigma$ and $\gamma$ are swapped with respect to the ordering $\langle \gamma, \sigma\rangle$ in the Kronecker index.
\end{definition}

\begin{remark}
A choice of orientation can also be made by a nowhere-vanishing top differential form.  Indeed, consider such a differential form $\phi$ which in a coordinate chart $\alpha$ with coordinates $x^\alpha$ can be written $\phi = \phi_\alpha \rd x_1^\alpha \wedge \dotso \wedge \rd x_n^\alpha$.  Then, one can check that $o_\alpha = \operatorname{sgn} \phi_\alpha$ transforms in the right way to define an orientation.
\end{remark}

\begin{remark}
A complex manifold has a canonical choice of orientation that is invariant under holomorphic changes of coordinates.  Indeed we can define such an orientation by the differential form $\rd \Re x_1 \wedge \rd \Im x_1 \wedge \dotso \wedge \rd \Re x_n \wedge \rd \Im x_n$.
\end{remark}

Let us compute a Kronecker index of two special $n$-dimensional cycles in a $n$-dimensional complex space.  We can arrange so that the cycles intersect at the origin in a convenient coordinate frame.  In a small neighborhood of the origin we take the first cycle, $e$, to be defined by $\Im x_1 = \dotso = \Im x_n = 0$ so it is parametrized by coordinates $(\Re x_1, \dotsc, \Re x_n)$.  We can then choose a system of indicators
\begin{equation}
    \frac{\partial}{\partial \Re x_1}, \dotsc, \frac{\partial}{\partial \Re x_n}
\end{equation}
for $e$.  We take the second cycle, $h$, to be defined by $\Im x_i = \lambda \Re x_i$ for $i = 1, \dotsc, n$.  When represented as a vector field, this is a radial vector field which vanishes at the origin.  It is a source if $\lambda > 0$ and a sink if $\lambda < 0$.

The tangent space to $h$ at $0$ is generated by
\begin{equation}
    \frac{\partial}{\partial \Re x_1} + \lambda \frac{\partial}{\partial \Im x_1}, \dotsc,
    \frac{\partial}{\partial \Re x_n} + \lambda \frac{\partial}{\partial \Im x_n} \, ,
\end{equation}
and we choose the orientation of $h$ to be described by this system of indicators.  To finish the computation we need to compare the orientation of the concatenation of these systems of indicators to that defined by
\begin{equation}
    \frac{\partial}{\partial \Re x_1}, \frac{\partial}{\partial \Im x_1}\dotsc, \frac{\partial}{\partial \Re x_n}, \frac{\partial}{\partial \Im x_n}.
\end{equation}
One can see now that there are two contributions to the sign which compares the two orientations.  The first is from $(\operatorname{sgn} \lambda)^n$ and the second is from sorting the vectors to agree with the orientation of $X$ as a complex manifold which yields $(-1)^{\frac {n (n + 1)} 2}$.

In conclusion, we have shown that
\begin{equation}
    \langle e, h\rangle = \begin{cases}
        (-1)^{\frac{n (n + 1)}{2}}, & \qquad \lambda > 0, \\
        (-1)^{\frac{n (n - 1)}{2}}, & \qquad \lambda < 0.
    \end{cases}
\end{equation}

\begin{remark}
From these results it is easy to prove Cartan's theorem on the Kronecker index of a sphere $S$ with itself.  Indeed, we can deform one of the cycles by a vector field which has a source at the North pole and a sink at the South pole.  Then we can apply the result above in a patch around the North pole and in a patch around the South pole to find
\begin{equation}
    \langle S, S\rangle = (-1)^{\frac{n(n-1)}{2}} + (-1)^{\frac{n(n+1)}{2}} = (-1)^{\frac{n(n-1)}{2}} (1 + (-1)^n).
\end{equation}
The self-intersection number vanishes for odd-dimensional spheres, it is $+2$ if the dimension is divisible by $4$ and $-2$ if the dimension is even but not divisible by $4$.
\end{remark}
   \section{Poincar\'e Duality}
\label{sec:signs}
In this appendix, we work out the Poincar\'e duality that was needed to derive Cutkosky's formula in Eq.~\eqref{IGmform3}.

Consider a real $n$-dimensional manifold $X$, without boundary, which contains a submanifold $N$ of real codimension one.  Suppose that $N$ is given by a global equation $f(x) = 0$ and that $f(x) \geq 0$ defines a compact manifold $M$ with boundary $N$. Then, we have
\begin{equation}
  \int_M \omega = \int_X \theta(f) \omega,
\end{equation}
where $\theta$ is the Heaviside function $\theta(x) = 1$ if $x \geq 0$ and $\theta(x) = 0$ if $x < 0$.  If $\omega$ is an exact form and can be written as $\omega = \rd \rho$, then
\begin{equation}
  \int_M d \rho = \int_X \theta(f) \rd \rho = \int_X \Bigl(\rd (\theta(f) \rho) - (\rd \theta(f)) \wedge \rho\Bigr).
\end{equation}
The first term yields zero by Stokes theorem since $X$ has no boundary.  The second term localizes the integral on $N = \partial M$ since $\frac {d \theta}{d x} = \delta(x)$.  Indeed, since
\begin{equation}
  \rd \theta(f) = \delta(f) \rd f,
\end{equation}
we have
\begin{equation}
  \int_M \rd \rho = -\int_X \delta(f) \rd f \wedge \rho = \int_N \rho,
\end{equation}
where the second equality follows from Stokes theorem.  The statement of Stokes theorem assumes that an orientation on $N$ is induced from an orientation on $M$ and we will always follow this convention.

The identity
\begin{equation}
  \label{eq:stokes_with_theta}
  \int_{\partial M} \rho = -\int_X \delta(f) \rd f \wedge \rho
\end{equation}
will be important for us.  In this example, the integration over $X$ can be replaced by an integration over a domain that contains $N = \partial M$.  The differential form $-\delta(f) \rd f$ (or, more precisely, the current) is the Poincar\'e dual to $N$ (see page~51 of Ref.~\cite{MR658304}, but note that our definition of the Poincar\'e dual differs from theirs by a sign).

Let us illustrate this with an example, in which we take $X = \mathbb{R}$ and $M = [a, b]$, so $N$ consists of the two points $a, b \in \mathbb{R}$ with $a < b$.\footnote{In this case $X$ has a boundary, so it does not satisfy the requirements in the discussion above; however, in this example we will integrate only forms of compact support where the boundary plays no role, so we ignore this subtlety.}  We then choose $f(x) = (b - x)(x - a)$.\footnote{Note that this choice $f$ is not unique; in particular, we can multiply $f$ by any positive function.} If we pick a test function $\phi$ on $\mathbb{R}$, then we have
\begin{equation}
  \left\langle \frac{d \theta(f)}{d x}, \phi\right\rangle =
  -\left\langle \theta(f), \frac {d \phi}{d x}\right\rangle =
  -\int_a^b \frac {d \phi}{d x} = -\phi(b) + \phi(a) =
  \langle -\delta(x - b) + \delta(x - a), \phi\rangle.
\end{equation}
Since this holds for all test functions $\phi$, we have that
\begin{equation}
  \frac{d \theta((b - x)(x - a))}{d x} = -\delta(x - b) + \delta(x - a).
\end{equation}
On the other hand, using $\rd \theta(f) = \delta(f) \rd f$ we have
\begin{equation}
  \rd \theta\bigl((b - x)(x - a)\bigr) = \delta\bigl((b - x)(x - a)\bigr) \rd \bigl((b - x)(x - a)\bigr).
\end{equation}
Using $\delta(f(x)) = \sum_{x_i \mid f(x_i) = 0} \frac {\delta(x - x_i)}{\lvert f'(x_i)\rvert}$, we can rewrite 
\begin{equation}
  \delta\bigl((b - x)(x - a)\bigr) = \frac {\delta(x - b)}{\lvert b - a\rvert} + \frac {\delta (x - a)}{\lvert b - a\rvert}.
\end{equation}
Combining this with the fact that
\begin{equation}
  \rd \bigl((b - x)(x - a)\bigr) = (-(x - a) + (b - x)) \rd x,
\end{equation}
we find
\begin{multline}
  \delta\bigl((b - x)(x - a)\bigr) \rd \bigl((b - x)(x - a)\bigr) = \\
  \Bigl(\frac {\delta(x - b)}{\lvert b - a\rvert} + \frac {\delta (x - a)}{\lvert b - a\rvert}\Bigr) (-(x - a) + (b - x)) \rd x = \\
  \frac {\delta(x - b)}{\lvert b - a\rvert} (-(x - a)) \rd x +
  \frac {\delta (x - a)}{\lvert b - a\rvert} (b - x) \rd x = \\
  -\delta(x - b) \rd x + \delta(x - a) \rd x.
\end{multline}
Higher-dimensional examples work in a similar way.

Applying the Eq.~\eqref{eq:stokes_with_theta} repeatedly, we thus find that
\begin{equation}
  \label{eq:stokes_with_theta_iter}
  \int_{\partial_m \dotso \partial_1 e} \omega = (-1)^m \int_U \delta(s_1) \rd s_1 \wedge \dotso \wedge \delta(s_m) \rd s_m \wedge \omega,
\end{equation}
where $U$ is a space containing $\partial_m \dotso \partial_1 e$.

   \section{Sketch of Proof of Theorem~\ref{thm:pham}}
\label{sec:sing_absorption}
We here sketch the proof of Theorem~\ref{thm:pham}, that the singularities of the absorption integral $\cA_G^\kappa$ only occur on the Pham loci $\P_\kappap$ associated with graphs $G^\kappap$ that dominate $G^\kappa$.  The discussion in this appendix follows section II.2.1 of Ref.~\cite{pham}.  

We start by considering a Feynman integral $I = \int_h \omega$, where the contour $h$ is almost real and the form $\omega$ is real for real momenta. Computing the monodromy of this integral around the Pham locus $\P_\kappa$ that is associated with the contraction $\kappa \colon G^\kappa \twoheadrightarrow G_0$ 
leads to the absorption integral
\begin{equation}
    \label{eq:sing_absorption}
    \cA_G^\kappa = (2 \pi i)^m \int_{\partial_m \cdots \partial_1 e} \frac{s_1 \dotso s_m \omega}{\rd s_1 \wedge \dotso \wedge \rd s_m},
\end{equation}
where $s_1, \dotsc, s_m$ are propagators corresponding to the edges $e_m \in E(\ker \kappa)$ and $\partial_m \cdots \partial_1 e$ is the vanishing sphere.
In Sec.~\ref{sec:landau_review}, we saw that the Pham locus corresponding to the contraction $\kappa$ can be written in differential form as
\begin{equation}
    \label{eq:sing_absorption_dell}
    \rd \ell(\kappa) = \alpha_1 \rd s_1 + \dotsc \alpha_m \rd s_m = \sum_{e \in E(\ker \kappa)} \alpha_e \rd s_e.
\end{equation}
The constraint $\rd \ell=0$ encodes the content of Lemma~\ref{lem:crit}, that the Pham locus corresponds to critical points of the projection map from $\S(G^\kappa)$ to $S(G_0)$.

To see where $\cA_G^\kappa$ can become singular, we first rewrite the absorption integral $\cA_G^\kappa$ as
\begin{equation}
    \cA_G^\kappa = (2 \pi i)^m \int_{\partial_m \cdots \partial_1 e} \frac{\prod_{c \in \Chat(G)}\rd^{d} k_c}{\rd s_1 \wedge \dotso \wedge \rd s_m \, s_{m+1} \dotso s_{\nint}} \,.
\end{equation}
We now apply the Landau equations to this integral. To do so, we note that the integration contour for $\cA_G^\kappa$ is the vanishing sphere $\partial_m \cdots \partial_1 e$, which corresponds to the surface $s_1=\cdots= s_m=0$. Thus, all the singularities of $\cA_G^\kappa$ must correspond to points where all these $s_i$, and possibly more, vanish. The singularities of $\cA_G^\kappa$ will thus be described by graphs $G^\kappap$ that dominate $G^\kappa$. This motivates constructing the diagram
\begin{equation}
\begin{tikzcd}
G \arrow[r,twoheadrightarrow, ""] &
G^\kappap \arrow[r, twoheadrightarrow,"\kappas"] \arrow[dr,twoheadrightarrow,"\kappap"{below}] & G^\kappa \arrow[d,twoheadrightarrow, "\kappa"] \\
&& G_0
\end{tikzcd}
\end{equation}
where we have introduced the contraction $\kappas \colon G^\kappap \twoheadrightarrow G^\kappa$ and the composition $\kappap = \kappa \circ \kappas$. 

We denote by $s_{m+1}, \dotsc, s_{m + m'}$ the propagators corresponding to the edges in $E(\ker \kappas)$. These are a subset of the uncut denominators that remain in $\omega$.
The integrand in Eq.~\eqref{eq:sing_absorption} will become singular when these denominators vanishes, and when $\rd s_1 \wedge \dotso \wedge \rd s_m = 0$. The last condition can be equivalently written as $\rd \ell(\kappa) = 0$.
The analog of the pinching condition for Feynman integrals then becomes the condition that there exist $\alpha_0, \alpha_{m+1}, \dotsc, \alpha_{m + m'}$, not all vanishing, such that
\begin{equation}
    \label{eq:sing_absorption_pinching}
    \alpha_ 0 \rd \ell(\kappa) + \alpha_{m+1} \rd s_{m+1} + \dotso + \alpha_{m + m'} \rd s_{m + m'}\Big|_p =0
\end{equation}
where $|_p$ means external momenta are held fixed.

Combining Eq.~\eqref{eq:sing_absorption_pinching} with Eq.~\eqref{eq:sing_absorption_dell} we have that
\begin{equation}
    \sum_{e \in E(\ker \kappas) \cup E(\ker \kappa)} \alpha_e \rd s_e \Big|_p =0
    \label{eq:absorption_landau}
\end{equation}
for a singularity of $\cA_G^\kappa$. Since $E(\ker \kappa) \cup E(\ker \kappas) = E(\ker \kappap)$, we see that this condition is exactly the condition for being on the Pham locus $\P_\kappap$. We conclude that the absorption integral $\cA_G^\kappa$ can only have singularities on Pham loci $\P_\kappap$ where $\kappap \colon G^\kappap \twoheadrightarrow G_0$. 
 
 Note that if $\kappap$ is of codimension two, then one potential solution to the Landau equation in Eq.~\eqref{eq:absorption_landau} is that $\alpha_e=0$ for all $e \in E(\ker \kappa)$, so the absorption integral is potentially singular at the location of the Pham locus $\P_\kappas$. This type of singularity, for a contraction whose target is not the elementary graph, plays an important role in our analysis of codimension-two Pham diagrams in Sec.~\ref{sec:codim2}.

\paragraph{Example}

For a concrete example, consider the triangle absorption integral of the box:
\begin{equation}
\resizebox{!}{1cm}{
\begin{tikzpicture}[baseline=(current bounding box.center),
    line width=1.5,scale=0.3]
    \draw[black] (-4,-4)   -- (-2,-2)  node[midway,below] {$~~~p_1$};
    \draw[black] (-4,4) -- (-2,2) node[midway,above] {$~~~p_2$}; 
    \draw[black] (4,4) -- (2,2) node[midway,above] {$p_3~~$};
    \draw[black] (4,-4) -- (2,-2)  node[midway,below] {$p_4~~$};
    \draw[darkorange] (2,2) -- (2,-2)  node[midway,right] {$q_4$};
    \draw[olddarkgreen] (2,-2) -- (-2,-2)   node[midway,below] {$q_1$};
    \draw[newdarkblue2] (-2,-2) -- (-2,2)   node[midway,left] {$q_2$};
    \draw[darkred] (-2,2) -- (2,2)   node[midway,above] {$q_3$};
\end{tikzpicture}
}
 \xrightarrowdbl{~~\kappas~~}
\resizebox{!}{1cm}{
\begin{tikzpicture}[baseline=(current bounding box.center),
    line width=1.5,scale=0.7]
    \draw[black] (-4,1.5) --  (-2,1)  node[midway,above] {$p_2$};
    \draw[black] (-4,-1.5) -- (-2,-1) node[midway, below] {$p_1$};
    \draw[newdarkblue2] (-2,-1) --  (-2,1) node[midway,left] {$q_2$};
    \draw[darkred] (-2,1) --  (0,0) node[midway,above] {$~~~~q_3$};
    \draw[olddarkgreen] (0,0) -- (-2,-1) node[midway,below] {$~~~~~q_1$};
    \draw[black] (0,0) --  (1,0.5)  node[midway,above] {$p_3$};
    \draw[black] (0,0) --  (1,-0.5)  node[midway,below] {$p_4$};
    \end{tikzpicture}
}
 \xrightarrowdbl{~~\kappa~~}
\resizebox{!}{0.8cm}{
\begin{tikzpicture}[baseline=(current bounding box.center),
    line width=1.5,scale=0.5]
    \draw[black] (-2,2) -- (0,0)  node[midway,above] {$~~p_2$};
    \draw[black] (2,2) --  (0,0)  node[midway,above] {$p_3~~$};
    \draw[black] (2,-2) --  (0,0)  node[midway,below] {$p_4~~$};
    \draw[black] (-2,-2) -- (0,0)  node[midway,below] {$~~p_1$};
    \end{tikzpicture}
}
\label{boxtotriangle}
\end{equation}
The box integral is given by
\begin{equation}
    I_\squaretcolii = \int_h \frac {\rd^4 k}{s_1 s_2 s_3 s_4}\, ,
\end{equation}
while the triangle monodromy of the box integral is
\begin{equation}
A_{\squaretcolii}^{\tritcol} = 
(\bbone - \mathscr{M}_{\tritcol}) I_{\squaretcolii} =
\int_{\partial_3 \partial_2 \partial_1 e}
\frac {\rd^4 k}{(\rd s_1 \wedge \rd s_2 \wedge \rd s_3) s_4}.
\end{equation}
Here the vanishing sphere  $\partial_3 \partial_2 \partial_1 e$ is the surface $s_1=s_2=s_3=0$. This can have singularities when $s_4=0$, which puts the fourth propagator on-shell and leads to the Pham locus $\P_\squaretcolii$. Or it can have singularities when $\rd s_1 \wedge \rd s_2 \wedge \rd s_3=0$, which is the triangle singularity again.

Both $\P_\squaretcolii$ and $\P_\tritcol$ dominate $\P_\tritcol$, consistent with Thm.~\ref{thm:pham}.
Another possibility is that $\rd s_1$ and $\rd s_2$ in the denominator $\rd s_1 \wedge \rd s_2 \wedge \rd s_3$ of the $A_\squaretcolii^\tritcol$ integrand become proportional.  This looks like a bubble singularity, $\P_{\bubrgT}$. However, it is not exactly a bubble since we still have the extra on-shell condition $s_3 = 0$ from the integration domain. This is the bubble-like singularity contained within $\P_\tritcol$ and is codimension-two because of the extra on-shell constraint. A codimension-one bubble-like singularity is discussed in Sec.~\ref{sec:icecreamtobubbleandsun}.
   \section{Critical Points and Hessian Matrices}
\label{sec:hessians}
In Sec.~\ref{sec:principal}, we argued that since the Hessian matrix of $\ell(p)=\sum_e \alpha_e (q_e^2-m_e^2)$ at a principal singularity $p^\ast$ was negative definite, the corresponding vanishing cell entering the Picard--Lefschetz theorem was real. In this appendix, we show why the Hessian matrix of $\ell(p)$, when differentiating with respect to the loop momenta $k_{c_i}^\mu$, is the quantity that leads to a bounded vanishing cell. In particular, we show that the on-shell space for real momenta close to the principal singularity at $p^\ast$ is bounded. We first discuss some general features of a constrained extremization problem, before applying it to Feynman integrals below.

Consider a function $f \colon \mathbb{R}^n \to \mathbb{R}$ that we want to minimize subject to some constraints $g_1 = \dotso = g_m = 0$, where $g_i \colon \mathbb{R}^n \to \mathbb{R}$ for $i = 1, \dotsc, m$.  For this purpose we can use the method of Lagrange multipliers. Thus, we define an auxiliary function $F \colon \mathbb{R}^n \times \mathbb{R}^m \to \mathbb{R}$ by
\begin{equation}
    F(x, t) = f(x) - \sum_{i = 1}^m t_i g_i(x).
\end{equation}
The critical point conditions for $F$ read
\begin{gather}
    \label{eqs:lagrange_critical_point}
    \frac{\partial f(x)}{\partial x_a} - \sum_{i = 1}^m t_i \frac{\partial g_i(x)}{\partial x_a} = 0, \\
    g_i(x) = 0\, ,
\end{gather}
where the second set of conditions ensure that the solution for $x$ satisfies the original constraints.

Now, let us study the conditions under which a point that satisfies the conditions above is a minimum.  Let us consider a curve $x(u)$ inside the intersection of the constraint hypersurfaces, such that $x^* = x(0)$ is the critical point.  This means that $g_i(x(u)) = 0$ for $i = 1, \dotsc, m$.  Taking the derivatives with respect to $u$ we find
\begin{gather}
    \frac{\partial g_i(x(u))}{\partial x_a} \frac{\partial x_a(u)}{\partial u} = 0, \\
    \frac{\partial^2 g_i(x(u))}{\partial x_a \partial x_b} \frac{\partial x_a(u)}{\partial u} \frac{\partial x_a(u)}{\partial u} + \frac{\partial g_i(x(u))}{\partial x_a} \frac{\partial^2 x_a(u)}{\partial u^2} = 0.
\end{gather}
Multiplying by $t_i^*$, summing over $i$, and using the critical point condition we find
\begin{equation}
    \sum_{i = 1}^m t_i^* \frac{\partial^2 g_i(x^*)}{\partial x_a \partial x_b} \frac{\partial x_a(0)}{\partial u} \frac{\partial x_b(0)}{\partial u} + \frac{\partial f(x^*)}{\partial x_a} \frac{\partial^2 x_a(0)}{\partial u^2} = 0.
\end{equation}
The function $f$ has a local minimum at $x^*$ if the function $f(x(u))$ has a local minimum at $u = 0$ for all curves $x(u)$ satisfying the constraints $g_i(x(u)) = 0$ for $i = 1, \dotsc, m$.  We have
\begin{gather}
    \frac{\partial f(x(u))}{\partial u} = \frac{\partial f(x(u))}{\partial x_a} \frac{\partial x_a(u)}{\partial u}, \\
    \frac{\partial^2 f(x(u))}{\partial u^2} = \frac{\partial^2 f(x(u))}{\partial x_a \partial x_b} \frac{\partial x_a(u)}{\partial u} \frac{\partial x_b(u)}{\partial u} + \frac{\partial f(x(u))}{\partial x_a} \frac{\partial^2 x_a(u)}{\partial u^2}.
\end{gather}
Setting $u = 0$ and using a result above we find
\begin{equation}
    \frac{\partial^2 f(x(0))}{\partial u^2} = \Bigl(\frac{\partial^2 f(x^*)}{\partial x_a \partial x_b} - \sum_{i = 1}^m t_i^* \frac{\partial^2 g_i(x^*)}{\partial x_a \partial x_b}\Bigr) \frac{\partial x_a(0)}{\partial u} \frac{\partial x_b(0)}{\partial u},
\end{equation}
where $t_i^*$ are the values of the Lagrange multipliers at the critical point.\footnote{The solution for $t^*$ does not have to be unique.  Higher-dimensional solution spaces for $t^*$ arise for higher codimension Landau singularities.}  Since this must hold for all velocities $\frac{\partial x(u)}{\partial u}$ satisfying the condition $\frac{\partial g_i(x^*)}{\partial x_a} \frac{\partial x_a(0)}{\partial u} = 0$ then we must have that the quadratic form
\begin{equation}
    \frac{\partial^2 f(x^*)}{\partial x_a \partial x_b} - \sum_{i = 1}^m t_i^* \frac{\partial^2 g_i(x^*)}{\partial x_a \partial x_b}
\end{equation}
should be positive-definite when restricted to such velocities.

Hence, we have reduced the problem of deciding if a critical point is minimal to the question of whether a quadratic form $A(X) = X^t A X$ restricted to the kernel of a linear operator $B X = 0$ is positive-definite or not.  Sometimes it is inconvenient to solve for the kernel of $B$.  In those cases, we can apply the following result.

If $A$ is an $n \times n$ symmetric real matrix and $B$ is a $k \times n$ real matrix with $k < n$ and $\operatorname{rank} B = k$, then the quadratic form $A$ restricted to $\ker B$ is positive definite when the following conditions are satisfied.  We first define the $(n + k) \times (n + k)$ \emph{bordered Hessian matrix}
\begin{equation}
    H = \begin{pmatrix}
    0 & B \\
    B^t & A
    \end{pmatrix} \, ,
\end{equation}
and $H_l$ to be the $l \times l$ minor obtained by erasing the last $n + k - l$ rows and columns. We then have that $A\rvert_{\ker B} > 0$ if and only if $(-1)^k \det H_l > 0$ for $l = 2 k + 1, \dotsc, n + k$~\cite{10.2307/2324784}.  Sometimes in the literature only the last condition $l = n + k$ is stated explicitly, as in Ref.~\cite{Hwa:102287}.

Now we can apply the results above to the Feynman integral in section~\ref{sec:principal}. In our application of finding critical points of Feynman integrals, the constraints $g_i$ are the on-shell conditions, the Lagrange multipliers $t_i$ are the variables $\alpha_e$, and the variables $x$ are $k_c$ and $p_i$.  It is less obvious what $f$ should be.  To identify it we write the critical point condition in differential form
\begin{equation}
    \rd f = \sum_{e = 1}^m \alpha_e \rd s_e,
\end{equation}
where the $\rd$ in the right-hand side acts on both $k$ and $p$.  This means we can identify $f(p) = \ell(p)$, where $\ell(p) = 0$ is the local equation for the Landau variety. This choice of $f$ ensures that Eq.~\eqref{eqs:lagrange_critical_point} is satisfied.

The derivatives with respect to the $x$ variables become derivatives with respect to the momenta $k_c$ and $p_i$.  If we restrict to derivatives with respect to $k_c$, the quadratic form becomes
\begin{equation}
- \sum_{e = 1}^m \alpha_e \frac{\partial^2 s_e(k^*, p^*)}{\partial k_{c_1}^\mu \partial k_{c_2}^\nu} = -2 \sum_{e = 1}^m \alpha_e b_{c_1 e} b_{c_2 e} \eta_{\mu \nu}
\end{equation}
since the derivative of $f$ with respect to $k$ vanishes at the singular locus. If we remove the global minus sign we obtain a negative-definite quadratic form. When contracting this equation with the vectors $X_{c_1}^\mu$ and $X_{c_2}^\nu$ from section~\ref{sec:principal} and summing over $c_1$ and $c_2$, the right-hand side becomes
\begin{equation}
A(X_c^\mu) = \sum_{\substack{c_1, c_2 \in \Chat(\ker \kappa) \\ e \in E(\ker \kappa)}} \alpha_e b_{c_1 e} b_{c_2 e} X_{c_1} \cdot X_{c_2} \,,
\end{equation}
which is precisely Eq.~\eqref{eq:quad_form}.

Looking at the double derivatives with respect to the $p_i$, we obtain a quadratic form involving the second derivative of $f$ that does not seem to have been considered in the literature before. The extra matrix elements involving mixed derivatives with respect to $k_c$ and $p_i$ can be seen to play a role in the transversality condition (see page~182 of Ref.~\cite{pham}) and the definition of \emph{exceptional critical points} (see page~183 of Ref.~\cite{pham}).
The case where the critical point equations~\eqref{eqs:lagrange_critical_point} have a higher dimensional space of solutions in $t$ is more complicated, but the corresponding equations can also be worked out in this case.

   \section{Homotopy of Tangential Intersections}
\label{sec:homotopy}

In this appendix, we work out the first homotopy group of $\mathbb{C}^2 \setminus (\P_1 \cup \P_2)$, where $\P_1$ and $\P_2$ are two Pham loci that intersect tangentially at a point with a quadratic contact.  Without loss of generality, we take $\P_1$ and $\P_2$ to be given by the equations $y = 0$  and $y = x^2$, where $(x, y) \in \mathbb{C}^2$ are complex coordinates. The resulting space can be seen as a fibration over the base $B = \mathbb{C}^*$; for every value of $y$ there are two values of $x$ such that $y = x^2$, so the fiber is $F_y = \mathbb{C} \setminus \{+\sqrt{y}, -\sqrt{y}\}$.  Since we have excluded the value $y = 0$, the topology of the fiber is that of the complex plane minus two points.  We take $E$ to be the total space of the fibration, which is the space $\mathbb{C}^2 \setminus (\P_1 \cup \P_2)$ we wanted to study.

At this point we have a fibration $E \to B$ with fiber $F$.  Next, we apply the long exact sequence for the homotopy of fibrations, which reads (see, for example, Thm.~4.41 of Ref.~\cite{MR1867354})
\begin{equation}
\cdots \to \pi_n(F) \to \pi_n(E) \to \pi_n(B) \to \pi_{n-1}(F) \to \cdots \to \pi_0(E) \to 1,
\end{equation}
where we have used $1$ to denote the trivial group (consisting only of the neutral element $1$). We have $\pi_0(E) = 1$ since the space is path connected, and $\pi_2(B) = 1$ since $\mathbb{C} \setminus \{-\sqrt{y}, \sqrt{y}\}$ does not have a nontrivial second homotopy group.  So the long exact sequence becomes a short exact sequence
\begin{equation}
1 \to \pi_1(F) \to \pi_1(E) \to \pi_1(B) \to 1.
\end{equation}
The group $\pi_1(F)$ is the first homotopy of the complex plane minus two points, so it is generated by two paths, one going around $+\sqrt{y}$ and the other around $-\sqrt{y}$.  Let us call these generators $\eta_{\pm}$.  Also, the base $B = \mathbb{C}^*$, so $\pi_1(B)$ is generated by a path  $\gamma$ going around $y = 0$.

Thus, we have that $\pi_1(F) \cong \mathbb{Z} * \mathbb{Z}$,  where $G * H$ is the free product of the groups $G$ and $H$, which is the group generated by arbitrary elements $g_1 h_1 g_2 h_2 \cdots$.  If we pick a generator $\eta_+$ for the first $\mathbb{Z}$, and a generator $\eta_-$ for the second $\mathbb{Z}$, then the elements of $\mathbb{Z} * \mathbb{Z}$ are $\eta_+^{n_1} \eta_-^{m_1} \eta_+^{n_2} \eta_-^{m_2} \cdots$ where $n_1, m_1, n_2, m_2, \dots \in \mathbb{Z}$. Here, the multiplication corresponds to concatenation followed by the grouping of common terms.  We also have that $\pi_1(B) \cong \mathbb{Z}$ with generator $\gamma$.

We are now faced with the problem of finding the middle group from a short exact sequence of groups.  We start by considering a general short exact sequence
\begin{equation}
1 \to N \xrightarrow{\iota} G \xrightarrow{\pi} Q \to 1.
\end{equation}
We can think of the elements of the fundamental group $G$ as pairs $(n, q) \in N \times Q$ and the maps in the short exact sequence being
\begin{equation}
\iota(n) = (n, 1), \qquad
\pi(n, q) = q. 
\end{equation}
Clearly $\pi \circ \iota = 1$ (where $1$ is the group homomorphism that sends everything to the identity). To fully identify the group $G$, we need to define the group law on its elements $(n, q) \in G$, such that $\iota$ and $\pi$ are homomorphisms.  We already know that
\begin{gather}
(n, 1) (n', 1) = (n n', 1), \qquad
(1, q)(1, q') = (f(q, q'), q q'). \label{short_sequence_constraints}
\end{gather}
The term $f(q, q')$ appears because the only information we have is the action of $\pi$, and that erases the first term.

In the present case, we want to solve the constraints in~\eqref{short_sequence_constraints} for the case where $N = \mathbb{Z} * \mathbb{Z}$ and $Q = \mathbb{Z}$.  These constraints are difficult to solve in general and the solution is \emph{not} unique.  However, there are two obvious solutions, one of which is the case where $\pi_1(E) \cong (\mathbb{Z} * \mathbb{Z}) \times \mathbb{Z}$.  The second solution is a semidirect product defined as follows.  We have a group homomorphism $\phi \colon Q \to \operatorname{Aut}(N)$, where $\operatorname{Aut}(N)$ is the automorphism group of $N$, that is the group of bijective homomorphisms from $N$ to itself.  Then the group law in $G$ can be written as
\begin{equation}
    (n, q) (n', q') = (n \phi(q)(n'), q q').
\end{equation}
This group is written $N \rtimes_\phi Q$.

We will now use the extra information that $\pi_1(B)$ has an action $\phi$ on $\pi_1(F)$. In the special case we are interested in, where $Q \cong \mathbb{Z}$ and $N \cong \mathbb{Z} * \mathbb{Z}$, we take
\begin{gather}
    \phi(\gamma)(\eta_+) = \eta_-, \qquad
    \phi(\gamma)(\eta_-) = \eta_+,
\end{gather}
which is an automorphism of $N$.  In general we have $\phi(\gamma^{2 n})(\eta_\pm) = \eta_\pm$ and $\phi(\gamma^{2 n + 1})(\eta_\pm) = \eta_\mp$. Using the group law described above, we have
\begin{gather}
    (\eta_\pm, 1) (1, \gamma) = (\eta_\pm, \gamma), \\
    (1, \gamma) (\eta_\pm, 1) = (\eta_\mp, \gamma).
\end{gather}
If we identify $(\eta_\pm, 1)$ with $\eta_\pm$ in $\pi_1(E)$ and $(1, \gamma)$ with $\gamma$ in $\pi_1(E)$, then these two equalities can be written as the relation $\eta_\pm \circ \gamma = \gamma \circ \eta_\mp$.

To summarize, the group $\pi_1(E)$ can be written as a semidirect product as $\pi_1(E) \cong (\mathbb{Z} * \mathbb{Z}) \rtimes_\phi \mathbb{Z}$.  It also has a presentation 
\begin{equation}
    \pi_1(E) = \langle \eta_+, \gamma \mid \gamma^2 \circ \eta_+ \circ \gamma^{-2} \circ \eta_+^{-1} = 1\rangle
\end{equation}
in terms of generators and relations, where we have solved for $\eta_-$.

\bibliographystyle{jhep}

\bibliography{all_massive_pham}

\end{document}